# ALICE EMCal
# Physics Performance Report

for submission to the U.S. Department of Energy

November 2009

U. Abeysekara[4], J. Allen[5], T. Aronsson[17], T. Awes[9], A. Badalá[24], S. Baumgart[17], R. Bellwied[16], L. Benhabib[10], C. Bernard[8], N. Bianchi[5] , F. Blanco[12], Y. Bortoli[10], B. Boswell[1], G. Bourdaud[10], O. Bourrion[8], B. Boyer[8], C.R. Brown[1], E. Bruna[17], J. Butterworth[4], H. Caines[17], D. Calvo Diaz Aldagalan[29], G.P. Capitani[5], Y. Carcagno[8], A. Casanova Diaz[5], M. Cherney[4], G. Conesa Balbastre[5], T.M. Cormier[16], M.R. Cosentino[22], L. Cunqueiro Mendez[5], H. Delagrange[10], M. Del Franco[5], M. Dialinas[10], P. Di Nezza[5], A. Donoghue[1], M. Elnimr[16], A. Enokizono[9], M. Estienne[10], J. Faivre[8], A. Fantoni[5], B. Fenton-Olsen[6,19], F. Fichera[24], M.A.S. Figueredo[23], B. Foglio[8], S. Fresneau[10], J. Fujita[4], C. Furget[8], S. Gadrat[8], I. Garishvili[14], M. Germain[10], N. Giudice[25], Y. N. Gorbunov[4], A. Grimaldi[24], R. Guernane[8], C. Hadjidakis[10], J. Hamblen[14], J. W. Harris[17], D. Hasch[5], M. Heinz[17], B. Hicks[17], P.T. Hille[17], D. Hornback[14], R. Ichou[10], P. Jacobs[6], S. Jangal[27], K. Jayananda[4], T. Kalliokoski[13], Y.Kharlov[18], J.L. Klay[1], A. Knospe[17], S. Kox[8], J. Kral[13], P. Laloux[10], S. LaPointe[16], P. La Rocca[25,26], S. Lewis[1], Q. Li[16], F. Librizzi[24], R. Ma[17], D. Madagodahettige Don[12], Y. Mao [8], C. Markert [15], I. Martashvili[14], B. Mayes[12], T. Milletto[10], J. Mlynarz[16], V. Muccifora[5], H. Mueller[3], M.G. Munhoz [23], J.F. Muraz[8], J. Newby[7], C. Nattrass[14], F. Noto[25], N. Novitzky[13], B.S. Nilsen[4], G. Odyniec[6], A. Orlandi[5], A. Palmeri[24], G.S. Pappalardo[24], A. Pavlinov[16], W. Pesci[5], V. Petrov[16], C. Petta[25], P. Pichot[10], L. Pinsky[12], M. Ploskon[6], F. Pompei[16], A. Pulvirenti[25], J. Putschke[17], C.A. Pruneau[16], J. Rak[13], J. Rasson[6], K.F. Read[14], J.S. Real[8], A.R. Reolon[5], F. Riggi[25], J. Riso[16], F. Ronchetti[5], C. Roy[27], D. Roy[10], M. Salemi[24], S. Salur[6], M. Sano[6,11], R.P. Scharenberg[20], M. Sharma[16], D. Silvermyr[9], N. Smirnov[17], R. Soltz[7], S. Sorensen[14], V. Sparti[24], B.K. Srivastava[20], J.-S. Stutzmann[10], J. Symons[6], A. Tarazona Martinez[28], L. Tarini[16], R. Thomen[4], A. Timmins[16], A. Turvey[4], M. van Leeuwen[21], R. Vieira[5], A. Viticchié[5], S. Voloshin[16], R. Vernet[2], D. Wang[30], Y. Wang[30], R.M. Ward[1]

[1] *California Polytechnic State University, San Luis Obispo, CA 93407, USA*
[2] *CCIN2P3, Lyon, France*
[3] *CERN, European Organization for Nuclear Research, 1211 Geneve, Switzerland*
[4] *Creighton University, Omaha, NE 68178, USA*
[5] *INFN Laboratori Nazionali di Frascati, 00044 Frascati, Italy*
[6] *Lawrence Berkeley National Laboratory, Berkeley, CA 94720, USA*
[7] *Lawrence Livermore National Laboratory (LLNL), Livermore, CA 94550, USA*
[8] *LPSC, University Joseph Fourier Grenoble, CNRS/IN2P3, Grenoble, France*
[9] *Oak Ridge National Laboratory (ORNL), Oak Ridge, TN 37831, USA*
[10] *SUBATECH, Ecole des Mines, Universite de Nantes, CNRS/IN2P3, Nantes, France*
[11] *Institute of Physics, University of Tsukuba, Japan*
[12] *University of Houston, Houston, TX 77204, USA*
[13] *University of Jyvaskyla, Jyvaskyla, Finland*
[14] *University of Tennessee, Knoxville, TN 37996, USA*
[15] *University of Texas at Austin, Austin, TX 78702, USA*
[16] *Wayne State University, Detroit, MI 48201, USA*
[17] *Yale University, New Haven, CT 06511, USA*
[18] *IHEP, Protvino, Russia*
[19] *Niels Bohr Institute, Copenhagen, Denmark*
[20] *Purdue University, West Lafayette, IN 47907, USA*
[21] *NIKHEF, Amsterdam, Netherlands*
[22] *Universidade Estadual de Campinas, Campinas, Brazil*
[23] *Universidade de Sao Paulo, Sao Paulo, Brazil*
[24] *Sezione INFN, Catania, Italy*
[25] *Universita Catania e Sezione INFN, Catania, Italy*
[26] *Museo Storico della Fisica e Centro Studi e Ricerche Enrico Fermi, Roma, Italy*
[27] *IPHC, Universite Louis Pasteur, CNRS-IN2P3, Strasbourg, France*
[28] *Valencia Polytechnic University, Valencia, Spain*
[29] *University of Valencia, Valencia, Spain*
[30] *Hua–Zhong Normal University, Wuhan, China*

# Contents

# Chapter 1

# Executive Summary

The ALICE detector at the LHC (A Large Ion Collider Experiment) will carry out comprehensive measurements of high energy nucleus-nucleus collisions, in order to study QCD matter under extreme conditions and the phase transtion between confined matter and the Quark-Gluon Plasma (QGP). ALICE contains a wide array of detector systems for precise measurements of hadrons, leptons, photons, and their correlations, over a very broad kinematic range. The ALICE detector is shown in Fig. 1.1. Discussion of the full ALICE physics program can be found in [1, 2].

This report presents our current state of understanding of the Physics Performance of the large acceptance Electromagnetic Calorimeter (EMCal) in the ALICE central detector. The EMCal enhances ALICE's capabilities for jet measurements. The EMCal enables triggering and full reconstruction of high energy jets in ALICE, and augments existing ALICE capabilities to measure high momentum photons and electrons. Combined with ALICE's excellent capabilities to track and identify particles from very low $p_T$ to high $p_T$, the EMCal enables a comprehensive study of jet interactions in the medium produced in heavy ion collisions at the LHC. The interaction and energy loss of high energy partons in matter provides a sensitive tomographic probe of the medium generated in high energy nuclear collisions. The EMCal has previously been presented in [3, 4].

The remainder of this chapter provides an overview of the physics of the EMCal. Chapter 2 describes the detector layout, technical specification, and technical performance of the EMCal, based on test beam and cosmic ray data as well as simulations. Chapter 3 details the triggering capabilities enabled by the EMCal. Chapter 4 presents the motivation and present understanding of jet physics in relativistic heavy ion collisions at RHIC and the LHC. Chapters 5,6,7, and 8 present the anticipated physics performance for jet reconstruction, photons, heavy flavor production and particle identification in jets. Chapter 9 is a summary of this report.

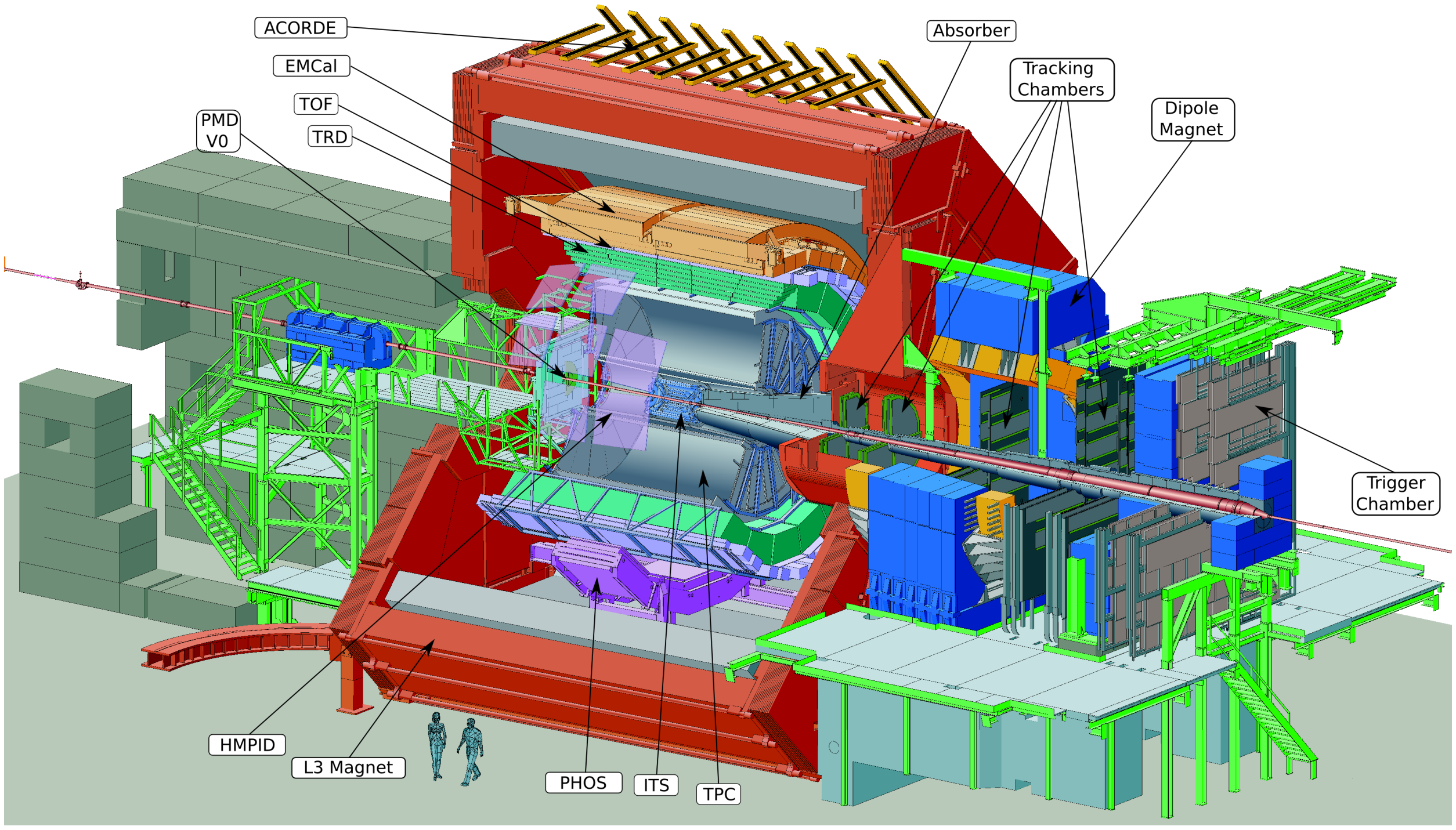

**Figure 1.1:** The ALICE detector in the L3 magnet.

## 1.1 Overview of ALICE EMCal Physics

Hard (high $Q^2$) probes will play a leading role in the LHC heavy ion program [5, 6]. Figure 1.2 shows the yields for various hard processes within the EMCal acceptance, for one running year of $\sqrt{s_{NN}}$=5.5 TeV Pb–Pb minimum bias collisions. There is a large increase in the rate of these processes relative to RHIC, due to the much higher collision energy at the LHC. Statistically significant measurements can be made for jets with $E_T$ greater than 200 GeV, for $\pi^0$ with $p_T$ greater than 80 GeV, for electrons with $p_T$ greater than 50 GeV, and for $\gamma$ with $p_T$ greater than 60 GeV. This enormous kinematic reach enables qualitatively new experimental observables of the interaction of hard probes with matter, including the QCD evolution of such interactions. Measurements will be made in Pb–Pb collisions but also in p–p and other reference systems ($p - A$, lighter nuclei), to isolate the effects of interactions in the Quark Gluon Plasma from nuclear effects in the initial state.

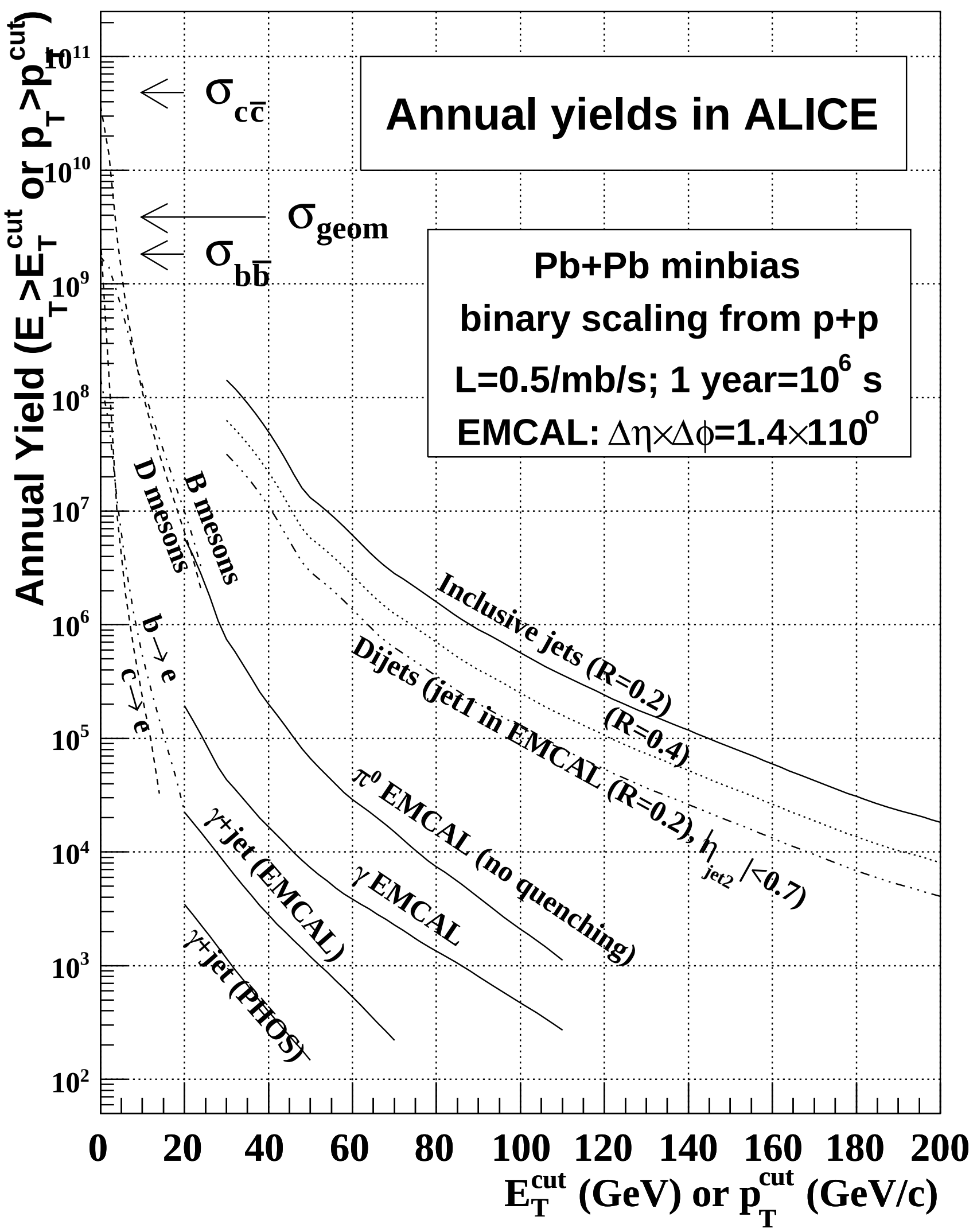

**Figure 1.2:** Annual yields in ALICE with the EMCal for various hard processes as a function of energy or $p_T$ threshold for $\sqrt{s_{NN}} = 5.5$ TeV Pb–Pb minimum bias collisions.

### 1.1.1 Jets

The EMCal, coupled with ALICE tracking detectors, enables ALICE to reconstruct large transverse momentum jets. Most quantitative studies of jet quenching to date have relied on observables of high $p_T$ hadrons and their correlations, i.e. leading fragments of jets, in order to suppress the large underlying event backgrounds in heavy ion collisions. While this approach has provided significant insights into the physics of jet quenching, it is limited by its intrinsic bias: such leading hadrons are preferentially from jet fragmentation of hard partons that have interacted *least* in the medium, in particular those generated at the surface of the fireball and headed outward. Full jet reconstruction in heavy ion collisions is essential to overcome such biases and to exploit fully the kinematic reach of the LHC.

In addition, jet quenching is intrinsically a partonic process, and its measurement in terms of hadronic observables introduces complex questions of hadronization that may mask essential physics. The measurement of jet structure and its modification in terms of *energy flow* rather than *hadron distributions* promises a much closer connection to underlying theory [7]. There is a wide array of jet quenching observables made possible by full jet reconstruction, discussed in detail in Chapter 4.

While accurate full jet reconstruction in a high background environment is a complex task, significant recent progress has been made in both theory and experiment, and we discuss the current status in Chapter 4. Chapter 5 discusses in detail the capabilities of ALICE+EMCal for full jet measurements.

The structure of jets is expected to be sensitive to parton energy loss [8, 9]. In addition to energy flow measurements, this can be observed as a modification of jet fragmentation patterns or the broadening of jets due to interactions in matter. Softening of fragmentation due to parton energy loss will lead to suppression of hadrons with a large momentum fraction $x$ of the jet and enhancement of the yield of soft, low x hadrons. Jet "heating" from the energy loss can be measured in the hadron momentum distributions perpendicular to the jet axis, and through jet shapes and energy clustering within the jet, utilizing variables such as $j_T$ and $k_T$.

A fast and unbiased jet trigger is essential for ALICE to exploit fully the kinematic range of jets shown in Fig. 1.2. Chapter 3 discusses the implementation of such a trigger, which in heavy ion collisions utilizies both the EMCal and the High Level Trigger to provide the required rejection.

### 1.1.2 $\pi^0$ and $\gamma$

The EMCal has a six-fold larger acceptance for measurements of photons and neutral mesons ($\pi^0$, $\eta$,...) than the highly granular PHOS detector. Discrimination of $\gamma$ and $\pi^0$ using EM shower shape characteristics is possible in the EMCal up to $p_T \sim 30$ GeV/c, while additional techniques, in particular isolation cuts, can be used for photon measurements to higher $p_T$.

The inclusive direct photon cross section is a key test of pQCD calculations. The coincidence of a direct photon with a recoiling jet or hadron (jet fragment) is the cleanest way to measure jet fragmentation in hadronic collisions, since at leading order the photon energy equals that of the recoiling jet. Furthermore, the dominant process for producing photon+jet pairs is Compton scattering in which the outgoing jet is a light quark, whereas di-jet generation at the LHC is dominated by gluon jets. Because the photon does not carry color charge, $\gamma$+hadron measurements have been proposed as a sensitive probe of jet quenching [10], and results from RHIC of this observable are now emerging [11, 12]. Such measurements will be a major area of focus for the EMCal, in conjunction with other subdetectors in ALICE.

### 1.1.3 Heavy Flavor

The measurement of heavy flavor production at high $p_T$ provides unique observables of jet quenching. The suggestion that massive quarks experience reduced energy loss due to the suppression of forward radiation ("dead cone effect" [13]) has not been borne out by RHIC measurements [14, 15], leading to signficant theoretical and experimental activity to elucidate the underlying physics of this process. Individual charm and beauty measurements at high $p_T$ will also be a focus at the LHC. At high $p_T$, semi-leptonic decay channels (branching ratio $\sim$10% for both B and D mesons) are favorable for heavy flavor measurements because they can be triggered, but also require good hadron rejection. While ALICE has extensive capabilities for electron measurements, via both TPC energy loss and the TRD, the EMCal is the primary tool for electrons at $p_T$>10 GeV/c, providing both an efficient and fast trigger and sufficient hadron rejection. Secondary vertexing provides additional discrimination, and ALICE with the EMCal can measure b-jet production in Pb–Pb collisions up to $E_T$$\sim$80 GeV. At these high energies the "dead cone" effect will be negligible, and such high energy quark jets will provide the cleanest measurement of the color-charge dependence of partonic energy loss [16].

### 1.1.4 Identified Particles

ALICE's capabilities for particle identification are unique at the LHC. Measurement of the spectra and yields of a large variety of hadrons, hadronic resonances, and their correlations in triggered jet events will constrain medium properties as well as hadronization and energy loss models. Gluon splitting is expected to be the dominant mechanism for partonic energy loss in the medium. Certain theories [17] predict that this could lead to a hadron specific fragmentation function modification, which would be measurable in high momentum hadronic yields and ratios in reconstructed jets. The spectra of hadrons in jets will be measured out to $p_T$ = 30 GeV/c to extract differences in the energy loss of quarks (from heavy-flavor and photon tagged jets) and gluons to study the path-length dependence of parton propagation

in the medium and the impact of the color factor on parton energy loss. The apparent lack of sensitivity to the color factor in the RHIC data has been interpreted as evidence that flavor conversion of partons traversing the medium may be a dominant effect in the $p_T$ range around 10 GeV/c [18] . $\mathrm{R}^{\gamma}_{AA}/\mathrm{R}^{\pi}_{AA}$ or $\mathrm{R}^{p}_{AA}/\mathrm{R}^{\pi}_{AA}$ are expected to be sensitive to the flavor conversion probability. Evidence for flavor conversion would alter our understanding of the relationship between final hadronic cross sections and the initial parton flavor.

The decay properties of strongly decaying hadronic resonances with lifetimes comparable to the lifetime of the dense medium will likely be sensitive to the properties of the medium. The enhanced yield of high momentum resonances accessible in the EMCal triggered jet sample will enable a detailed comparative study of in-medium modification of resonances in jets and bulk matter.

# Chapter 2

# EMCal Detector Layout and Performance

## 2.1 Design Overview

The EMCal is located inside the large, ambient-temperature solenoidal magnet of ALICE. The EMCal occupies a cylindrical integration volume approximately 110 cm deep in the radial direction, with front face $\sim$ 450 cm from the beam line. This volume is sandwiched between the ALICE space-frame, which supports the entire central detector, and the magnet coils. The PHOton Spectrometer (PHOS) carriage below the ALICE TPC and the the High Momentum Particle IDentifier (HMPID) above the ALICE TPC, define the azimuthal EMCal coverage to be 107 degrees. In the longitudinal direction the EMCal has a length $\sim$ 700 cm, covering $|\eta| < 0.7$.

The EMCal detector design is based on the Shashlik technology, as implemented in the PHENIX experiment [1] at RHIC, HERA-B [2] at HERA, and LHCb [3] at CERN. The detector is a layered Pb-scintillator sampling calorimeter with a longitudinal pitch of 1.44 mm Pb and 1.76 mm scintillator with longitudinal wavelength shifting fiber light collection. The scope and basic design parameters of the proposed calorimeter have been chosen to match the physics performance requirements of the proposed ALICE high $p_T$ physics program.

Figure 2.1 shows the EMCal Super Modules, the basic structural units of the calorimeter, mounted in their installed positions on the support structure. A continuous arch of Super Modules, each spanning $\sim$20 degrees in azimuth, is indicated. The EMCal is positioned to provide partial back-to-back coverage with the PHOS calorimeter.

The detector is segmented into 12,288 towers, each of which is approximately projective in $\eta$ and $\phi$ to the interaction vertex. The towers are grouped into Super Modules of two types. There are 10 full size and 2 one-third size Super Modules in the full detector acceptance (see Fig. 2.1). The full size modules span $\Delta\eta$ =0.7 and $\Delta\phi = 20°$, whereas the 1/3 modules span

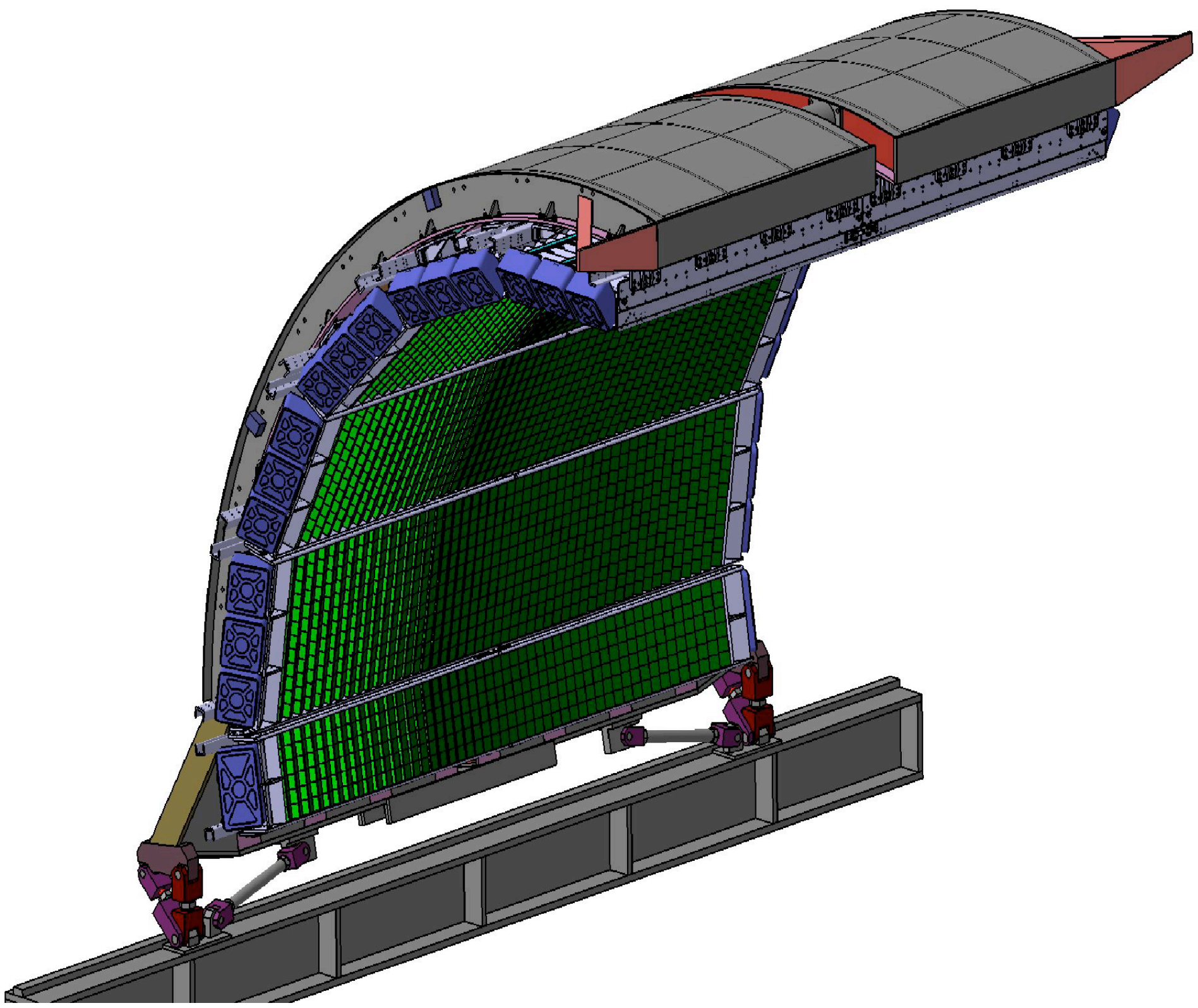

**Figure 2.1:** The array of Super Modules shown in their installed positions on the support structure.

a smaller azimuthal range of $\Delta\phi = 7°$.

Each full-sized Super Module is assembled from $12 \times 24 = 288$ modules arranged in 24 strip modules of $12 \times 1$ modules each. Each one-third size Super Module is assembled from $4 \times 24 = 96$ modules. Each module has a fixed width in the $\phi$ direction and a tapered width in the $\eta$ direction with a full taper of 1.5°. The resultant assembly of stacked strip modules is approximately projective in $\eta$ with an average angle of incidence at the front face of a module of less than 2° in $\eta$ and less than 5° in $\phi$.

A module is a single self-contained detector unit. All modules in the calorimeter are mechanically and dimensionally identical. Each module comprises four independent detection channels/towers giving a total of 1152 towers per full sized Super Module.

The calorimeter design incorporates on average a moderate active volume density of $\sim$5.68 g/$cm^2$ which results from a $\sim 1 : 1.22$ Pb to scintillator ratio by volume. This results in a compact detector consistent with the EMCal integration volume at the chosen detector thickness of $\sim$ 20 radiation lengths. In simulations, this number of radiation lengths gives a maximum deviation from linearity (due mainly to shower leakage) of $\sim$2.8% for the most probable energy response in the range up to 100 GeV photons.

The physical characteristics of the EMCal are summarized in Table 2.1. An exploded view

drawing of the module showing all single components is shown in Fig. 2.2

**Table 2.1:** The EMCal Physical Parameters.

| Quantity | Value |
|---|---|
| Tower Size (at $\eta$=0) | $\sim$6.0 $\times$ $\sim$6.0 $\times$ 24.6 $cm$ (active) |
| Tower Size | $\Delta\phi \times \Delta\eta = 0.0143 \times 0.0143$ |
| Sampling Ratio | 1.44 mm Pb / 1.76 mm Scintillator |
| Number of Layers | 77 |
| Effective Radiation Length $X_o$ | 12.3 mm |
| Effective Moliere Radius $R_M$ | 3.20 cm |
| Effective Density | 5.68 g/$cm^2$ |
| Sampling Fraction | 10.5 |
| Number of Radiation Lengths | 20.1 |
| Number of Towers | 12,288 |
| Number of Modules | 3072 |
| Number of Super Modules | 10 full size, 2 one-third size |
| Weight of Super Module | $\sim$7.7 metric tons (full size) |
| Total Coverage | $\Delta\phi$ = 107$^o$, -0.7 < $\eta$ < 0.7 |

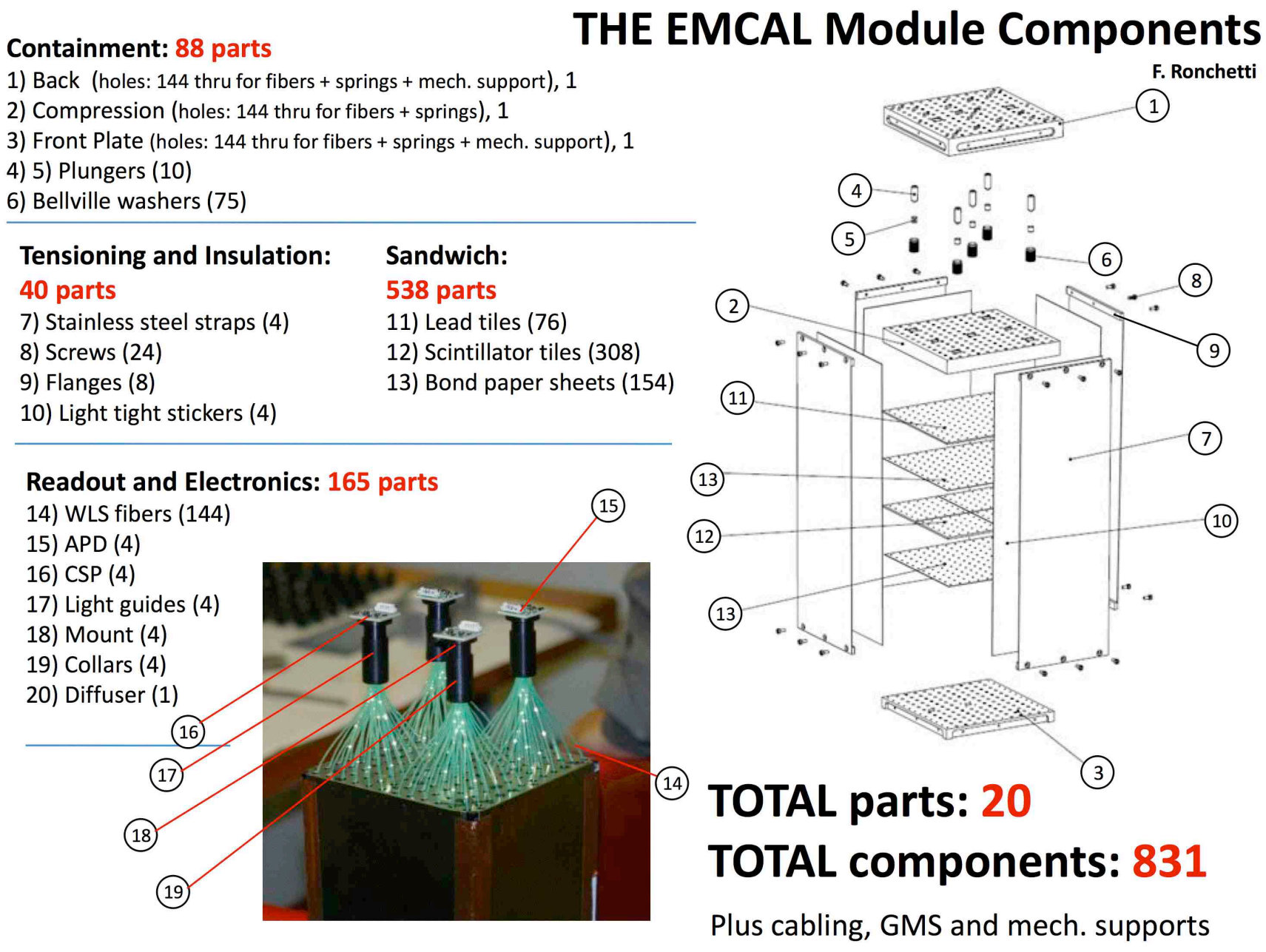


**Figure 2.2:** Exploded view drawing of EMCal module showing all components.

The contribution to the integrated radiation length (material budget) in front of the EMCal is shown in Fig 2.3.

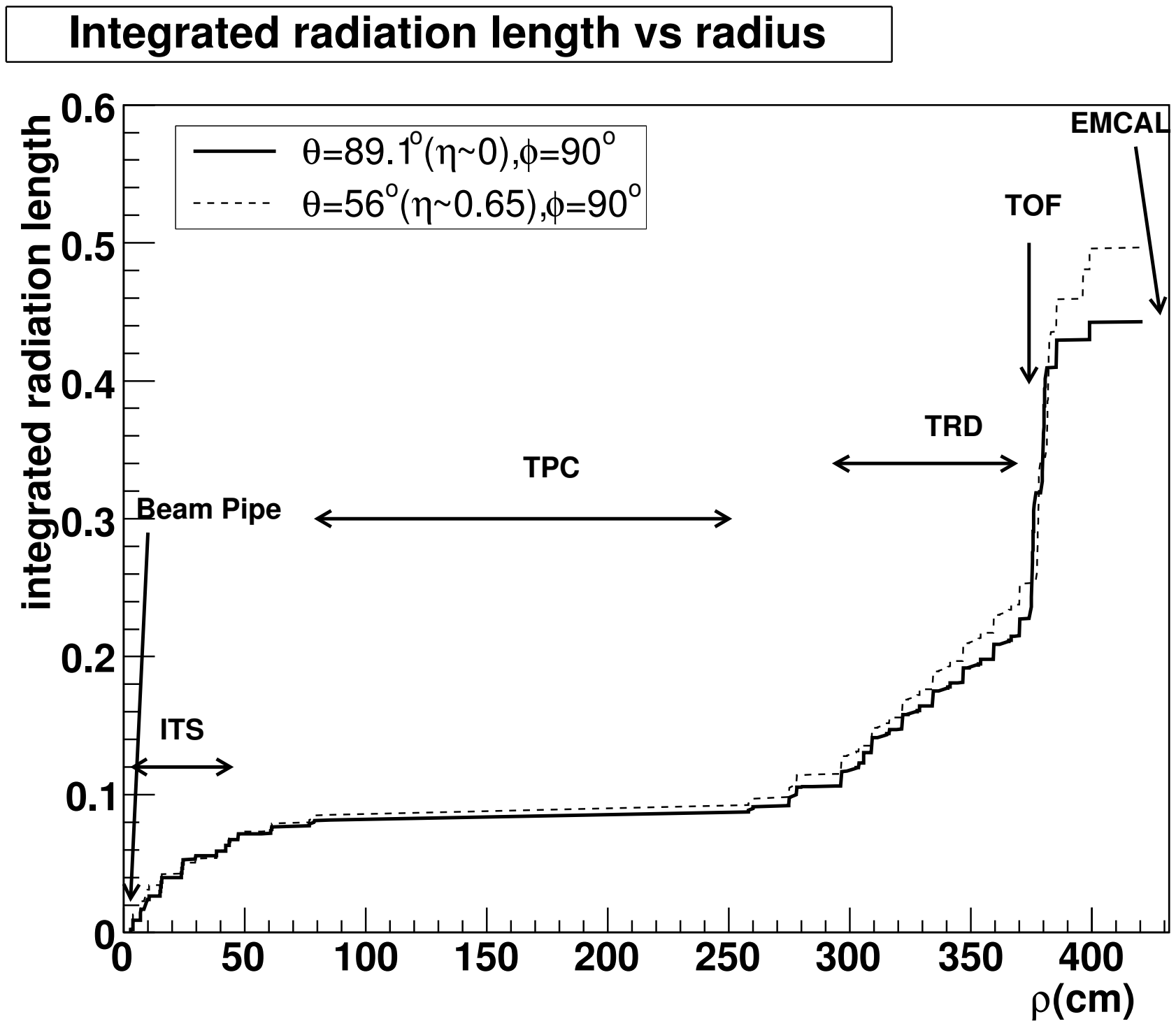


**Figure 2.3:** Integrated radiation length vs. distance from the beam line at two values of $\eta$ in ALICE.

## 2.2 Electronics

The active readout element of the EMCal detector is a radiation hard $5 \times 5$ mm$^2$ active area Avalanche PhotoDiode (APD), with high quantum efficiency, low dark current, and very good stability and reliability. The APDs are operated at moderate gain for low noise and high gain stability in order to maximize the energy resolution. With a nominal APD gain of M=30, about 132 electrons are generated in the APD per MeV of energy deposited by showering electromagnetic particles (4.4 $e^-$/MeV) .

The APD is connected directly to the back of a Charge Sensitive Preamplifier (CSP) which integrates the charge output from the APD over a 1 pF capacitor into a voltage step pulse. This step pulse is conditioned using a CR-2RC Gaussian shaper prior to digitization with the ALICE TPC ReadOut (ALTRO) chip [5]. The output of the shaper is amplified with two different gains (high and low) in order to cover the large dynamic range of the EMCal.

The Front End Electronic Card (FEEC) contains 32 remotely controlled precision High Voltage (HV) bias regulators [4], 64 shapers and digitizers, a board controller, and a power

regulation system which prevents noise coupling between digital or High Voltage sections and the analog signal section. Four ALTRO chips are required, each containing 16 10-bit flash ADCs and internal multi-event buffers, for a total of 32 high gain and 32 low gain channels per FEE card. The ALTRO ADCs sample the output signals from the shapers at 10 MHz (programmable up to 20 MHz).

The choice of the ALTRO chip, combined with a board controller FPGA similar to the board controller used by the TPC, allows PHOS and EMCal to re-use the readout back-end protocol of the ALICE TPC via an external Readout Control Unit (RCU) [6].

### 2.2.1 Readout

The readout of the FEE of one Super Module is performed by two RCUs, each RCU controlling the readout of 18 FEECs. Each RCU is connected to the Data Acquisition System (DAQ) via a dedicated Detector Data Link (DDL). The DDL is common to all detectors writing data to storage in the ALICE experiment and comprises a Source Interface Unit (SIU), which for EMCal is connected directly to the RCU, a Destination Interface Unit (DIU) situated in the DAQ counting room, and an optical fiber connecting the SIU with the DIU.

### 2.2.2 Data Volume and Bandwidth

Each RCU will transfer data from 18 FEECs for a total of $32 \times 2 \times 18 = 1152$ channels (576 towers) of 10-bit data. There are 10 SM + 2 one-third-size SM planned for the full EMCAL for a combined total of 24576 readout channels (12288 towers).

The data volume per channel is a function of how many ADC time-samples are read out from the ALTRO chip. With the 200 ns shaping time for the EMCal shaper and a 10 MHz sampling frequency, about 10 samples would adequately sample the peak. In the current default configuration we include up to 15 data samples per channel. Each sample is a 10-bit ADC word, and the data is formatted into 40-bit words in the ALTRO chip.

The plan is to always operate with zero suppression, so that only channels with data above certain threshold over pedestal are kept. For every channel that is not suppressed there is also an additional 4 bytes with channel address and payload information, needed for data unpacking. Currently, the data volume for a channel with all 15 samples kept is 24 bytes.

The tower hit rate for the full 8 kHz min bias Pb–Pb collision rate is estimated to be 2000 Hz based on HIJING simulations . This would correspond to a maximum data transfer rate of $2000 \times 2 \times 24$(Bytes) $\simeq$ 94 kBytes/s/tower, or 26 MBytes/s/GTL with zero suppression (no data from towers without hits). This is below the DDL data transfer limit of 200 MBytes/s, thus allowing the EMCal to operate without deadtime at the full Pb–Pb min bias trigger

rate. The data volume for readout of the full EMCal would be (2 gain ranges) × (12288 towers) × 24(Bytes) = 576 kBytes per event. The 2 kHz tower hit rate corresponds to a 40% average EMCal occupancy for Pb–Pb collisions at 8 kHz. The corresponding EMCal average total event size would be ∼ 230 kBytes per event. This is much smaller than the ∼ 75 MBytes size of the average TPC event for Pb–Pb collisions. For p–p collisions and cosmics, the EMCal average event size is of course suppressed/reduced much further.

## 2.3 Trigger

The ALICE Trigger System is made up of two independent components: a Central Trigger Processor (CTP) and a Trigger Distribution Network. The CTP provides the trigger decision logic, generating triggers for the readout detectors by evaluating inputs from triggering detectors. The Trigger Distribution Network delivers these triggers to the detectors.

The earliest trigger decision (Level 0, or L0) is issued 1.2 $\mu$s after the event, L1 is issued at 6.5 $\mu$s, and L2 is issued at 88 $\mu$s. L1 and L2 decisions provide rejection of L0 triggers. The CTP checks the events for pile-up, and acceptance at L2 implies 'free of pile-up'. The High Level Trigger (HLT) may also provide pile-up rejection [7].

### 2.3.1 Level 0/1 EMCAL Trigger

The L0/1 EMCal trigger is implemented in two hierarchically configured layers:

- Lower layer: each EMCal 32 channel FEE card forms 8 analog charge sums of 2 × 2 adjacent towers, for fast L0 and deadtime-less L1 triggers. The fast-OR signals are sent to Trigger Region Units (TRU), which receives 96 analog sums from 12 FEE cards. The TRU digitizes the sums using a commercial FADC and inputs the full space and time image of all channels to a single FPGA. This configuration is identical to that used by PHOS. Only decisions based on the local region of the EMCal accessed by a single TRU can be made, appropriate for photon ($\pi^0$, $\eta$,...) and electron triggers.
- The upper layer is implemented as a single, global trigger unit (Summary Trigger Unit, or STU) that receives trigger data from all TRUs via LVDS cables clocked at 40 MHz, and inputs them into a single FPGA.

Since QCD jets subtend an area larger than that accessed by a single TRU, L1 jet triggering is carried out in the STU. The jet patch trigger in the STU employs a simple jet-finding algorithm applied over the EMCal acceptance, in which the energy is integrated within a patch of defined size and the patch is stepped over the entire EMCal fiducial area. An L1 jet trigger signal is issued if a patch energy in the event exceeds a defined threshold.

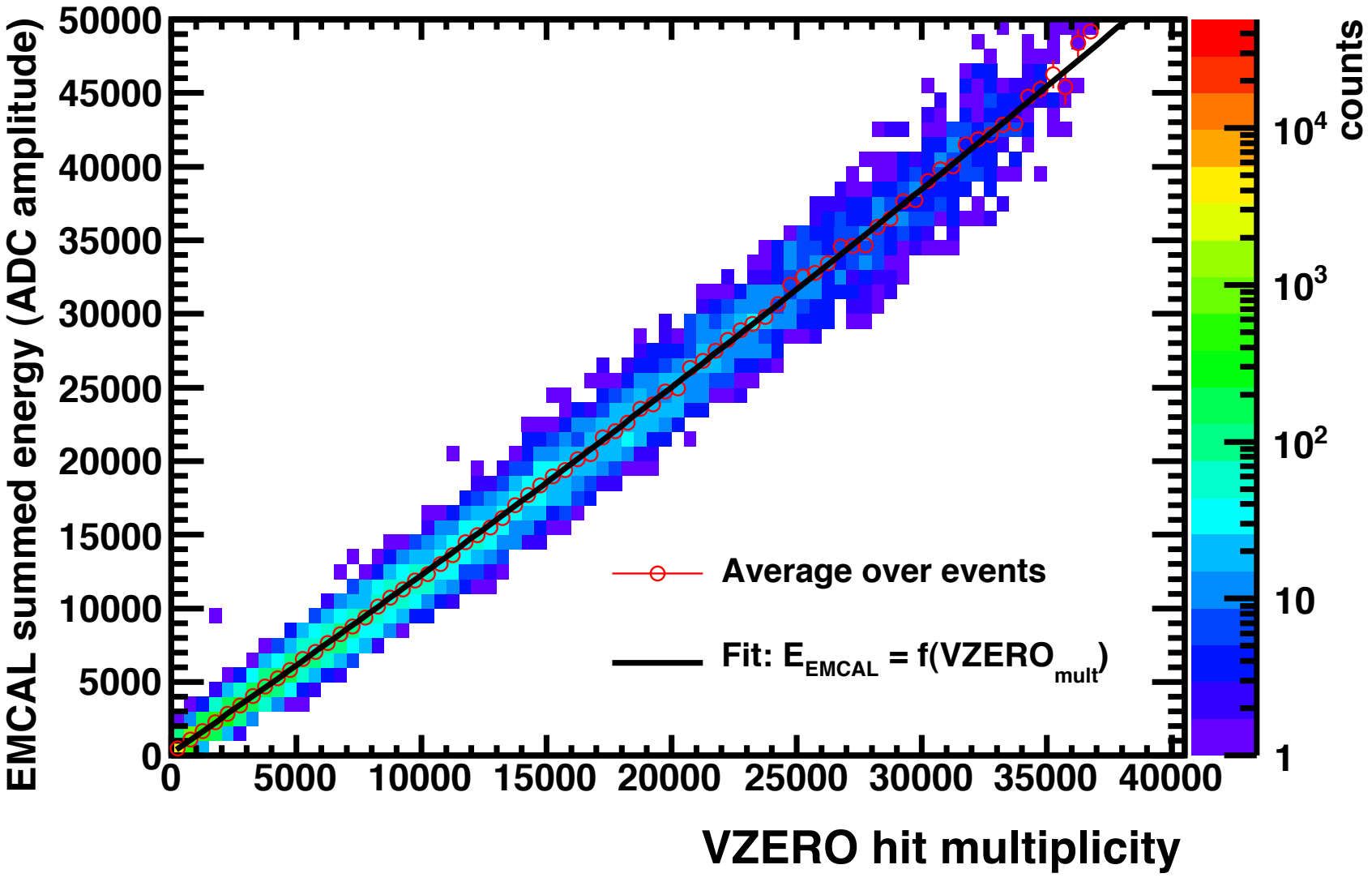


**Figure 2.4:** Correlation of integrated signal in V0 forward scintillators ($-3.7 < \eta < -1.7$, $2.8 < \eta < 5.1$) with total energy (summed ADC counts) in EMCal, for minimum bias Pb–Pb collisions at $\sqrt{s} = 5.5$ TeV. Simulation is HIJING events filtered through a detailed GEANT model of ALICE.

To suppress the heavy-ion event background in the jet patch the trigger requires a multiplicity input from the V0 forward scintillator detectors ($-3.7 < \eta < -1.7$, $2.8 < \eta < 5.1$) whose response is closely correlated with the event centrality and $E_T$ in the ALICE central region. Figure 2.4 shows the correlation between V0 multiplicity and total energy within the EMCal, simulated with HIJING events filtered through a detailed GEANT model of ALICE. The average of this correlation, indicated in the figure, is used at L1 to account for the average background energy accumulated within a given jet patch.

### 2.3.2 High Level Trigger

The HLT is an online event filter and trigger system, with input bandwidth 25 GB/s at event rates of up to 1 kHz. The system is designed as a scalable PC cluster, implementing several hundred nodes. The transport of data in the system is handled by an object-oriented data flow framework operating on the basis of the publisher-subscriber principle. The design is fully pipelined, with low processing overhead and communication latency in the cluster.

The HLT system is designed to increase the statistics of recorded physics events of interest by a factor of 10. The current tracking performance shows that a sufficient event reconstruction within the central Pb–Pb event rate of 200 Hz will be achievable for multiplicity densities of $dN_{ch}/d\eta \simeq 2000$. For higher densities, cluster deconvolution based on track parameters becomes necessary. The software framework allows to run similar (or exactly the same)

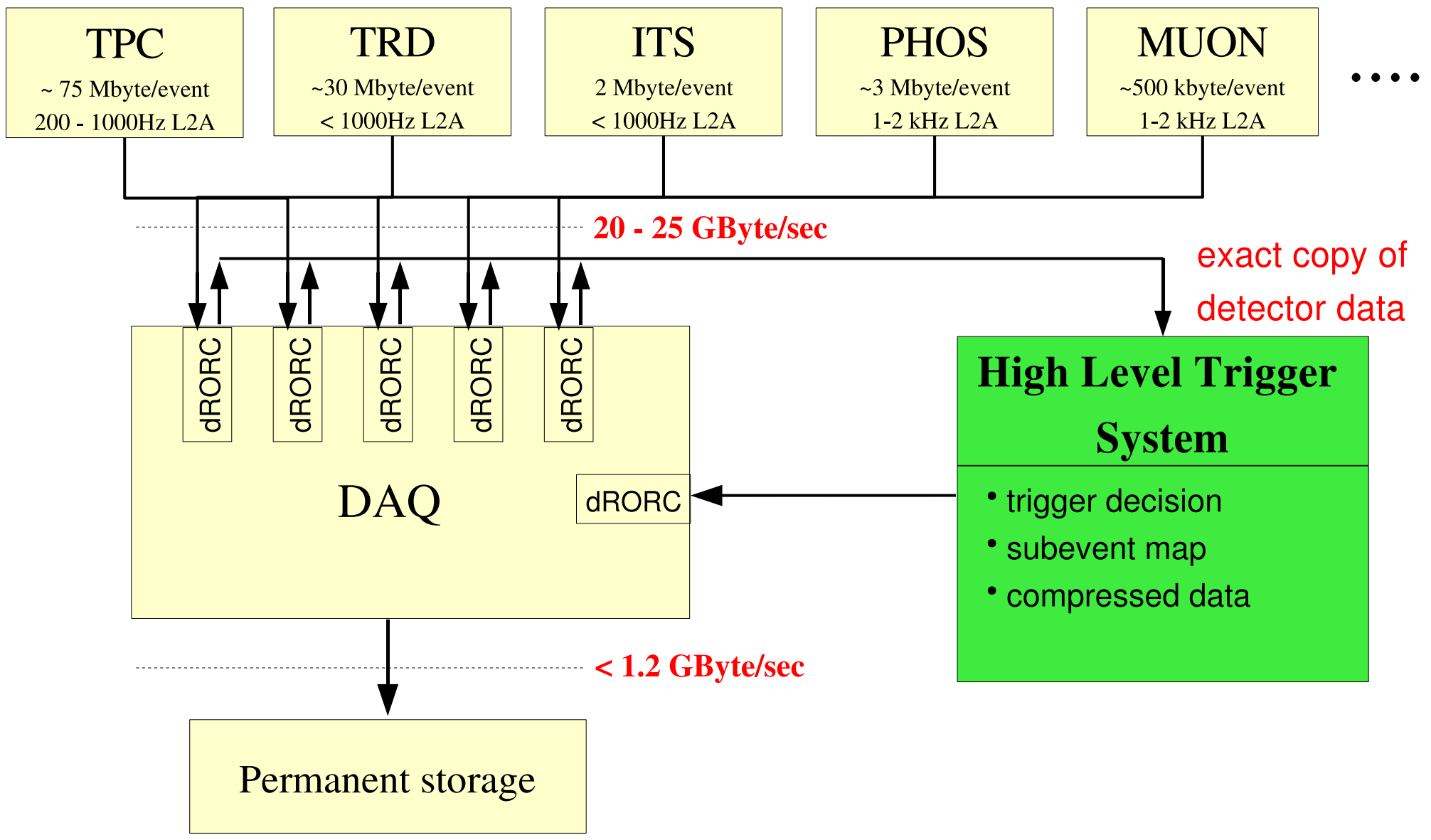


**Figure 2.5:** High-Level Trigger system in the ALICE data stream. The HLT receives a copy of the detector data and is treated by DAQ as an additional detector. The specied numbers are upper limits for the event size delivered by a sub-detector.

reconstruction algorithms to offline reconstruction software. The online and offline software performance have been shown to deliver similar physics response. The HLT is also calibration aware allowing for the *best-guess* online reconstruction of the events thanks to the interface to the Offline Condition Data Base and Detector Control System.

Figure 2.5 shows the integration of the HLT into the data flow of the ALICE experiment. The raw data are transferred via optical fibers from the detector front-end to the DAQ system. The Detector Data Link (optical link) is used commonly for data readout of all ALICE detectors. The data stream is received by the HLT RORC (H-RORC). In total, 454 DDLs are forwarded to HLT (200 MB/s each), including all relevant detectors. The trigger decision, reconstructed events, and compressed data are transferred back to the DAQ via the ALICE standard DDL.

### 2.3.3 Bandwidth Limitations

The trigger enhances the recorded sample of rare processes in minimum bias Pb–Pb and p–p collisions; for the EMCal these processes comprise high $p_T$ jets, $\pi^0$, photons, electrons. Rejection rates must satisfy the following bandwidth considerations:

- Collisions at ALICE are expected to occur with rates of about 4 kHz for heavy-ion and 400 kHz for p–p .

- The expected safe operation of the TPC allows for maximum gating frequency of about 500 Hz in p–p collisions and up to 100 Hz in Pb–Pb .
- The rejection rate at L1 is dictated by the input bandwidth of HLT. The total data volume from all ALICE detectors expected in minimum bias Pb–Pb is 20 MB. This amounts to 80 GB/s data flow which must be reduced to HLT input specifications.
- The HLT can handle up to 25 GB/s at event rates of up to 1 kHz.
- The HLT must provide additional rejection to fulfill the limitation of the recording bandwidth, which is 1.25 GB/s.

## 2.4 Detector Performance

The results presented here are based on two sets of test measurements. Early tests were performed in November 2005 at the Meson Test Beam (MTEST) at FNAL utilizing a stacked $4 \times 4$ array of *prototype* EMCal modules ($8 \times 8$ towers). Then, during a period of five weeks in autumn 2007, the final ALICE EMCal modules were tested in the CERN SPS and PS test beam lines. The test utilized a stacked $4 \times 4$ array of EMCal modules ($8 \times 8$ towers). All towers were instrumented with the full electronics chain with shapers and APD gains operated as planned in ALICE. The readout of the front end electronics used the full ALICE DAQ readout chain.

At the SPS, a primary proton beam of 400 GeV, with intensities up to $10^{12}$ particles per spill is incident on a primary target providing pion, electron, and muon secondary beams. The maximum achievable momenta at the three secondary beam lines are highly correlated. Thus the maximum electron energy was constrained to be between 5 and 120 GeV. The electron beam had a purity of better than 99% and a typical momentum spread of $\delta$p/p $\sim$ 1.3% (defined by the chosen aperture). The test measurements at Fermilab utilized a primary proton beam which generated mixed beams with good particle identification (e/$\pi$/p discrimination) over the full range of available momenta (3-33 GeV).

### 2.4.1 Energy resolution

#### 2.4.1.1 Contributions to the EMCal energy resolution

The energy resolution of an electromagnetic calorimeter can be parameterized as

$$\sigma/E = a/\sqrt{E} \oplus b \oplus c/E \; , \tag{2.1}$$

where E is the shower energy and the first term characterized by the parameter $a$ arises from stochastic fluctuations due to intrinsic detector effects such as energy deposit, energy sampling, light collection efficiency, etc. The constant term, $b$, arises from systematic effects, such as shower leakage, detector non-uniformity or channel-by-channel calibration errors. The third term, $c$, arises from electronic noise summed over the towers of the cluster used to reconstruct the electromagnetic shower. The three resolution contributions add together in quadrature as indicated in Eq. 2.1. Over the lower half of the energy range of interest in ALICE, the stochastic term dominates with the constant term increasing in significance only at the highest energies.

The energy resolution for a given sampling frequency in an electromagnetic calorimeter varies with the sampling frequency approximately as $\sigma/E \sim \sqrt{d_{\mathrm{Sc}}/f_{\mathrm{s}}}$ where $d_{\mathrm{Sc}}$ is the scintillator thickness in mm and $f_{\mathrm{s}}$ is the sampling fraction for minimum ionizing particles. For optimum resolution in a given physical space and total radiation lengths, there is thus a desire to have the highest possible sampling frequency. Practical considerations, including the cost of the total assembly labour, suggest reducing the total number of Pb/scintillator layers thus decreasing the sampling frequency. Using the final 1:1.22 Pb to scintillator ratio with a sampling geometry of Pb(1.44 mm)/Scint(1.76 mm), detailed GEANT3 simulations yield $a/\sqrt{E} \oplus b\%$ with fit results $a = (6.90 \pm 0.09)\%$ and $b = (1.44 \pm 0.03)\%$ over the range $p_T$ = 5 to 100 GeV/c. The simulation results are shown in Fig. 2.6. These results are based on energy deposition only and do not include photon transport efficiencies or the electronic noise contribution. Systematic contributions to the resolution arising from calibration and related systematic uncertainties are ignored.

Some increase in the constant $a$ is to be expected from photon transport and related effects. This has been studied in a series of test beam measurements of prototypes of this detector with various sampling frequencies including - Pb(1.6 mm)/Scint(1.6 mm) also shown in Fig. 2.6 - and preliminary results are consistent with a small increase in $a$.

The value of the constant term $b$ is dominated by shower leakage in these calculations. Other systematic effects which arise during detector fabrication and from the tower-by-tower calibration uncertainties will increase $b$. The latter effect is itself of the order of 1% typically.

The relative contributions of the electronics to the total EMCal energy resolution are shown in Fig. 2.7. The intrinsic energy resolution has been assumed to be $6.9\%/\sqrt{E} \oplus 1.4\%$, based on GEANT3 simulations for the production module. The digitization resolution has been assumed to be determined by a maximum energy scale set to 250 GeV with 10-bits of digitization resolution and dual gain ranges separated by a factor of 16. The constant energy contribution due to calibration errors has been assumed to be 1%. Finally, the electronics noise contribution has been conservatively assumed to be $\sigma_{ENC} = 2000\ e^-$ for an integration time of 100 ns. With a light yield of 4.4 $e^-$/MeV, a gain of 30, and $3 \times 3$ modules included in the energy sum, this corresponds to an electronics noise contribution to the resolution of $c = 48\mathrm{MeV}/E$ (Eq. 2.1). This contribution (dotted curve) is seen to be negligible compared

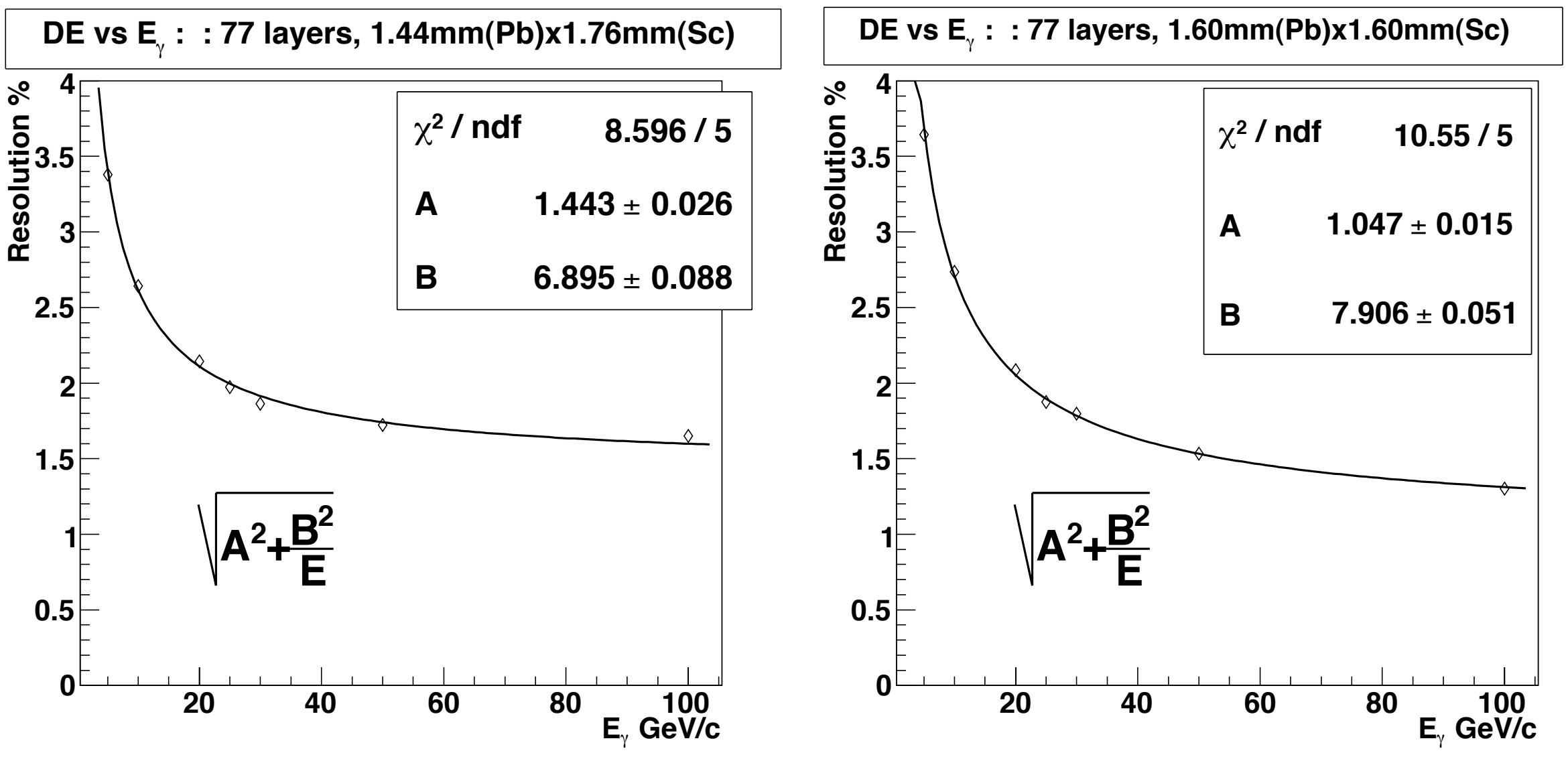


**Figure 2.6:** GEANT3 simulations of the EMCal module resolution for a proposed production module (left) and a prototype test module (right).

to the intrinsic noise contribution (solid dark curve) except at photon energies much below 1 GeV.

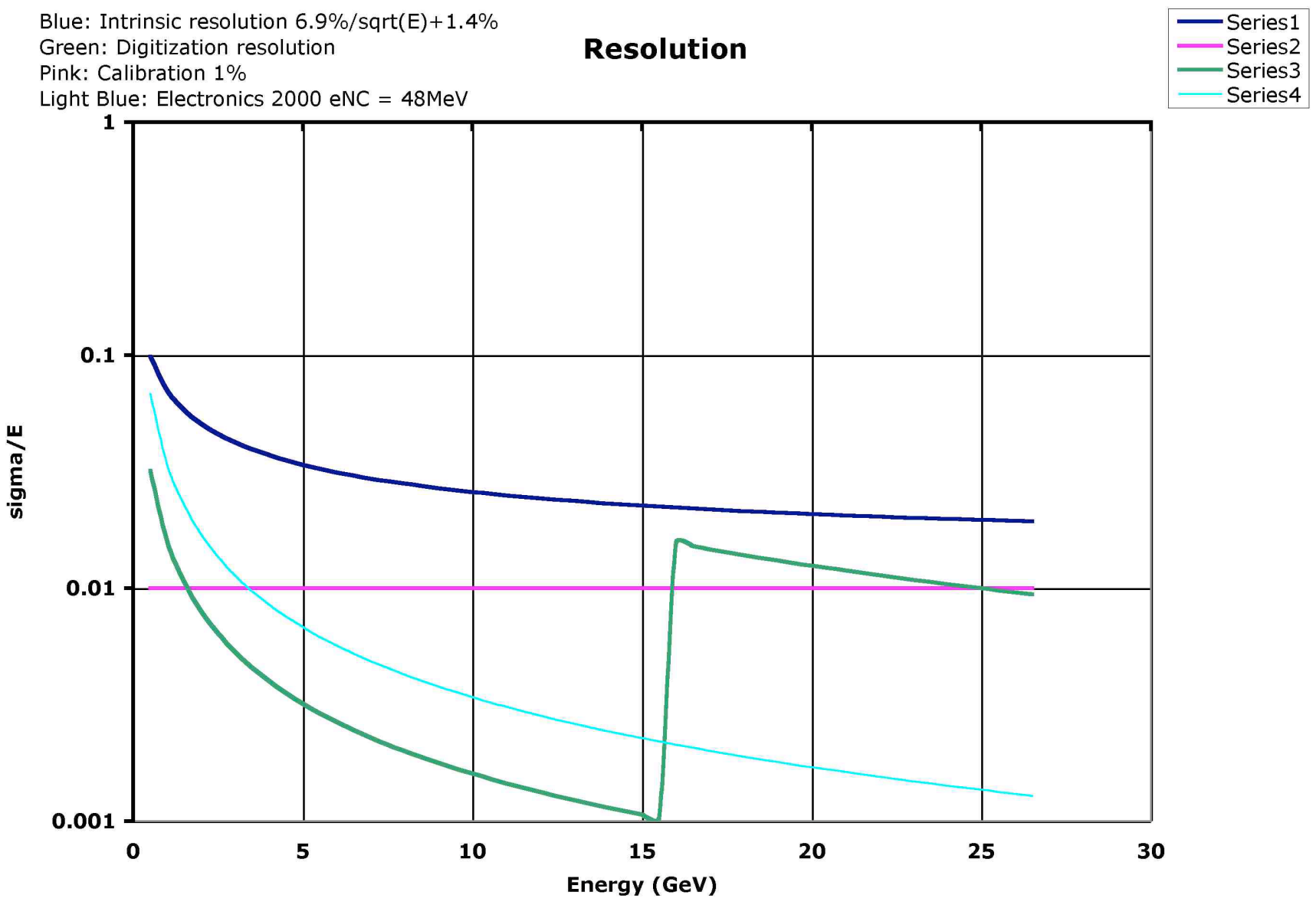


**Figure 2.7:** Contributions to the total EMCal photon energy resolution. Blue curve: $6.9\%/\sqrt{E} \oplus 1.4\%$ intrinsic contribution, Green curve: digitization contribution, Light blue curve: electronic noise contribution.

#### 2.4.1.2 Test beam results on energy resolution

In order to reach the design EMCal energy resolution for high energy electromagnetic showers, a tower-by-tower relative energy calibration of about 1% has to be obtained and maintained in the offline analysis. In addition, since analog tower energy sums provide the basis of the L0 and L1 high energy shower trigger input to the ALICE trigger decision, the EMCal should operate with APD gains adjusted to match online relative tower energy calibrations to better than about 5%.

A LED calibration system, in which all towers view a calibrated pulsed LED light source, has been successfully tested to track and adjust for the temperature dependence of the APD gains during operation. During the test beam the temperature varied by at most 3 degrees which led to a 3% variation in the LED amplitude. The temperature coefficients obtained from the fits of distributions were used to correct for the time dependence of the APD gain.

An absolute energy calibration of the test beam data was obtained from the known incident electron energy using an iterative procedure. An initial relative tower-by-tower calibration was performed using the MIP peak from the hadron beams. The LED calibration system was used to track and adjust for the time dependence of the calibration coefficients. The energy range of 0.5 - 100 GeV could be explored. Such energy scans were performed at several different positions, including tower and module edges. No systematic variation of the resolution depending on the position was observed. The resolution obtained at the different positions was combined and the average values as a function of the incident beam momentum are displayed in Fig. 2.8. For the SPS data, the momentum spread of the incident beam of typically 1.3% was subtracted in quadrature.

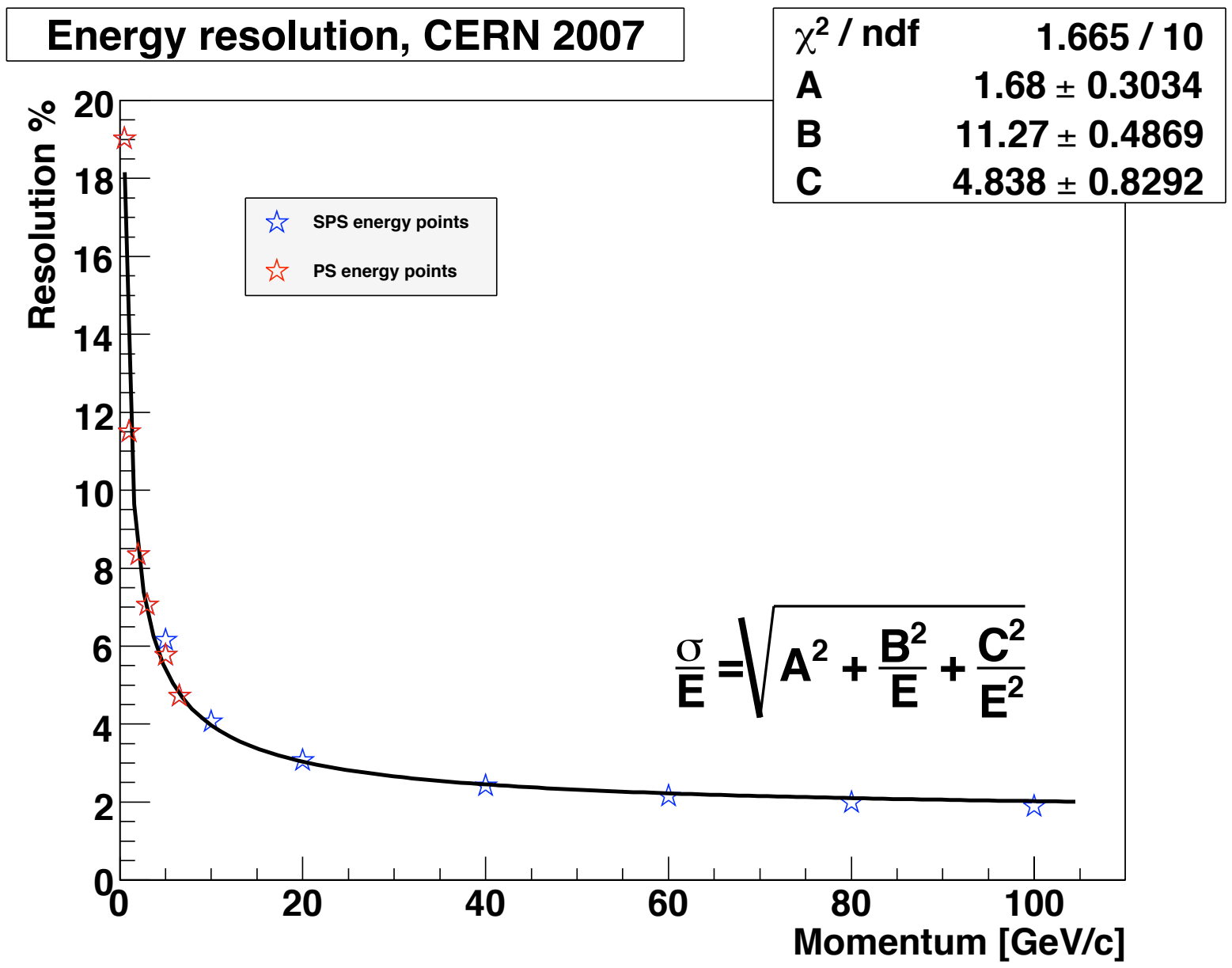


**Figure 2.8:** Energy resolution for electrons as a function of the incident beam momentum. The beam energy spread was subtracted from the measured result.

A fit to the energy resolution as a function of the incident energy is also shown in Fig. 2.8. The fit is made with the conventional constant, $\sqrt{E}$, and linear $E$ terms, added in quadrature. The constant and $\sqrt{E}$ terms, respectively $a = 1.7 \pm 0.3$ and $b = 11.3 \pm 0.5$, may be compared with the GEANT3 simulation result for the EMCal module geometry (without light transport included in the simulation) shown in Fig. 2.6 that gave $a = 1.44 \pm 0.03$ and $b = 6.9 \pm 0.1$. The performance is quite similar to the PHENIX EMCAL [1] with similar physical characteristics and better than the original requirements.

The impact of detector energy resolution on the proposed physics program has been studied. Given the nature of the proposed physics, and in particular, the main focus on jet physics, there is no sharp cutoff on the required energy resolution for isolated electromagnetic clusters. Simulations show that a resolution of the order of $\sim 15\%/\sqrt{E} \oplus 2\%$ is sufficient for the jet physics program and this is fixed as the minimum detector requirement. The electron and photon physics programs will benefit from better resolution. Based on simulations and test beam results it is expected that the EMCal minimum performance requirements will be readily met and we project an ultimate performance of better than $\sim 12\%/\sqrt{E} \oplus 1.7\%$.

### 2.4.2 Linearity and Uniformity of the Energy Response

The linearity of the energy response was investigated in conjunction with the energy resolution. Fig. 2.9 displays the average ratio of the reconstructed and incident beam energy as a function of the incident beam energy obtained by combining the measurements at different detector positions. A very good linearity is observed.

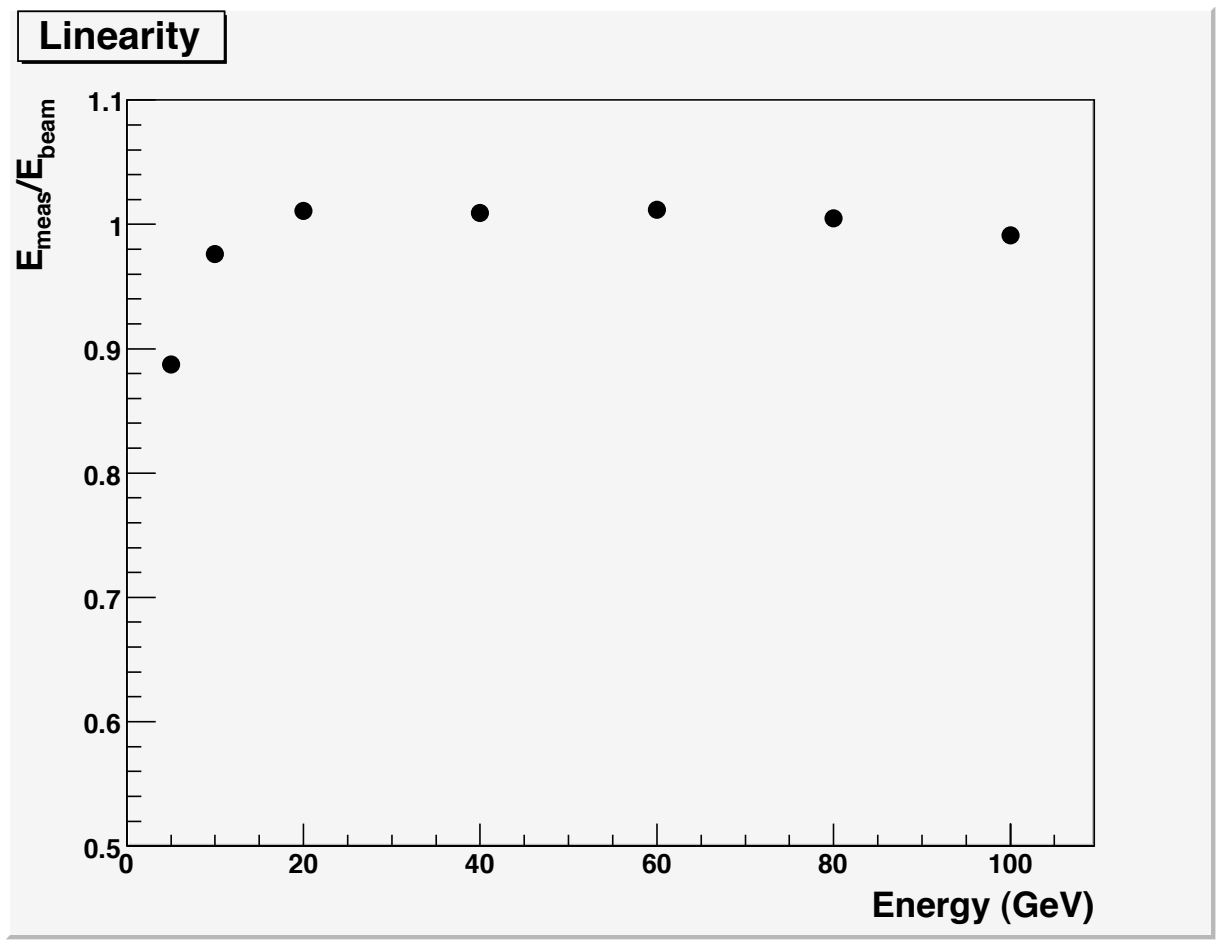


**Figure 2.9:** Average value of the measured to incident beam energy as a function of the incident beam energy for several different positions taken at the SPS.

Deviations of this ratio from unity were expected at high energies due to leakage, but have not yet been observed in the data. At very low energies, threshold effects might be non-negligible compared to the total energy and light transmission losses might have an impact.

In fact, the reconstructed energy is systematically lower than the incident one for energies equal or below 5 GeV. A drop of $\sim 10\%$ is observed at 5 GeV.

The uniformity of the energy response was studied under several different conditions. All module centers and a major part of tower centers were scanned using 80 GeV electrons. In addition, data were taken across tower and module borders as well as for tilted or recessed modules, as used at the large Z end of a Super Module. The latter allow to study the uniformity of the response for incidence on towers in different locations within the Super Module as installed in ALICE. Figure 2.10 shows the reconstructed energy (upper panel) and resolution (lower panel) for the different configurations as indicated in the figure. At the corners and towers located at the edges (marked with red in the figure) significant losses are expected (which however are partially compensated by the calibration constants).

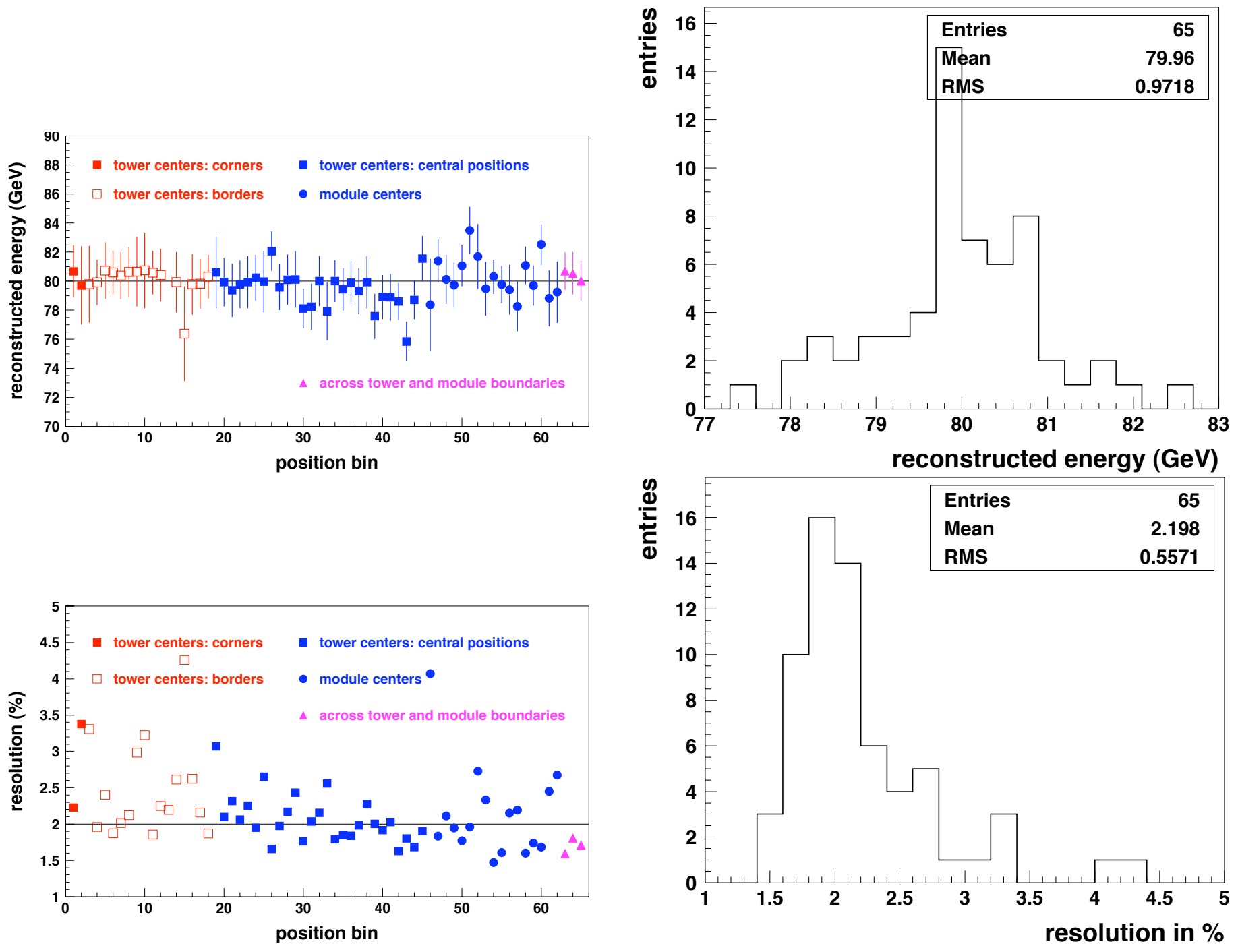


**Figure 2.10:** Upper panel: reconstructed energy for 80 GeV electrons at different positions as indicated in the figure and described in the text. Lower panel: the corresponding energy resolution for 80 GeV electrons at different positions as indicated in the figure.

No correction for time dependent effects have been applied yet, which might be important as data taken in very different time periods are compared. The preliminary results give a response of the EMCal with an RMS better than 1 GeV, for 80 GeV incoming electrons (see upper, right plot of Fig. 2.10). This result implies a very good uniformity of the EMCal construction and readout. In particular, no systematic effect is observed for positions across tower or module boundaries. The still large variations of the energy resolution at different positions, varying from 1.5% to 4.3%, underline the need for a full tower-by-tower inter-calibration.

The energy resolution was also studied for different incidence locations corresponding to the Super Module as installed in ALICE. A configuration where the beam hits the EMCal modules perpendicularly, was compared to configurations where $\phi$ was tilted by 6 deg or 9 deg at different surface positions. No significant difference with the average resolution at zero position was observed.

### 2.4.3 Position Resolution

The FNAL test beam data were also analyzed using the incident beam location projected from the tracking information from the MWPCs to investigate the position resolution of the EMCal. The $x$ and $y$ positions in the EMCal are calculated using distribution of energies in the towers of the cluster. The coordinate locations are calculated using a logarithmic weighting [8] of the tower energy deposits. The $x$ and $y$ position resolution as a function of incident momentum for electrons is shown in Fig. 2.11. As expected, no significant difference between the $x$ and $y$ position is observed. The electromagnetic shower position resolution is seen to be described as 1.5 mm + 5.3 mm/$\sqrt{E_{Deposit}}$.

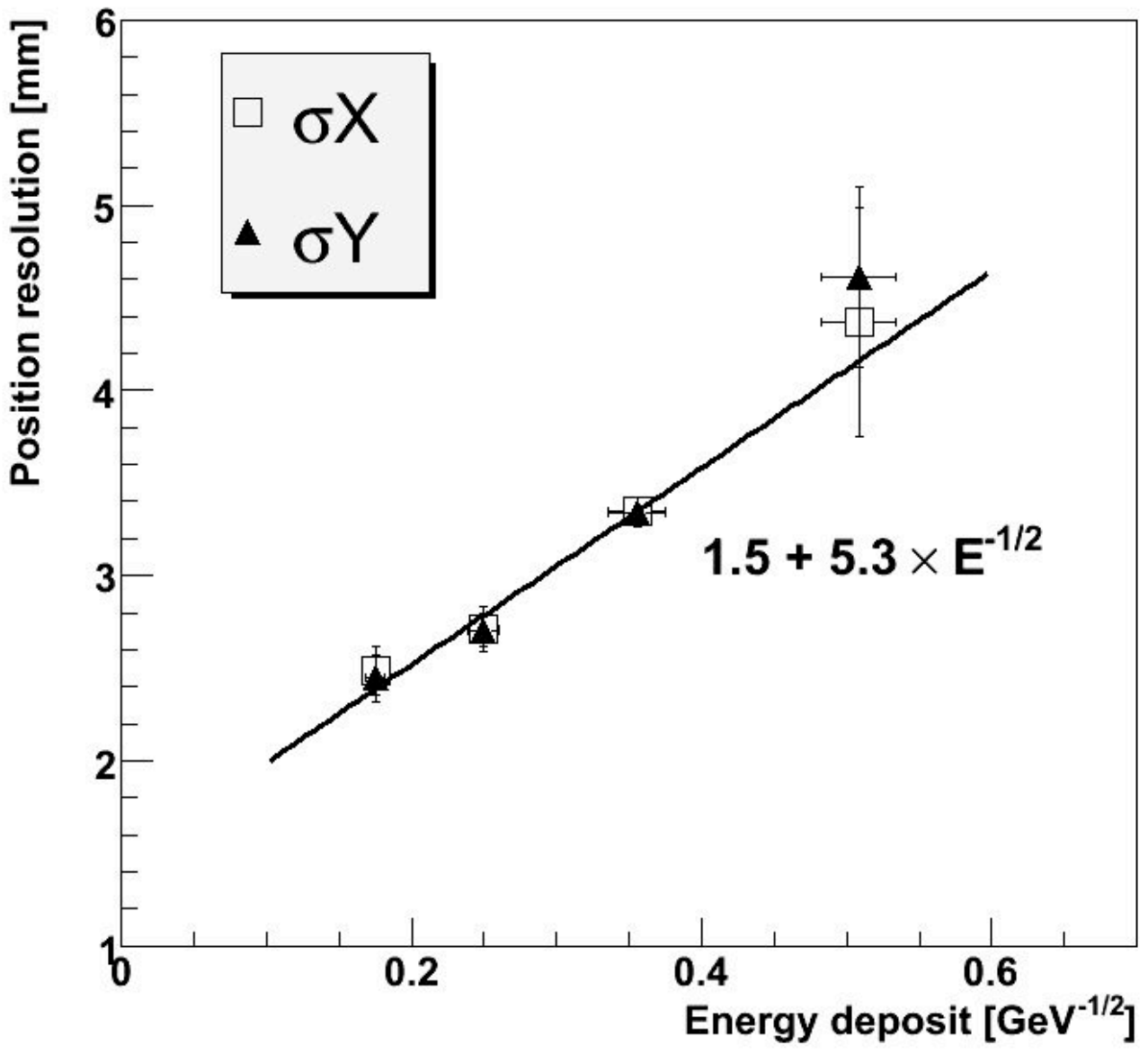


**Figure 2.11:** Dependence of the position resolution as a function of $1/\sqrt{E}$ (GeV) for electrons. The curve shows the best fit result.

### 2.4.4 In-situ calibration improvements

In order to maintain an optimal EMCal energy resolution the tower-by-tower relative energy calibration has to be held to better than 1%. Several in-beam analyses are planned, which

will improve on the results achieved through gain monitoring, the pre-calibration of the Avalanche Photo Diodes (APD), and the calibration with cosmic muons.

High statistics MIP data enable us to adjust the relative gain factors between towers. Identified electrons will be used to check the EMCal energy scale compared to the electron momentum measured by tracking in ALICE. Most importantly, the $\pi^0$ invariant mass spectrum will be used to confirm the absolute EMCal energy calibration. Overall the in-situ calibrations based on data are expected to improve the tower by tower energy alignment and the linearity by an order of magnitude compared to the test beam results.

With high statistics p–p or peripheral Pb–Pb collision data, two-photon invariant mass spectra can be accumulated for each tower if either of the two photons is centered in the tower. The position of the observed $\pi^0$ peak can be used to improve the tower energy calibration [9].

Full simulations were used to quantify the improvement of the relative calibration of the EMCal towers based on $\pi^0$ reconstruction. The same calibration constant was used for all towers. A 10% de-calibration was then applied to the tower calibration coefficients as expected from a pre-calibration based on cosmic muon data.

For the ideal case, the reconstructed two-photon invariant mass for the full Super Module shows a peak which lies at the mass of the $\pi^0$ with a width corresponding to the intrinsic resolution of the electromagnetic calorimeter. For the 10% de-calibration case, the width of the two-photon invariant mass peak increases by 50%. The calibration coefficients were then corrected tower by tower using the two-photon invariant mass spectra. For each tower $i$, the two-photon combinations were selected only if one of the clusters deposits at least 50% of its energy in the tower. The resulting two-photon invariant mass distribution was fitted and the extracted mean value $m_i$ was used to correct each tower calibration coefficient $cc_i$ using:

$$cc_i^{corr} = cc_i \cdot (1 + k_i^2)/2 \tag{2.2}$$

where $k_i = m_{\pi^0}/m_i$.

The procedure was repeated several times in order to obtain an invariant mass distribution centered at the mass of the pion for each tower. Figure 2.12 shows the de-calibrated $\pi^0$ peak on the left and the result after seven iterations on the right.

Figure 2.13 shows the calibration coefficients distribution for one Super Module for the ideal case (blue line), for 10% de-calibration (dashed line) and after calibration coefficients correction (full line).

In the latter distribution, the towers which lie at the edge of the Super Module were excluded. The corrected distribution is centered at the value of the ideal calibration with a 1% accuracy. The green and brown distributions show the corrected coefficient calibrations for the case of the edge towers: they are different from the inner towers due to shower leakage at the boundaries of the Super Module. The resulting final two-photon invariant mass for the full

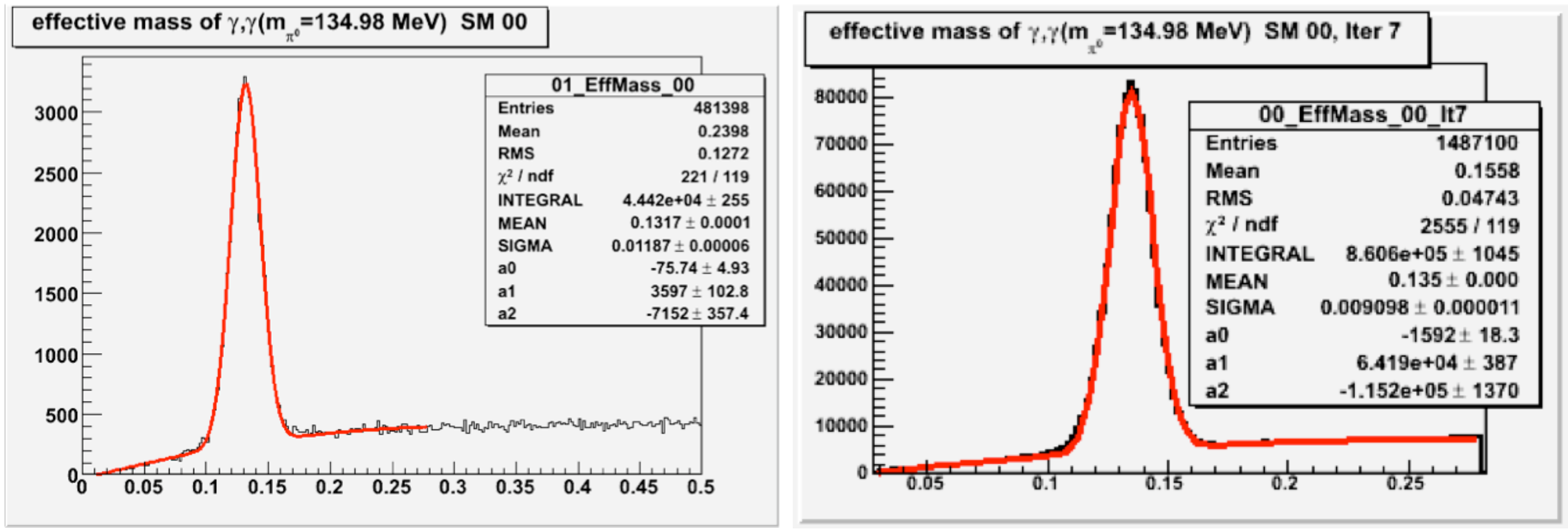


**Figure 2.12:** Left panel: reconstructed $\pi^0$ peak after 10% de-calibration (represents the level of accuracy based on cosmics calibration). Right panel: reconstructed $\pi^0$ peak after seven in-situ calibration iterations.

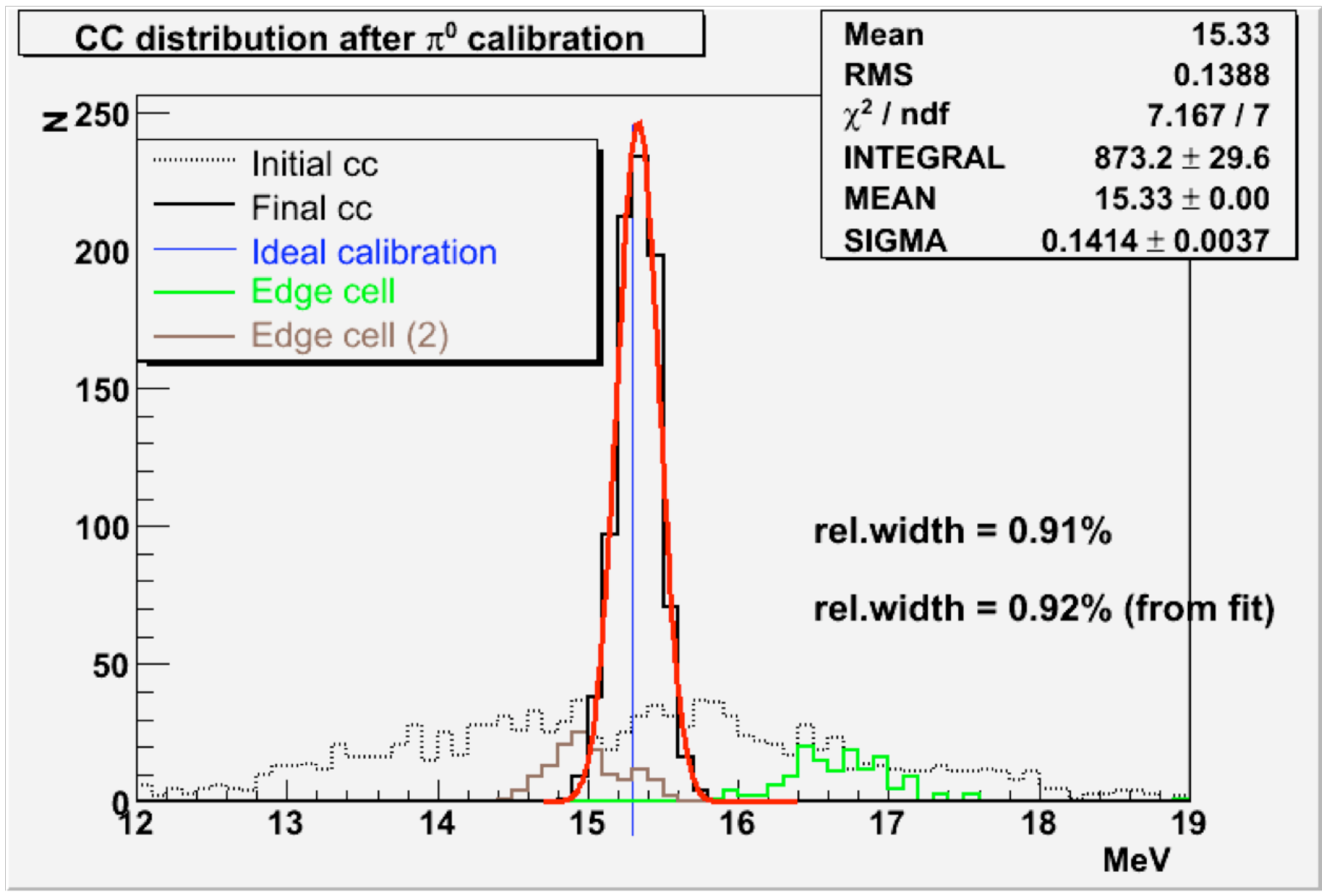


**Figure 2.13:** Distribution of tower relative calibration coefficients for one Super Module for an ideal calibration (blue line), for 10% de-calibration (dashed line), after corrections as explained in the text (full line). The latter distribution is fitted by a gaussian (red line). The green and brown lines shows the corrected calibration coefficients for the edge towers case.

Super Module after applying the corrected calibration coefficients was found to be centered at the pion mass with a width only 5% larger than that of the ideal case.

This study demonstrates that a cosmics calibration with 10% accuracy, can be improved to a calibration uncertainty in the range of 1% using the $\pi^0$ invariant mass in p–p running. The statistics needed to obtain such accuracy is approximatively 700 $\pi^0$ per tower. Photon pairs with the higher energy photon centered on the tower should be used.

# Chapter 3

# EMCal Trigger System

This section presents the EMCal triggering strategy and physics performance.

The architecture of the ALICE trigger system and the EMCal-specific trigger hardware is presented in Chapter 2.3. The goal of the EMCal Trigger system is to enhance the kinematic reach of recorded data for hard probes such as high-$p_T$ $\pi^0$, $\gamma$, electrons, and jets, within the overall trigger rate and bandwidth constraints of ALICE. These constraints, presented in Chapter 2.3.3, dictate the event rate reduction and the data rate reduction that must be achieved by the EMCal Trigger System. We assess its physics performance in terms of the reduction factors, trigger bias, and efficiencies.

The main focus of this chapter is on the jet trigger, which requires the most sophisticated trigger strategy because jets subtend a large phase space area. The high and fluctuating background in heavy ion collisions must be accounted for to obtain an efficient, unbiased jet trigger. The design of the fast (Level 1) jet trigger hardware allows for exploration and optimization of the jet area integration ("patch size") and background correction, and this chapter evaluates several possible scenarios.

We assess the physics performance of all triggers at rejection rates that we regard as realistic, in that they generate data rates that are a limited fraction of the overall ALICE bandwidth and are therefore candidates for a realistic run plan.

This chapter is organized as follows: discussion of trigger requirements for p–p and Pb–Pb collisions; description of the simulations framework for L0/L1 and HLT triggers; performance evaluation of selected triggers and strategies in p–p and Pb–Pb; and summary.

### 3.0.5 Trigger Requirements

The key requirements of the triggering system are shown schematically in Fig. 3.1.

The L1 trigger has different requirements in p–p and Pb–Pb collisions, due to the different

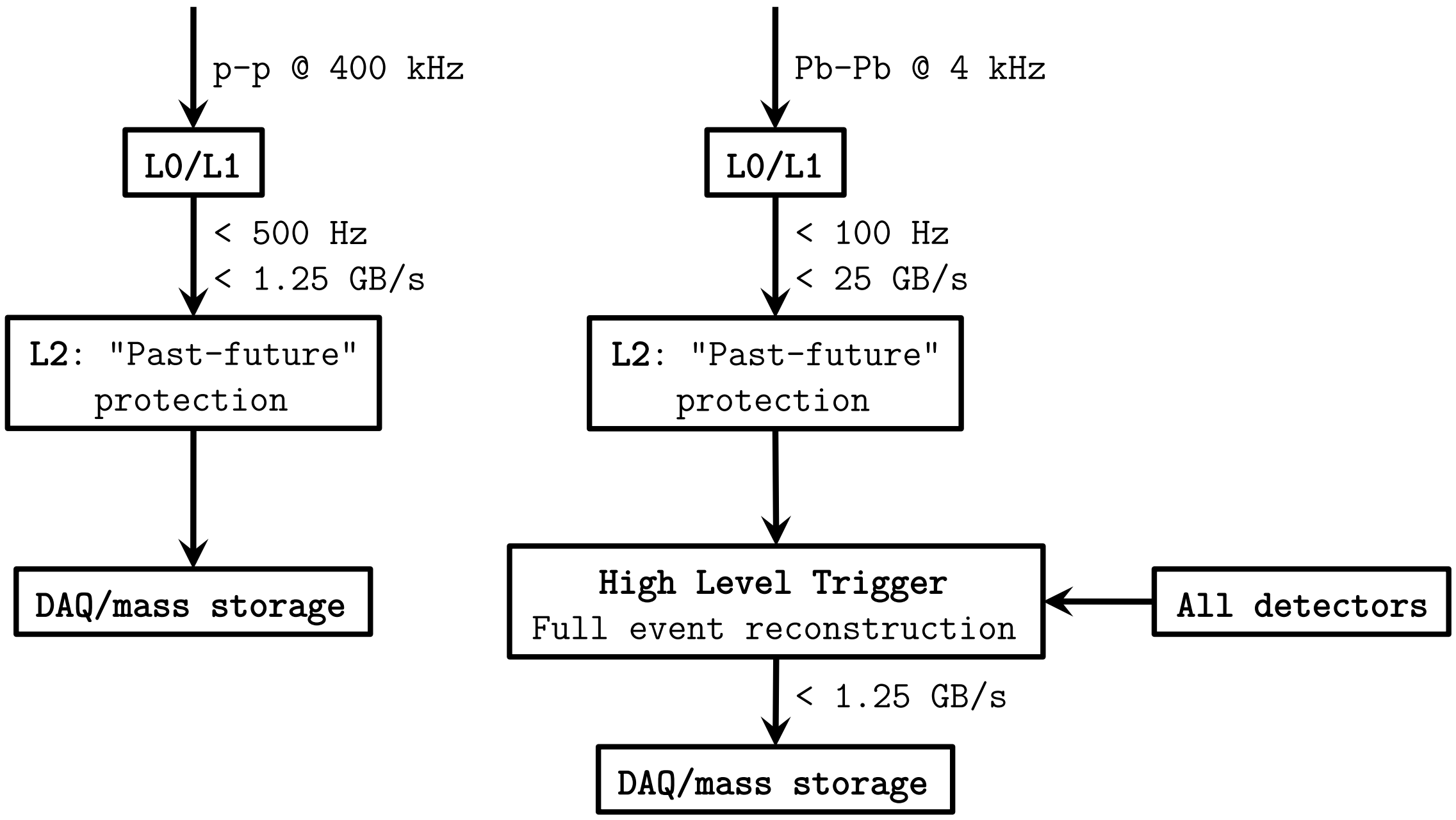


**Figure 3.1:** A schematic view of the trigger data flow with the essential information and requirements in p–p collisions (left) and Pb–Pb collisions (right).

interaction rates and event sizes:

- The Level 1 (L1) trigger accept rate must be less than the safe gating frequency of the TPC, which is 500 Hz in p–p collisions and 100 Hz in Pb–Pb collisions.
- p–p collisions: interaction rate is $\sim$ 400 kHz with event size of a few MB. Given the maximum TPC gating rate of 500 Hz, broadly speaking all p–p events for which the TPC has been gated can be written to tape without further rejection. The full rejection in p–p must therefore occur at Level 1, with rejection rates around 400 kHz/500 Hz $\sim$ 1000 or greater. The HLT does not provide significant rejection in p–p collisions.
- Pb–Pb collisions: interaction rate is $\sim$ 4 kHz, with average event size $\sim$ 20 MB for minimum bias and $\sim$ 80 MB for central events. L1 data rate reduction must be $\sim 5-10$ to satisfy the TPC gating ($\sim$ 100 Hz) and HLT input bandwidth limitations (<25 GB/s, see Sect. 2.3.2). The HLT input event rate must also be less than 1 kHz, which is however less constraining than the TPC maximum gating frequency. The overall trigger rejection must ensure a data rate to tape below the DAQ limit of 1.25 GB/s.

The jet trigger should generate a bias-free jet population above some energy threshold. Optimization of the trigger strategy is complex due to several factors, however. At L1 the jet trigger operates only on the EMCal response (neutral energy plus an admixture of charged particle energy) and a full energy measurement is possible only at HLT, where the charged

particle momenta become available as input to the trigger algorithms. Additionally, jet quenching in Pb–Pb collisions may broaden jets via large angle radiation, requiring larger integration area to measure a given fraction of jet energy, while larger intergration area includes more background and consequently larger fluctuations.

Optimization of these competing effects requires a flexible trigger system and experience with real data. In this section we investigate several trigger scenarios, and demonstrate the variation in physics performance with variation in trigger parameters.

### 3.0.6 Simulation Framework

The trigger response was simulated within AliROOT, the ALICE simulation and reconstruction software package. Events generated with PYTHIA (p–p) and HIJING (Pb–Pb) where transported through the detector using a detailed GEANT model, and reconstructed with AliROOT.

The background of heavy-ion events was simulated using HIJING (unquenched). The L0 and L1 trigger response were emulated using digitized (ADC) signals of the individual EMCal towers.

HLT algorithms have been run on the fully reconstructed tracks within the TPC acceptance $|\eta| < 0.9$ and clusters reconstructed within the EMCal.

#### 3.0.6.1 Cluster trigger

The EMCal trigger at Level 0 and Level 1 is described in Chapter 2.3 (see also [1]). The cluster trigger searches for high $p_T$ showers from $\gamma$ ($\pi^0$, $\eta$,...) and electrons.

The L0 algorithm identifies the shower energy above threshold in the local region of a Trigger Region Unit (TRU). The L0 provides a pre-trigger signal for the L1 trigger in p–p collisions, enabling measurement of (for instance) an unbiased inclusive $\pi^0$ spectrum.

The cluster trigger at L1 is evaluated $6.5\mu s$ after the interaction. The energy is summed over a sliding window of $4 \times 4$ towers and compared to a threshold above noise, as in L0, but the L1 decision is evaluated within the Summary Trigger Unit (STU), allowing a scan over the entire EMCal surface.

We simulate the cluster trigger decision through a sliding window corresponding to the cluster size of $4 \times 4$ towers. The algorithm utilizes the digitized EMCal tower signals.

#### 3.0.6.2 L1 jet trigger

Jets are extended objects. A jet trigger that is efficient and unbiased requires integration over a phase space region larger than that subtended by a single TRU. The L1 jet trigger therefore must be processed in the Summary Trigger Unit (STU, see Section 2.3.1), which has access to the entire EMCal energy distribution.

The jet trigger decision at L1 is evaluated using a "patch" trigger. A single patch is composed of a number of adjacent $n \times n$ towers. We simulate the trigger response using various patch sizes in $\Delta\eta \times \Delta\varphi$ over the full detector acceptance.

In heavy ion collisions there is a large and fluctuating underlying event. The L1 jet trigger algorithm must accommodate the event-wise variation in this background in order to provide an efficient and unbaised trigger for jets in both peripheral and central collisions. The total energy accumulated within the EMCal is correlated with the V0 multiplicity (see Section 2.3.1). The integrated V0 signal is available to the STU at L1, enabling the correction of each patch energy for the background $E_{\rm patch}^{\rm background}$. We find that the resolution of the energy deposited in the EMCal $E_{\rm EMCal}$ obtained via the correlation with V0 allows for accurate ($\sigma(E_{\rm EMCal})/E_{\rm EMCal}$ ~5–10%) estimation of the energy per unit area $\rho = E_{\rm EMCal}/A_{\rm EMCal}$, where $A_{\rm EMCal}$ is the area of the EMCal in radians ($\Delta\eta \times \Delta\varphi$ (rad)). We use the value of $\rho$ for the calculation of the background energy within a patch with a given area according to $E_{\rm patch}^{\rm background} = \rho \times A_{\rm patch}$.

The patch with maximum energy found within the fiducial acceptance of the EMCal, after adjustment for underlying background via the V0 signal, is selected for comparison with the threshold.

To explore the trigger response and identify the optimum trigger configuration, several patch sizes have been considered ($0.1 \times 0.1$, $0.2 \times 0.2$, $0.3 \times 0.3$ and $0.4 \times 0.4$). CDF measurements of charged jet profiles in p–p collisions show that for ~50–100 GeV jets about 80% of the jet energy is contained within $R < 0.15$ [2]. Maximizing the patch size will reduce the trigger bias; however, patch sizes larger than $0.3 \times 0.3$ are susceptible to larger background fluctuations in heavy-ion collisions and as such provide poor performance.

Implementation of this L1 algorithm in the FPGA of the STU shows that the timing of the algorithm is well within the timing constraints of 5 $\mu s$.

#### 3.0.6.3 Jet trigger in HLT

The events accepted by L1 are filtered by HLT [3] to provide additional rejection. HLT provides full event reconstruction at a level close to that achieved offline.

The HLT response has been evaluated using offline tracking algorithms. The agreement between the offline and online tracking has been established at the percent level in p–p

and central Pb–Pb via simulations. Excellent agreement between online and offline tracking has also been observed in real life conditions during the recent cosmic tests (Fall 2009). A cosmic muon passing close to the TPC center in the transverse plane may leave two long, symmetric tracks in the upper and lower sections of the detector. Figure 3.2 shows the relative transverse momentum resolution of two such track segments produced by cosmic rays traversing the TPC. The relative resolution is defined as $2 \times |\frac{1/p_T^{'} - 1/p_T^{''}}{1/p_T^{'} + 1/p_T^{''}}|$. Figure 3.2 compares the distribution of relative resolution of a population of such cosmic muons for HLT tracking and offline reconstruction; good agreement is seen.

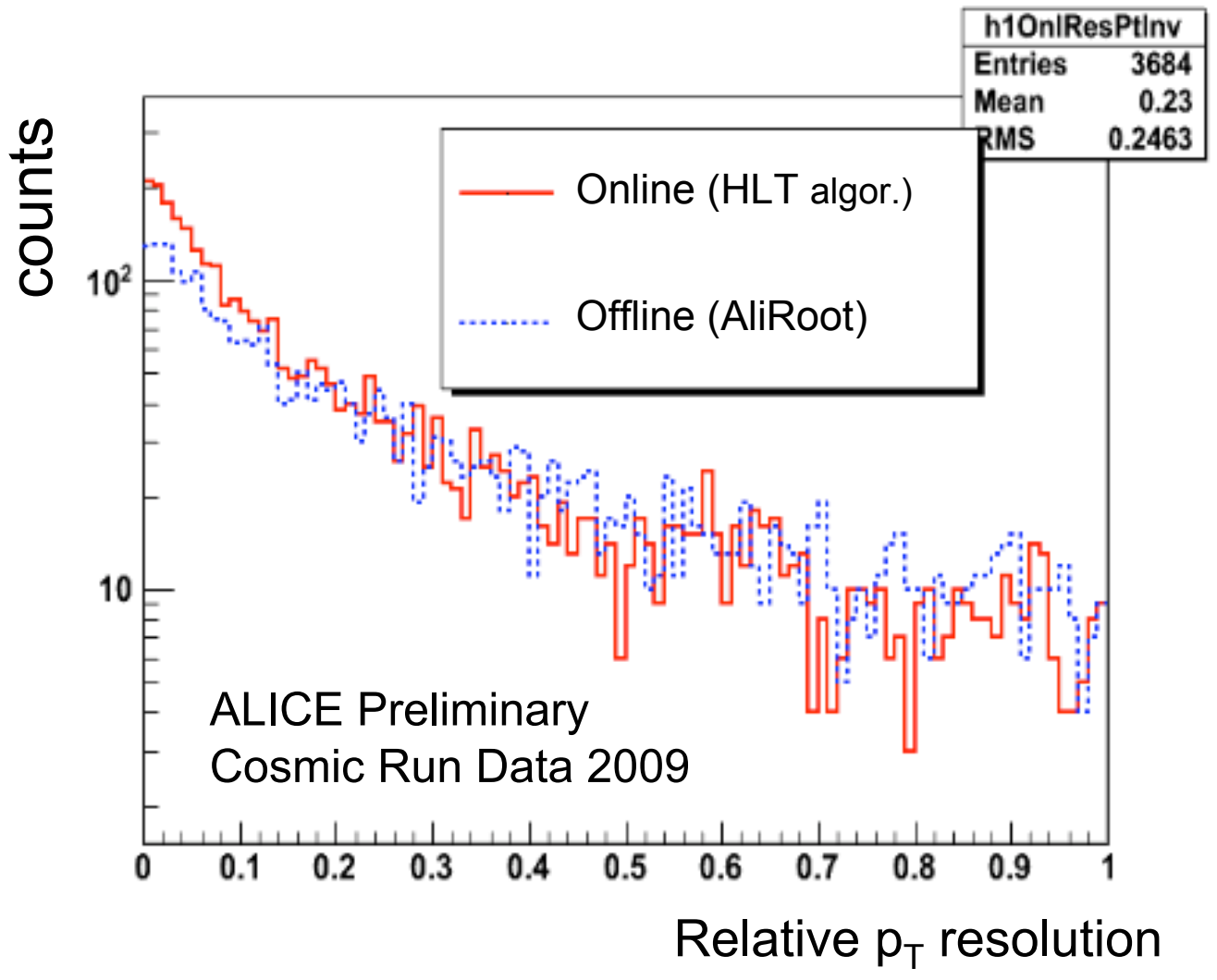


**Figure 3.2:** Distribution of relative $p_T$ resolution for two segments of the same cosmic muon track passing close to the TPC center, for both HLT and Offline reconstruction.

Full event reconstruction in the HLT enables the use of sophisticated jet finders in the trigger decision. As a first application we run the anti-$k_T$ jet finder (see Chapter 5), using charged particle tracks and reconstructed EMCal clusters. Jet $p_T$ is corrected for underlying event contributions according to the average background energy per unit area calculated event by event within the EMCal acceptance. The HLT Trigger decision is based on the maximum jet energy found by the anti-$k_T$ jet finder with $R = 0.4$ within the EMCal fiducial acceptance.

The algorithms within HLT must run at the frequency of the L1/L2 trigger (for L2 definition see Section 2.3). It is found that HLT timing requirements can in general be satisfied by parallelization of the HLT components.

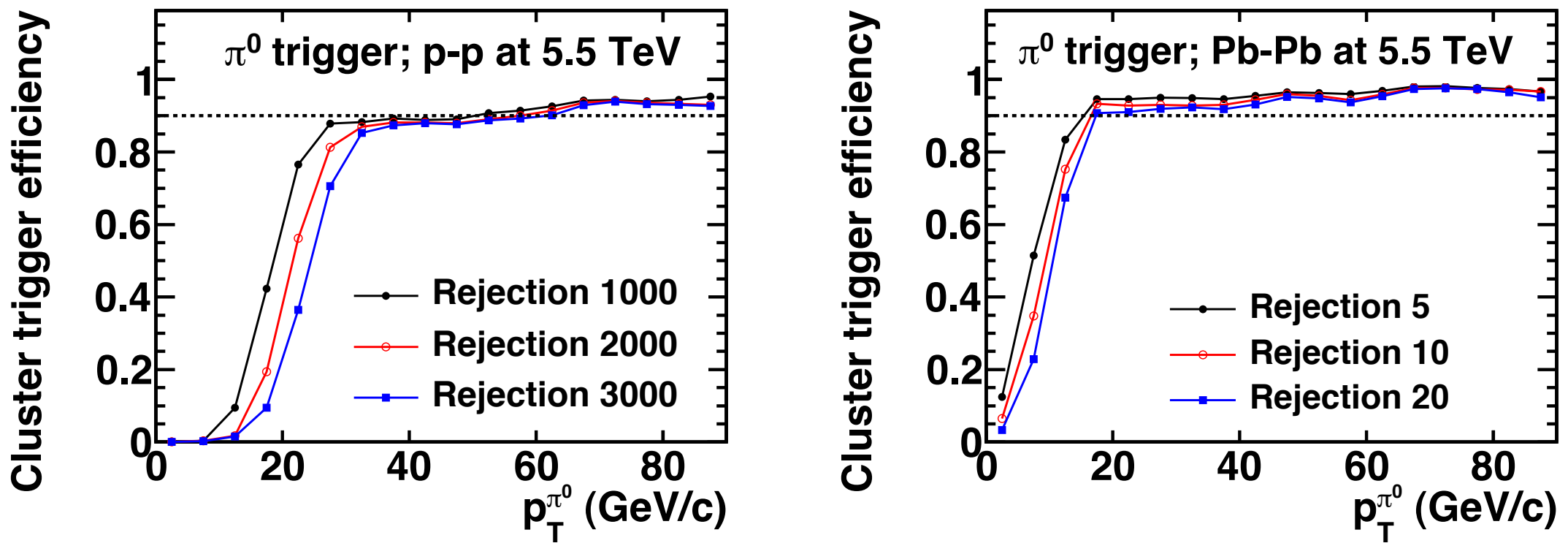


**Figure 3.3:** Cluster trigger efficiencies for $\pi^0$ in p–p and Pb–Pb collisions at 5.5 TeV for three rejection factors. Cluster size $0.04 \times 0.04$. The horizontal dashed line marks 90% efficiency.

# 3.1 Trigger performance

## 3.1.1 Cluster trigger in p–p

In practice the cluster trigger will be implemented as a hierarchy of three triggers with different scale-downs. Here we evaluate the performance of the cluster trigger with the largest rejection by considering $\pi^0$ mesons within the EMCal acceptance. Figure 3.3 shows the cluster trigger efficiencies exclusively for $\pi^0$. For the three different rejection rates the cluster trigger performs very well with efficiencies reaching a plateau of 90% at $p_T^{\pi^0} > 50$ GeV/$c$ in p–p collisions and 30 GeV/$c$ in Pb–Pb collisions. Similar triggering strategies are expected for high-$p_T$ electrons and photons.

## 3.1.2 Jet trigger in p–p

As outlined above, due to the ALICE bandwidth constraints the L1 EMCal trigger is required to achieve an event rate rejection greater than 1000 for p–p collisions, to limit the event frequency to the acceptable rate for TPC gating of 200–500 Hz. Figure 3.4 shows the expected L1 accept rate as a function of the L1 jet patch threshold. At thresholds of 600–800 (11–15 GeV) the data rate rejection reaches 1000–3000, depending on the patch size. For the jet patch size of $0.4 \times 0.4$ and threshold of 600 ($\sim$11 GeV) the L1 trigger accept rate goes down to $\sim$ 50 Hz and data rate reduction reaches 3000. Assuming the maximum gating frequency for the TPC of 500 Hz, the EMCal trigger with patch size of $0.4 \times 0.4$ and rejection of 1000 ($\sim$170 Hz event rate) would occupy 40% of the total ALICE data bandwidth written to tape; rejection of 3000 ($\sim$50 Hz event rate) corresponds to 10% of the recording bandwidth.

Figure 3.5 shows the inclusive jet spectrum in the EMCal acceptance (upper panels) and

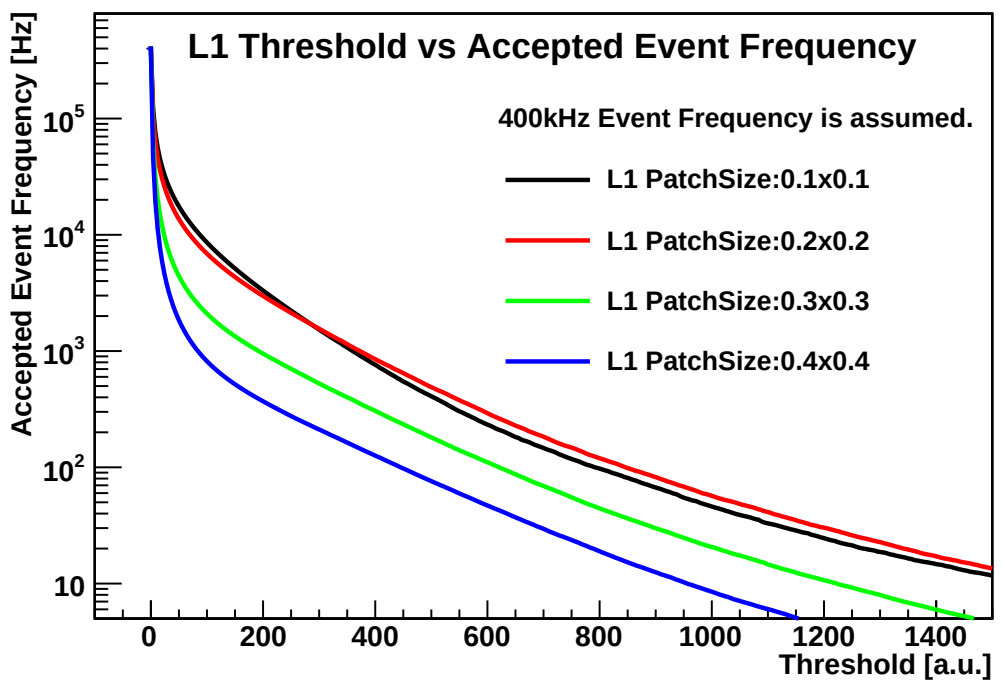


**Figure 3.4:** L1 accept frequency as a function of L1 threshold for min bias p–p at $\sqrt{s_{\rm NN}} = 5.5$ TeV.

corresponding trigger efficiencies (lower panels) for two selections of the patch size and two rejection factors of 1000 and 3000. The trigger response is calculated using a detailed GEANT model of the the ALICE detector ("Detector Level" simulation) and comparison to threshold.

For clarity, we report the trigger efficiency as a function of jet energy determined rather at the "Particle Level", i.e. the output of the event generator without considering detector response, using the anti-$k_T$ algorithm with $R = 0.4$ for jet reconstruction. This choice decouples the detector effects related to the jet reconstruction (jet energy scale, energy resolution, tracking efficiency, etc.) from the physical jet efficiency calculation which is relevant here. The trigger efficiency at a given Particle Level jet energy is defined as the ratio of jet yield with patch trigger rejection applied to the yield for minimum bias events.

The anti-$k_T$ jets used for the efficiency calculation are fully contained within the EMCal acceptance such that the jet centroid is at least a distance $R$ from the edges of the detector. The largest jet patch $0.4\times0.4$ shows good (>90%) jet efficiency at $p_T^{\rm jet} > 30$ GeV for rejection 1000. For rejection 3000 the trigger is >90% efficient at $p_T^{\rm jet} > 60$GeV . As expected, the largest patch introduces the least bias on the recorded jet population.

### 3.1.3 Jet trigger in Pb–Pb

In Pb–Pb collisions, the total event size is strongly correlated to the track multiplicity in the TPC, therefore the charged particle multiplicity is used as an estimator of the data volume for calculating the rejection factors. Figure 3.6, left panel, shows the data volume rejection at L1 for minimum bias HIJING events of Pb–Pb collisions at $\sqrt{s_{\rm NN}} = 5.5$ TeV as a function of the threshold applied, for four L1 trigger patch sizes. Background rejection of 10 means a ten-fold reduction in data volume at L1, i.e. 8 GB/s for minimum bias Pb–Pb events, which is $\sim 30\%$ of the total HLT input bandwidth.

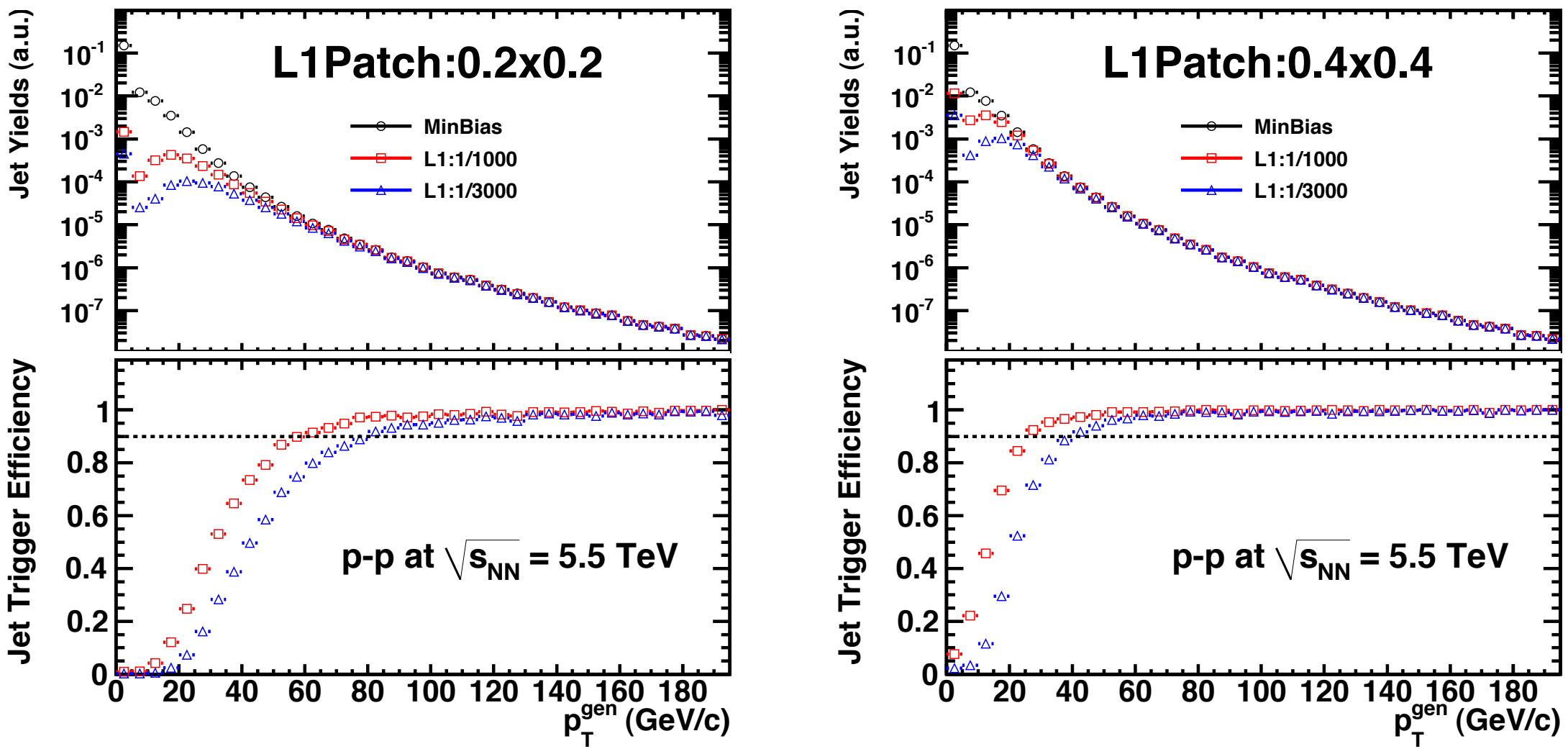


**Figure 3.5:** L1 jet trigger performance: Jet yields (*upper panel*) and trigger efficiencies (*lower panel*) in p–p collisions at $\sqrt{s_{\rm NN}} = 5.5$ TeV with L1 rejection rates of 1000 and 3000. For two different L1 patch sizes: $0.2 \times 0.2$ and $0.4 \times 0.4$. The horizontal dashed line marks 90% efficiency.

As discussed in Section 2.3.3 the input event rate at L1 must be limited to allow for a safe gating frequency of the TPC. Figure 3.6, right panel, shows the frequency of the L1 trigger as a function of the data volume rejection factor for the four different jet patch trigger selections. A rejection of 5 at L1 brings the L1 frequency to 200 Hz and 100 Hz at a rejection of 10.

As discussed in sections 2.3 and 3.0.6.2, event-wise fluctuations correlated with the centrality of the heavy ion collisions are corrected on average using the total charged multiplicity measured in the forward V0 detectors. Fig. 2.4 shows the correlation between V0 multiplicity and total energy in the EMCal. The mean of this distribution, indicated by the line in the figure, is used in the trigger simulation to adjust the L1 threshold for event centrality.

To satisfy the DAQ data recording bandwidth limit of 1.25 GB/s, the HLT must further reduce the data rate. In the following sections we demonstrate the trigger capabilities with the HLT data reduction of 40, which brings the data volume rate to 400 MB/s for L1 rejection of 5 (32% of total DAQ bandwidth) and 200 MB/s for L1 rejection of 10 (16% of total DAQ bandwidth).

The jet trigger efficiency is calculated as described above for p–p collisions. The inclusive spectrum of (particle-level) jets is reconstructed with anti-$k_T$ for $R = 0.4$ within the EMCal acceptance, and the efficiency is defined as the ratio of yields with and without the trigger rejection applied. To model jet production in Pb–Pb events, the spectrum of jets generated by PYTHIA (p–p at 5.5 TeV) is merged with a sample of minimum bias HIJING events. Only one Pythia event was embedded into a given HIJING event.

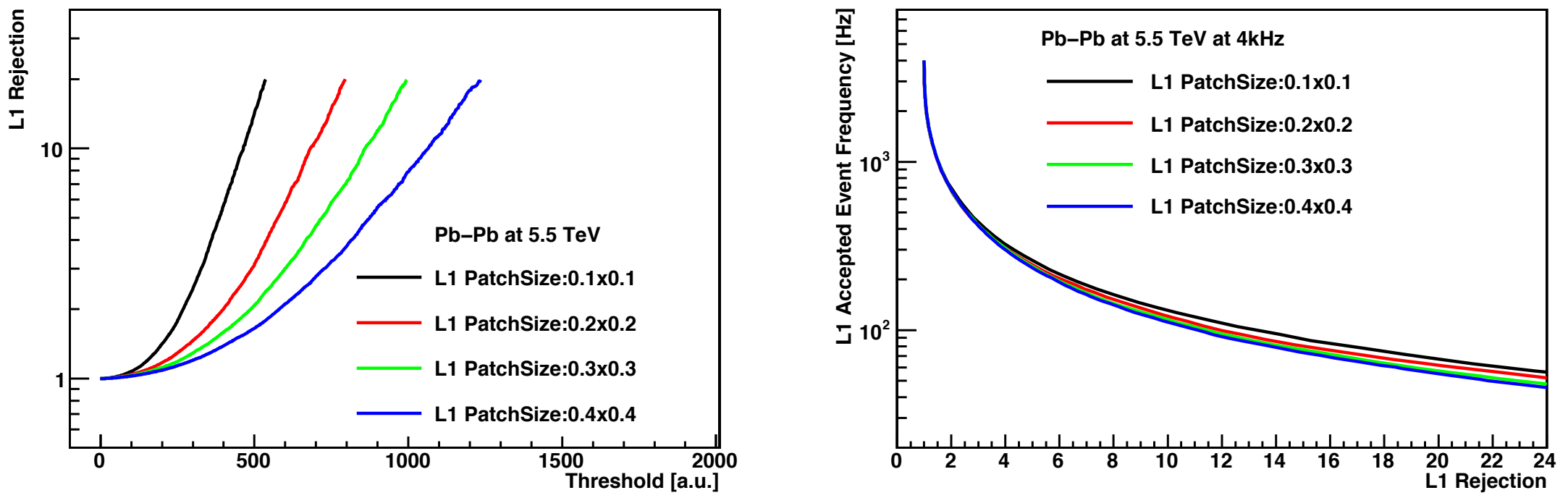


**Figure 3.6:** Data volume rejection as a function of L1 threshold (left) and L1 accept frequency as a function of L1 data volume rejection (right) for minimum bias Pb–Pb collisions at $\sqrt{s_{\mathrm{NN}}} = 5.5$ TeV.

The upper panels in Figure 3.7 show the jet spectrum for three event selections: minimum bias, L1 accepted, and L1+HLT accepted. The different panels are for two different patch sizes but with the same rejection of 5 at L1 and 40 at HLT, corresponding to a TPC gating frequency of 200 Hz. The lower panels show jet trigger efficiencies. The trigger efficiency at L1 reaches 90% at 80 GeV for the smallest patch and 60 GeV for the $0.3 \times 0.3$. After HLT rejection, all trigger selections show a similarly sharp turn-on, with very low efficiency for jets below 30 GeVand greater than 90% efficiency above 80 GeV.

Figure 3.8 shows similar curves as Fig. 3.7 but with data-volume rejection at L1 changed from 5 to 10, corresponding to a TPC gating frequency of 100 Hz. For this rejection at L1 the trigger curves have slower turn on, however. For the smallest patch the trigger efficiency reaches 90% at 90 GeV, whereas for the largest patch considered the 90% efficiency is reached at 70 GeV. For a combined trigger of L1 and HLT the efficiency is negligible below 40 GeV and is greater than 90% above 80 GeV.

### Pb–Pb : Jet trigger in central and peripheral events

Figure 3.9 shows the trigger efficiency for central (0–20%) and peripheral (60–80%) Pb–Pb collisions, for a data rate reduction of 10 at L1 and 40 at HLT. We observe a larger L1 trigger efficiency in central events compared to peripheral events (left panel), which results from the specific implementation in this calculation of the correction for underlying background based on the V0 signal. However, as seen in the right panel the difference is negligible after HLT filtering. There is also a minor difference in the efficiency curves for patches $0.1 \times 0.1$ and $0.3 \times 0.3$, indicating that the gain in jet energy resolution for the latter relative to the former is offset by the increased contribution of background fluctuations. The performance of the trigger with $0.3 \times 0.3$ patch size is satisfactory.

Jet broadening due to quenching may modify these conclusions significantly. Progress has been made towards Monte Carlo implementations of quenching (see Section 4.5), though an accurate assessment of quenching effects on trigger performance is not yet achievable.

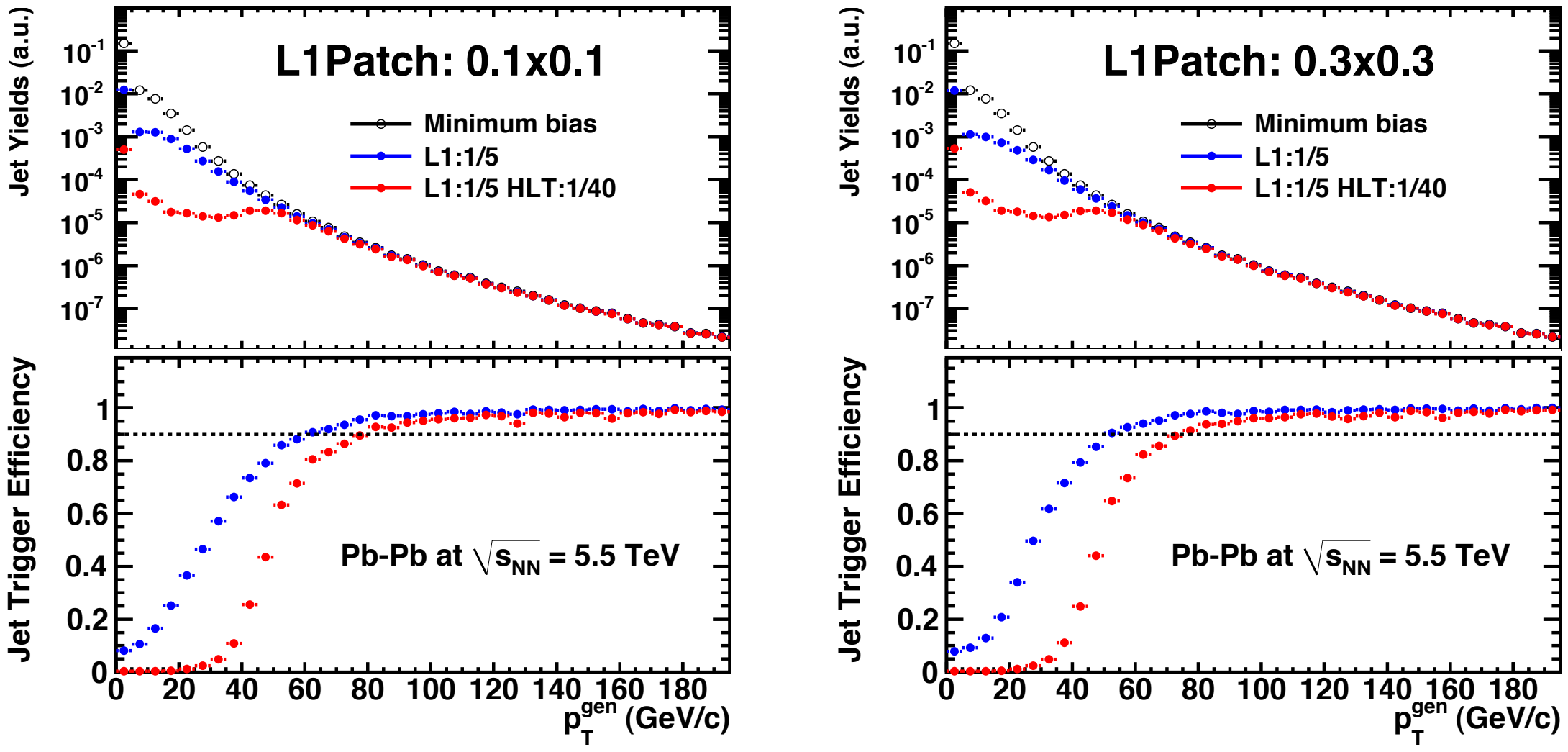


**Figure 3.7:** Jet trigger efficiency in minimum bias Pb–Pb at $\sqrt{s}$= 5.5 TeV, with total rejection of 200. Upper panels are Jet yields, lower panels are trigger efficiencies. L1 and L1+HLT are shown with rejection rate of 5 at L1 and 40 at HLT, for two different patch sizes at L1: $0.1 \times 0.1$ (left) and $0.3 \times 0.3$ (right). See text for details.

This work is in progress. In any case, the ALICE+EMCal trigger architecture has sufficient flexibility to optimize the trigger efficiencies once these issues are better understood.

### Measurement of trigger bias

The trigger bias can be estimated using simulations, but accurate measurement of the bias requires sufficient minimum bias event statistics for normalization in the region where the trigger is expected to be highly efficient (i.e. kinematic reach). We estimate the minimum bias data set required for better than 5% statistical precision for normalization in the plateau region. For each system we select the trigger settings where the performance at the threshold is optimum, and then require sufficient minimum bias statistics such that there are 1000 jets above the threshold.

For the 5.5 TeV p–p case the jet patch trigger of $0.4 \times 0.4$ reaches 90% efficiency for $p_T^{\rm jet}$ $> 50$ GeV. ALICE must record $\sim 25$ M minimum bias events to measure 1000 jets above 50 GeV. One p–p running year in ALICE is $\sim 6 \times 10^6$ seconds, meaning that sufficient minimum bias statistics to normalize the jet trigger can be accumulated by running minimum bias at $\sim 25 \times 10^6 \sim 4$ Hz. This is a modest fraction of the ALICE DAQ bandwidth, and is achievable in practice. There is no dependence on luminosity in this estimate.

We perform a similar calculation for Pb–Pb. We select the trigger patch at L1 of $0.3 \times 0.3$ which yields a reduction of 10. For this selection the efficiency of 90% is reached at $p_T^{\rm jet}$ $\sim 90$ GeV. About $2 \times 10^6$ minimum bias events are required to record 1000 jets above this threshold. Given a Pb–Pb running year of $10^6$ seconds, these statistics will be accumulated

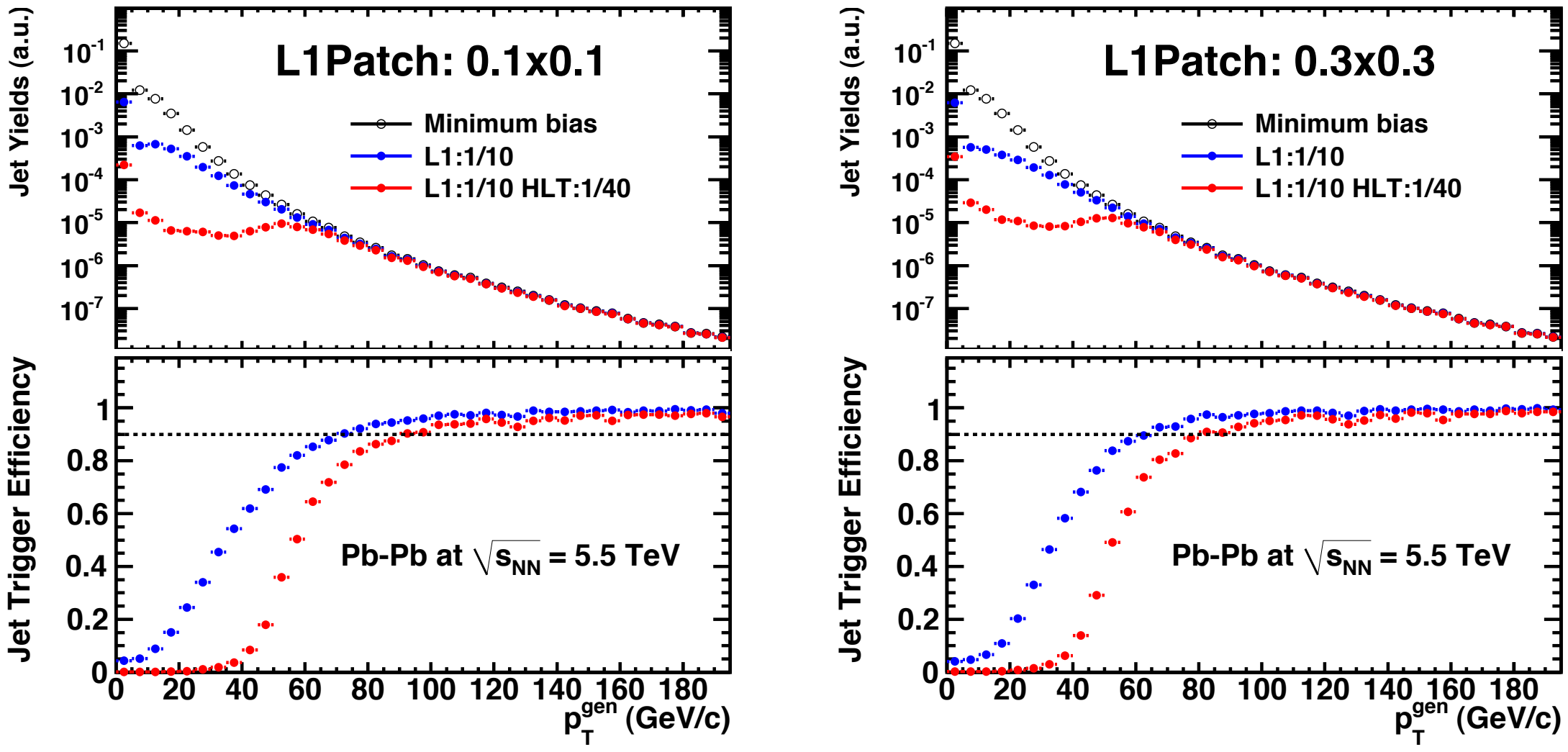


**Figure 3.8:** Jet trigger efficiency in minimum bias Pb–Pb at $\sqrt{s}$= 5.5 TeV, with total rejection of 400. Upper panels are Jet yields, lower panels are trigger efficiencies. L1 and L1+HLT are shown with rejection rate of 5 at L1 and 40 at HLT, for two different patch sizes at L1: $0.1 \times 0.1$ (left) and $0.3 \times 0.3$ (right). See text for details.

with a minimum bias trigger running at 2 Hz. This is likewise a modest and achievable goal.

### 3.1.4 Enhancement factors

Table 3.1 shows the expected enhancement in recorded jet statistics due to the EMCal jet trigger, compared to ALICE capabilities without the EMCal. The assumed luminosity and effective running time are shown. The comparison is between the rate of jets triggered by and reconstructed with ALICE+EMCal, and jets recorded using geometrical triggers only (minimum bias Pb–Pb or p–p, central Pb–Pb). The enhancement calculation takes into account the EMCal jet trigger acceptance relative to charged particle jets (17.3%) and trigger efficiency and live-time (combined factor 80%), together with event rate limitations due to DAQ bandwidth and maximum TPC gating frequency. The enhancement factors in Table 3.1 are applicable for $E_{jet} > 100$ GeV, where the trigger is close to 100% efficient (see Figures 3.7 and 3.8). The enhancement factor is seen to be greatest for the smallest collision systems (p–p, peripheral Pb–Pb). This is because the smallest systems have the largest interaction rate, while the ALICE DAQ rate is limited in all cases to 500 Hz.

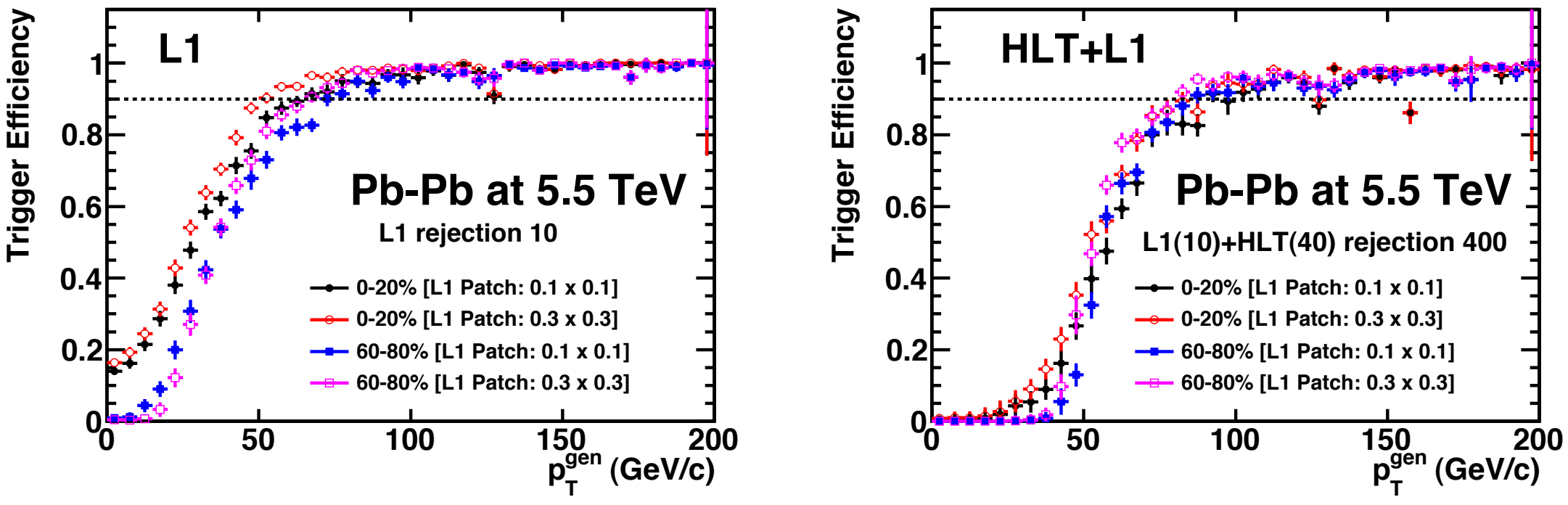


**Figure 3.9:** Jet trigger efficiencies as in Fig. 3.7 and 3.8 for central and peripheral Pb–Pb collisions at 5.5 TeV. L1 (left) and L1+HLT (right) with rejection rates of 10 at L1 and 40 at HLT for two different patch sizes at L1: $0.1 \times 0.1$ and $0.3 \times 0.3$. HLT runs anti-$k_T$ with R=0.4. The horizontal dashed line marks 90% efficiency.

## 3.2 Summary

This chapter has presented the triggering strategies for the EMCal and an assessment of the trigger performance. Calculations are based on a complete simulation of detector response. While the conclusions have some model dependence on specific physics event generators, the triggering strategies appear to be robust and these calculations provide initial guidelines for real data-taking.

This chapter has established the following:

- The L0 cluster trigger provides an efficient pre-trigger for $\pi^0, \gamma$ and high-$p_T$ electrons.
- The V0 multiplicity, used to estimate the transverse energy $E_T$ within the EMCal, provides sufficient precision for a centrality-dependent jet trigger threshold correction in Pb–Pb collisions.
- The L1 jet trigger investigated with several patch size configurations provides the required rejection rates and satisfactory efficiencies for jets with $p_T > 100$ GeV/c in both p–p and Pb–Pb collisions. A final evaluation of the patch sizes and geometries will be possible only with real ALICE data. The patch size is driven by the requirement of unbiased triggering in light of potentially large quenching effects, but limited from above by increasing background for larger patch size. The proposed L1 hardware (STU) provides the necessary flexibility to implement the proposed patch geometries.
- The application of the HLT with filtering algorithms based on fully reconstructed jets in heavy-ion collisions allows for additional data rate reduction (improves rejection of the low energy jets and trigger on-set as compared to L1) which satisfies the DAQ limitations and does not impose additional trigger bias above $p_T^{\text{jet}} > 100$ GeV.

**Table 3.1:** EMCal jet trigger enhancement factors: Gain in recorded jet statistics for various systems due to the EMCal Jet Trigger, together with assumed mean luminosity, annual running time and ALICE DAQ rate.

| System | $\sqrt{s}$(TeV) | $L_{mean}(cm^{-2}s^{-1})$ | Time (s) | DAQ rate (Hz) | Gain at L1 |
|---|---|---|---|---|---|
| p–p | 5.5 | $5{\times}10^{30}$ | $10^5$ | 500 | 110 |
| p–p | 14 | $5{\times}10^{30}$ | $10^7$ | 500 | 500 |
| Pb–Pb Centrality | | | | | |
| min. bias | 5.5 | $5{\times}10^{26}$ | $10^6$ | 20 | 21 |
| min. bias | 5.5 | $5{\times}10^{26}$ | $10^6$ | 50 | 9 |
| min. bias | 5.5 | $5{\times}10^{26}$ | $10^6$ | 100 | 4 |
| 0–10% | 5.5 | $5{\times}10^{26}$ | $10^6$ | 50 | 5 |
| 0–10% | 5.5 | $5{\times}10^{26}$ | $10^6$ | 100 | 2 |
| 60–80% | 5.5 | $5{\times}10^{26}$ | $10^6$ | 50 | 12 |
| 60–80% | 5.5 | $5{\times}10^{26}$ | $10^6$ | 100 | 6 |

- Based on the simulation results presented in this chapter we find that in order to correct for the trigger bias and recover the full inclusive jet cross sections from the triggered samples with a 5% systematic uncertainty, ALICE will need to record statistics of 25 M minimum bias p–p events and 2 M minimum bias Pb–Pb events. This can be easily accommodated within the ALICE DAQ bandwidth by downscaled minimum bias triggers.

# Chapter 4

# Jet Measurements in Heavy Ion Collisions

## 4.1 Introduction

The interaction of high energy partons with a colored medium (*jet quenching*) provides a unique set of penetrating probes of the structure and dynamics of the Quark Gluon Plasma [1, 2]. In QCD, as in QED, energy loss will be dominated in the high energy limit by bremsstrahlung (medium-induced soft gluon radiation). In this limit, jet quenching manifests itself as the *medium-induced modification of jet fragmentation*; in other words, modification of the internal structure of a jet.

A number of experimental observables have been proposed to measure jet structure in heavy ion collisions. Striking effects of jet modification have been observed in high energy nuclear collisions at RHIC [3]. Jet production will be an even more dominant feature of nuclear collisions at the LHC due to the large increase in $\sqrt{s}$, and jet quenching is expected to play a central role in the study of QCD matter at the LHC.

The measurement of jets in heavy ion collisions is a major area of research focus, with substantial recent progress both in theory (algorithms to reconstruct jets accurately in the presence of large background [4]; implementation of jet quenching in theoretically well-motivated Monte Carlo event generators [5, 6, 7, 8]) and in experiment (first application of full jet reconstruction to RHIC heavy ion collisions [9, 10, 11]). Understanding of the manifestations of jet quenching, and how to measure it in experimentally robust ways, is evolving rapidly.

This chapter provides an overview of jet quenching theory and experiment, giving a context for the ALICE-specific jet measurement studies presented in Chapter 5. The emphasis here is on full jet reconstruction and on the observables utilizing fully reconstructed jets that have been studied in most detail at RHIC, in particular the inclusive jet production cross section.

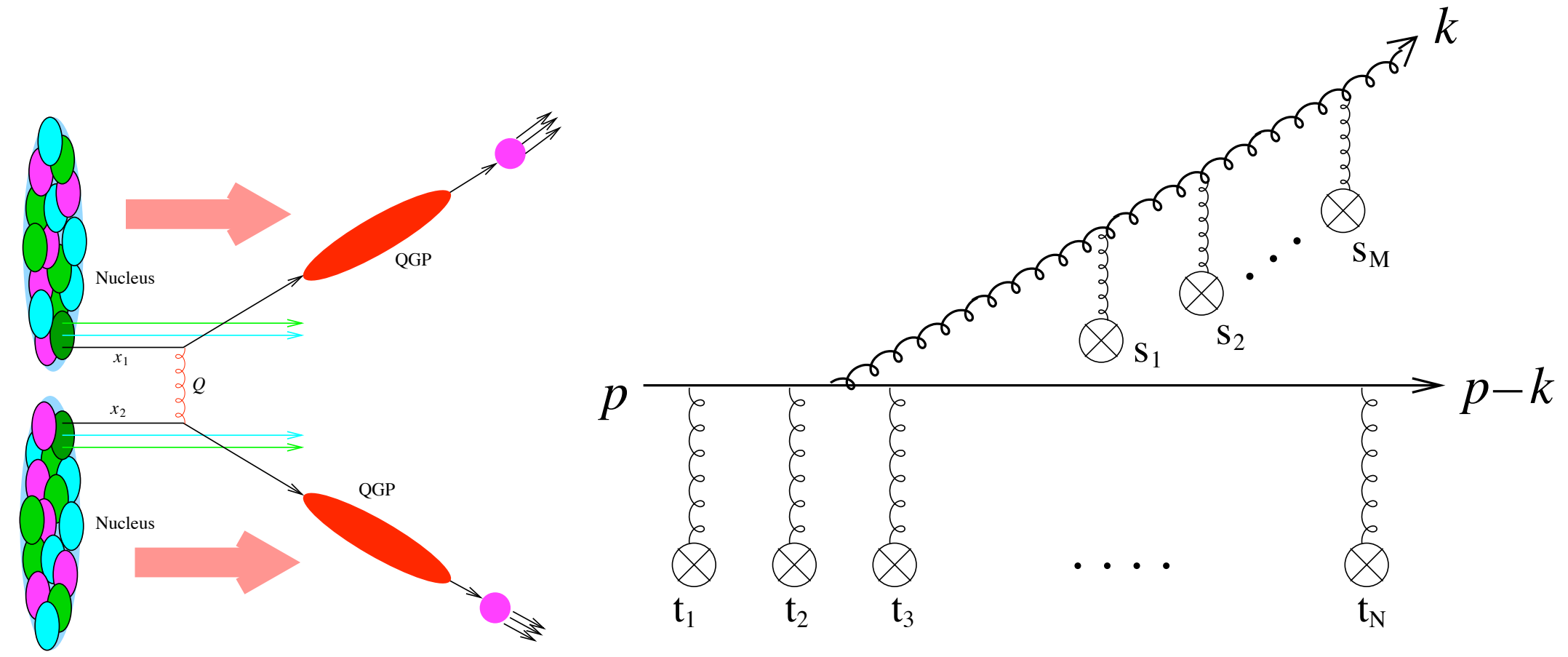


**Figure 4.1:** Theoretical picture of jet quenching in heavy ion collisions. Left: factorization of medium effects. Right: QCD bremsstrahlung in matter. Figures from [21].

We also present predictions for jet quenching at the LHC from two different theoretical approaches: analytical calculations based on the GLV formalism [12, 13], and qPYTHIA [5], one of the new Monte Carlo generators that incorporate quenching into the PYTHIA or HERWIG frameworks.

## 4.2 Jet quenching overview

Jet measurements in heavy ion collisions are challenging, due to the large and complex background whose magnitude and fluctuations must be accurately accounted for. Initial jet quenching studies in heavy ion collisions at RHIC minimized such background effects by focusing on measurements of high $p_T$ hadrons, to take advantage of the substantially harder $p_T$ distribution of jet fragments relative to the soft underlying background. A dramatic suppression (factor $\sim 5$) has been observed in nuclear collisions at high $p_T$ of both inclusive hadron yields [14, 15] and the rate of azimuthally back-to-back hadron pairs [16, 17]. A related semi-inclusive measurement, in which there is no constraint on the momentum of the recoiling hadron, shows qualitatively a softening and azimuthal broadening of the recoiling jet [18, 19]. Taken together, these observations confirm the overall picture of a strong disruption of jet structure relative to vacuum fragmentation, due to the propagation of the color charged parton in a dense, colored medium.

### 4.2.1 Jet quenching: theory

Fig. 4.1, left panel, illustrates the factorized approach underlying most theoretical calculations of jet quenching. Jet production arises from a momentum transfer $Q^2$ that is large, in

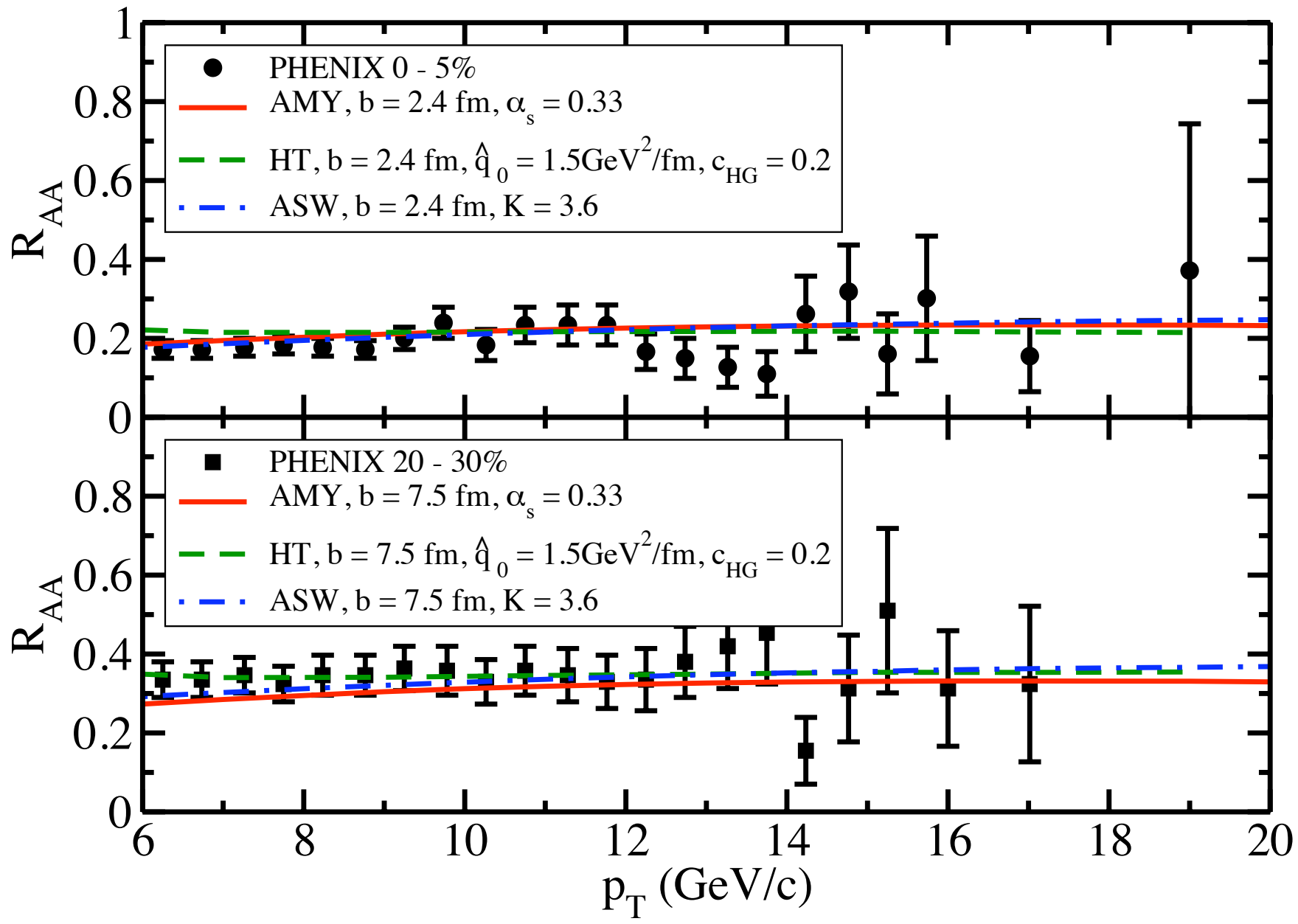


**Figure 4.2:** High $p_T$ pion suppression measured in 200 GeV Au–Au collisions [23] compared to several jet quenching calculations embedded in the same hydrodynamic model [20]. Model parameters are fit to central collision data (upper panel); pion suppression for non-central collisions are then predicted (lower panel).

the sense that its time scale ($\sim 1/Q \sim 0.2\ fm/c$ for $Q \sim 10$ GeV) is short compared to the formation time of the hot QCD medium, so that the two processes effectively decouple. The influence of the QGP on jets is therefore a final-state effect that modifies jet fragmentation, indicated by the red blobs on the outgoing partons. The dominant energy loss mechanism in the high jet energy limit is QCD bremsstrahlung, as illustrated in Fig. 4.1, right panel. This differs from QED bremsstrahlung due to the non-Abelian nature of QCD, indicated in the figure by the additional interactions of the radiated gluon with the colored medium.

There are various approximations for calculating QCD bremsstrahlung, all based on the assumption, implicit in the figure, that hard partons couple weakly (in the pQCD sense) with the medium so that perturbative tools can be applied (for recent reviews see [1, 2, 21, 22]). All calculations take into account coherence effects between multiple scattering centers $t_1, t_2, \ldots$ or $s_1, s_2, \ldots$ (LPM effect). The first theoretical approaches treated the scattering centers as randomly distributed, static (massive), and thermally screened, with no correlation between successive radiations (Poisson distribution of medium-induced gluons). More recent analytical calculations relax these assumptions, and include dynamic scattering centers with resummed interactions.

Heavy ion collisions are highly dynamic, with the jets propagating through a medium that is itself rapidly evolving and expanding. Comparison of jet quenching calculations to experimental observables therefore requires the coupling of energy loss calculations to a detailed model of the nuclear collision. Fig. 4.2 shows a set of such calculations, utilizing ideal relativistic hydrodynamics, for one of the primary jet quenching signatures, the suppression of

high $p_T$ pions in 200 GeV Au–Au collisions. The measured data [23] are compared to several jet quenching calculations coupled to the same hydrodynamic model of the fireball evolution [20]. Three different jet quenching formalisms, based on very different assumptions (Eikonal approximation - BDMPS/ASW [24]; higher twist formalism (HT) [25]; finite temperature field theory - AMY [26]), are compared.

All three formalisms predict correctly the $p_T$-dependence of $R_{\mathrm{AA}}$ (upper panel, the ratio of inclusive pion yield per central Au–Au collision to the binary collision-scaled yield measured in NSD p–p collisions), as well as the suppression for non-central collisions (lower panel) based on a model parameter fit to the central collision data. However, when the fit to data is expressed in terms of a transport coefficient of the hot QCD medium, $\hat{q}$ ($= \mu^2/\Lambda$, where $\mu$ is the RMS momentum transfer per interaction and $\Lambda$ is the gluon mean free path), the extracted values of $\hat{q}$ vary by a factor five or more between the models [20].

This large discrepancy has its origin in the collinear approximation for induced radiation made by all formalisms, which is violated in practical calculations, leading to large theoretical "systematic uncertainties". A major recent advance in the theory sector is the incorporation of quenching effects into the well-established Monte Carlo generators, PYTHIA and HERWIG [21, 5, 6, 7, 8]. This approach enables the modeling of energy-momentum conservation at every vertex, overcoming the essential limitation of the analytical approaches, but at the cost of introducing interference effects in a probabilistic way.

## 4.3 Jet reconstruction algorithms

While a great deal has been learned about jet quenching from the study of high $p_T$ hadrons and correlations, such an approach is limited by its intrinsic bias: the observed hadrons are the leading fragments of those jets that interacted *least* in the medium. A more in-depth exploration of jet quenching therefore requires full reconstruction of the jet signal, including the soft radiation whose hadronic fragments cannot be distinguished from the underlying background by means of kinematic cuts. Such an approach requires a full characterization of the underlying background, including fluctuations, and its influence on jet measurements.

Conceptually, a QCD jet is the hadronic final state of a parton shower generated by a highly virtual quark or gluon. Experimentally, a jet manifests itself in the detector as a correlated spray of hadrons (Fig. 4.3). Theoretically, the shower can be calculated perturbatively to NLO or NNLO, while calculation of experimental observables requires the modeling of further shower development and hadronization (e.g. PYTHIA and HERWIG).

A jet reconstruction algorithm specifies a clustering procedure that can be applied in a consistent fashion to both (N)NLO perturbative calculations and to experimental measurements, allowing robust comparison between them without a strong dependence on modeling or detailed experimental cuts. This requires the algorithm to be both infrared and collinear safe,

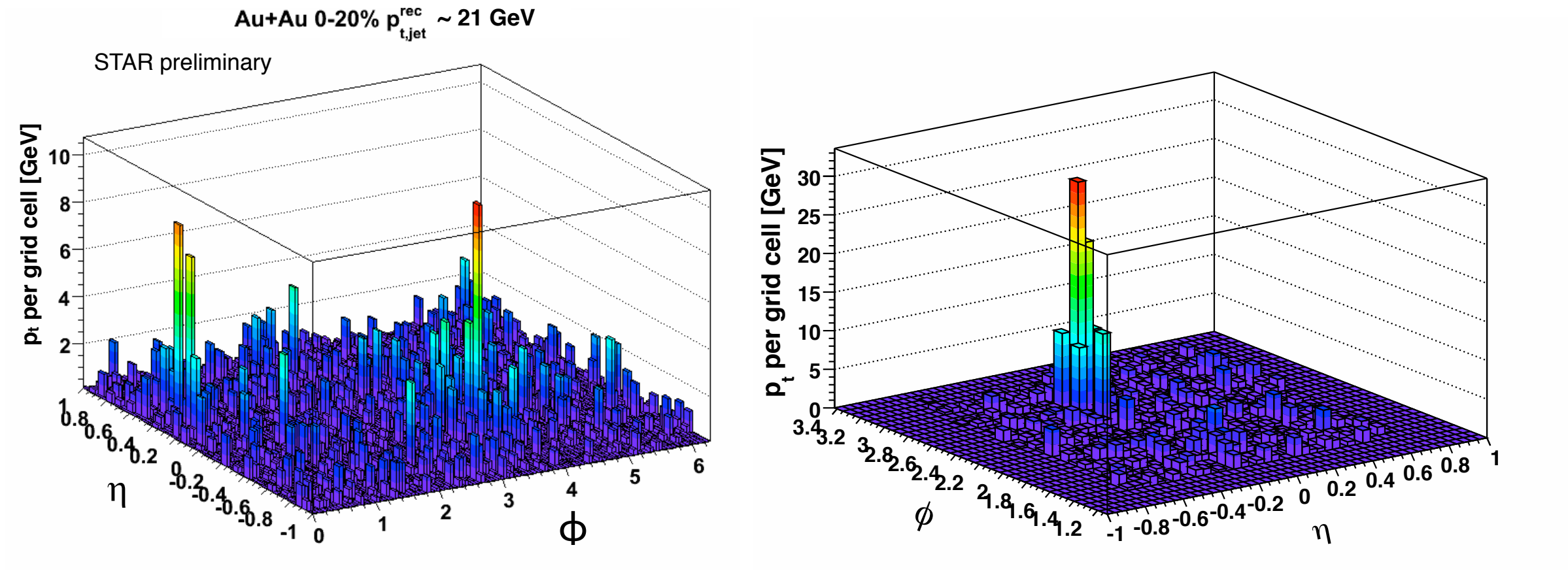


**Figure 4.3:** Single-event displays of jets in heavy ion collisions. Left: measured di-jet ($E_T \sim 21$ GeV) in a central Au–Au collisions at $\sqrt{s_{\rm NN}}$=200 GeV at STAR. Right: simulated jet (PYTHIA, $E_T \sim 120$ GeV) in ALICE(+EMCal), in a central Pb–Pb collision at $\sqrt{s_{\rm NN}}$=5.5 TeV.

with an implementation for data analysis that is unbiased (seedless) and does not require inordinate CPU time [27].

There are two broad classes of jet reconstruction algorithms that meet these criteria: modified cone, and sequential recombination. Fig. 4.4 shows a measurement at RHIC of the inclusive jet production cross section in 200 GeV p–p collisions using a mid-point cone algorithm, compared to an NLO calculation (left), and to similar measurements using sequential recombination algorithms (right). Agreement is found within experimental uncertainties in all cases. Differences between the algorithms may arise when there is large underlying event background, discussed further below.

There has been significant recent progress in jet reconstruction algorithms, most notably as encoded in the FastJet package [4]. FastJet includes all modern jet reconstruction algorithms, such as seedless cones and various sequential recombination approaches. From an experimental standpoint there are two main advances in the FastJet implementation [4]:

- algorithmic developments resulting in a large reduction in processing time for sequential recombination algorithms, whose application to the hadronic collider environment was previously very limited for computational reasons.
- the rigorous definition of *jet area* for any infrared safe algorithm[30], which is crucial for an accurate correction for underlying event effects. Jet area is measured by seeding an event with a large number of very soft "ghost particles", uniformly distributed in rapidity $\eta$ and azimuth $\phi$, counting the number of ghost particles swept up in each reconstructed jet, and accounting for the ghost-particle phase-space density.

The development of the jet area definition was initially driven by the need for precise jet

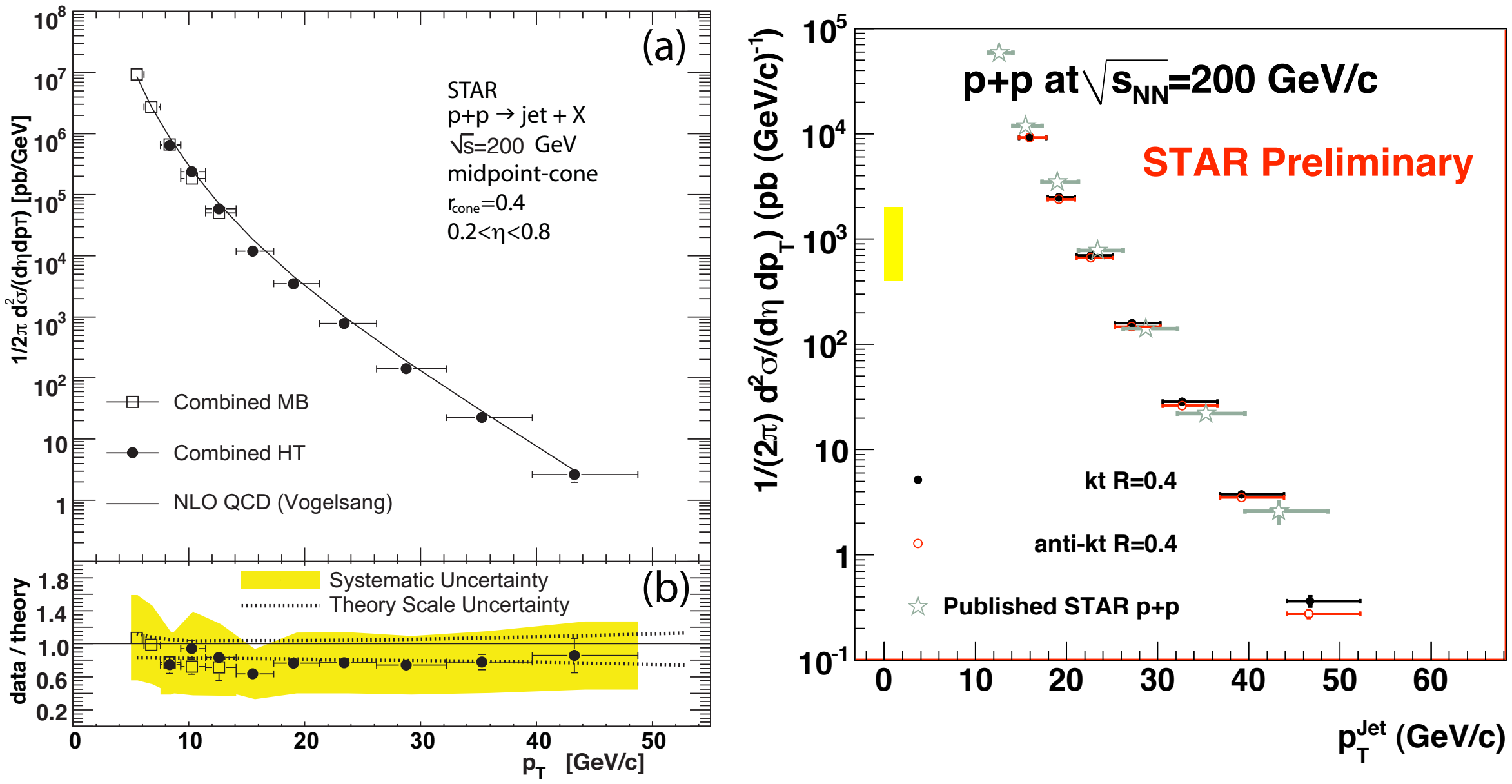


**Figure 4.4:** Inclusive jet differential cross section in $\sqrt{s}$=200 GeV p–p collisions, measured by STAR. Left: Published spectrum using a Mid-Point Cone algorithm [28]. NLO calculation from [29]. Right: Preliminary spectra using FastJet $k_T$ and anti-$k_T$ algorithms [11], compared to published distribution from left panel.

measurements at the TeV scale in the search for BSM physics in 14 TeV p–p collisions at LHC design luminosity $(10^{34}/cm^{-2}/s)$, where there will be substantial pile-up due to multiple interactions per bunch crossing. However, it was quickly realized that the same approach could be applied to heavy ion events, where the origin of the background underlying the jet is the event itself. In both cases, accurate measurement of the jet area from FastJet, combined with accurate characterization of the unbiased background, provides the optimal correction for underlying event energy on a jet-wise basis, and thereby the most accurate measurement of jet observables. Characterization of the background in heavy ion events is complex, and we defer discussion of it to section 5.2.9 below.

The essential parameter in each algorithm is denoted $R$. For a simple cone algorithm without splitting, $R$ is simply the cone radius $R = \sqrt{\Delta\phi^2 + \Delta\eta^2}$, but for more sophisticated algorithms $R$ is better thought of as the resolution scale at which the structure of the event is being measured. We refer to $R$ below by the more generic term "resolution scale". The choice of $R$ in heavy ion events is limited by the large underlying background, which contributes to the jet energy as $R^2$. The crucial issue in this regard is the precision with which such background can be corrected, and the resultant systematic uncertainty on the jet observable in question. We address background correction and its systematics later in the report. At present we limit our studies of jet reconstruction in heavy ion collisions to $R = 0.4$ and smaller.

As remarked above, algorithms vary in their response to large underlying event background,

and not all choices are equally good in this case. The well-known $k_T$ algorithm clusters first the softest radiation in the event, thereby building up jets from the tail inward to the core. In contrast, the FastJet anti-$k_T$ algorithm [31] reverses the ordering of the clustering by a minor algorithmic change (which leaves the algorithm infrared safe), beginning with the hardest radiation in the event. Studies with both simulations [31] and real data show that anti-$k_T$ jet reconstruction is much more robust than $k_T$ against disruption by large underlying background, providing more stable and interpretable results. Fig. 4.5 and 4.6 show the response of $k_T$ and anti-$k_T$ in central Au–Au collisions at RHIC; the difference between them is largest at the lowest jet energies, where the effects of background are expected to be largest.

Based on RHIC experience and the model studies in [31], our investigation of jet measurements in ALICE concentrates on the anti-$k_T$ algorithm, which at present appears to be the optimal choice for heavy ion collisions.

## 4.4 Jet reconstruction in heavy ion collisions: experimental results

We now discuss in more detail the application of jet algorithms to heavy ion collisions. Jets are clearly visible to the eye in both panels of Figure 4.3, but for quantitative measurements of jet energy and the modification of jet structure it is necessary to go beyond the simple identification of the presence of a jet, with accurate assessment of the effects of background and its fluctuations on jet reconstruction.

Figure 4.4 shows the inclusive differential jet production cross section in $\sqrt{s}$=200 GeV p–p collisions, measured by STAR using several algorithms. The measurements are carried out using charged particle tracking and electromagnetic calorimetry, with corrections that are very similar to those required for jet measurements in ALICE (see Chapter 5). The analysis in the left panel utilizes the Mid-Point Cone algorithm [28]. The systematic uncertainty, dominated by the calibration uncertainty in the STAR EMC Barrel, corresponds to a fully-correlated 50% uncertainty in the cross section (it is expected that this uncertainty will be significantly smaller in future measurements). An NLO calculation [29] is seen to agree with the measurement within uncertainties. The right panel utilizes the FastJet $k_T$ and anti-$k_T$ which are compared to the Mid-Point Cone analysis from the left panel. In all cases a resolution scale of $R = 0.4$ is used. The results of all three algorithms are seen to agree within uncertainties for p–p collisions at RHIC.

Figure 4.5, left panel, shows the equivalent measurement of inclusive jet cross section in central 200 GeV Au–Au collisions [11], utilizing the FastJet $k_T$ and anti-$k_T$ algorithms and the same correction scheme utilized for the FastJet measurements in Fig. 4.4. In this case, correction for the underlying background and its fluctuations is also a large systematic un-

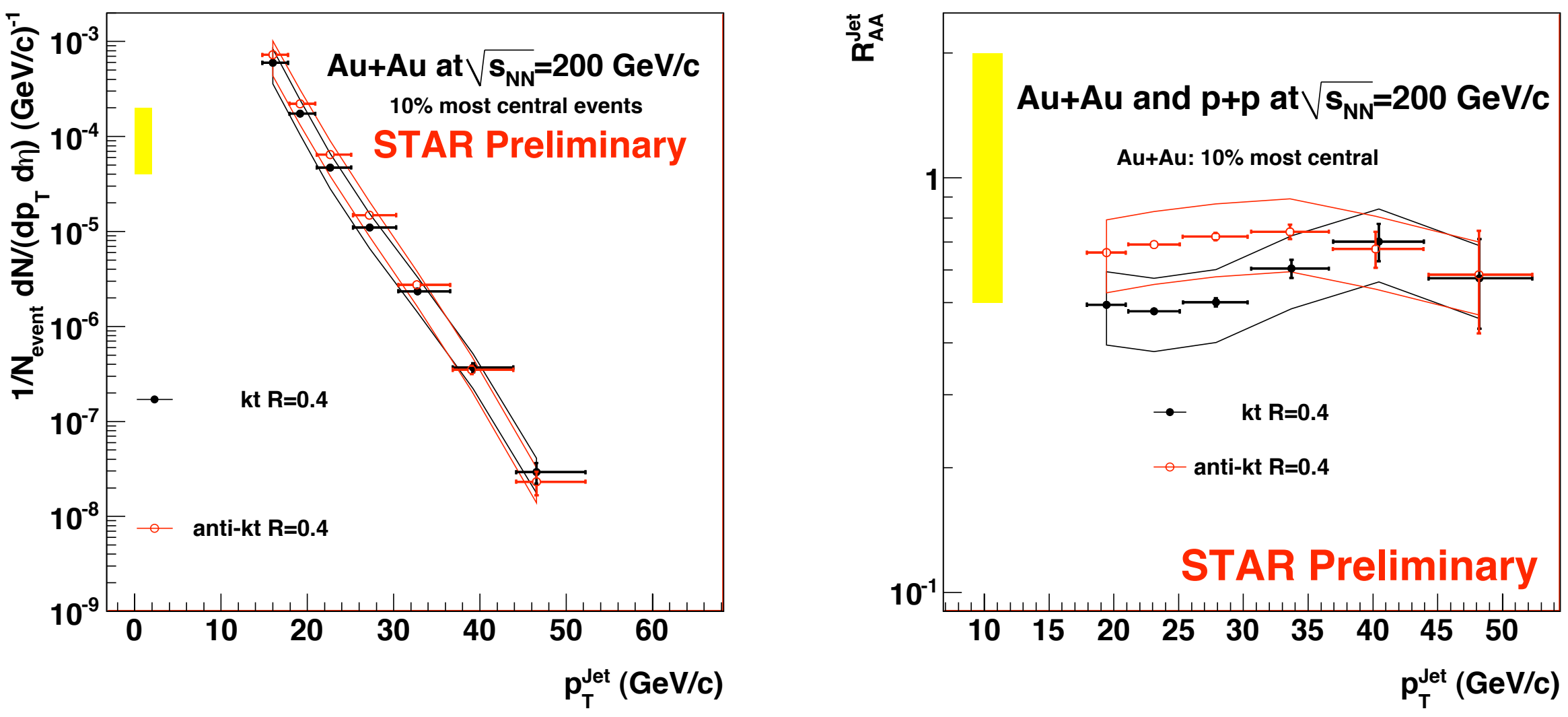


**Figure 4.5:** Full jet reconstruction in heavy ion collisions at RHIC, from STAR. Left: Inclusive cross section. Right: Jet $R_{AA}$ from [11].

certainty, expressed as the band around the data points. The techniques used to characterize the background and unfold its effects on the spectrum are the same as those proposed for ALICE EMCal measurements, discussed in detail in Chapter 5.

Figure 4.5, right panel, shows the jet $R_{AA}$, based on the above measurements of the inclusive jet production cross section in p–p and Au–Au collisions at 200 GeV [11]. Unbiased jet reconstruction, in which the full jet production cross section in heavy ion collisions is recovered even in the presence of strong quenching effects, would correspond to $R_{AA}$=1. The figure shows that the measured value of $R_{AA}$ is indeed consistent with unity when the (correlated) systematic uncertainty is taken into account (note the yellow band centered at unity). However, this uncertainty is large at present, due to the complexities inherent in forming the ratio of measurements from two different datasets. A factor $\sim 2$ deficit relative to the unbiased cross section is also consistent with the measurement. Such a deficit in measured cross section would indicate that not all jets have their energy fully reconstructed in heavy ion collisions, due to quenching effects (in particular, medium induced large angle radiation).

Prior to the measurements in Fig. 4.5, jet reconstruction in the heavy ion environment was thought to require kinematic cuts to reduce background, specifically the exclusion of low $p_T$ charged tracks and low energy calorimeter towers from the input to the jet reconstruction algorithms. An earlier version of this analysis [9] showed that such cuts result in a substantial cross section deficit, which arises from significant reconstruction biases *against* highly quenched jets. The analysis shown in Figure 4.5 does not apply such kinematic cuts, with only minimal tracking ($p_T > 200$ MeV/c) and tower energy cuts ($E_{tower} > 200$ MeV) applied

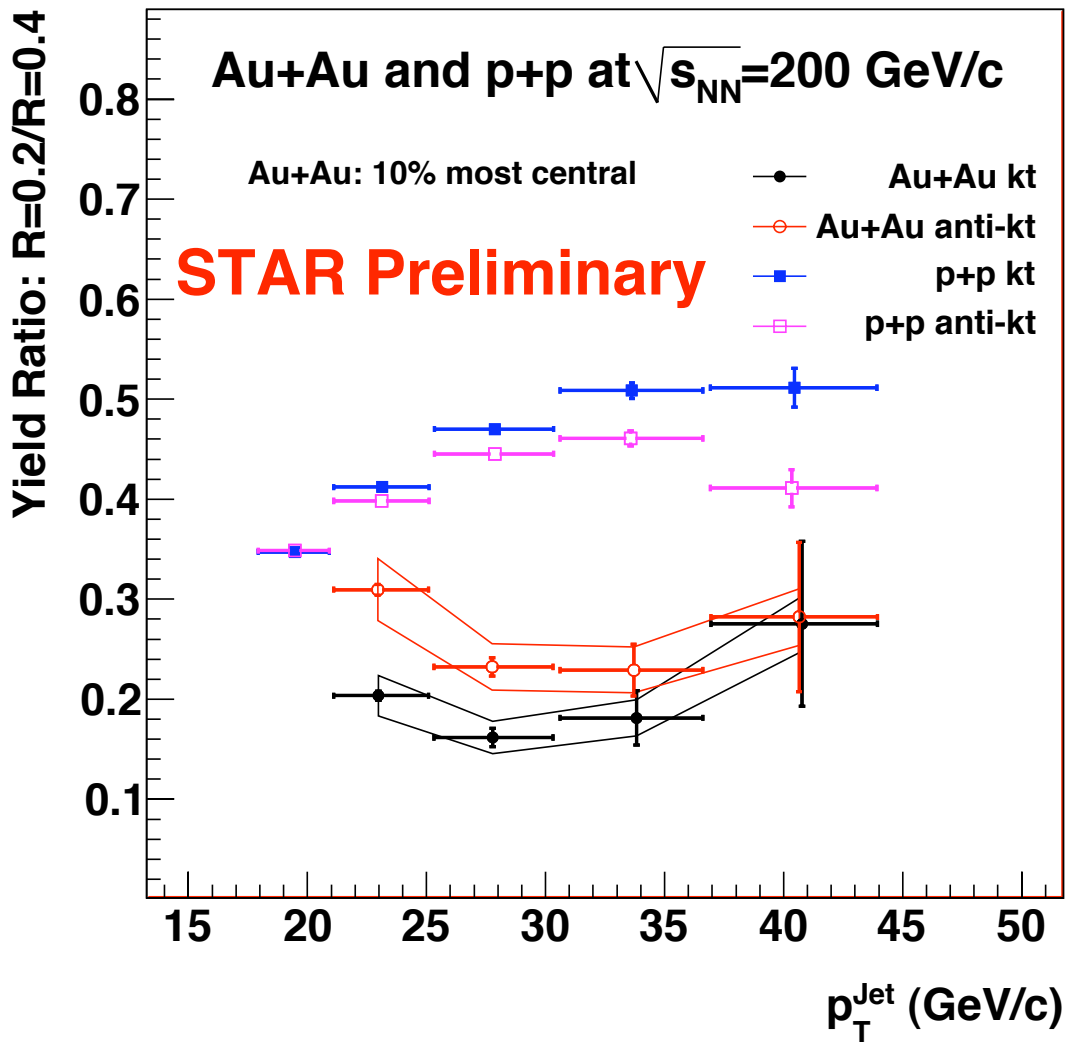


**Figure 4.6:** Ratio of inclusive differential jet cross sections at two resolution scales, $d\sigma(R = 0.2)/d\sigma(R = 0.4)$, for 200 GeV p–p and Au–Au collisions [11].

to eliminate noise and ensure analysis quality. This approach is intended to preserve the signal for highly quenched jets, at the cost of increasing substantially the complexity of the analysis, due to the markedly larger background and associated fluctuations. We will take the same approach for jet reconstruction in ALICE.

Other jet observables at present have smaller systematic uncertainties than $R_{AA}$, with correspondingly greater potential sensitivity to quenching effects. Fig. 4.6 shows the measurement of one such observable, the ratio of inclusive differential jet cross sections measured with two different resolution scales ($d\sigma(R = 0.2)/d\sigma(R = 0.4)$). Many systematic effects cancel in this measurement. The figure shows the ratio for both p–p and central Au–Au collisions. The cross section ratio for p–p is well-described by PYTHIA (see Fig. 4.9 below), whereas the ratio for central Au–Au is significantly suppressed. Bearing in mind that the observable is the ratio of cross sections at the same measured jet energy, this suppression indicates a substantial angular broadening of jets due to quenching. This is the first direct observation of jet energy re-distribution due to quenching. We discuss below this measurement in comparison to theoretical calculations.

## 4.5 Jet quenching calculations

In this section we discuss two specific implementations of the theoretical approaches sketched in section 4.2.1, to obtain guidance for jet quenching effects at the LHC: the analytic GLV

formalism, based on an opacity expansion in the number of in-medium scatterings [12, 13], and qPYTHIA [32], a modification of the standard PYTHIA Monte Carlo particle generator [33], in which medium-induced quenching effects are introduced as an additive correction to the standard vacuum splitting functions in the parton shower. The modification is based on the BDMSP multiple-soft scattering approximation [34].

Application of analytical approaches to the description of RHIC data was discussed in section 4.2.1 (see also [35]). While there are several new Monte Carlo generators under development that include quenching effects, qPYTHIA is at present the only such code that is available for general use. We compare qPYTHIA predictions to existing RHIC data, to explore its current capabilities.

### 4.5.1 qPYTHIA vs. RHIC Data

We stress that this is not an exhaustive study of qPYTHIA physics capabilities - indeed, this is the first time that comparisons to RHIC data have been made at this level of detail. This section should be regarded simply as an indication of what can be learned from such theory/experiment comparisons.

Figure 4.7, left panel, shows the inclusive $\pi^0$ production cross section in p–p collisions at $\sqrt{s}$=200 GeV measured by PHENIX [36], compared to a qPYTHIA calculation with $\hat{q} = 0$. Agreement is seen to be good between calculation and data. Note that qPYTHIA in vacuum is not identical to PYTHIA, in that it utilizes the full splitting function whereas PYTHIA takes the splitting function in the limit $P(z \rightarrow 1)$. The difference in choice of splitting function does not affect leading $\pi^0$ production ($z \rightarrow 1$), but modifies the jets in such a manner that they become collinear.

Quenching effects are quantified for inclusive cross sections via the ratio $R_{\mathrm{AA}}$, defined as:

$$R_{AA} = \frac{d^2N_{AuAu}/dydp_T}{\langle T_{AB}\rangle d^2\sigma_{pp}/dydp_T}, \tag{4.1}$$

where $\langle T_{AB}\rangle \simeq 22\ \mathrm{mb}^{-1}$ accounts for the nuclear geometry; $R_{\mathrm{AA}}$=1 if there are no nuclear-specific effects in the hard production process (shadowing, quenching).

Nuclear geometry is modeled in qPYTHIA using the PQM prescription [38], which defines a $\hat{q}_{local}$ and calculates an effective $\hat{q}$ for a given jet trajectory in the medium via an appropriately weighted average of $\hat{q}_{local}$. Figure 4.7, right panel, shows $\pi^0$ $R_{\mathrm{AA}}$ measured by PHENIX in 200 GeV central Au–Au collisions [36] compared to qPYTHIA for various values of $\hat{q}$. The well-known suppression factor ∼5 requires $\hat{q} \simeq 6\ \mathrm{GeV}^2/\mathrm{fm}$, qualitatively consistent with the analysis in Fig. 4.2. Substantially larger values of $\hat{q}$ are disfavored by the data.

Turning now to jet observables, Fig. 4.8, left panel, shows the inclusive jet spectrum for

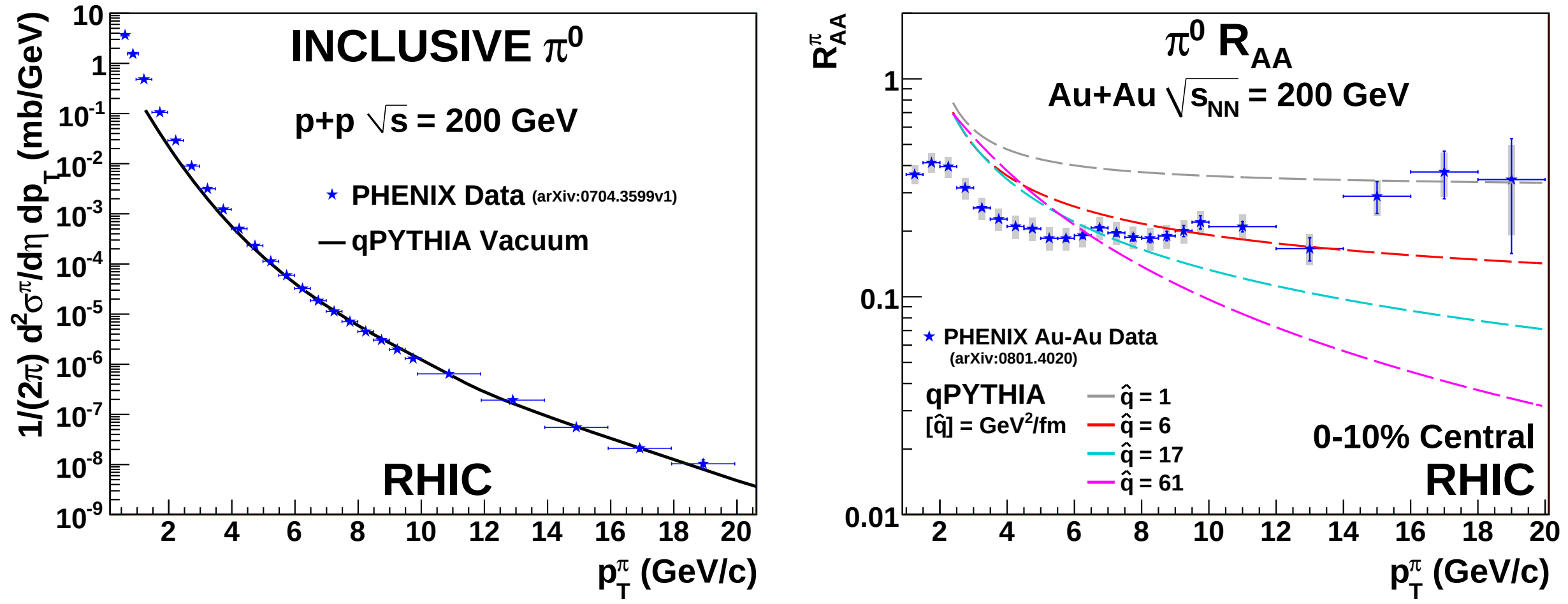


**Figure 4.7:** Inclusive $\pi^0$ production at $\sqrt{s_{NN}}$=200 GeV. Left: p–p collisions at $\sqrt{s}$=200 GeV [36], compared to qPYTHIA calculation in vacuum. Right: $\pi^0$ $R_{\mathrm{AA}}$ [37] compared to qPYTHIA calculation with several values of $\hat{q}$.

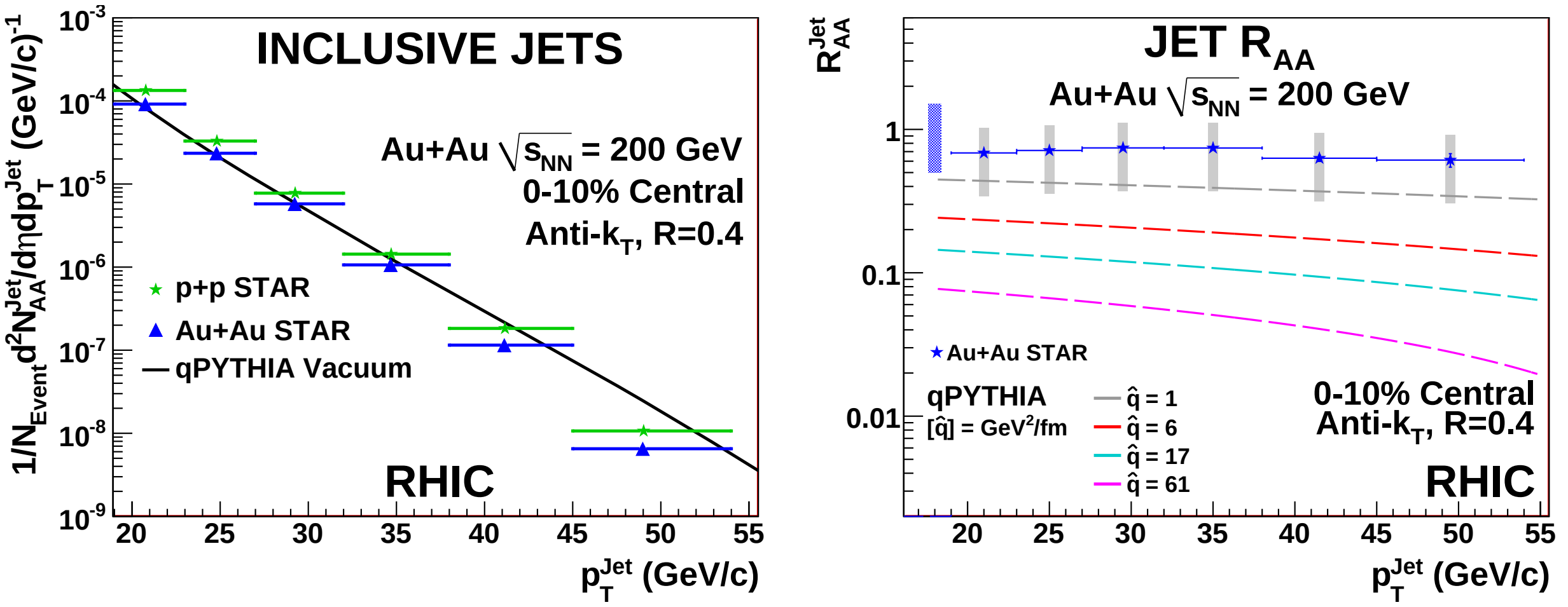


**Figure 4.8:** Inclusive jet production at $\sqrt{s_{NN}}$=200 GeV, using FastJet anti-$k_T$ $R = 0.4$. Left: Inclusive jet cross section for p–p and Au–Au collisions [11], compared to a qPYTHIA vacuum calculation. Right: Jet $R_{\mathrm{AA}}$ for 200 GeV 0-10% central Au–Au collisions [11]. The solid box centered at unity shows the correlated systematic uncertainty, dominated by the calorimeter calibration uncertainty; grey boxes show point-to-point systematic uncertainties; curves show qPYTHIA calculations for several values of $\hat{q}$.

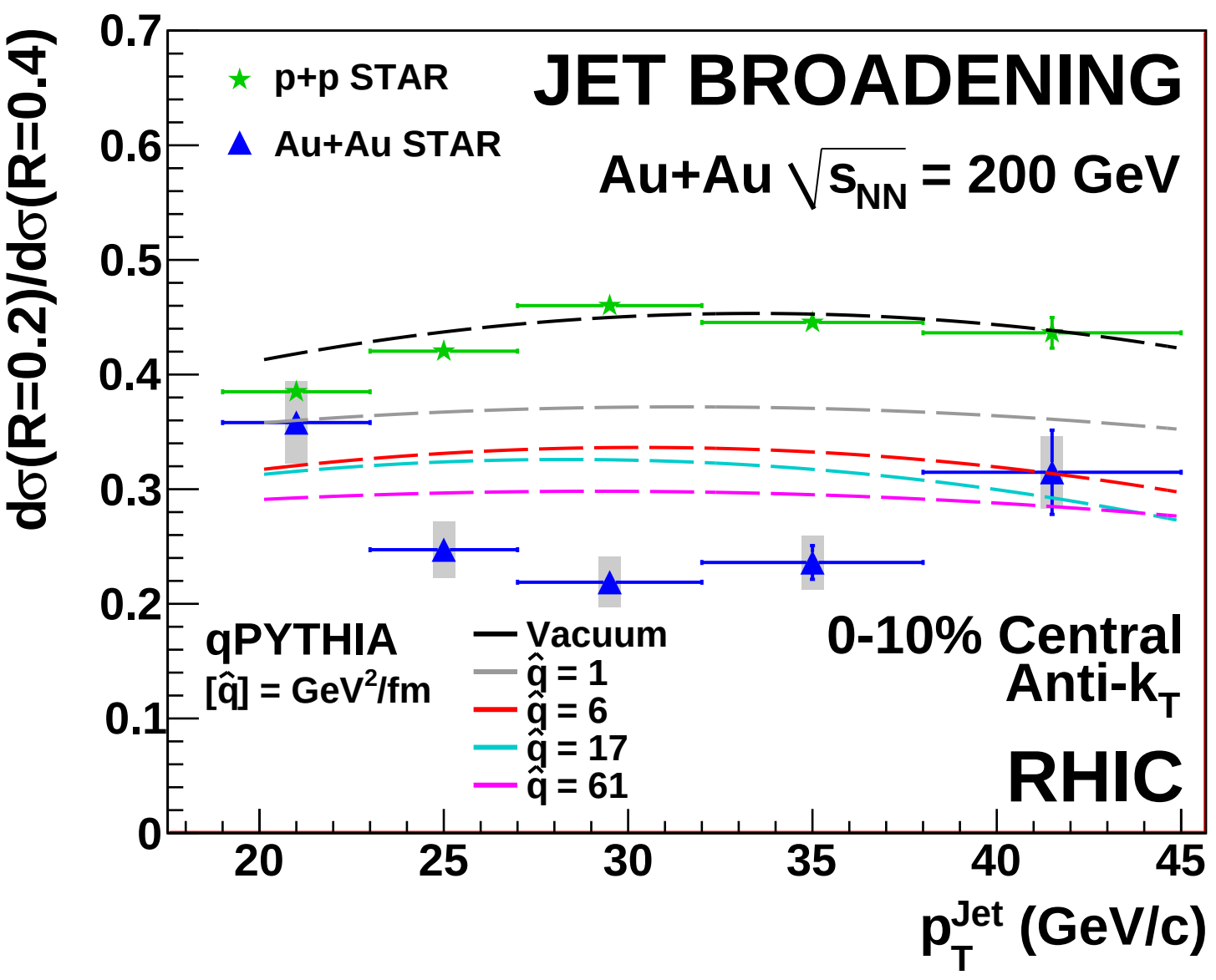


**Figure 4.9:** Ratio of inclusive jet differential cross sections $d\sigma(R=0.2)/d\sigma(R=0.4)$ for 200 GeV p–p and Au–Au collisions [11], compared to qPYTHIA calculations for various $\hat{q}$. Data are the same as Fig.4.6.

200 GeV p–p collisions (cross section scaled by $\langle T_{\rm AB}\rangle$) and central Au–Au collisions [11], compared to a qPYTHIA calculation for jet production in vacuum ($\hat{q}$=0). Figure 4.8, right panel, shows the measured jet $R_{\rm AA}$ from these spectra, compared to qPYTHIA with various $\hat{q}$. While qPYTHIA predicts substantially more suppresssion for $\hat{q}$=6 GeV$^2$/fm than the central value of the measurement, the data and calculation are compatible at the limit of the large systematic uncertainties.

Comparison of the inclusive jet production cross section for different resolution scales $R$ is sensitive to the distribution of energy within the jet. Figure 4.9 shows the measured ratio of inclusive jet cross section (anti-$k_T$) for $R=0.2$ and $R=0.4$ in 200 GeV p–p and central Au–Au collisions [11]. The significant difference in the ratio between p–p and Au–Au collisions suggests a broadening of the jet structure due to quenching. The figure also shows qPYTHIA calculations of this ratio for both systems. Reasonable agreement is seen for jets produced in vacuum (p–p collisions), although better agreement is achieved with standard PYTHIA, which utilizes a slightly different splitting function. The quenched data (Au–Au collisions) are not described by qPYTHIA for any value of $\hat{q}$, which predicts significantly less broadening than seen in the data. This disagreement indicates that qPYTHIA does not yet capture fully the physics of quenching.

### 4.5.2 LHC predictions

We now discuss theoretical predictions of jet quenching observables for 5.5 TeV Pb–Pb collisions, using qPYTHIA and the GLV formalism. While section 4.5.1 has pointed out some

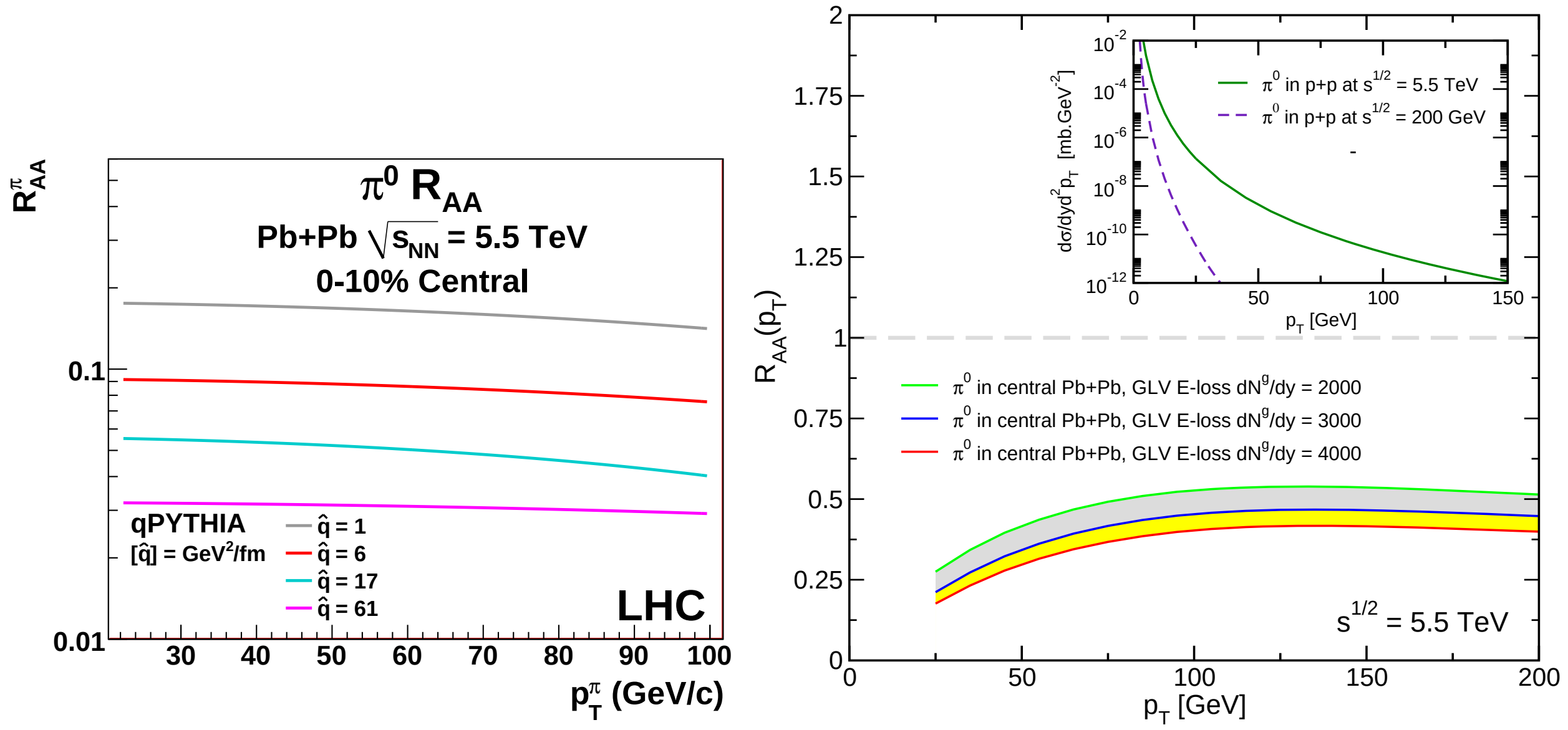


**Figure 4.10:** $\pi^0$ $R_{\rm AA}$ predictions for 5.5 TeV central Pb–Pb collisions. Left: qPYTHIA [5]. Right: GLV-based analytic approach (I. Vitev in [39])

limitations of qPYTHIA in its current implementation for describing RHIC jet measurements, it is nevertheless the most mature Monte Carlo generator incorporating quenching effects that is currently publicly available. Its predictions for quenching effects at the LHC can give guidance for the statistical and systematic sensitivity required in quenching measurements.

The extrapolation from RHIC measurements to quenching parameters at the LHC has large uncertainties. Current estimates of $\hat{q}$ at RHIC cover a broad range $\hat{q} \simeq 1-6$ GeV$^2$/fm (see section 4.2.1). $\hat{q}$ is expected to scale with the entropy density, reflected in the charged particle multiplicity, which is expected to be a factor $\sim 5$ larger in nuclear collisions at LHC than at RHIC. Thus, we project $\hat{q} \simeq 25$ GeV$^2$/fm at the LHC. This estimate is only rough guidance, with a large and otherwise unknown error bar.

Figure 4.10 shows predictions for $\pi^0$ $R_{\rm AA}$ in 5.5 TeV central Pb–Pb collisions from qPYTHIA (left) and the GLV analytic formalism (right). The model parameter for the GLV calculation is not $\hat{q}$ but gluon multiplicity; nevertheless, the choice of parameters in the right panel also corresponds to a scaling of multiplicity a factor $\sim 3-5$ relative to RHIC, so the predictions are directly comparable. For values of $\hat{q}$ expected at the LHC, qPYTHIA predicts $R_{\rm AA}$ substantially less than 0.2 at $p_T$=100 GeV/c, whereas GLV predicts $R_{\rm AA} \sim 0.35-0.5$.

Figure 4.11 shows predictions for jet $R_{\rm AA}$ for 5.5 TeV central Pb–Pb collisions from qPYTHIA (left) and the GLV approach (right; cone algorithm, various $R$) [13]. Comparison for $R=0.4$ at $E_T$=150 GeV reveals a factor $\sim 3$ larger jet $R_{\rm AA}$ for GLV than qPYTHIA, suggesting relatively more collimated quenched jets.

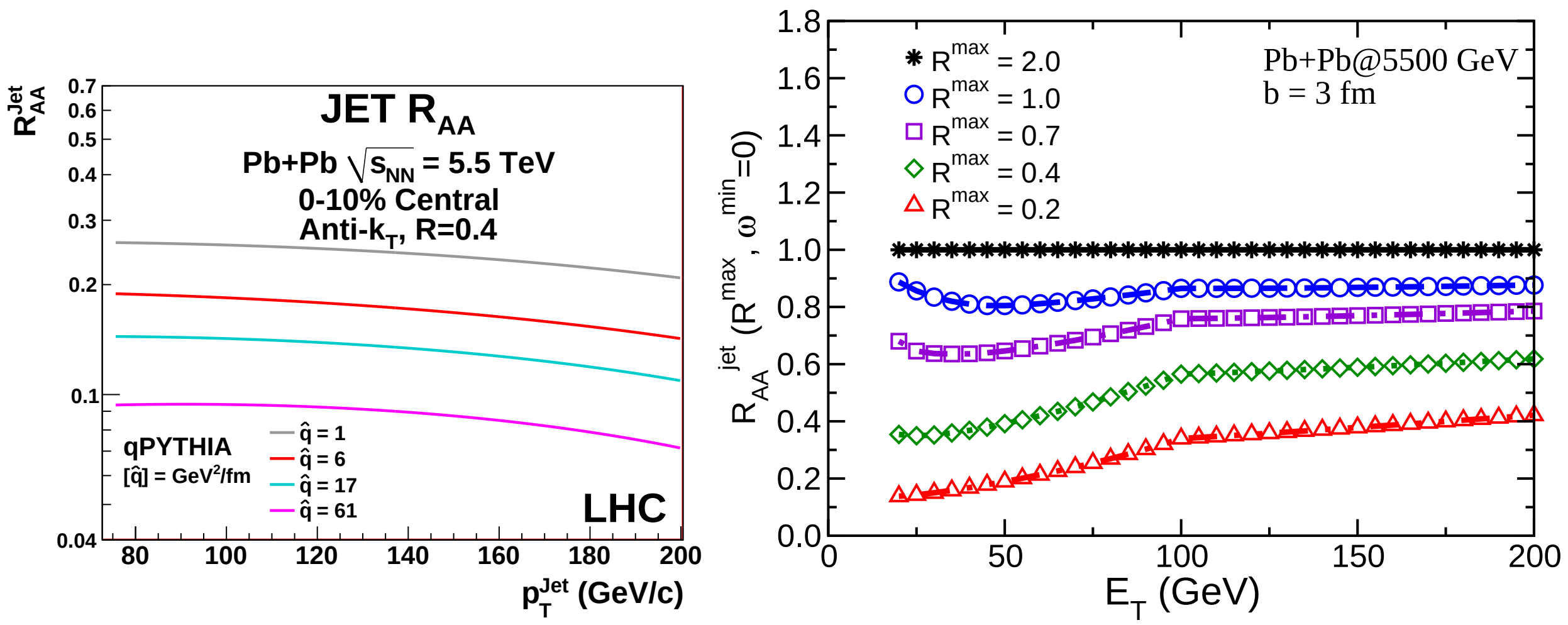


**Figure 4.11:** Jet $R_{\rm AA}$ for 5.5 TeV central Pb–Pb collisions. Left: qPYTHIA predictions for various $\hat{q}$. Jet algorithm is anti-$k_T$, $R = 0.4$. Right: GLV predictions for various gluon densities [13]. Jet algorithm is modified cone, various $R$.

Figure 4.12, left panel, shows the qPYTHIA prediction for the ratio of jet cross sections $\sigma(R = 0.2)/\sigma(R = 0.4)$, sensitive to jet broadening. Significant suppression is predicted for $\hat{q} \sim 6$ GeV$^2$/fm relative to the vacuum case, though the broadening has little sensitivity to $\hat{q}$ beyond that value.

Figure 4.12, right panel, presents the GLV calculation of the event-wise measurement of the differential jet energy profile,

$$\psi(r; R) = \frac{d\Psi_{int}(r; R)}{dr}, \tag{4.2}$$

where

$$\Psi_{int}(r; R) = \frac{\sum_i (E_T)_i \Theta(r - R_{jet,i})}{\sum_i (E_T)_i \Theta(R - R_{jet,i})}, \tag{4.3}$$

$R$ is the jet resolution scale, $i$ runs over all particles in the jet, and $R_{jet,i} = \sqrt{\Delta\eta_i^2 + \Delta\phi_i^2}$ is the distance from the jet centroid to the particle. This distribution has been measured by CDF at the Tevatron [40], though not yet in nuclear collisions. The figure shows substantial broadening of jets in medium relative to vacuum, though the observable distribution (labelled "Total" in the figure) shows no significant broadening relative to vacuum. The authors of [13] attribute this to the effect of experimental cuts due to the $R$ parameter: there is significant radiation at angles larger than $R$, meaning that quenched jets falling into a given window of reconstructed $E_T$ originate at larger jet energy $E_T{}'$ and are thereby more collimated, offsetting the effects of jet broadening.

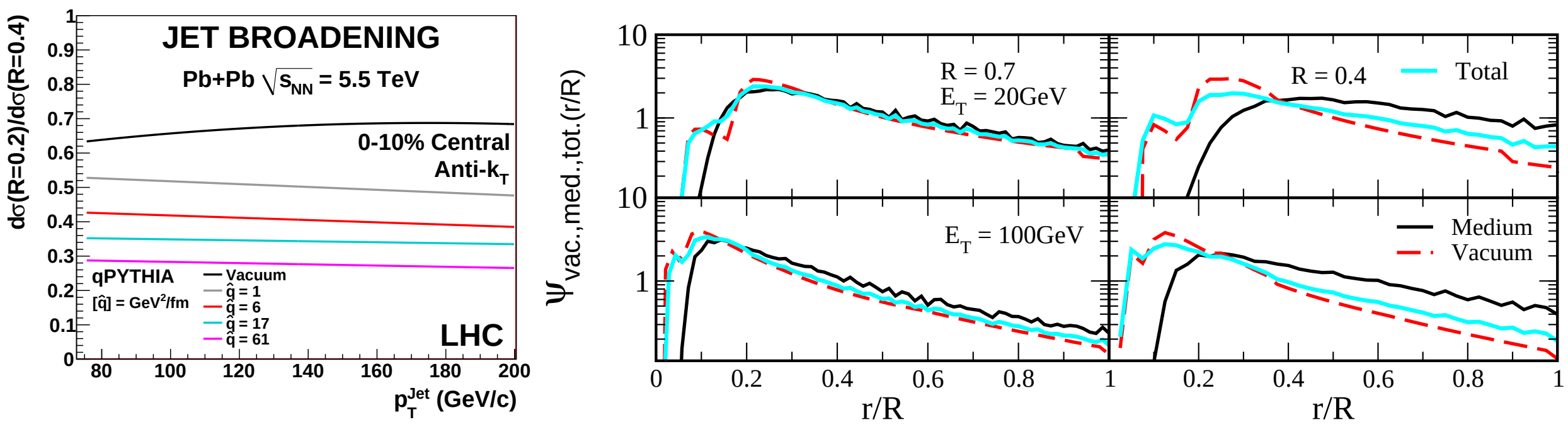


**Figure 4.12:** Quenched jet shapes at LHC. Left: qPYTHIA calculation of the ratio of inclusive jet differential cross sections $d\sigma(R = 0.2)/d\sigma(R = 0.4)$. Right: event-wise jet energy profile (eq. 4.2) from GLV [13].

In section 5.4 we return to these predictions and compare them to the expected sensitivity of ALICE jet measurements, in order to quantify the jet physics performance of ALICE with the EMCal.

# Chapter 5

# Jet Measurements in ALICE

This chapter presents the physics performance of ALICE with the EMCal for jet measurements in p–p and Pb–Pb collisions. We do not present a wide variety of jet observables, focusing rather on jet measurements in ALICE that are most advanced at RHIC, in particular the inclusive jet cross section. The cross section analysis raises many issues that are common to more differential jet measurements, and therefore provides an instructive worked example.

We present our current understanding of systematic effects, drawing extensively from experience with the same measurement in STAR. We estimate systematic uncertainties for jet cross section measurements in ALICE, together with systematic uncertainties for cross section ratios that are sensitive to quenching.

Section 5.4 compares the expectations from qPYTHIA presented in Chapter 4 to the projected systematic uncertainties in the inclusive spectrum measurements. This gives the best available quantitative assessment of the ALICE+EMCal physics performance for jets, showing the projected sensitivity of ALICE in the context of the expected magnitude of jet quenching signals.

## 5.1 Methodology

ALICE takes the same approach to jet measurements as STAR, using Electromagnetic Calorimetry (EMCal) combined with precise charged particle tracking in a moderate-strength magnetic field (0.5 T). Fast (Level 0/1) triggering is provided by the EMCal. This technique differs from the more common approach in collider experiments, in which a purely calorimetric jet measurement is made via both EM and Hadronic calorimeters.

The choice to utilize tracking rather than hadronic calorimetry was taken to optimize jet reconstruction in the complex environment of heavy ion collisions, where detailed and fine-

grained characterization of the large event background and its fluctuations are required for systematically well-controlled jet measurements. Such an approach introduces additional systematic effects that are not present in a purely calorimetric measurement, however, leading to the following requirements in ALICE:

- precise control over tracking efficiency, momentum resolution and scale, up to very high $p_T$ ($\sim 100$ GeV/c), including the effects of the local tracking environment in the core of a hard jet;
- correction for the double-counting of energy deposited by charged tracks showering in the EMCal (both electrons and hadrons);
- correction for un-measured neutral hadrons, with the dominant effect due to neutrons and $K^0_L$.

A proof of principle of this approach has been provided by the STAR experiment, which also measures jets utilizing tracking and EM Calorimetry (see Figs. 4.4 and 4.5).

Other approaches to jet reconstruction in the heavy ion environment apply kinematic cuts on the tracks or calorimeter towers utilized in jet reconstruction in order to suppress background fluctuations [1], or have such kinematic cuts imposed by hardware cutoffs arising from a strong magnetic field [2]. Preliminary study of jet reconstruction in heavy ion collisions by the STAR experiment shows that such kinematic cuts introduce strong biases against quenched jets [3], and that unbiased jet quenching measurements require the avoidance or minimization of such cuts. This can be accomplished in ALICE with high precision tracking and the highly granular EMCal, which impose very low instrumental cuts on tracks and towers ($\sim 100 - 200$ MeV), in contrast to other LHC experiments [2].

The main systematic effects in the ALICE measurement of the inclusive differential jet cross section are listed in Table 5.1.

## 5.2 Jet measurements: systematic effects and uncertainties

In this section we discuss the systematic effects listed in Table 5.1, and estimate data-driven correction factors and corresponding systematic uncertainties for the inclusive jet differential cross section measurement.

The systematic effects are assessed in part via event simulations, with a jet population generated by PYTHIA and 5.5 TeV Pb–Pb background events generated by HIJING (unquenched). We utilize simulated data of two kinds:

| **Effect** | **Distribution or shift** | **Estimation/ correction** | **Where correction applied** | **Sys. uncertainty** |
|---|---|---|---|---|
| **Common in p–p and A–A** | | | | |
| *Hadronic and electron energy double counting* | Energy shift | Remove fraction of EMCal energy matched to track | Before jetfinding | error band |
| *Tracking efficiency* | Energy shift | MC studies | spectrum | error band |
| *Instrumental resolution (momentum, EMCal res.)* | Distribution | MC studies | spectrum deconvolution | error band |
| *TPC space charge* | Energy shift | systematic studies, limit track $p_T$ | before jet finding | error band |
| *EMCal energy scale* | Energy shift | calibration/ systematic studies | spectrum | error band |
| *Unobserved neutral energy* | Energy shift | MC studies | spectrum | error band |
| *Hadronization* | Energy shift | not corrected | not corrected | none |
| *Recombination scheme* | Energy shift | Algorithm choice | Algorithm | none |
| **A–A specific: Underlying heavy-ion event** | | | | |
| *Underlying event fluctuations* | Energy shift and fluctuations | multiple schemes and estimates | event-wise shift; spectrum deconvolution | error band |
| *False jets* | shift in yield | multiple schemes | spectrum | error band |

**Table 5.1:** Systematic effects in the measurement of the inclusive differential jet cross section. Column 2 describes how the spectrum is affected; column 3 decribes how the effect is estimated or corrected; column 4 describes where the correction is applied; and column 5 describes how the systematic uncertainty is expressed. Details of each correction are found in the text.

- **Particle level:** This refers to distributions of particles produced by the physics event generator (PYTHIA, qPYTHIA), without instrumental effects such as interactions in matter and finite decay length. All stable particles except neutrinos are included in the analysis.

- **Detector level:** This refers to generated events or single particles filtered through a detailed GEANT simulation of the ALICE apparatus, to account for interactions in matter, finite decay length, and instrumental response. Neutral, long-lived particles (neutrons, $K_L^0$) generally do not register at the detector level.

In this analysis we utilize EMCal clusters, formed from contiguous towers above a threshold and assigned a single space-point. A minimum cluster energy cut of 200 MeV for p–p and 500 MeV for Pb–Pb was applied, which will be reduced in future analyses. Standard ALICE

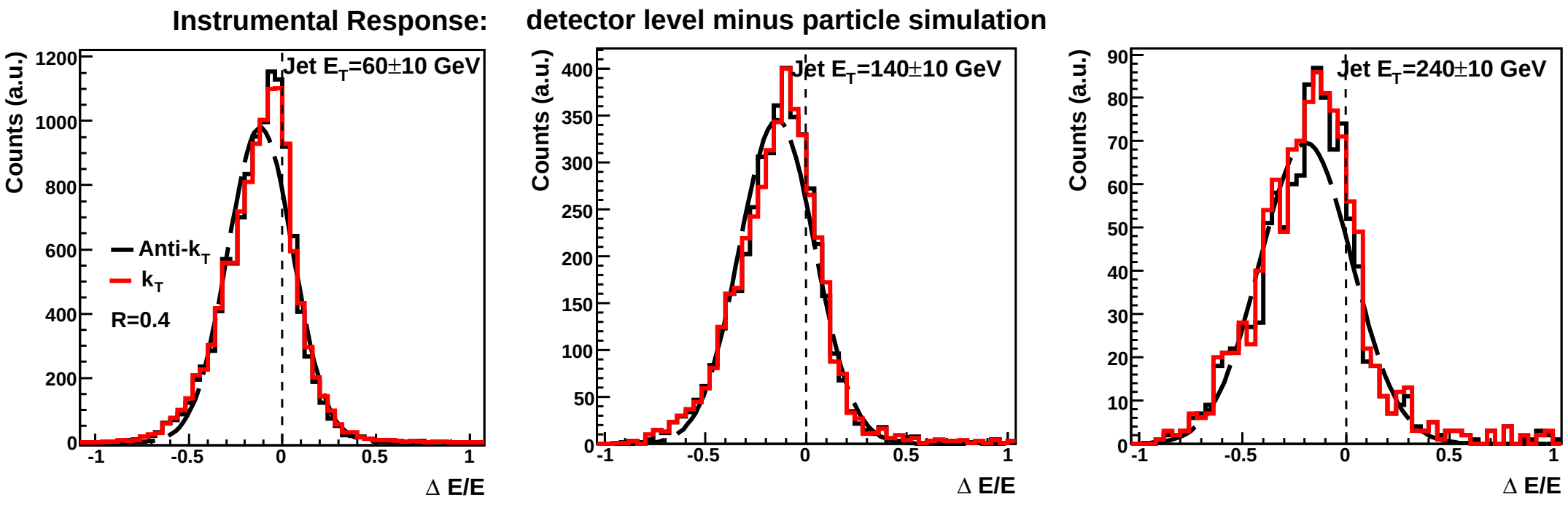


**Figure 5.1:** Instrumental effects on jet reconstruction, for jets from 5.5 TeV p–p collisions in the EMCal acceptance. $\Delta E$ is the event-wise difference between jets reconstructed on the Detector minus Particle level, for selected intervals in Particle level jet energy $E$. Algorithms are FastJet $k_T$ and anti-$k_T$ $R = 0.4$. Solid curve is Gaussian function fit to anti-$k_T$ distribution.

quality cuts were applied to the charged tracks [4], with no minimum $p_T$ cut on the tracks contributing to jet reconstruction.

### 5.2.1 Jet energy measurement: instrumental effects

The overall instrumental effects on jet reconstruction are shown in Figs. 5.1 and 5.2, which compare jets reconstructed (anti-$k_T$ $R$=0.4) within the EMCal acceptance on the detector vs. particle level, for PYTHIA 5.5 TeV p–p events. The only correction applied is for EMCal energy deposited by charged hadrons and electrons, to avoid double counting (see section 5.2.2). This simulation accounts for the variation in tracking resolution and efficiency in the dense core of high energy jets, among other effects.

Figure 5.1 shows the event-wise relative difference in jet energy (detector level minus particle level), for selected bins in particle-level reconstructed energy. This comparison shows the cumulative effect of experimental cuts, undetected neutral particles, tracking efficiency (which together generate a systematic energy shift), and tracking and calorimeter resolution. The *instrumental* jet energy resolution is defined as the width of a Gaussian function fit to this distribution.

Figure 5.2 shows the dependence of both the relative shift (left panel) and relative resolution (right panel), as a function of the particle-level reconstructed jet energy. For jets between 50 and 200 GeV, the mean energy deficit at the detector level is 11-20% and resolution is 18-22%. The dominant contributions to these effects are due to undetected neutral particles and charged particle tracking resolution.

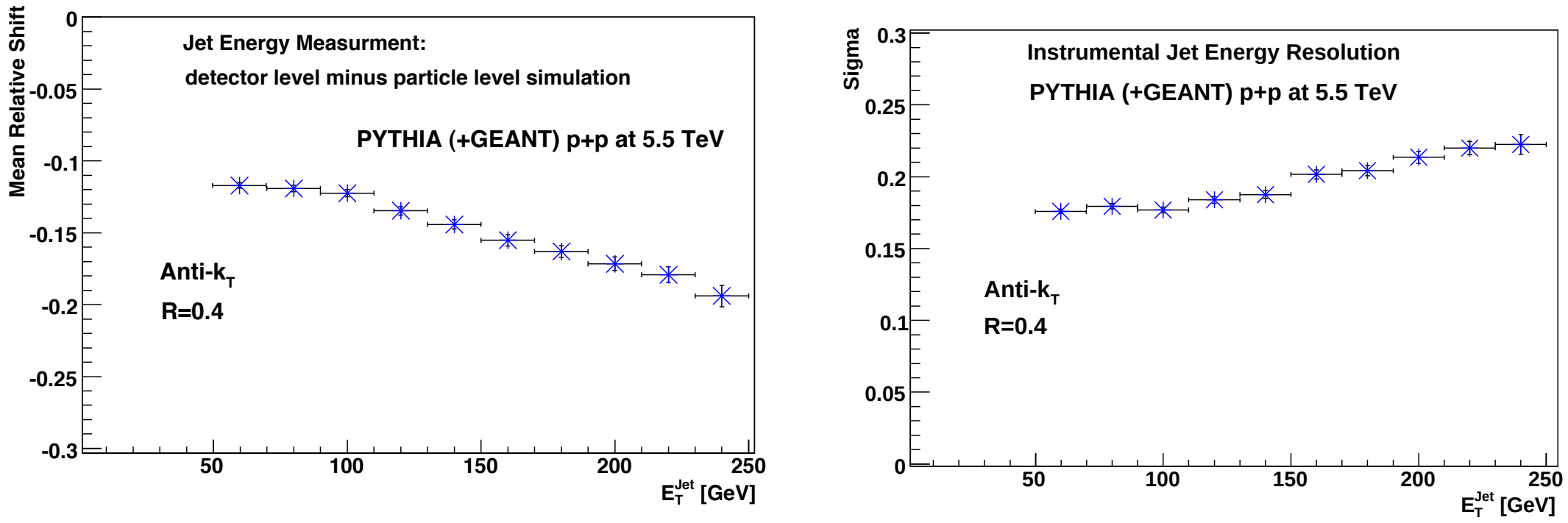


**Figure 5.2:** Instrumental effects on jet reconstruction: jet $E_T$ dependence of mean and width of Gaussian function fit to distributions in Fig. 5.1. Left: shift in mean $\Delta E/E$ (detector - particle level). Right: width from Gaussian fit (instrumental jet energy resolution).

### 5.2.2 Hadronic and electron energy double counting

The EMCal response includes electromagnetic showers from photons (due dominantly to $\pi^0$ and $\eta$ decay), EM showers from electrons, MIP energy deposition, and hadronic showers from long-lived hadrons.

Electron and charged hadron momenta are also measured via charged tracking, and corrections for these contributions to the EMCal response must be applied in order to isolate the neutral component of the EM response and to avoid double counting of energy.

The response of the EMCal to charged hadrons is shown in Fig. 7.7. Correction for this energy deposition in the EMCal can only be applied on an average basis. Specifically,

$$E_{cluster}^{corr.} = E_{cluster}^{rec} - \sum_{i=0}^{N_{matched}} f \times p_{h^{\pm}}^{i}, \tag{5.1}$$

with $f$ the fraction of the charged track momentum $p_{h^{\pm}}$ to be subtracted, and $N_{matched}$ the number of charged tracks matched to a reconstructed EMCal cluster. It is required that $E_{cluster}^{corr.} \geq 0$.

Studies of jet reconstruction in STAR indicate that subtracting 100% of the track momentum (i.e. $f = 1$) is optimal for hadrons. Since $f = 1$ is also appropriate for electrons, this correction is applied for all charged tracks matched to EMCal clusters.

The systematic uncertainty on the jet energy scale originating from this correction procedure is estimated by varying $f$ by 25%.

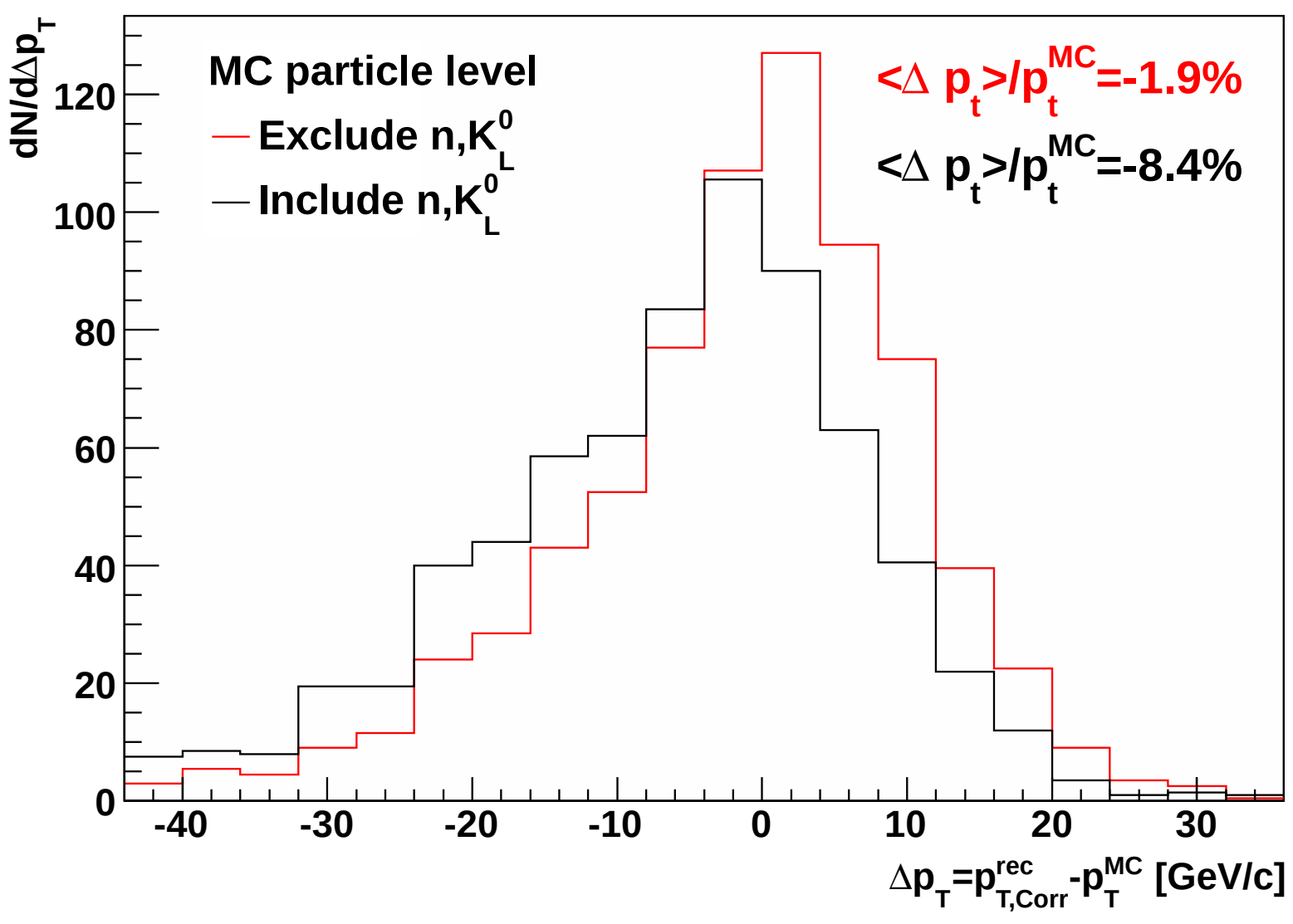


**Figure 5.3:** Jet energy for detector minus particle level after correction for tracking efficiency. See text for details.

### 5.2.3 Tracking efficiency

The charged jet energy component $E^{\mathrm{Ch}}_{\mathrm{Jet,rec}}$ must be corrected for the $p_T$ -dependent tracking efficiency $\epsilon(p_T)$. The ALICE design tracking efficiency is ∼90% for $p_T >\sim 1$ GeV/c, with weak $p_T$ dependence [4]. The $p_T$ dependence of the efficiency may neverthless have a finite effect on reconstructed jet energy. An effective tracking efficiency is therefore calculated, taking into account the charged track $p_T$ distribution in the observed jet population for a given jet energy selection:

$$\epsilon_{eff} = \sum_{i=0}^{N} p^i_{T,h^\pm} / \left( \sum_{i=0}^{N} p^i_{T,h^\pm} \times 1/\epsilon(p_T) \right), \tag{5.2}$$

where $N$ is the number of the charged jet constituents from all jets in the selected ensemble of jets.

Figure 5.3 illustrates the result of the tracking efficiency corrections for jets with $E_T \simeq 75$ GeV from 5.5 TeV p–p collisions, generated by PYTHIA. The event-wise difference of the reconstructed jet energy at the detector minus particle level is shown, similar to Fig. 5.1, but corrected in addition for tracking inefficiencies. Two cases are shown, in which neutron and $K^0_L$ energies are included or excluded in the particle-level jet energy. The mean jet energy after correction agrees within 2% with the particle-level jet energy excluding neutrons and $K^0_L$. The 2% systematic underestimation is due to $\gamma$ conversion in the ITS.

A systematic uncertainty of 5% is assigned to the tracking efficiency, based on STAR studies

[5].

### 5.2.4 Charged particle tracking performance

High $p_T$ tracking will ultimately be limited by systematic distortions, both static (due to misalignment) and dynamic (luminosity-dependent). We address here only the dynamic distortions. This is of particular importance for rare processes such as high $p_T$ jets, since the bulk of the statistics for these observables will come from high luminosity running.

An example of a luminosity-dependent distortion is the accumulation of space charge in the TPC, which distorts the drift field and thereby the momentum reconstruction. Various physics processes may be utilized to monitor the luminosity-dependent tracking performance. However, for the highest energy jets measurable by ALICE, "high $p_T$" means tracks with $p_T \simeq 50 - 100$ GeV/c, which restricts the candidate processes to those with statistically significant yield in that range.

Space charge distortion generates anomalous transport of drifting electron clusters in the TPC, modifying the measured sagitta and thereby shifting the measured momenta up or down, depending on the charge sign of tracks. Given the sharply falling distribution of the charged particle cross section with increasing $p_T$, this momentum shift is observable via luminosity-dependent relative yields of various species of positive and negative particles.

Charged pions have substantial yield at high $p_T$ and are therefore primary candidates to monitor such distortions. While the ratio of yields of positive and negative pions at high $p_T$ is not *a priori* known, it must of course be independent of luminosity. Fig. 5.4, left panel, illustrates the modification of the $\pi^+/\pi^-$ ratio as a function of $p_T$, due to a luminosity-dependent shift in pion momentum by a charge sign-dependent amount $\pm\delta$ (e.g. arising from residual space charge effects). The bands show the 1-$\sigma$ statistical error on the $\pi^+/\pi^-$ ratio, calculated using pion statistics in $10^7$ minimum bias events of 5.5 TeV Pb–Pb. The low luminosity sample, assumed in this example to comprise 10% of the data set, corresponds to no momentum shift ($\delta = 0$, or "truth"), with statistical precision represented by the yellow band around unity. Momentum shifts due to spatial distortion in the High Luminosity sample (90% of the dataset), of magnitude $|\delta| =1$, 2, and 3 GeV/c, are shown by the colored bands, whose widths likewise indicate the statistical precision of that data sample. While this worked example is somewhat arbitrary, it suggests that pion statistics will be sufficient to detect a distortion in momentum scale of magnitude $\delta \simeq 1$ GeV/c at $p_T \simeq 70$ GeV/c. A more realistic study requires real ALICE data.

To illustrate the effects of limited tracking performance on jet measurements, Fig. 5.4, right panel, shows the bias in the jet population if charged tracking is truncated above a certain threshold. The figure shows the ratio of inclusive jet cross sections in 5.5 TeV p–p collisions, where the numerator comprises only those jets whose leading charged particle has $p_T < 50$ or 75 GeV/c, and the denominator is the unbiased population. A strong bias above jet

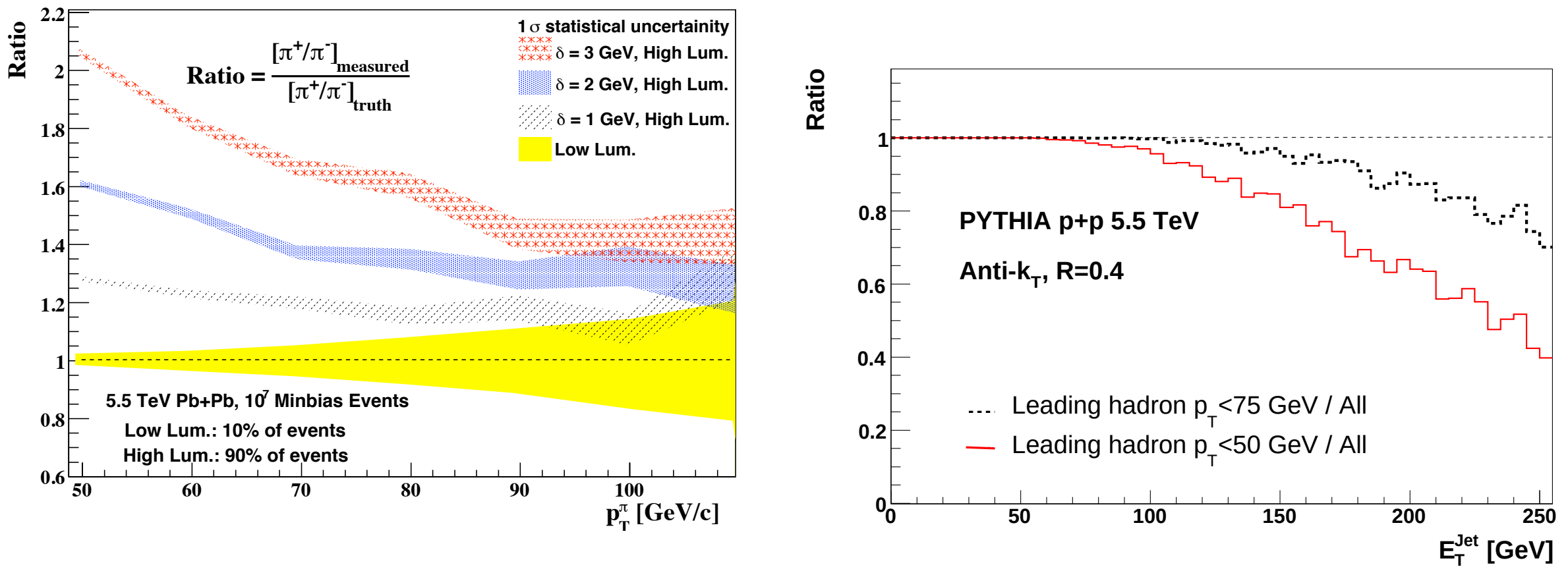


**Figure 5.4:** High $p_T$ charged tracking for jet measurements. Left: sensitivity of modification of $\pi^+/\pi^-$ ratio due to momentum-scale distortions (see text). Right: bias of jet population for 5.5 TeV p–p collisions, due to truncation of tracking at high $p_T$(see text).

$E_T$=150 GeV is induced for truncation at $p_T$=50 GeV/c, less so for truncation at $p_T$=75 GeV/c. ALICE charged tracking is expected to perform well beyond $p_T$=75 GeV/c.

Fig. 5.4, right panel, is for 5.5 TeV p–p collisions. The jet bias in central Pb–Pb collisions will be substantially less than indicated in the figure for a given threshold, due to jet quenching. It is however difficult to quantify that effect at present.

### 5.2.5 EMCal energy scale

The uncertainty on the neutral energy scale, driven by the precision of the EMCal calibration, is one of the dominant contributions to the systematic uncertainties of jet cross-section measurements. The uncertainty on the neutral energy scale reported by STAR is around 5% [6]. Substantially better calibrations were achieved by CDF, which quotes calibration uncertainties below 1% [7], demonstrating that with sufficient effort one can reduce this uncertainty to a "minor" contribution. For the purpose of this document we assume a conservative uncertainty of 5% on the neutral energy scale.

### 5.2.6 Unobserved neutral energy

Neutral, long-lived particles (neutrons, $K_L^0$) do not register as a charged track or MIP in the EMCal, and are unlikely to develop a significant shower in the EMCal. Consequently, a correction must be applied for their unobserved energy, $\Delta E_{jet}^{unobserved}$.

The baseline correction scheme for this effect utilizes PYTHIA to simulate the unobserved neutral energy in both p–p and heavy-ion collisions. While such an approach is well defined

for p–p collisions, it does not account for possible modifications to hadron production in heavy ion collisions, in particular the baryon enhancement at intermediate $p_T$ that is now well-established (though not yet fully understood) at RHIC [8, 9]. The missing neutral energy can be estimated on an inclusive basis through the measurement of identified particles in jets (kaons, protons) and invoking isospin symmetry. How significant such effects are, and how well they can be determined in practice, requires experience with ALICE measurements of real LHC collisions. These corrections are not considered here.

From PYTHIA simulations $\Delta E_{jet}^{unobserved}$ is approximately 6-8% (see Fig. 5.3) with an assigned systematic uncertainty of 3%.

### 5.2.7 Hadronization

The process of hadronization introduces an uncertainty in the measurement of jet energy, as gauged on the partonic level. Hadronization corrections to jet measurements at RHIC and the Tevatron are made by comparing the PYTHIA and HERWIG hadronization schemes, and are assigned typical uncertainties of a few percent of the jet energy [7].

Hadronization in heavy ion collisions is known from RHIC data to differ dramatically from that in p–p collisions in the region $p_T < 5$ GeV/c [8, 9]. It is not yet known whether such an effect persists in the hadronization of the parton shower in a jet, but given the dramatic effects seen on the inclusive level it is unclear whether PYTHIA or HERWIG hadronization schemes apply to jet production in heavy ion collisions.

A better understanding of these effects requires ALICE measurements of real LHC events. Consequently, we choose not to make any hadronization correction in ALICE jet measurements at present, and do not assign systematic uncertainties due to hadronization.

### 5.2.8 Recombination scheme

Any sequential recombination algorithm combines pairs of kinematic objects (tracks, calorimeter towers or clusters, proto-jets) into a new, merged object. There is an ambiguity in forming the momentum four-vector of the merged object, depending upon whether energy and momentum are required to be conserved in the sum (“E-scheme”) or whether the merged object is required to be massless (“p-scheme”) [10].

Experience with STAR analysis suggests that differences due to the choice of recombination schemes are small compared to other uncertainties in the jet measurements. This will need to be studied with ALICE data, but we have not made estimates here of such effects.

### 5.2.9 Underlying heavy-ion event

The effects of the underlying event and its fluctuations generate the largest systematic corrections and uncertainties in the measurement of jet observables in the heavy ion environment. In a central Pb–Pb collision at 5.5 TeV (unquenched HIJING), a cone of radius $R$=0.4 contains on average 0.2 TeV of background energy, with a complex (and *a priori* unknown) fluctuation spectrum that is correlated on multiple length scales. The effect of the fluctuations depends on the observable being measured, and cannot be accounted for simply in terms of a correction to the overall jet energy scale. The fluctuation spectrum must be accurately characterized and unfolded from the measured distribution (which we illustrate in this section using the inclusive differential cross section). Given the complexity of this procedure, it is necessary to characterize the background and its effect on the jet spectrum with multiple, independent techniques.

Schematically, the correction for the underlying background can be written as:

$$E_{jet} = E_{jet}^{rec} - \rho \times A \pm \sigma \times \sqrt{A}. \tag{5.3}$$

Operationally, $\rho$ is measured on an event-wise basis and is used to correct the energy of each jet candidate. $\sigma$ characterizes the fluctuations about $\rho$, and its effects on the measured distribution of the observable can only be corrected on an inclusive basis via unfolding. The dependence of fluctuations on $\sqrt{A}$, where $A$ is the jet area, is likewise schematic since it implies an uncorrelated background, which may not apply in practice. We comment on these points further below.

| **R** | **$\langle\rho\rangle$ [GeV/unit area]** | |
|---|---|---|
| | Random Cones | FastJet $k_T$ |
| 0.2 | 198 ± 1.6 | 202 ± 1.7 |
| 0.3 | 200 ± 1.6 | 202 ± 1.7 |
| 0.4 | 199 ± 1.6 | 203 ± 1.7 |

**Table 5.2:** Median background energy density $\rho$ in central unquenched HIJING Pb–Pb events for different resolution scales $R$, determined by various algorithms (statistical errors only). See text for details.

The background energy density, $\rho$, for a given event is determined by applying a jet finding algorithm from the FastJet suite ($k_T$ is most commonly used [11]) or by the average energy from non-overlapping randomly distributed cones of fixed radius. For each event the algorithm returns a set of jet energies $\{E_i\}$ and areas $\{A_i\}$, where $i$ enumerates the jets in the event. We define the set of ratios $\{r_i\}$, where $r_i = E_i/A_i$. Following the FastJet approach [11], a robust estimator of the background energy per unit area in the event is obtained by

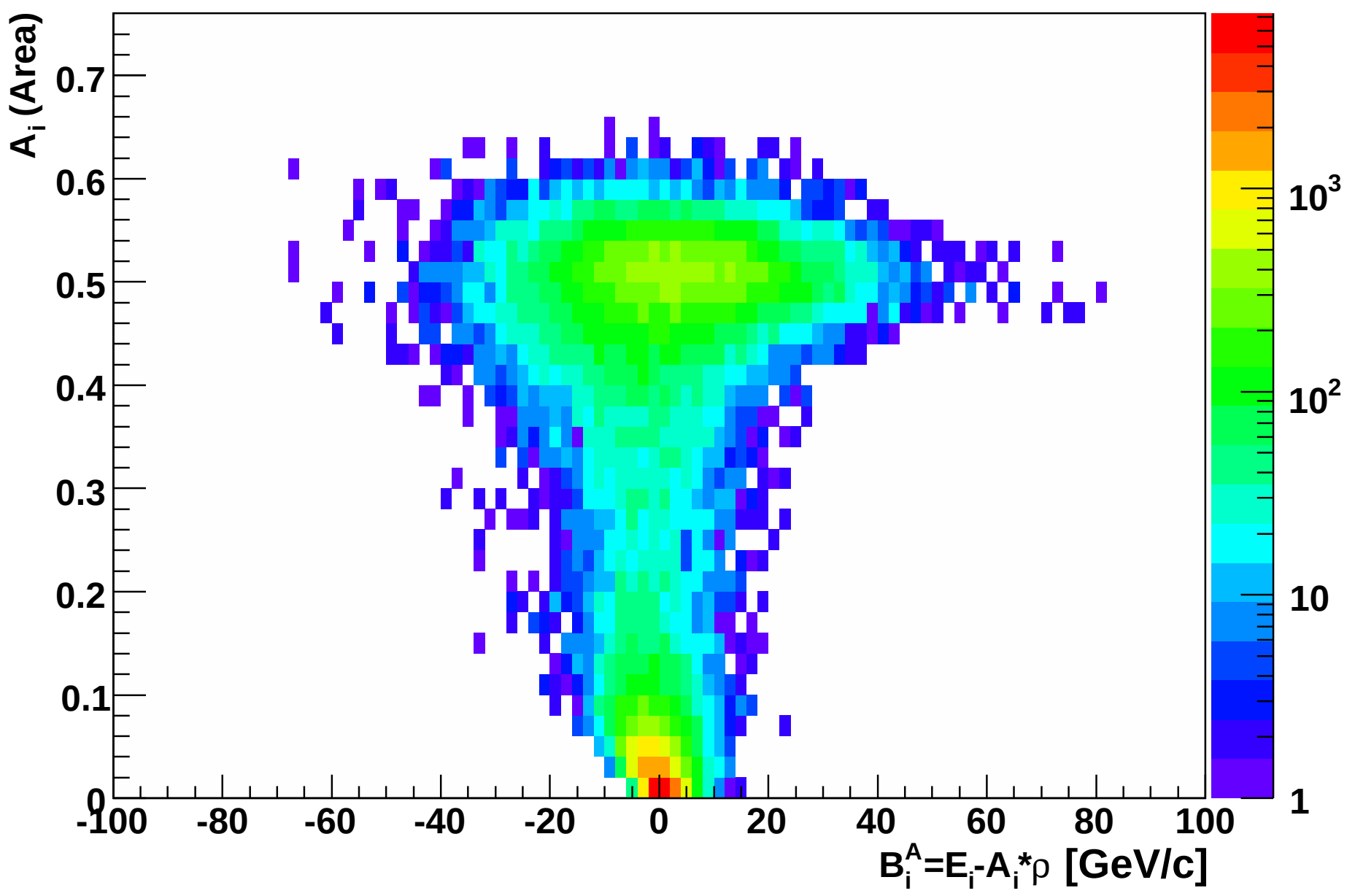


**Figure 5.5:** $B_i^A$ for central 5.5 TeV Pb–Pb events, calculated with HIJING (unquenched).

the median of the $r_i$ distribution:

$$\rho = median\{r_i\} \tag{5.4}$$

Table 5.2 lists the median background energy density $\rho$ within the EMCal acceptance for central (unquenched) HIJING events for various resolution scales (or cone radii) $R$, utilizing different algorithms. The random cones sample the event uniformly, whereas the $k_T$ algorithm clusters according to a distance and energy-weighted metric, generating different areas and shapes and leading to a non-uniform sampling of the event. The methods agree within $\sim 4$ GeV/unit area for the range of $R \sim 0.2 - 0.4$.

The distribution of the fluctuations around the event-wise estimate of background energy density is defined as:

$$B_i(A_i) = r_i - \rho \tag{5.5}$$

The FastJet algorithms utilize the $B_i$ distribution to characterize the background fluctuations by extracting the width of the second quartile of the distribution. This is intended to suppress the hard tail due to hard scattering. In practice, the background energy density is always scaled by jet area, so the spectrum of background fluctuations is seen most clearly via:

$$B_i^A(A_i) = E_i - \rho A_i, \tag{5.6}$$

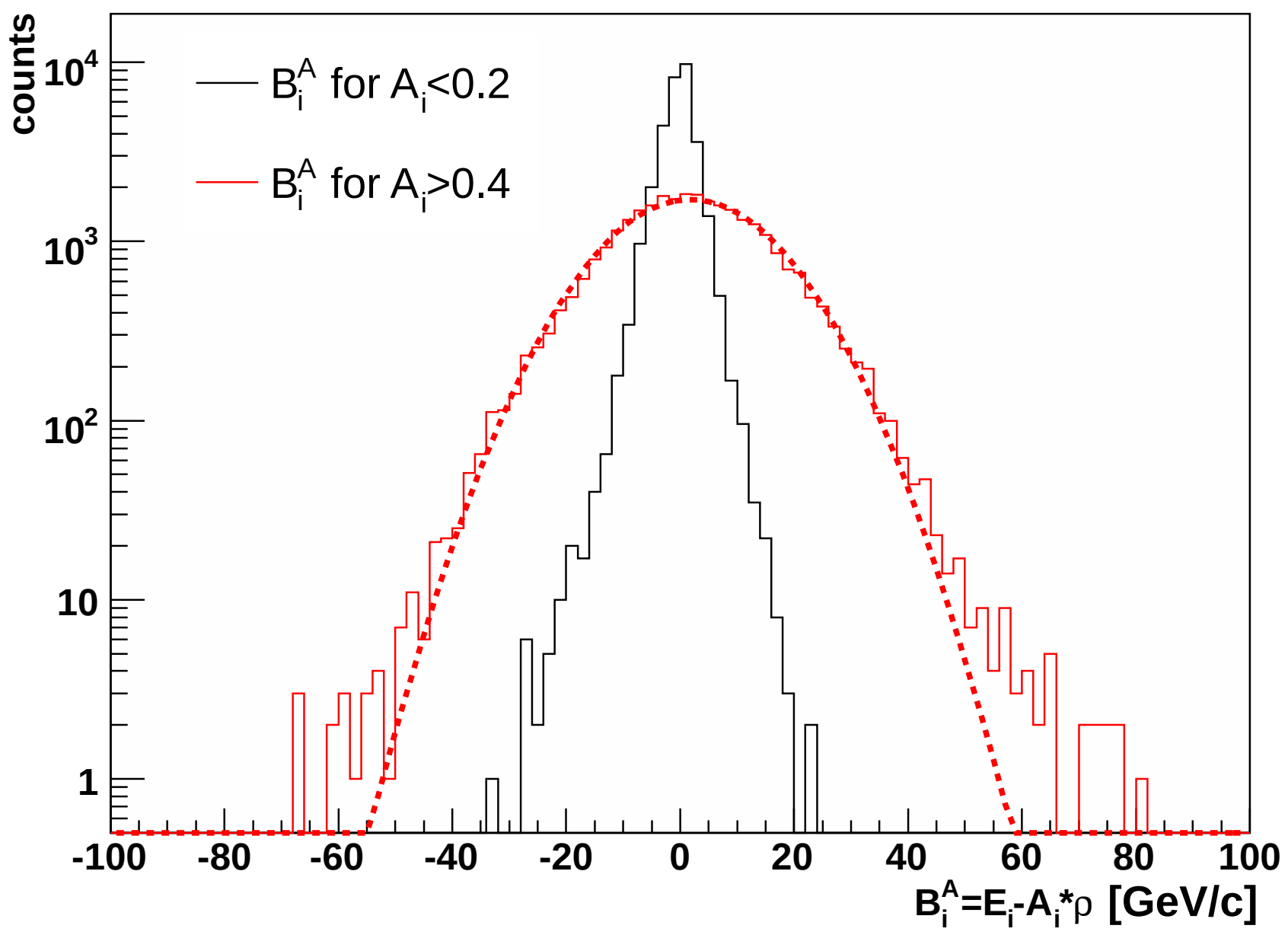


**Figure 5.6:** $B_i^A(A_i)$ for central Pb–Pb events (HIJING), for two separate ranges of jet area.

where the subscript $i$ runs over all jet candidates found in each event. Figure 5.5 shows the distribution of $B_i^A(A_i)$ as a function of jet area, for central 5.5 TeV Pb–Pb collisions simulated by HIJING.

Fig. 5.6 shows the $B_i^A(A_i)$ for two separate ranges of jet area. The Gaussian function fit to the larger area selection ($A_i > 0.4$, corresponding roughly to $R = 0.4$) shows that the $B_i^A$ distribution is approximately Gaussian, with an asymmetric tail towards the right that arises from hard scatterings present in the simulated HIJING events. The presence of such an asymmetric tail can have significant influence on the shape of the inclusive jet spectrum.

While the $B_i^A$ distribution can be used as the data-driven input to the unfolding procedure for correcting the inclusive jet spectrum, in practice such a procedure requires care to ensure a stable and robust result. An alternative approach, based on current STAR analysis, is to embed a spectrum of simulated jets into a background distribution of otherwise-unbiased central Pb–Pb events, and extract the smearing parameter required for the unfolding. While this approach depends on modeling the jet fragmentation, and is thereby less "data-driven" than applying the $B_i^A$ distribution directly, its model dependencies can nevertheless be checked by varying the fragmentation model and including jet quenching effects. In this report we use the correction derived from PYTHIA embedding, and continue to develop our understanding of unfolding corrections based directly on the $B_i^A$ distribution.

Fig. 5.7 shows the $B_i^A$ distribution for central Pb–Pb collisions, determined by several methods:

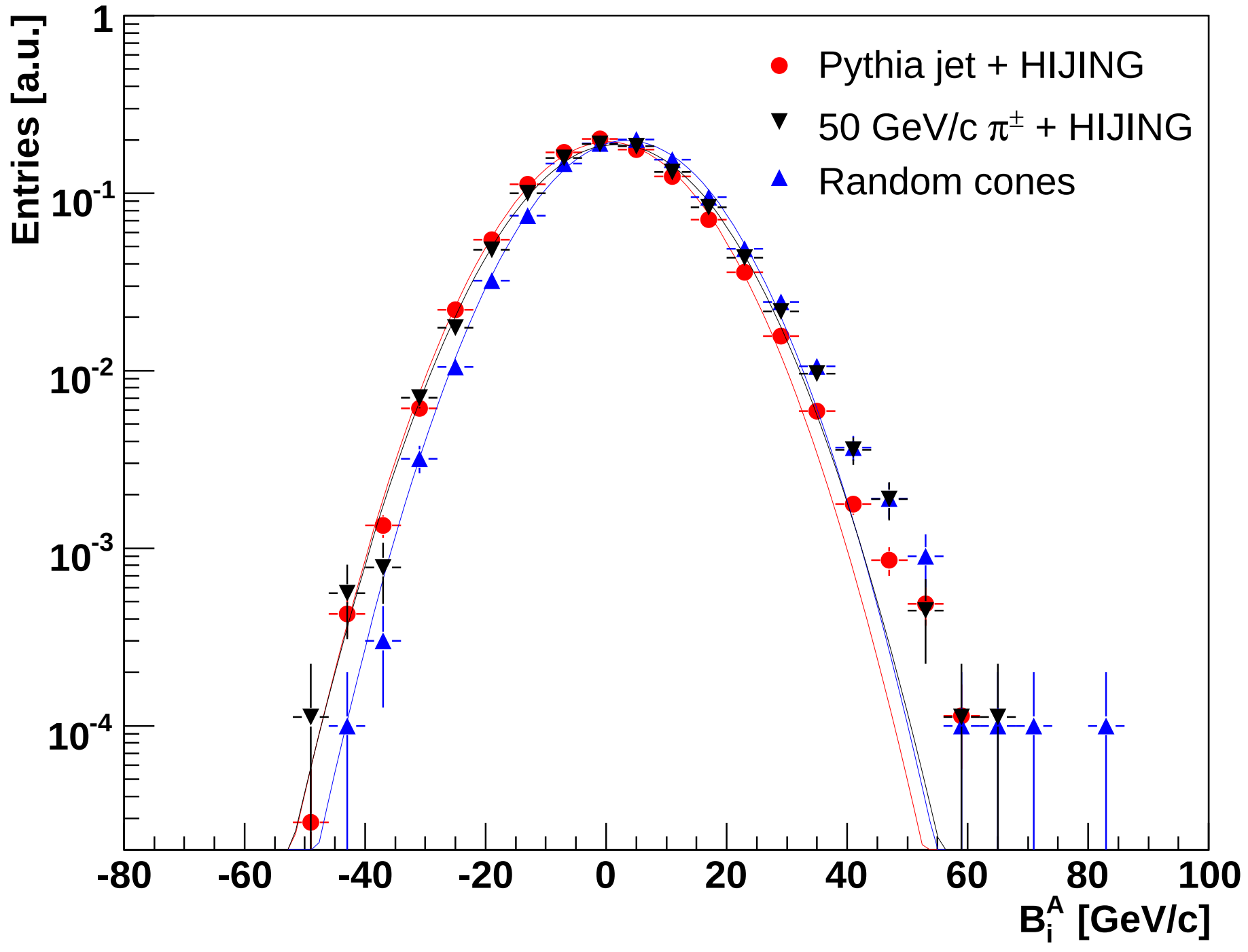


**Figure 5.7:** Comparison of $B_i^A$ distribution in central Pb–Pb collisions (HIJING), determined by different methods: embedding of single 50 GeV/c pions, embedding of single PYTHIA jets, randomly distributed cones.

| Method | $\sigma$ GeV/c |
|---|---|
| Random Cones | 11.8 ± 0.1 |
| Single $\pi$ | 12.5 ± 0.1 |
| PYTHIA jet | 12.2 ± 0.05 |

**Table 5.3:** Gaussian $\sigma$ from a fit to the $B_i^A$ distributions for different methods: embedding of single 50 GeV/c pions, embedding of single PYTHIA jets and randomly distributed cones (see Fig. 5.7). Statistical errors only.

- the embedding of single high $p_T$ pions (50 GeV/c) into HIJING-generated background events, jet reconstruction using FastJet anti-$k_T$ $R = 0.4$, and comparison to input momentum;
- same as first bullet but using a single PYTHIA jet instead of a single pion;
- distributing non-overlapping cones with $R = 0.4$ randomly in the event and integrating the energy in each of them.

The distributions are approximately Gaussian, with an asymmetric hard tail to the right. Widths from fits of Gaussian functions to each distribution are given in Table 5.3. The widths, which characterize the background fluctuations measured by the different methods, vary by less than 1 GeV/c. This variation provides a systematic uncertainty estimate for the value of $\sigma$ used to unfold the inclusive jet spectrum for effects due to background fluctuations.

## 5.3 Inclusive jet spectrum

To illustrate the influence of these systematic effects on physics observables, we now discuss in detail the measurement of the inclusive jet production cross section in both 5.5 TeV p–p and Pb–Pb collisions, and estimate their systematic uncertainties based on the foregoing discussion.

### 5.3.1 Deconvolution of fluctuations

A significant systematic uncertainty in the measurement of the inclusive jet cross section in heavy ion collisions arises from the large underlying background and its fluctuations, as discussed above. Correction for this effect is applied on an inclusive basis, via a deconvolution (or "unfolding") procedure. The effect of finite jet energy resolution due to instrumental effects, also present in p–p collisions, is likewise corrected via deconvolution.

Schematically, the distortion of the true inclusive jet spectrum, $d\sigma(E_{Jet})/dE_{Jet}$, can be expressed via smearing functions due to the instrumental jet energy resolution $g_{instr}$, unobserved neutral energy $h_{\Delta E^{unobserved}}$ and background fluctuations $f_{bkgd}$:

$$\frac{d\sigma(E_{Jet})}{dE_{Jet}} = \frac{d\sigma(dE'_{Jet})}{dE'_{Jet}} \otimes \left(f^{-1}_{bkgd} \otimes g^{-1}_{instr} \otimes h^{-1}_{\Delta E^{unobsoreved}}\right), \tag{5.7}$$

where $d\sigma/dE'_{Jet}$ is the measured jet cross-section at energy $E'_{Jet}$, corrected for hadronic/electron double counting and tracking efficiency. The smearing matrix $f_{bkgd}$ characterizes the background and does not depend on jet energy.

Figure 5.7 and Table 5.3 indicate that the background fluctuations in a central Pb–Pb collision at 5.5 TeV can be characterized by a Gaussian function with width $\sigma \sim 12$ GeV/c. Fig. 5.8 illustrates the unfolding procedure based on this background model, for a spectrum of jets at 5.5 GeV generated by PYTHIA embedded into central Pb–Pb events generated by HIJING. The unfolded (corrected) spectrum is within 20% of the generated spectrum over a broad jet energy range. The shaded band indicates the variation in the unfolded spectrum with a variation of the background fluctuation estimate of $\sigma = 12.0 \pm 1.5$ GeV/c, which brackets the width determined by the different methods, given in Table 5.3.

It is notable that the spectrum deviations arising from variations in the background fluctuation parameterization is only a few percent at 150 GeV/c and above. This may initially be surprising, since the background energy density is large. However, distortion of the spectrum is driven by the background fluctuations, which are estimated to be small relative to the total jet energy. When combined with the relatively "flat" spectral shape, due to the large LHC collision energy, the net result is only a mild spectrum distortion at very high $p_T$, enabling a precise measurement of the jet cross section even in the presence of complex heavy ion event background.

### 5.3.2 "False jet" rates

Fluctuations of soft processes in a heavy ion collision may generate accidental correlations in the final-state distribution of hadrons which mimic the morphology of jets. The correction for background fluctuations should in principle remove such false jets. To the extent that this is not fully accomplished, the inclusive jet spectrum will contain a background contribution from soft processes that does not originate in hard QCD scatterings. In order to assess the rate of such "false jets" we apply two independent techniques.

The first technique randomizes each event to destroy all phase space correlations and reruns the jet finder. Since high $p_T$ tracks (above $p_T \sim 3 - 5$ GeV/c) are themselves likely to be jet fragments and jets built around them would not be entirely "false", an iterative step can be taken to remove the leading particle of each jet from such a randomized event, or to

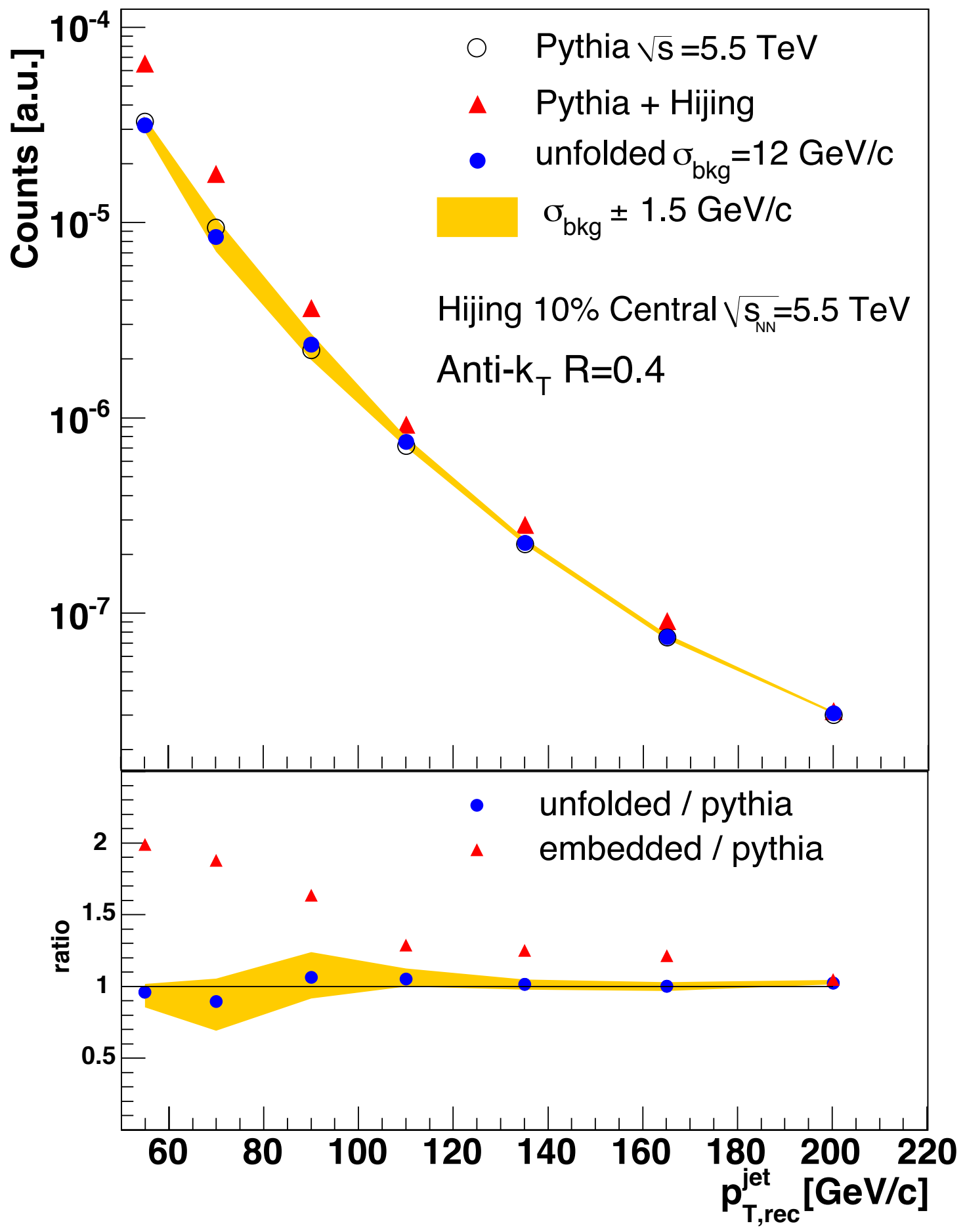


**Figure 5.8:** Spectrum distortion due to background fluctuations and correction via unfolding. Black: inclusive jet cross section from 5.5 TeV p–p collisions from PYTHIA. Blue: jet population embedded into central 5.5 TeV Pb–Pb events from HIJING. Red: embedded jet population after unfolding with background fluctuation width $\sigma = 12$ GeV. Yellow band indicates systematic variation in unfolded distribution due to variation in background fluctation $\sigma = 12 \pm 1.5$ GeV. Lower panel shows ratio of both smeared and unfolded spectra to the undistorted PYTHIA spectrum.

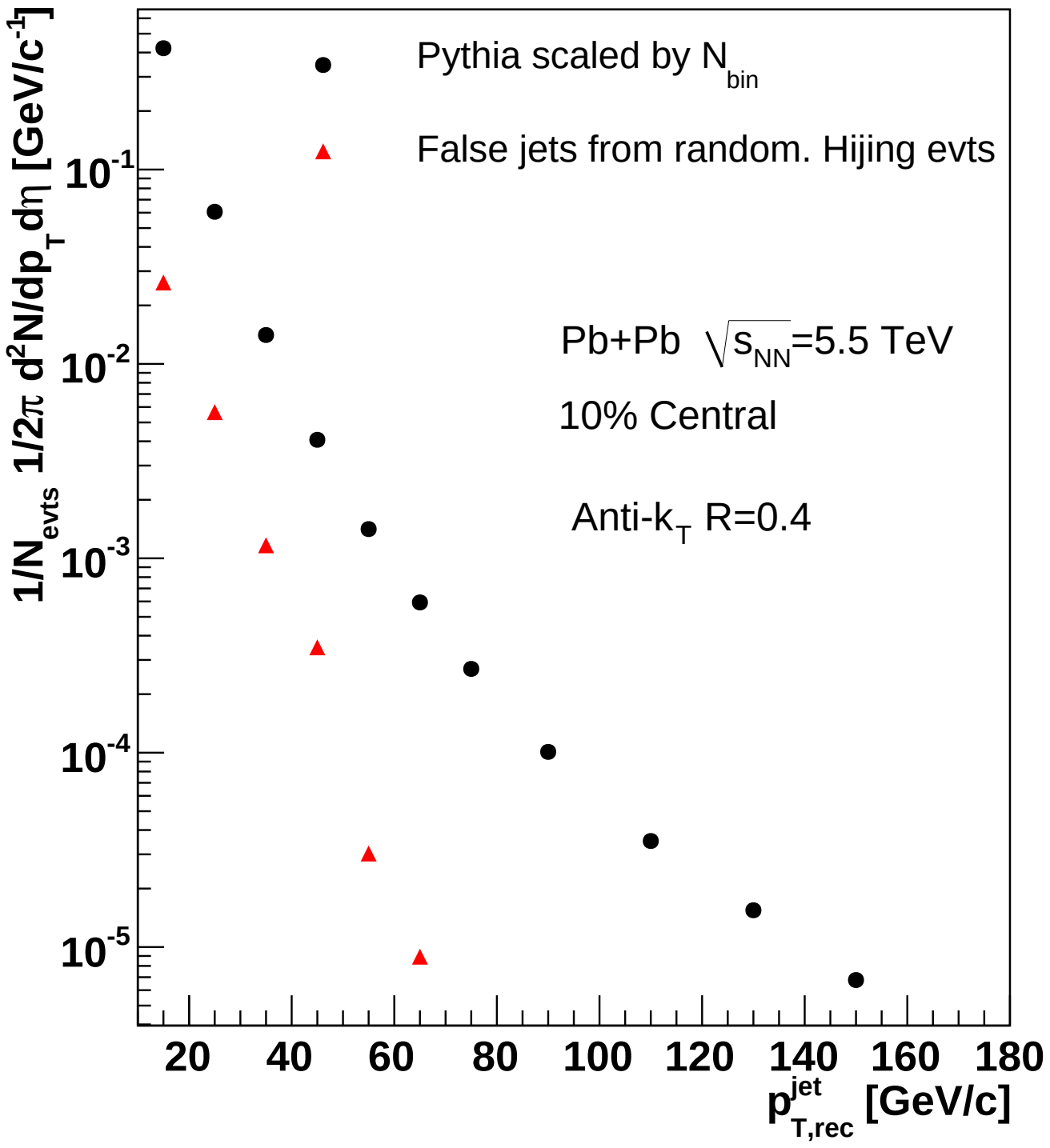


**Figure 5.9:** "False jet" spectrum in 5.5 TeV Pb–Pb events estimated via randomized central HIJING events (see text), compared to signal jet spectrum corresponding to $N_{bin}$-scaled 5.5 TeV p–p events from PYTHIA.

| **Systematic effect** | **Incl. cross section sys. uncert.** |
|---|---|
| **Common in p–p and A–A** | |
| *Tracking distortions (space charge etc.)* | unknown |
| *Tracking efficiency* | 1% |
| *Hadronic and electron energy double counting* | 3-4% |
| *EMCal energy scale* | 8-10% |
| *Unobserved neutral energy* | 13-15% |
| **Underlying event (central Pb–Pb)** | |
| *Fluctuations* | 20% (75 GeV/c), 3% (150 GeV/c) |
| *False Jets* | small (>50 GeV/c) |

**Table 5.4:** Estimated systematic uncertainties for the measurement of inclusive jet cross sections.

remove all charged particles and calorimeter clusters above some threshold. The inclusive jet spectrum from this procedure is shown in Fig. 5.9. The jet yield in randomized events falls well below the measured yield, indicating that the rate of such "false jets" arising purely from accidental correlations of background is negligible relative to the physical signal.

The second technique utilizes the back-to-back pairwise nature of jet production at leading order. In particular, we measure the rate of jets reconstructed in the EMCal in coincidence with a high-$p_T$ (charged) hadron trigger particle, measured in the ALICE tracking detectors. The true rate of this coincidence can be measured in p–p collisions. In central Pb–Pb collisions there will be additional coincidences due to multiple hard scatterings in the event, whose rate can be estimated from the inclusive jet production cross section. Any yield in excess of this rate is attributable to "false jets". The two methods should provide consistent false jet rate estimates.

### 5.3.3 Summary of systematic uncertainties

Table 5.4 presents systematic uncertainties specific to the inclusive spectrum measurement. In this report we do not consider more general systematic factors such as precision of integrated luminosity and finite trigger efficiency, which we expect will contribute an additional ~10% to the overall normalization uncertainty in cross section measurements.

Some systematic uncertainties, in particular those due to tracking distortions, cannot be accurately estimated using simulations and require experience with real ALICE data.

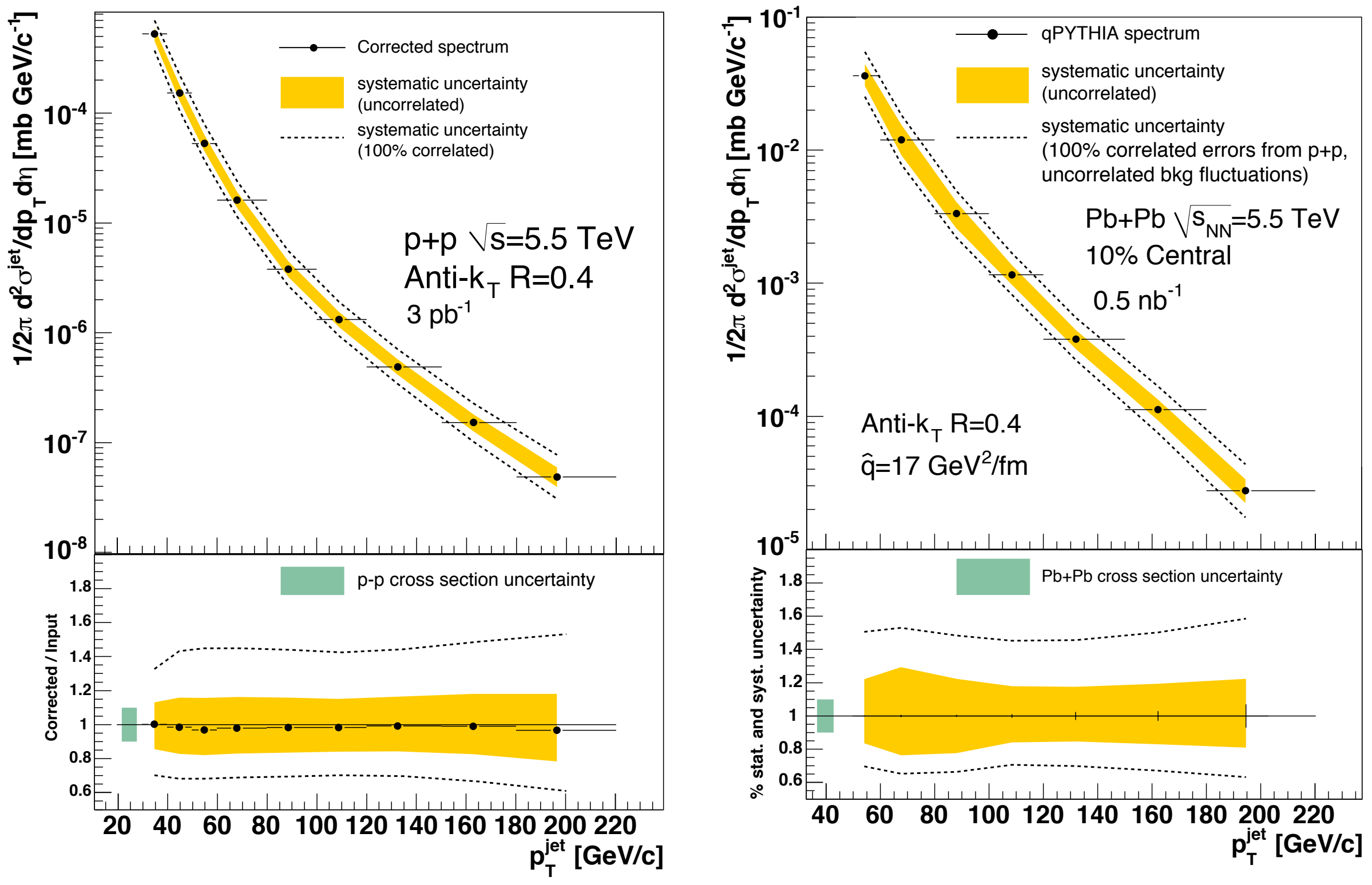


**Figure 5.10:** Final inclusive jet production cross section with systematic uncertainties (see text). Left: 5.5 TeV p–p. Right: 5.5 TeV central Pb–Pb.

### 5.3.4 Final result: inclusive jet spectra

Fig. 5.10 shows the final, corrected inclusive jet spectra in both p–p and Pb–Pb collisions at 5.5 TeV, with systematic uncertainties. Two uncertainty bands are given. The inner band corresponds to the assumption that all uncertainties are uncorrelated and are added in quadrature. The outer band is based on the conservative assumption that all uncertainties are highly correlated and should be added linearly, giving a measurement precision of about a factor two. We expect the achievable systematic uncertainty to be closer to the uncorrelated case.

## 5.4 Jet quenching measurements with the EMCal

We now return to the predictions for jet quenching effects at the LHC presented in Section 4.5.2, in particular jet $R_{AA}$ and the ratio of inclusive cross sections $d\sigma(R = 0.2)/d\sigma(R = 0.4)$. We assess the sensitivity of ALICE measurements with the EMCal to quenching effects in these observables, in light of the systematic uncertainties outlined in the previous section.

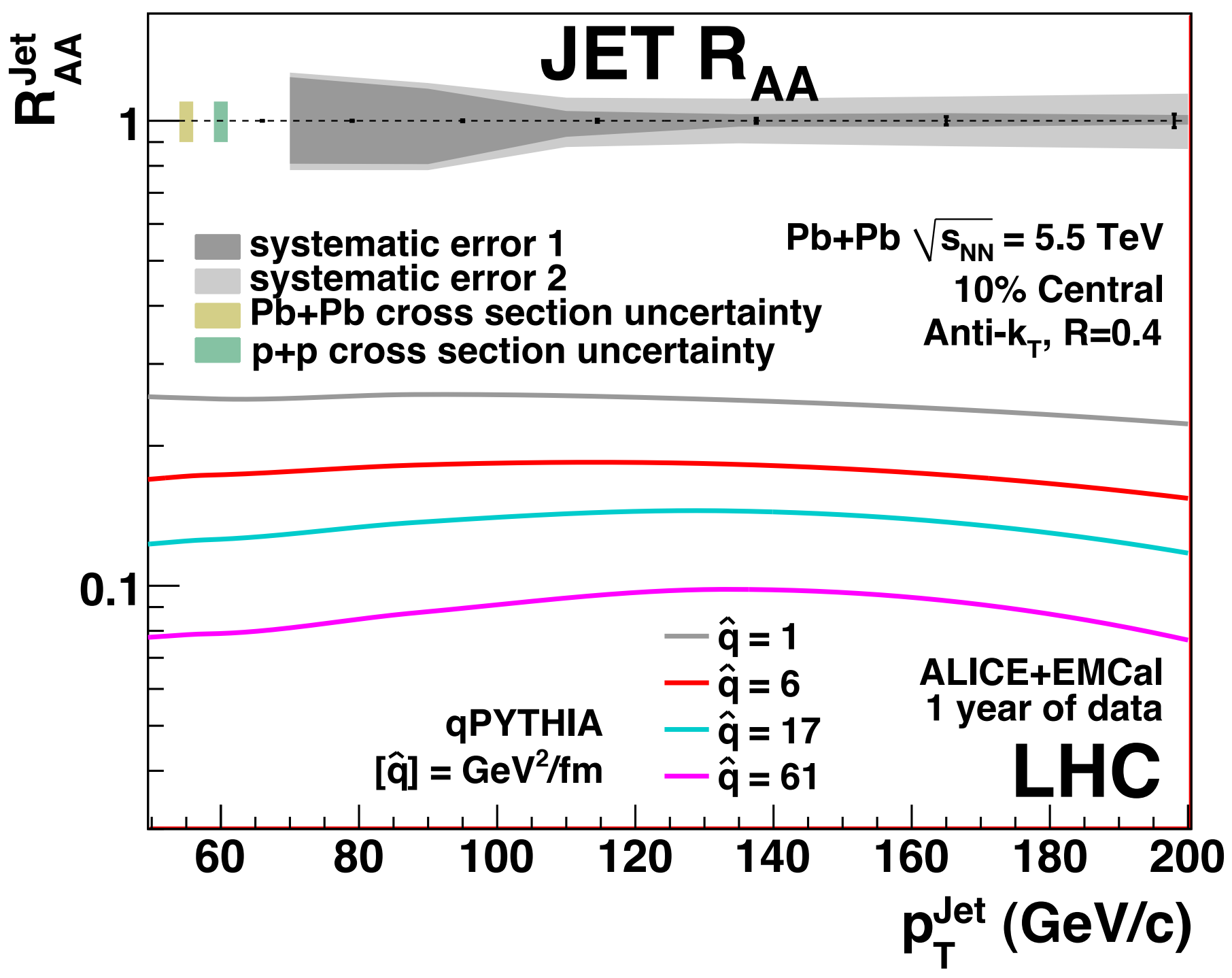


**Figure 5.11:** Jet $R_{\rm AA}$ for central Pb–Pb normalized by p–p for various $\hat{q}$ calculated by qPYTHIA. Systematic uncertainties in the ALICE measurement are centered at unity. For details see text.

Section 5.5 discusses additional quenching measurements possible with ALICE+EMCal.

### 5.4.1 Jet $R_{\rm AA}$

Figure 5.11 shows the qPYTHIA calculations from Fig. 4.11, left panel, for jet $R_{\rm AA}$ in central Pb–Pb collisions for various $\hat{q}$. The figure also shows the estimated statistical errors and systematic uncertainties for an ALICE+EMCal measurement from one heavy ion running period. (We assume that statistics for p–p data will be larger than for Pb–Pb. The bulk of p–p data measured by ALICE will be at larger $\sqrt{s}\sim\ 10-14$ TeV, which will require additional corrections for a $R_{\rm AA}$ measurement that we do not address here.)

The statistical errors are smaller than the systematic uncertainties, except for the highest $p_T$ bin. Several types of systematic uncertainty are specified. The p–p cross section uncertainty (10%) estimates the combined uncertainties in p–p integrated luminosity and trigger efficiency, as well as the uncertainty in Glauber scaling $\langle T_{AA}\rangle$. The Pb–Pb cross section uncertainty (10%) estimates the combined uncertainties in Pb–Pb integrated luminosity and trigger efficiency. The dominant systematic effect at jet energies below 120 GeV/c is the uncertainty in the background fluctuation estimate in the heavy-ion measurement. "Systematic error 1" is $p_T$-dependent and based on the assumption that all additional systematic

uncertainties cancel in the ratio of the two cross sections except for charged tracking efficiency, which is expected to differ between p–p and central Pb–Pb. In particular, the EMCal calibration is assumed to be the same for the two datasets. “Systematic error 2” is the same as “1” except for the assumption that the EMCal calibrations are uncorrelated between the p–p and Pb–Pb datasets, with a scale uncertainty in each case of 5%. We expect these assumptions to bracket the achievable systematic uncertainties.

Additional uncorrelated uncertainties may be present, for instance in the correction for unobserved neutral hadrons: the well-known baryon enhancement observed in RHIC heavy-ion collisions could lead to different systematics in the correction for unobserved neutral hadrons in Pb–Pb and p–p collisions. This effect is however impossible to assess quantitatively without real ALICE data.

The figure shows the state of the art in jet quenching modeling. While the $\hat{q}$-dependence of jet $R_{\rm AA}$ predicted by qPYTHIA requires further study, comparison with the systematic uncertainties achievable by ALICE+EMCal shows discriminating power in this observable. We refrain from extracting a specific numerical resolution in $\hat{q}$ from the figure, however, which would only parameterize (and possibly misrepresent) what is evident visually. The measurement and calculation in the figure are both at the forefront of jet quenching research, and we expect this study to spur additional investigation into the physics of this observable and the precision with which it can be measured.

### 5.4.2 Jet Broadening

Figure 5.12 shows the qPYTHIA calculations from Fig. 4.12, left panel, for the ratio of inclusive jet cross sections $d\sigma(R = 0.2)/d\sigma(R = 0.4)$ in central Pb–Pb collisions for various $\hat{q}$. The figure also shows the projected statistical errors and systematic uncertainties for an ALICE+EMCal measurement from one Pb–Pb running period.

As discussed above, this observable is a measurement of the jet energy profile. Since it is the ratio of two jet cross sections measured in the same dataset, many systematic uncertanties are expected to cancel, such as those due to integrated luminosity and EMCal calibrations. Other uncertainties will be highly, if not entirely, correlated, such as tracking efficiency and the correction for the unobserved neutral energy. Overall, we expect the systematic uncertainty of this observable to be markedly smaller than that of $R_{\rm AA}$, and dominated by the difference in background fluctuations due to the different jet areas for R=0.2 and 0.4, as shown by the shaded band in Fig. 5.12. The small uncertainty due to background fluctuations at jet energies $\sim$ 150 GeV/c and above will enable a precise measurement of the jet energy profile in terms of this ratio of cross sections.

As is apparent in the figure, qPYTHIA suggests a marked change in the jet energy profile from vacuum to moderate $\hat{q}$=6 GeV$^2$/fm, with little additional suppression in the ratio for larger $\hat{q}$. If this picture persists with further study and in other calculations, even such a

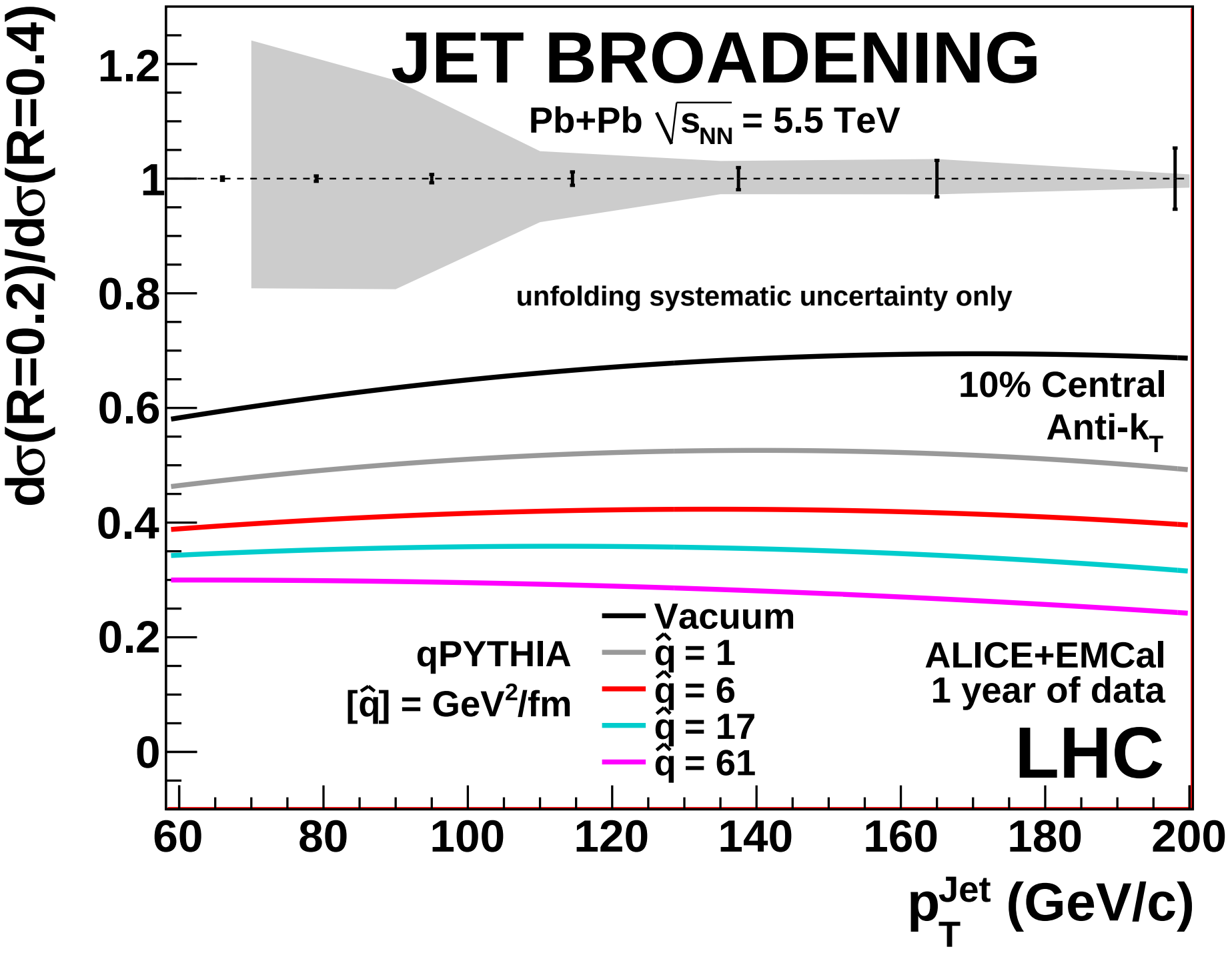


**Figure 5.12:** Ratio of inclusive jet differential cross sections $d\sigma(R = 0.2)/d\sigma(R = 0.4)$ for various $\hat{q}$, calculated by qPYTHIA. Systematic uncertainties in the ALICE measurement are centered at unity.

precise measurement as described in this section will have limited sensitivity to $\hat{q}$. We note, however, that the qPYTHIA predictions shown in the figure require additional investigation, taking into account the current disagreement with RHIC measurements (Fig. 4.9). This is a very active area of study, and the qPYTHIA results presented in Fig. 5.12 should be regarded as a work in progress.

## 5.5 Discussion of additional jet quenching measurements

The analysis of inclusive jet spectra in the previous sections demonstrates in detail the proposed correction scheme for jet analysis, and its resulting systematic uncertainties. With these well-controlled reference measurements and good understanding of corrections for background fluctuations caused by the underlying heavy-ion event, a variety of additional jet quenching observables become accessible.

### 5.5.1 Jet energy profile

Two additional measurements that are now readily accessible, given the understanding of inclusive spectrum analysis, are the jet broadening observable $\psi(r;R)$ (see section 4.5.2), and $R_{AA}^{Jet}(R_{max},\omega_{min})$ [12], measured as function of $R_{max}$ and $\omega_{min}$ (the minimum energy cut off for charged particles and neutral towers). ALICE is well suited to measure the $\omega_{min}$ dependence down to very low $p_T$, where jet-quenching effects are expected to be strongest, allowing a systematic study of this quantity over a broad phase-space.

### 5.5.2 Modified Fragmentation Function

The modification of the jet fragmentation function $\xi = \ln 1/z$ with $z = p_t^{hadron}/E_{Jet}$ due to partonic energy loss has been proposed as a key signature of jet quenching at the LHC [13].

Recently STAR made the first measurement of a modified fragmentation function (FF) [14] in the recoiling jet of a di-jet pair, designed to avoid bias in the fragmentation function caused by the online trigger. Fig. 5.13 shows this measurement, specifically the ratio of the measured FF in 200 GeV central Au–Au with respect to the p–p distribution. For high jet energies ($>$ 25 GeV/c), no significant modification in the di-jet FF is observed. This observation may be attributed to the strong jet broadening, resulting in a significant out-of-cone energy deposition with respect to p–p measurements, leading to an underestimation of the measured jet energy in heavy-ion collisions (see similar discussion of jet $R_{AA}$ at LHC energies in section 5.4). We note that this effect may complicate the interpretation of the FF measurements in heavy-ion collisions in general.

Trigger biases in the charged particle fragmentation can be minimized by studying the FF at sufficiently high jet energies (>100-120 GeV, see trigger chapter). Correction for detector effects and the jet energy scale uncertainty due to fluctuations of the underlying event require an unfolding procedure, similar to the inclusive spectrum. Resolution effects due to high-$p_T$ tracking must also be taken into account. While further studies are needed to assess the systematics of this measurement, we expect that the impact of background fluctuations on the FF measurement will be moderate for jet energies above 125 GeV/c, due to the hard jet spectrum at the LHC (see Fig. 5.8).

Overall, the FF measurements combined with the inclusive jet spectrum measurements will constrain quenching model calculations significantly, by measuring the internal jet structure in heavy-ion collisions in detail. In addition, particle-identified FF in heavy-ion collisions can be used to investigate novel hadronization effects in quenching.

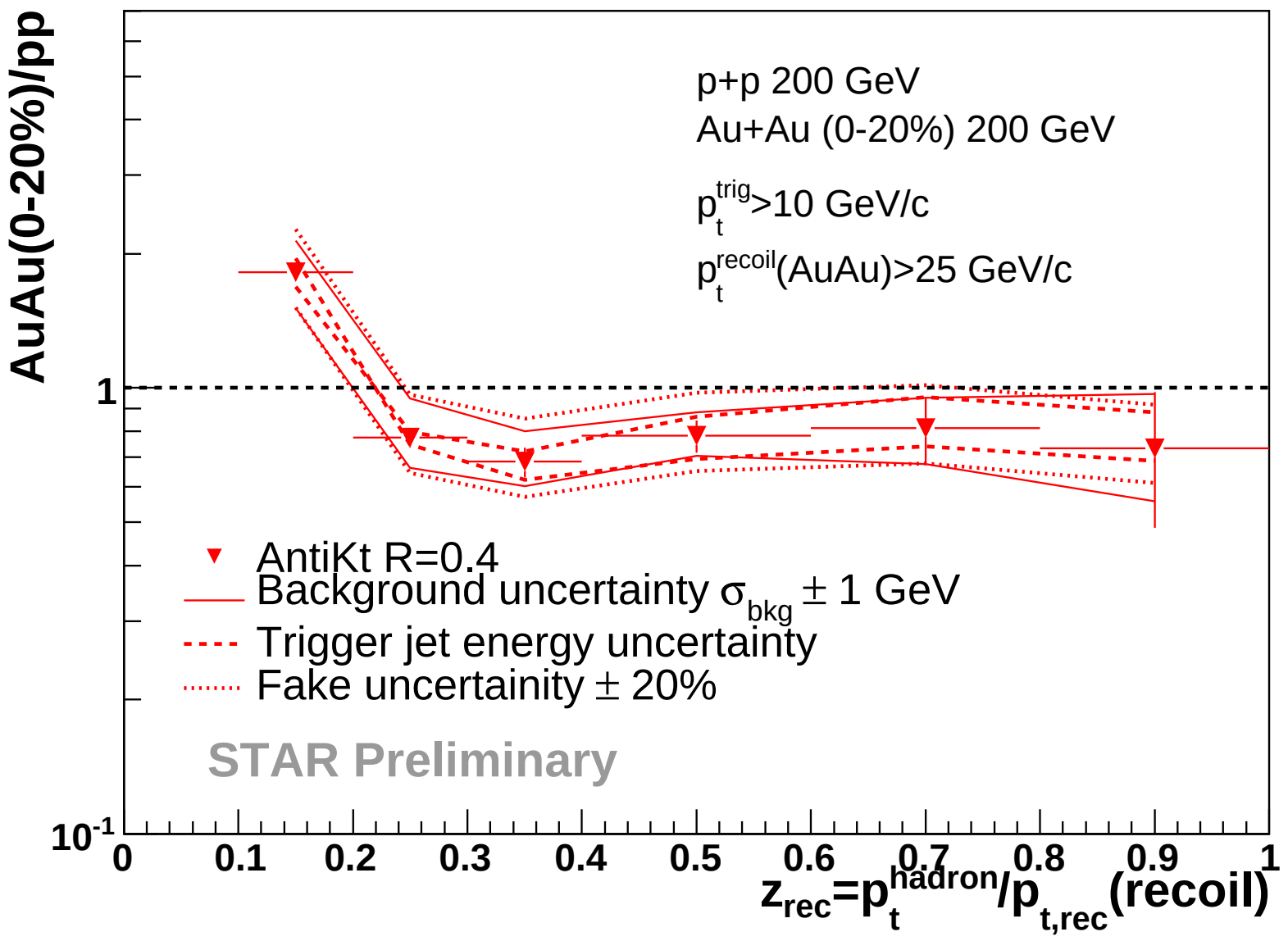


**Figure 5.13:** Recoil fragmentation function ratio Au–Au over p–p as a function of $z$ for $p_T^{recoil} > 25$ GeV/c (STAR) [14].

### 5.5.3 Sub-jets

A potentially robust measurement of the internal jet structure and its modification due to partonic energy loss utilizes the *n-jet fraction* or *sub-jet population* [15], which probes the modification of jet structure on various resolution scales.

### 5.5.4 Out-of-cone radiation

To further constrain and estimate the out-of-cone energy caused by jet quenching one can utilize jet-hadron correlations. The idea is to select highly biased jets in the EMCal, by requiring a highly energetic neutral particle and further applying a "high" $p_T$ cut on particles and towers ($p_T^{cut}$) to suppress background fluctuation contributions. Jets reconstructed in this way are expected to be biased towards the surface and/or small energy loss, allowing a direct comparison to p–p equivalent jets with reduced systematic uncertainties. With this jet selection one constructs the $\Delta\phi$ correlation of associated particles with respect to the reconstructed jet axis with different kinematic selections of the associated particles. The background is subtracted statistically.

Preliminary results of jet-hadron correlations from the STAR collaboration for $0.2 < p_T^{assoc} < 1$ GeV/c and $p_T^{assoc} > 2.5$ GeV/c for anti-$k_T$ jets ($R = 0.4$ and $p_T^{cut} > 2$ GeV/c) with a reconstructed energy above 20 GeV/c for central Au–Au and p–p collisions are shown in Fig. 5.14 [16]. The focus in this analysis is on the recoil side ($\Delta\phi = \pi$). One can clearly see a softening

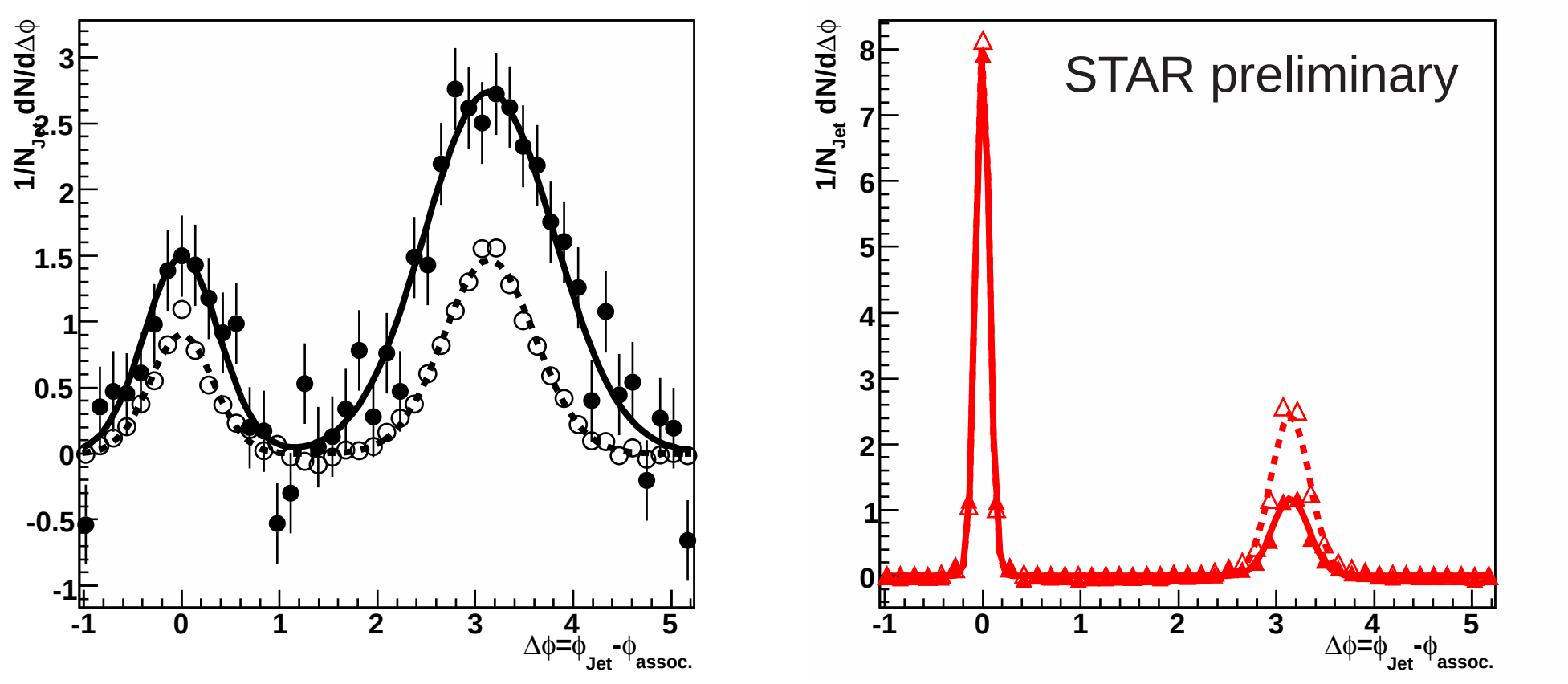


**Figure 5.14:** Jet-hadron correlations for 0.2< $p_T^{assoc}$ <1 GeV/c (left panel) and $p_T^{assoc}$ >2.5 GeV/c (right panel) for anti-$k_T$ jets ($R = 0.4$ and $p_T^{cut}$ >2 GeV/c) with reconstructed jet energy > 20 GeV/c for central Au–Au collisions (solid symbols) and p–p collisions (open symbols) (STAR) [16].

and broadening of the recoil jet for lower $p_T^{assoc}$ and a suppression at higher $p_T^{assoc}$, reflecting the expectation of a modification in the fragmentation function due to partonic energy loss. In addition one can on average estimate the additional out-of-cone energy beyond the resolution scale $R$ used in full jet-reconstruction with respect to the p–p reference measurements. This estimate could be used to "correct" for the jet broadening in the inclusive jet spectrum as well as in the FF measurements, allowing an estimation of the potential jet absorption fraction. A natural extension of this analysis would be to measure the pathlength dependence of jet quenching by measuring the jet-hadron correlations with respect to the event plane for different centralities.

## 5.6 Summary

This chapter presented a detailed discussion of inclusive jet cross section measurements using ALICE+EMCal, including evaluation of the proposed correction scheme and estimates of the associated systematic uncertainties. The data-driven correction scheme is independent of particular fragmentation or background models, and can be equally applied to p–p and heavy-ion jet measurements.

A notable finding is that the "flatness" of the jet spectrum at LHC energies, when compared to the significantly steeper spectrum at RHIC, reduces significantly the systematic uncertainties in the inclusive cross section measurements for heavy ion collisions. This is seen in the strikingly small systematic error arising from uncertainties in the background fluctuation estimate for jet energies above 150 GeV/c. The good systematic precision in the measurements presented in this chapter will allow ALICE+EMCal to explore the interaction of jets with hot QCD matter in unprecedented detail.

# Chapter 6

# Photon and Neutral Pion Measurements

## 6.1 Introduction

The EMCal capability to detect and identify energetic photons plays a crucial role in measurements of direct photon and $\pi^0$ production in p–p and Pb–Pb collisions in ALICE. In this report, we use the following terminology: *decay* photons are those coming from long-lived particle decays, in particular, $\pi^0$ decays; all other photons are called *direct* photons. Direct photons are further categorized as either *thermal* photons, carrying temperetature information from the expanding collsion system, *prompt* photons from initial hard parton Compton scattering or quark annihilation:

$$\begin{aligned} Compton: g + q &\rightarrow \gamma + q, \\ Annihilation: q + \bar{q} &\rightarrow \gamma + g, \end{aligned} \tag{6.1}$$

and *fragmentation* photons, produced by jet fragmentation. The study of thermal photons is one of the main physics goals of the ALICE-PHOS detector.

Since the largest background for direct photon measurements comes from decay photons, direct photon analyses require an accurate measurement of neutral meson production and good discriminative power between single photon and $\pi^0$ clusters. This discrimination becomes challenging at high transverse momentum, when clusters from decay photons merge. An example of an interesting direct photon measurement that requires good photon/$\pi^0$ discrimination is the nuclear modification factor. There are several different possible predictions that modify the yield of direct photons in Pb–Pb collisions, as shown in Fig. 6.1 [8]. The predictions include cases where prompt photons are not modified by the plasma, but fragmentation photons are, as well as effects of isospin and of cold nuclear matter on the yield of direct photons.

As shown in the following sections, the EMCal is capable to reliably reconstruct the neutral

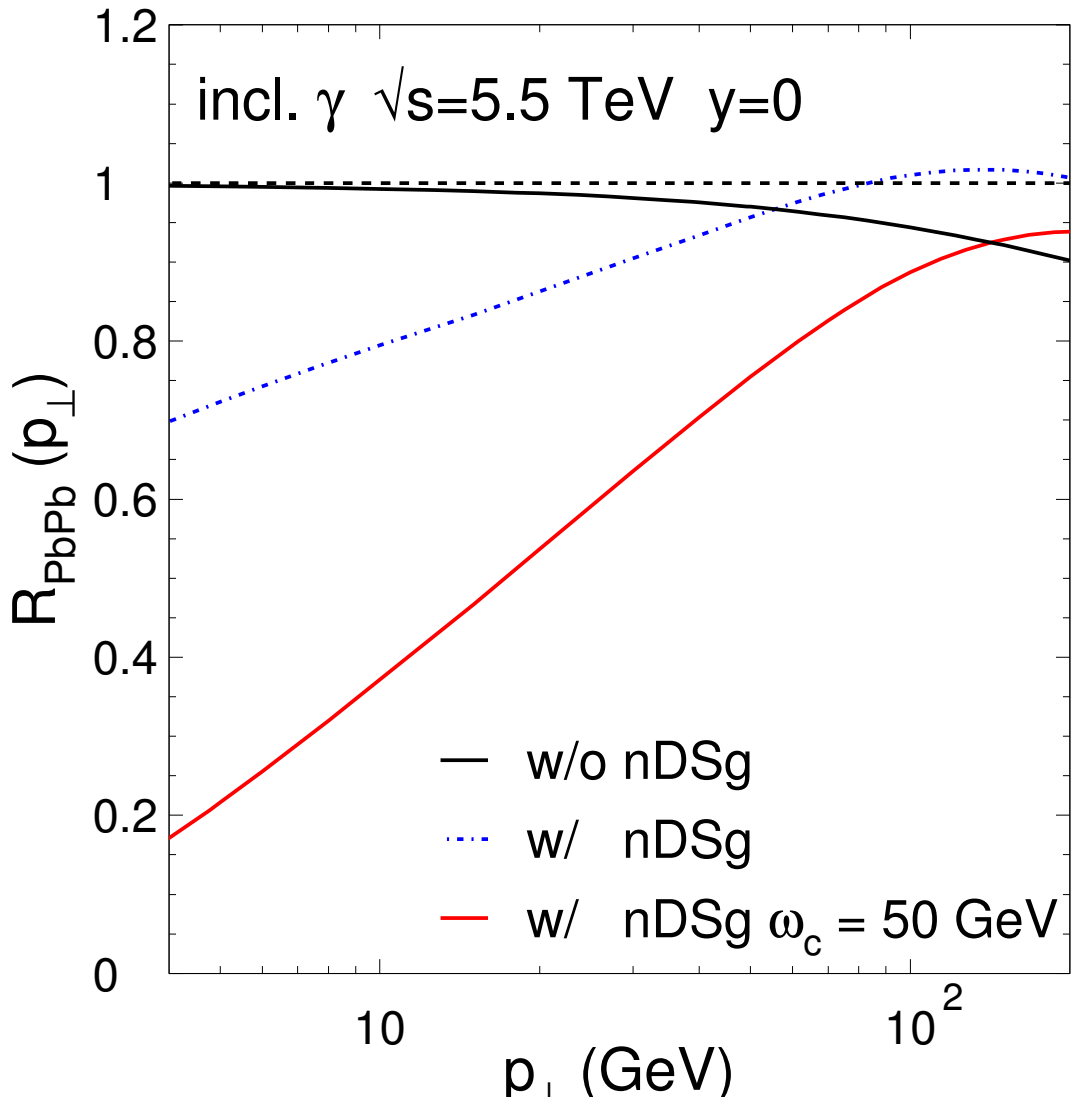


**Figure 6.1:** Nuclear modification factor as a function of transverse momentum for direct photon production at mid-rapidity in Pb–Pb collisions at $\sqrt{s_{\rm NN}}$= 5.5 TeV. The black line indicates suppression only due to isospin effect, while the blue and red lines include medium effects for two different cases. Extracted from [8]. nDSg is a particular parameterization of the parton distribution function in the nucleus (nPDF) [9].

pion spectra via an invariant mass analysis in the range of pion transverse momentum from about 1 to 10 GeV/$c$, when the two decay photons produce separate clusters in the EMCal, and from about 10 up to almost 50 GeV/$c$, when the two clusters merge, via cluster shape analysis. $\pi^0$ reconstruction via invariant mass is an important tool for precision calibration of the EMCal, as seen in Sec. 2.4.4.

In order to increase the purity of the identified prompt photon candidate sample, and further suppress the remaining background from $\pi^0$and other hadrons, and fragmentation photons, especially at high momenta, we use isolation cuts. In this way, we can select prompt photon candidates with high efficiency and purity in both p–p and Pb–Pb collisions for energies larger than 20 GeV. Measurements of the nuclear modification factor $R_{AA}$, i.e. the ratio of the prompt photon spectra in Pb–Pb and p–p collisions, scaled with the number of binary collisions, provides information on the initial parton distributions in colliding nuclei.

Finally, we show that accurate measurements of the prompt photon energy and the correlation with associated jets and/or hadrons would allow a precise measurement of the jet energy and medium modification of the jet fragmentation.

Most results presented in this chapter are based on simulations made with PYTHIA or modified PYTHIA with quenching models, qPYTHIA and PYQUEN. They were used to generate $\gamma$-jet and dijet (jet-jet) events. Dijet events with high momentum $\pi^0$'s represent the main background in the direct photon studies. Photon-jet events were generated with the photon within the EMCal acceptance and dijet events with one of the jets pointing toward

EMCal. To improve statistics at large transverse momenta, the simulations were performed for different parton transverse momenta ($\hat{p}_T$) bins (given in GeV/$c$), for $\gamma$-jet events: [5-10], [10-20], [20-30], [30-40], [40-50], [50-60], [60-70], [70-80], [80-90], [90-100], [$\hat{p}_T > 100$],

and for dijet events: [2-12], [12-16], [16-20], [20-24], [24-29], [29-35], [35-41], [41-50], [50-60], [60-72], [72-86], [86-104], [104-124], [124-149], [149-179], [179-215], [215-258].

## 6.2 Predictions and rates for photon and $\pi^0$ production

PYTHIA [1, 2] is used in this report as the main Monte Carlo generator. In this section, we compare its predictions with another Monte Carlo event generator, HERWIG [3], and with NLO theoretical calculations obtained with the INCNLO [4] program. Both PYTHIA and HERWIG are LO approximation generators but based on different theoretical models; PYTHIA is based on a string model and HERWIG on a clustering model. Both generators have been tuned to reproduce some NLO predictions. We use here the default settings for HERWIG and the *Tune A* settings for PYTHIA [5]. The INCNLO program (for inclusive NLO) allows to calculate perturbatively the photon production cross-section at the NLO accuracy in the QCD coupling $\alpha_s$. The calculation includes both the fragmentation and the direct photon components.

The results presented in this section have also been shown elsewhere [6]. We discuss here model predictions for the particle production in p–p collisions at $\sqrt{s}$=14 TeV up to $p_T$=100 GeV/$c$.

Figure 6.2 shows a comparison of the direct photon production cross sections obtained with the PYTHIA and HERWIG event generators and NLO theoretical calculations. The results from the event generators and the NLO calculations for direct photons are very similar.

Figure 6.2 also shows predictions for $\pi^0$ production in both generators. As expected, the pion yield is significantly higher than that of direct photons. The $\pi^0$ yield predicted by HERWIG is about 40% higher than in PYTHIA. This is mostly due to the fact that two generators consider different resonances (which can decay into $\pi^0$) as stable or unstable particles. This complicates the comparison between these generators and with NLO predictions unless all the particles are decayed, which was not done here. Figure 6.2 shows the NLO prediction for the yield ratio of direct photons over $\pi^0$ for different colliding systems. One can see that the $\pi^0$ production is one to two orders of magnitude larger than that of direct photons, depending on the colliding system, which emphasizes the need for good $\gamma/\pi^0$ discrimination capabilities for the direct photon analyses.

In Fig. 6.3 we show the cross sections and expected yields of prompt photons within the EMCal acceptance, obtained with PYTHIA for p–p collisions at $\sqrt{s}$= 14 TeV and Pb–Pb collisions at $\sqrt{s_{\rm NN}}$= 5.5 TeV. The cross sections for the Pb–Pb case are obtained from p–p results at $\sqrt{s_{\rm NN}}$= 5.5 TeV by scaling with the number of binary collisions, as described in

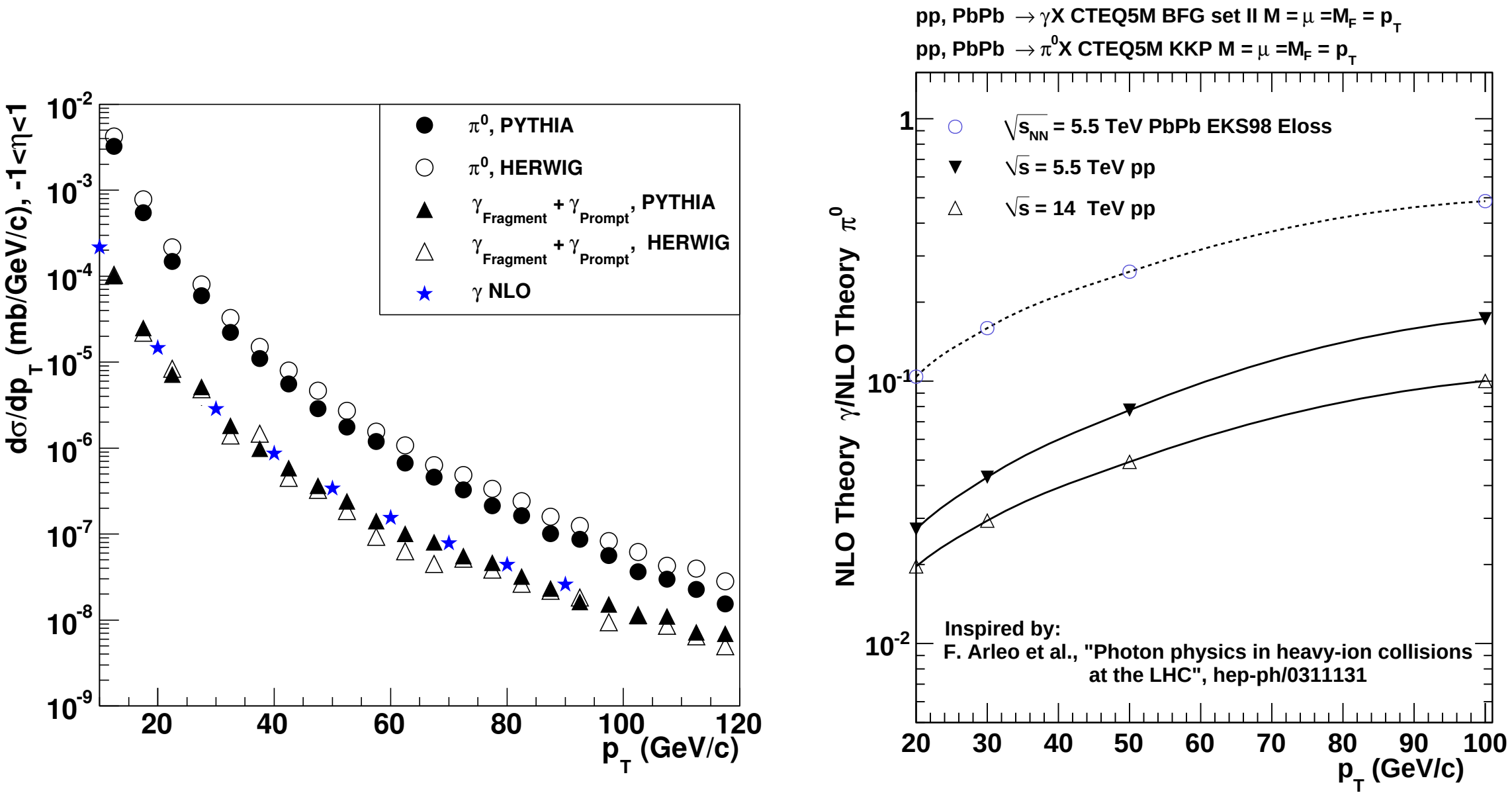


**Figure 6.2:** Left frame: Spectra of direct (the sum of prompt and fragmentation) photons (triangles) and $\pi^0$ (circles) produced with PYTHIA (solid markers), HERWIG (open markers), and INCNLO (stars) for p–p collisions at $\sqrt{s}$=14 TeV. Right frame: INCNLO calculations for the ratio of direct gamma to $\pi^0$ for p–p collisions at $\sqrt{s}$=14 TeV (triangles) and $\sqrt{s}$=5.5 TeV (inverted solid triangles), and Pb–Pb collisions at $\sqrt{s_{\rm NN}}$=5.5 TeV with jet quenching (circles) [7].

Refs. [7, 10, 11]. The annual prompt photon yields were estimated assuming the luminosity and running time values that are indicated in the figures. With these assumptions, we estimate that it should be possible to measure the yield of direct photons with reasonable statistics well above 100 GeV with the ALICE EMCal.

## 6.3 Photon and $\pi^0$ identification via shower shape analysis and a Bayesian approach

In this section, we present a particle identification (PID) method based on Bayes' theory of probabilities. In this approach, PID weights are assigned to every reconstructed particle in the EMCal acceptance. These PID weights can then also be combined with those derived in ALICE central tracking systems. As a result, for each EMCal cluster, the probabilities for being a particle of a certain type are calculated.

This technique is described in the ALICE-PPR [13, 14]. It was first developed for the ALICE PHOS detector and has been applied to the EMCal [15]. The method is based on a shower shape analysis, with the square of the main axis of the shower surface, $\lambda_0^2$ (in tower size units), taken as a discriminating parameter. The shower surface is defined by the intersection of

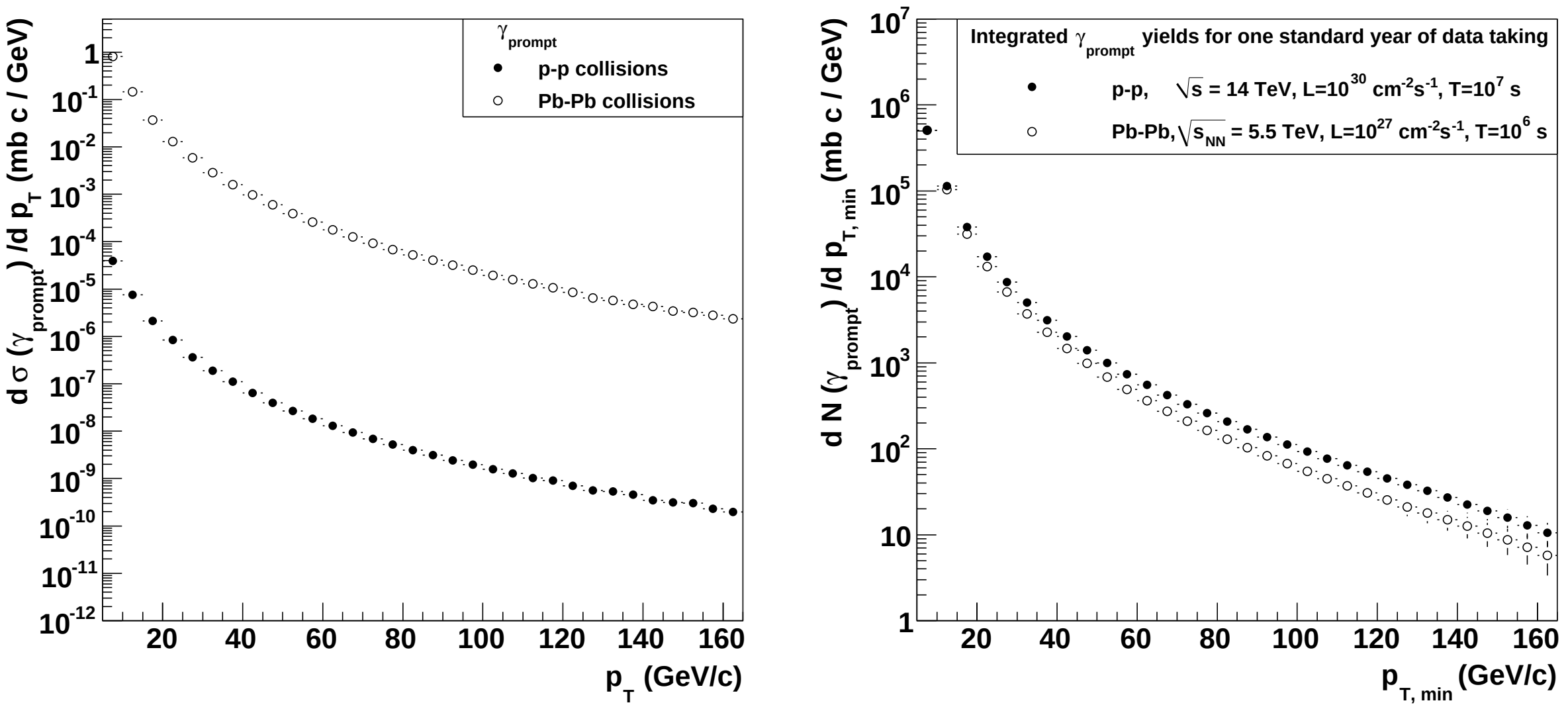


**Figure 6.3:** Left: Prompt photon production cross section as a function of $p_T$ within the EMCal acceptance for p–p collisions at $\sqrt{s}$=14 TeV (solid circles) and Pb–Pb collisions at $\sqrt{s_{\rm NN}}$= 5.5 TeV (open circles). Right: Integrated annual yield of direct photons with $p_T > p_{T,min}$ in EMCal acceptance in p–p collisions at $\sqrt{s}$= 14 TeV (solid circles) and Pb–Pb collisions at $\sqrt{s_{\rm NN}}$= 5.5 TeV (open circles). The error bars indicate the statistical error.

the cone containing the shower with the front plane of the calorimeter. Figure 6.4 shows the $\lambda_0^2$ distributions for photon, $\pi^0$, and charged hadron EMCal clusters, with a reconstructed energy of 20 and 40 GeV. Based on these distributions, the PID weights representing the probability that a reconstructed cluster is a photon, charged hadron, or $\pi^0$. are assigned for each cluster.

The weights are calculated as follows:

- Photons and high $p_T$ $\pi^0$ are generated with energies uniformly distributed in the 10 to 50 GeV range, and tracked through the EMCal acceptance. Above $p_T \simeq 50$ GeV/$c$ the showers generated by decay photons from $\pi^0$ are fully merged in EMCal, and the shower shape analysis can not be used to discriminate $\pi^0$ from photons[1].

  We also generate charged pions ($\pi^{\pm}$) and neutral hadrons ($K_L^0$) with a uniform energy distribution in the 10 to 100 GeV range. We extend the energy range in this case in order to obtain reasonable statistics for cluster energies up to 50 GeV. The reason for this is that these hadrons do not systematically develop electromagnetic showers in the EMCal, and thus deposit energy typically at the MIP energy, though the energy response is broad and extends up to the generated energy.

[1]For higher energies we can use isolation methods for photon discrimination, which are discussed in the next section.

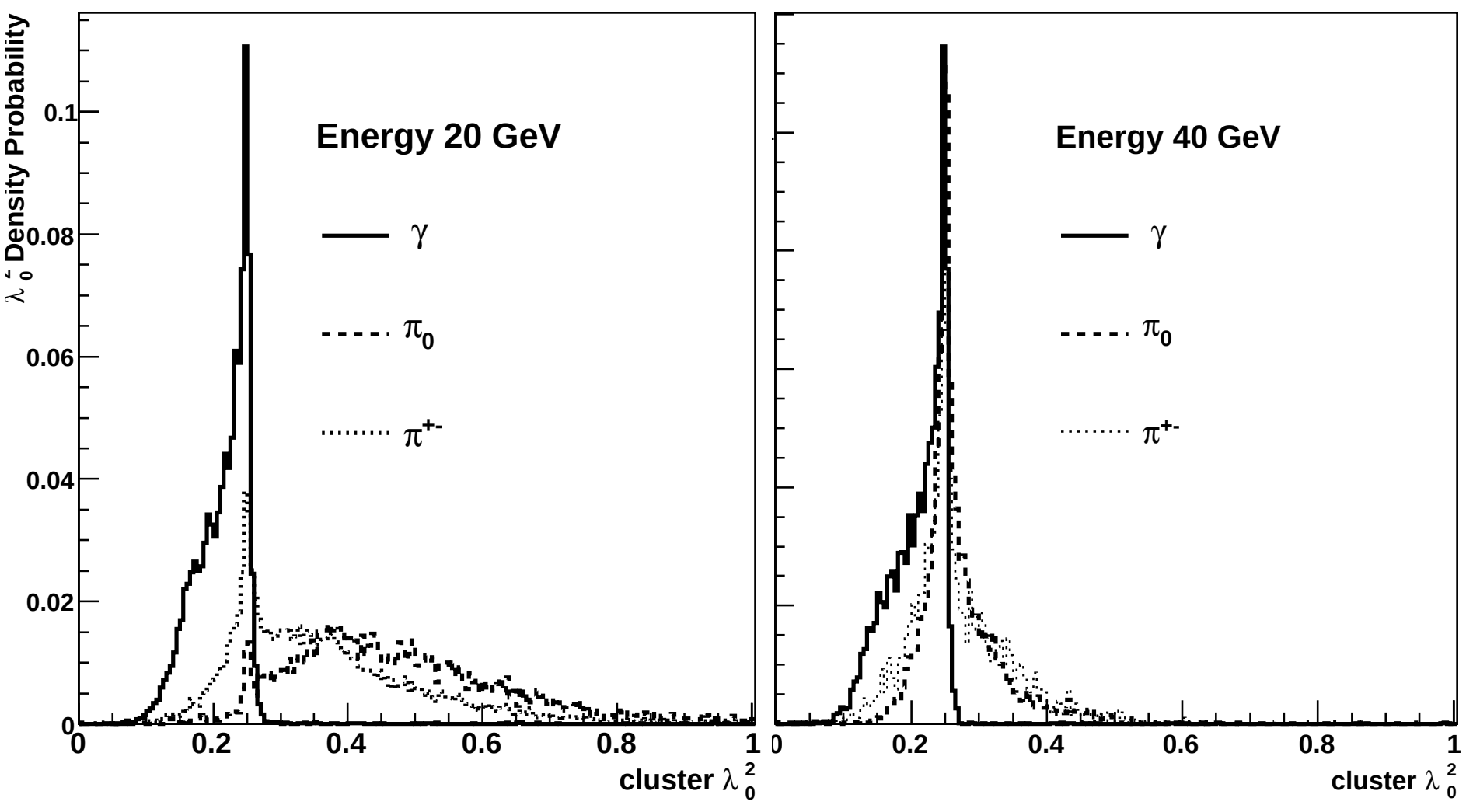


**Figure 6.4:** $\lambda_0^2$ distributions for photons, $\pi^0$ and charged hadrons for EMCal clusters with 20 and 40 GeV reconstructed energy.

- The probability density distributions $P(\lambda_0^2|i)$ for each particle type $i$ is calculated and parametrized by Landau and/or Gaussian distributions. If the number of EMCal towers contributing to the cluster is below two, the shape of the shower is not a significant discriminant between different particle types, which restricts the applicability of the method.

- The probability that a hadron develops a shower in EMCal is also taken into account. We define $P(cluster|i)$ as the probability that a particle generates a cluster for a given reconstructed energy. $P(cluster|i)$ for the case that $i$ represent hadrons has been parametrized by an exponential plus power law function, varying as a function of the reconstructed cluster energy. For photons and $\pi^0$ the $P(cluster|i)$ probability is set to 1, since both particles always develop an electromagnetic shower in EMCal.

- From the probability density distributions for each particle type ($\gamma$, hadrons, and $\pi^0$), a PID weight $W(i)$ is calculated for each reconstructed EMCal cluster, as a product of the two probability densities normalized to the sum of these products for all particle species:

$$W(i) = \frac{P(\lambda_0^2|i) \cdot P(cluster|i)}{\sum_s P(\lambda_0^2|s) \cdot P(cluster|s)} \tag{6.2}$$

We have applied this procedure to photons, $\pi^0$, and $\pi^\pm$, merged into PYTHIA p–p collisions at $\sqrt{s} = 14$ TeV[2] and HIJING Pb–Pb collisions at $\sqrt{s_{\mathrm{NN}}}= 5.5$ TeV with b < 5 fm. We define

[2] Jet-jet events with one jet fully in the EMCal acceptance, and $\hat{p}_T$ between 30 and 1000 GeV/$c$.

the PID efficiency as the probability to correctly identify a particle of type $i$, i.e. the ratio of reconstructed clusters generated by particles of type $i$ and identified with a PID weight $W(i)$ larger than a threshold value $W^{th}(i)$ (taken equal to 0.35), over all reconstructed clusters generated by the same type $i$ of particles (without a condition on $W(i)$). The PID purity is defined as the ratio of reconstructed clusters generated by particles of type $i$ and identified as type $i$, over all reconstructed particles identified as particle of type $i$.

The efficiency and purity of photons and $\pi^0$ identifications are presented in Fig. 6.5 for particles merged in p–p collisions and in Fig. 6.6 for particles merged in Pb–Pb collisions. The photon identification efficiency obtained in the p–p environment is very good, greater than 90%, over the whole energy range. The purity of identified photons in the p–p environment is low below 10 GeV because in that energy range photons from the decay of neutral pions produce separate showers and are then also identified in this method as photons instead of as $\pi^0$ś. For larger energies, between 10 and 50 GeV, the photons are well identified with a relatively high purity of between 60 and 80%. In the heavy-ion environment the photon efficiency is slightly lower (remaining around 80% over the whole energy range) with the same level of purity as in the p–p case.

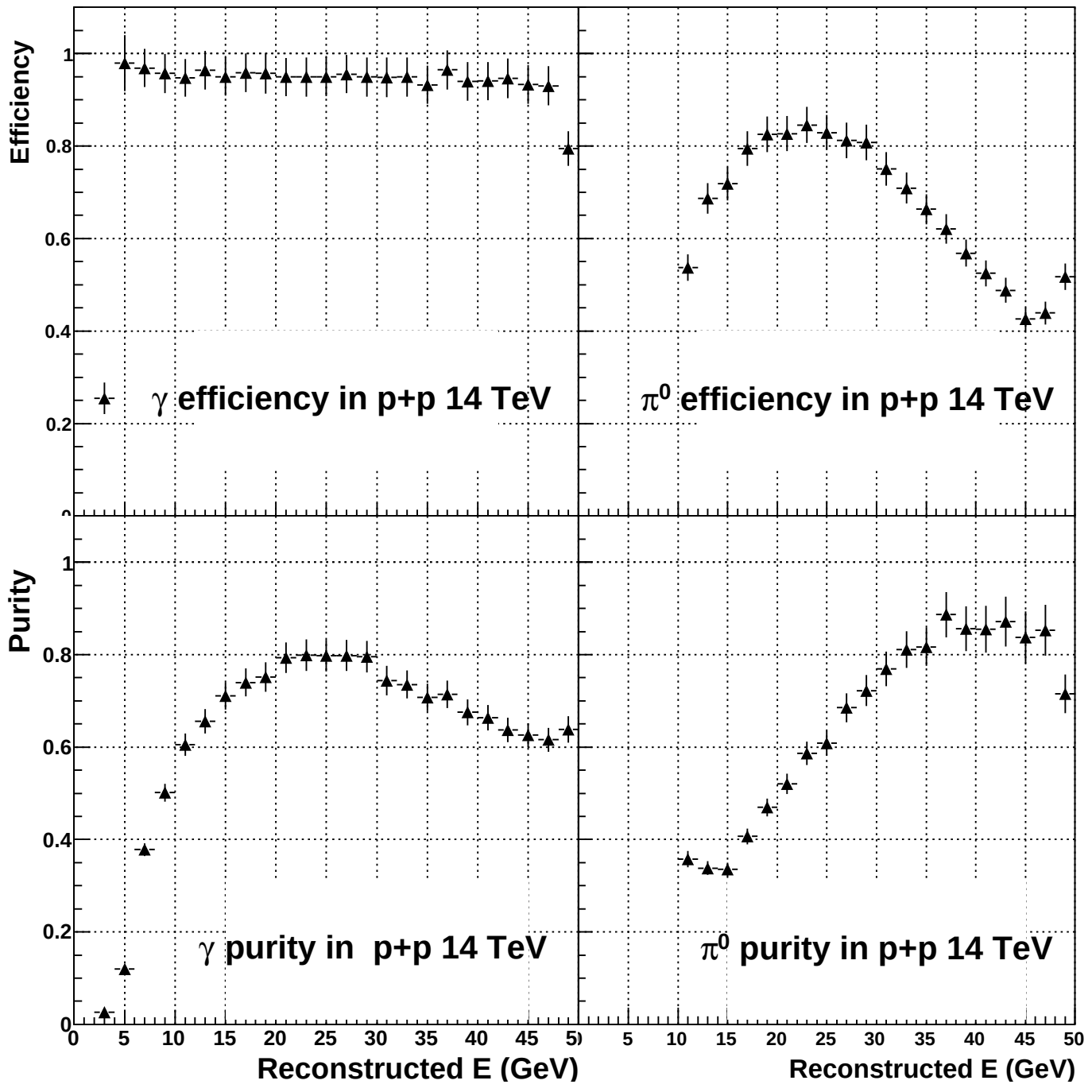


**Figure 6.5:** Particle identification efficiency and purity for photons and $\pi^0$ embedded into $\sqrt{s}$= 14 TeV p–p collisions simulated with PYTHIA.

The $\pi^0$ identification efficiency in p–p exhibits a maximum between 15 GeV and 30 GeV, reaching about 80%, in a range where the shower shape is expected to be an effective discriminant. Above 30 GeV, the efficiency decreases, which can be attributed to the fact that

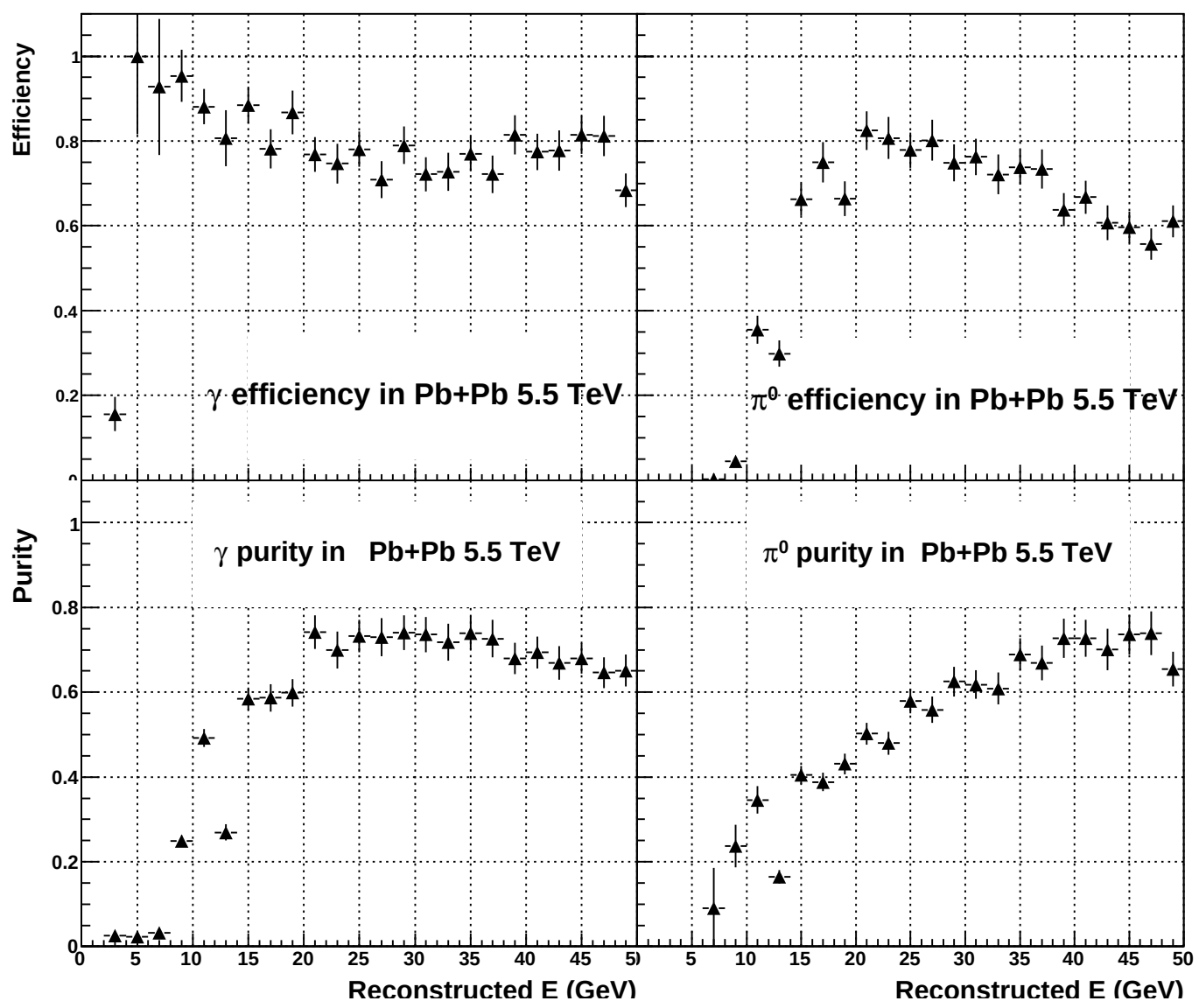


**Figure 6.6:** Particle identification efficiency and purity for photons and $\pi^0$ embedded into $\sqrt{s_{\rm NN}}$= 5.5 TeV Pb–Pb $b < 5$ fm, collisions simulated with HIJING.

with increasing energy the two decay photons are less separated.

In Pb–Pb simulations, a very similar behavior is observed in Fig. 6.6. In both cases a $\pi^0$ identification efficiency above 60% can be achieved for energies between 15 GeV and 35 GeV. We find that the purity of the identified $\pi^0$ increases with energy. The low values of $\pi^0$ purity at lower energies is explained by charged pions generating showers very similar to showers generated by $\pi^0$'s; and consequently identified as $\pi^0$'s. (In our simulations, the probability for charged pions to generate a shower with energy less than 20 GeV is not negligible.)

To summarize this section, we conclude that the EMCal detector provides a good capability to identify and discriminate between photons, $\pi^0$ś, and other hadrons using a Bayesian approach based on shower shape analysis. This method can be applied for energies in the range from 10 to 50 GeV. At lower energies, the shower shape analysis can still be used for discrimination between hadrons and photons. If we combine the EMCal information with the central detectors information (track matching, as seen in Sec. 7) to improve particle identification, the method can be extended to discriminate between electrons and photons, and also to identify charged hadrons. For photon identification at higher energies ($E > 50$ GeV), isolation methods have to be used, which are presented in section 6.5.

This particular analysis was tuned with PYTHIA events, where the relative particle abundances are known. In the real experiment, the relative particle yields for a given reconstructed energy bin have to be calculated first. For that, the shower shape analysis can also be used, but in that case on a statistical, not cluster-by-cluster, basis. This development is under investigation.

## 6.4 $\pi^0$ and $\eta$ identification with invariant mass analysis

In the transverse momentum range of $1 < p_T < 10$ GeV/$c$, $\pi^0$'s can be identified in the EMCal via invariant mass analysis of pairs of photons detected in the calorimeter. The upper limit of the accessible $p_T$ range is set by the EMCal spatial granularity, since at higher momenta clusters from $\pi^0$ decay photons are merged in EMCal, as discussed above. The ALICE-PHOS detector has smaller acceptance but better spatial resolution, which allows particle identification via invariant mass analysis up to about 3 times higher momenta. The larger EMCal acceptance is more favorable for statistics hungry analyses, e.g. ones based on multi-particle correlations. Also, $\pi^0$ reconstruction via invariant mass will be an important tool for precision calibration of the calorimeter (see Section 2.4.4).

Figure 6.7 presents the two-photon invariant mass distribution in two regions of pair transverse momentum together with a polynomial fit to the background. 34 million minimum bias PYTHIA p–p collisions at 10 TeV have been used in this study. The $\pi^0$ peak is well identified at lower pair momentum but is harder to identify at momenta greater than 10-12 GeV/$c$, as the two photon clusters begin to merge. The width of the Gaussian fit of the mass peak, shown in Fig. 6.8, is close to the expected value of about 10 MeV and determined mostly by the detector energy resolution and granularity. The transverse momentum dependence of the fit parameters is mostly due to to the change in the energy and position resolution.

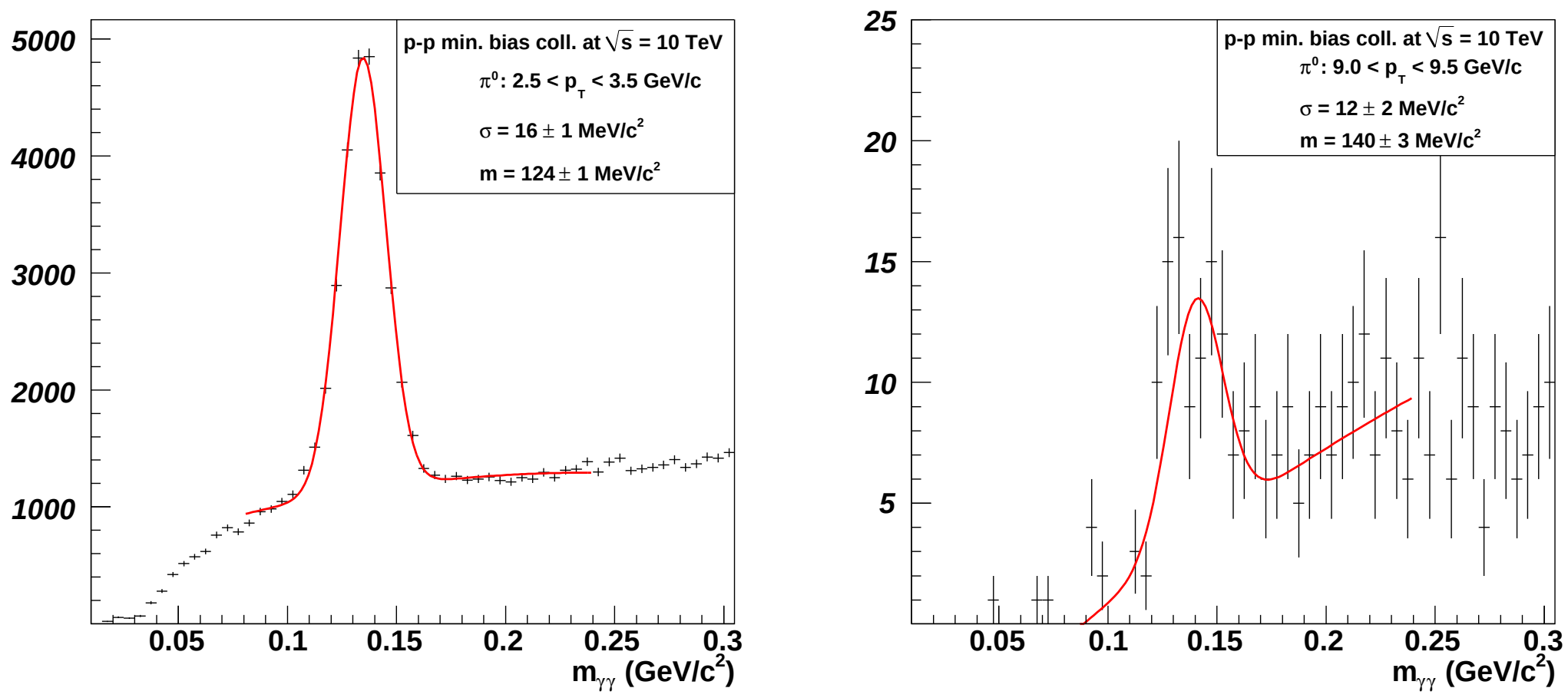


**Figure 6.7:** The two-photon invariant mass distribution for pair transverse momentum ranges $3 < p_T < 3.5$ GeV/$c$ (left) and $9 < p_T < 9.5$ GeV/$c$ (right) in minimum bias p–p collisions at $\sqrt{s}$=10 TeV.

The efficiency for the detection of both decay photons in the EMCal acceptance is about 60%; most losses of photons happen close to the EMCal edges. The reconstructed $\pi^0$ spectrum from this analysis is presented in Fig. 6.9-left.

Invariant mass analysis in the EMCal can also be used for detection of $\eta$ mesons, with an

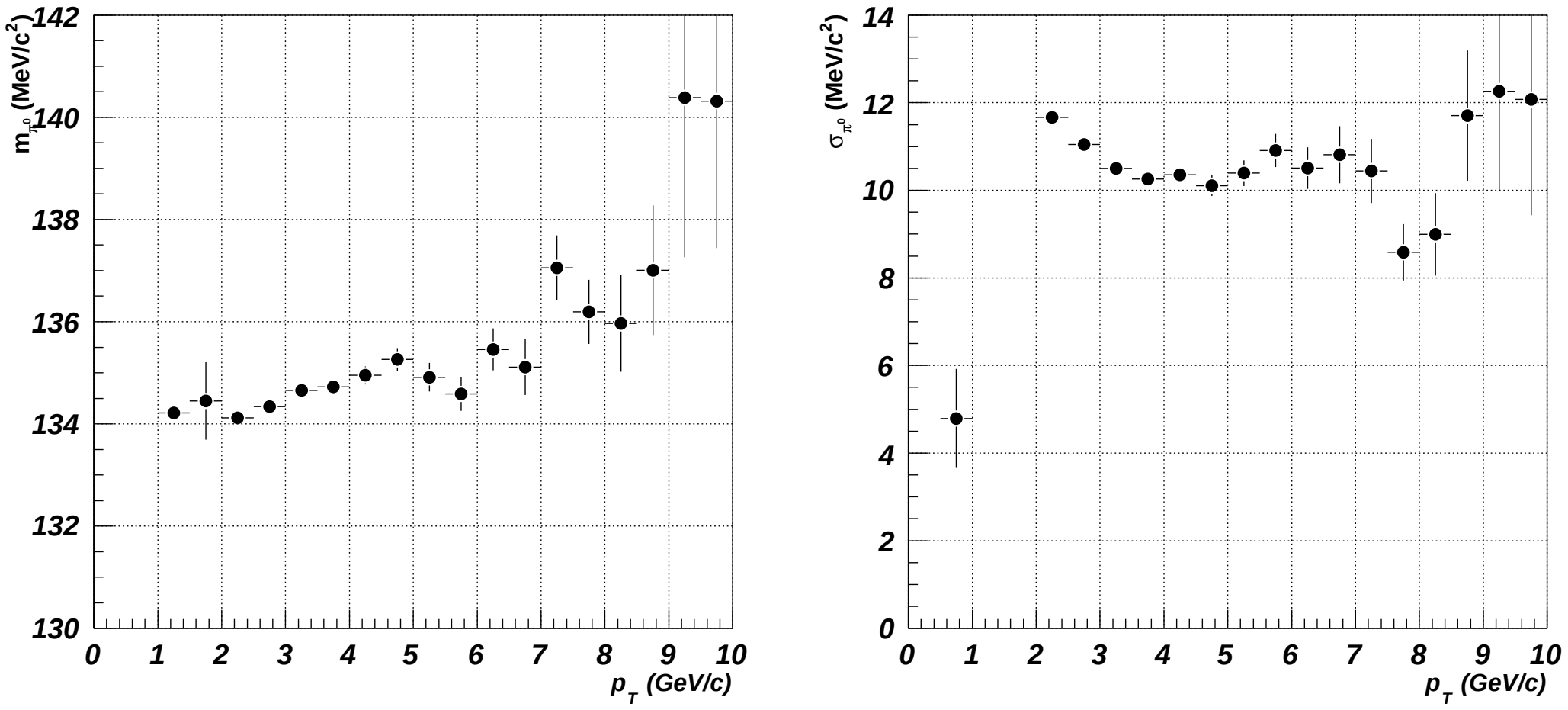


**Figure 6.8:** The mean and width values of Gaussian fit to the $\pi^0$ peak. PYTHIA minimum bias p–p collisions at $\sqrt{s}$=10 TeV.

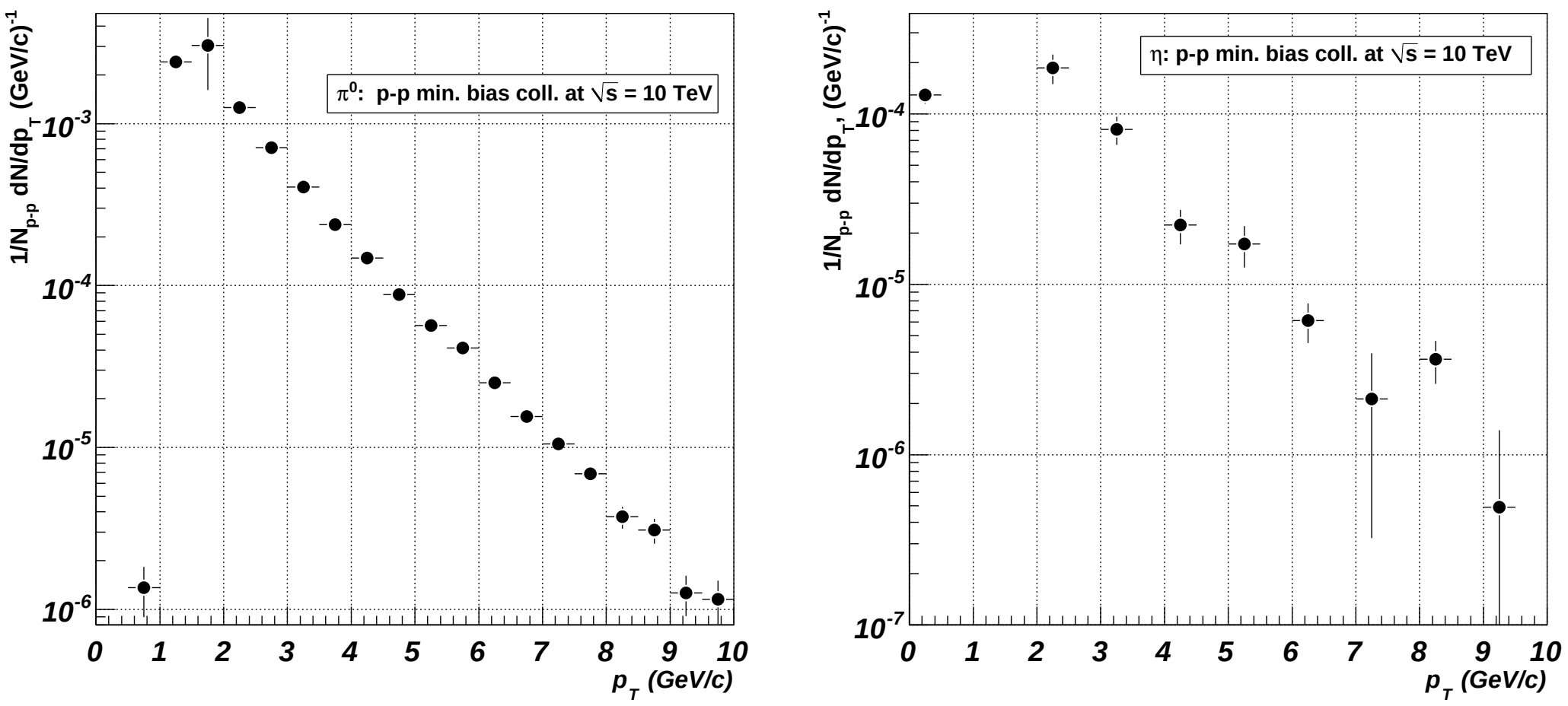


**Figure 6.9:** Transverse momentum distribution of reconstructed $\pi^0$ (left) and $\eta$ mesons (right). PYTHIA minimum bias p–p collisions at $\sqrt{s}$=10 TeV.

estimated accuracy comparable to PHOS. The reconstructed $\eta$-meson spectrum is shown in Fig. 6.9. The invariant mass distribution for two transverse momentum ranges are shown in Fig. 6.10. Typical width of the mass peak is about 30 MeV/$c^2$.

In Pb–Pb collisions, the combinatorial background of course increases dramatically, most notably in central collisions and at low $p_T$, with the Signal-to-Background (S/B) ratio being roughly inversely proportional to the number of participants. In addition, the photon energy resolution deteriorates due to increased occupancy of charged hadrons in the EMCal. With

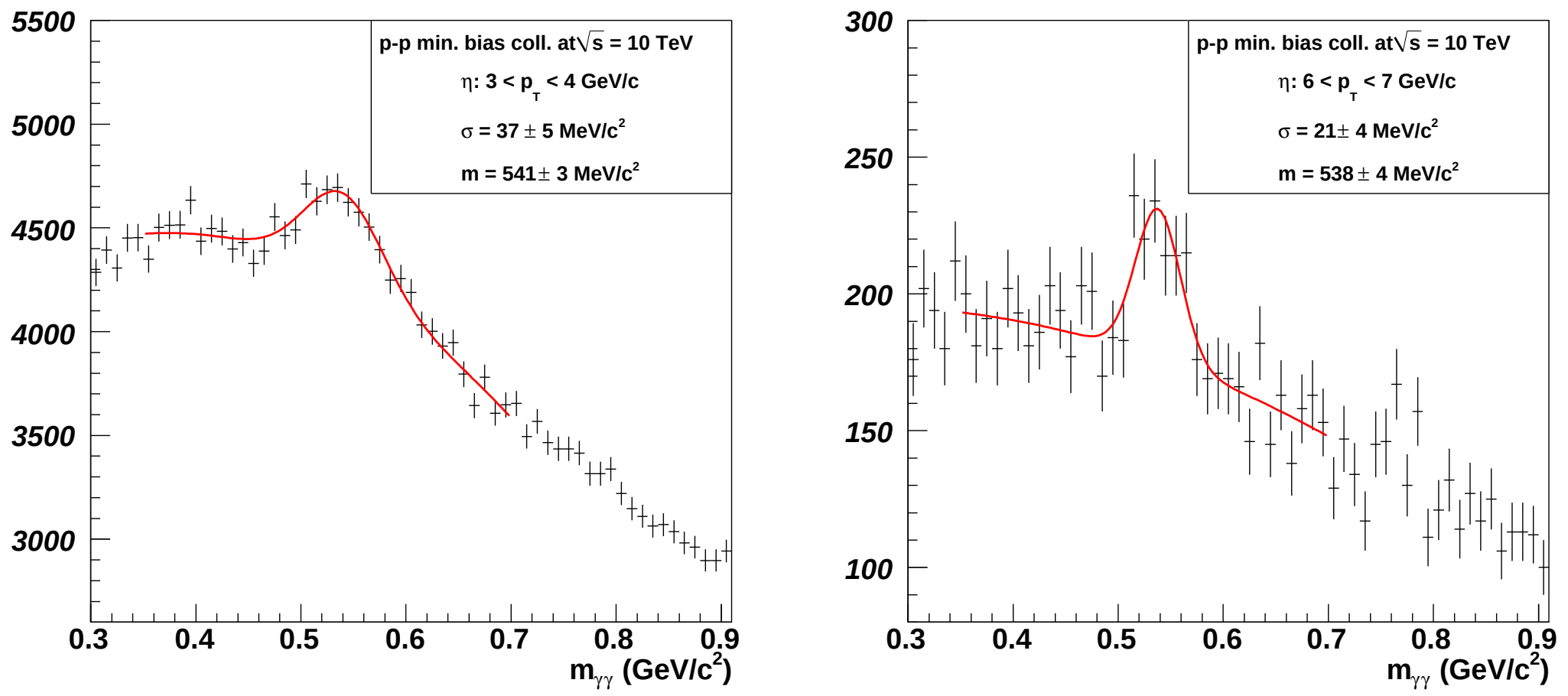


**Figure 6.10:** The two-photon invariant mass distribution for pair transverse momentum ranges $4 < p_T < 5$ GeV/$c$ (left) and $6 < p_T < 7$ GeV/$c$ (right) in minimum bias p–p collisions at $\sqrt{s}$=10 TeV.

about 100k minimum bias HIJING Pb–Pb events available in this analysis, the $\pi^0$ peak is clearly visible, see Fig. 6.11, but a larger number of events is needed for a more detailed study.

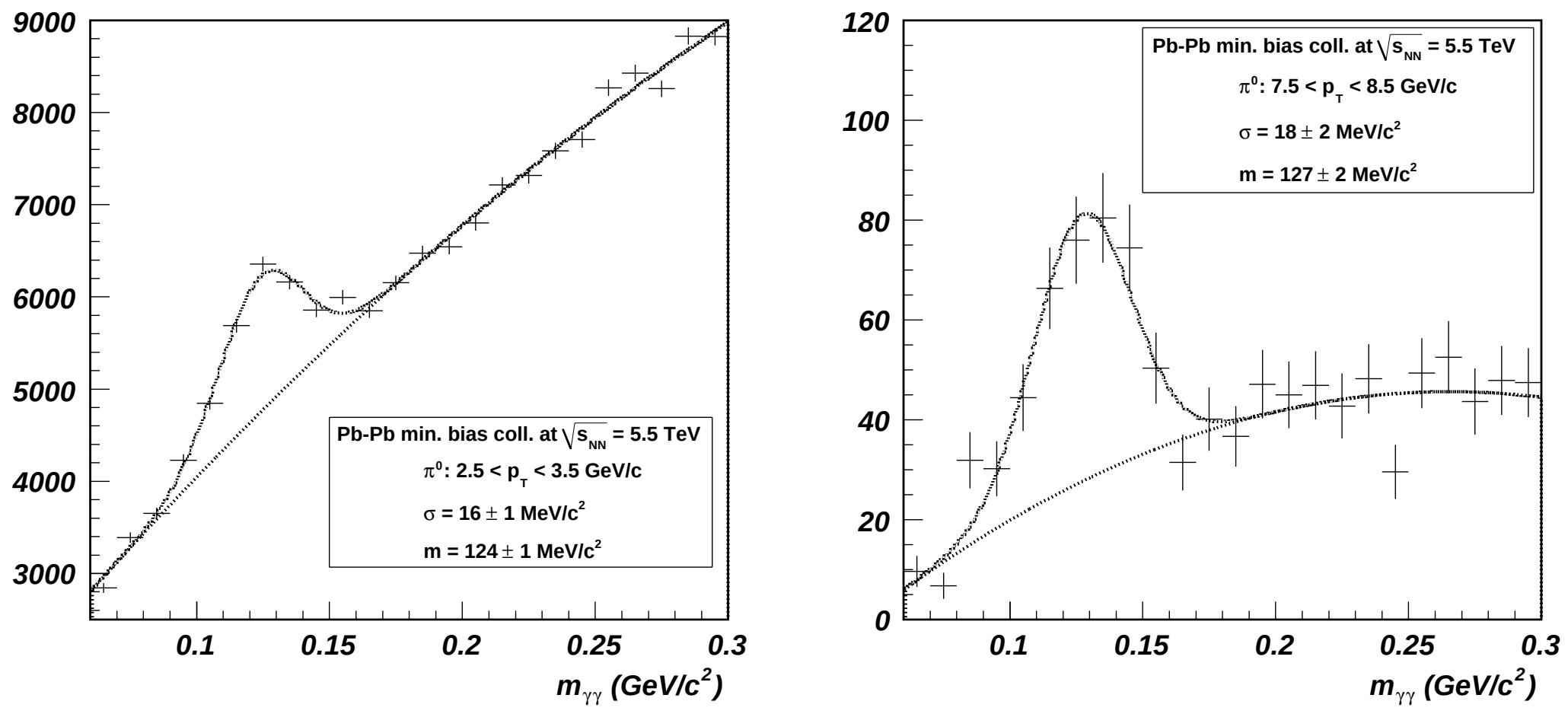


**Figure 6.11:** The two-photon invariant mass distribution for pair transverse momentum ranges $2.5 < p_T < 3.5$ GeV/$c$ (left) and $7.5 < p_T < 8.5$ GeV/$c$ (right) in minimum bias Pb–Pb collisions at $\sqrt{s_{\rm NN}}$= 5.5 TeV.

## 6.5 Direct photon identification with isolation cuts

### 6.5.1 Method description

Since there are no hadrons associated with the prompt photon produced in the Compton and annihilation process of Eq. ??, and the hadrons associated with the fragmentation of the recoiling parton are emitted in the direction opposite to the photon, an isolation cut can be used to enhance the selection of prompt photons. However, the underlying event, especially in the case of a heavy-ion collision, may perturb this ideal topology. To overcome this difficulty, different isolation cut methods may be used, as suggested in Refs. [11, 17]. These methods search for hadrons, either charged tracks measured in the central tracking system, or clusters in the calorimeter, inside a cone centered around the direction ($\eta_0$, $\phi_0$) of high-$p_T$ photon candidates.

The cone is defined by the radius $R = \sqrt{(\phi_0 - \phi)^2 + (\eta_0 - \eta)^2}$ in $\eta$-$\phi$ space. The multiplicity inside the cone depends on the cone size and the event type. For $\gamma$-jet events in $pp$ collisions, there is rarely a particle inside the cone, independent of the energy of the prompt photon. In contrast, for jet-jet events there is a clear increase in the multiplicity of particles within the cone with jet energy. Thus, applying $p_T$ cuts to the particles inside a cone around a photon candidate helps to distinguish between $\gamma$-jet and jet-jet events. The two methods used in Refs. [11, 17] to decide if a photon candidate is isolated and should be tagged as a prompt photon are defined as follows:

1. No hadron with $p_T$ above a given threshold is found within the cone. We denote this method ICM1.

2. The sum of the transverse momenta of all hadrons inside the cone is smaller than a certain $p_T$ threshold, $\sum_i p_{Ti} \leq p_{Tcut}$. Or, equivalently, using $\epsilon$, where $\sum_i p_{Ti} \leq \epsilon . p_{T\gamma}$ The sum of the transverse momenta of all hadrons inside the cone is smaller than a given fraction of the candidate $p_T$. We denote this method ICM2.

Here, we will use ICM1 for photon isolation in p–p and Pb–Pb events, because under the condition of large underlying event multiplicities in heavy-ion collisions, ICM1 has been shown to yield more consistent results [11, 17]. We will compare our result to a theoretical calculation using ICM2 for p–p collisions. Other possible isolation methods, which use jet reconstruction algorithms to study if a particle carries all or most of the jet energy, have been suggested but have not yet been applied in our simulations.

At present we calculate the isolation efficiency and purity based on Monte Carlo simulations, but in the future it will be necessary to perform a data driven analysis.

### 6.5.2 Comparison of different theoretical predictions

In this section, we compare the predictions from PYTHIA, HERWIG, and NLO calculations. The NLO calculations were obtained with the JETPHOX program (for JET-PHOton / hadron X-sections) [4], which is an improvement with respect to the INCNLO program since it allows to evaluate the effect of isolation cuts for prompt photon identification.

Based on the results in Ref. [17], we have selected a cone size $R = 0.5$ and a $p_T^{th} = 1$ GeV/$c$ as isolation parameters, which is a compromise between the optimized values for p–p and Pb–Pb collisions.

Fig. 6.12 shows the ratio of isolated particles of different types to all particles of the same type as a function of the particle momentum. Prompt photons are nearly all isolated as expected. However, the isolated fraction of prompt photons is 10% higher in HERWIG than in PYTHIA. This is understood by the observation that the particle multiplicity, from the underlying event, within the cone around the prompt photon (Fig. 6.13) is greater in PYTHIA than in HERWIG. Isolated hadrons are mostly leading particles with energies close to the jet energy. These hadrons are also more isolated in HERWIG. We observe that for PYTHIA, the fraction of fragmentation photon that are isolated is around 50%, slightly less than for HERWIG. We also observe that charged pions are more isolated than $\pi^0$ś, which can be explained by the fact that charged pions are produced in pairs with both hadrons likely to be within the isolation cone.

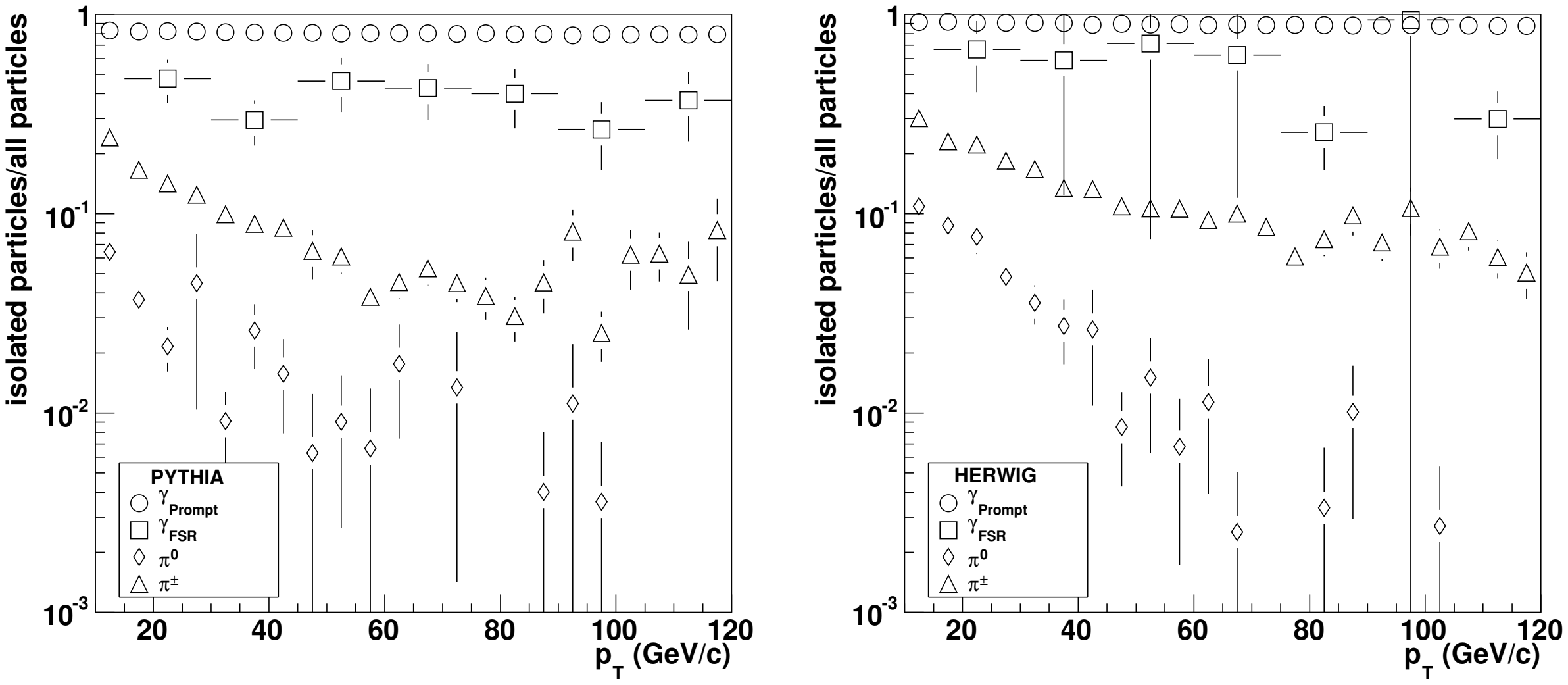


**Figure 6.12:** Fraction of particles that are isolated for different particle types as a function of their $p_T$ for PYTHIA (left) and HERWIG (right) for p–p collisions at $\sqrt{s}$=14 TeV. Isolation parameters: Cone size R=0.5 and $p_T$ $^{th}$=1 GeV/$c$.

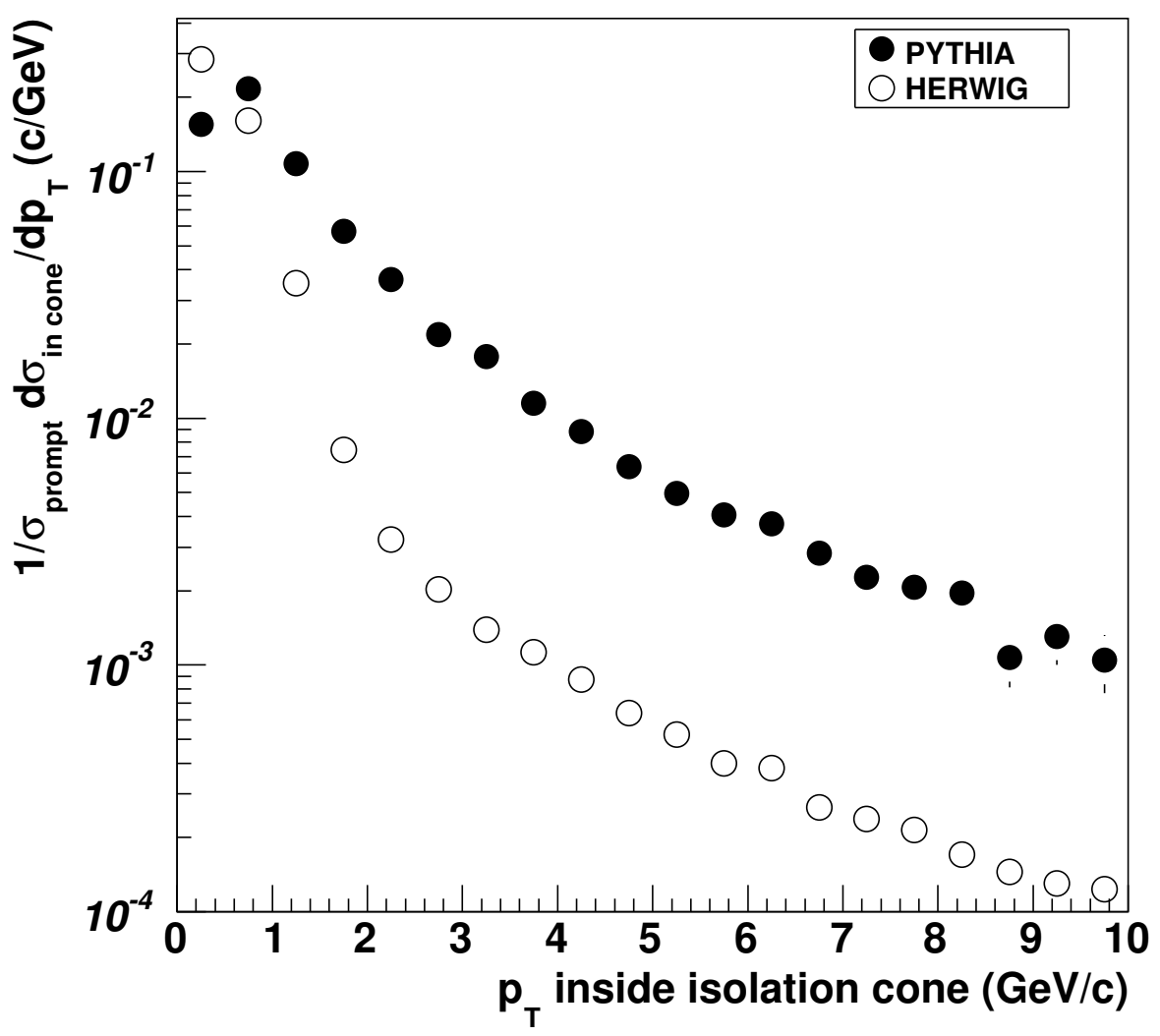


**Figure 6.13:** Single particle distribution in a cone around a prompt photon in HERWIG (open) and PYTHIA (full) events for p–p collisions at $\sqrt{s}$=14 TeV.

The JETPHOX program allows one to study combined samples of fragmentation and prompt photons. This simulation tool can then be used to optimize the parameters $\epsilon$ and $R$ for the isolation cut method ICM2. In order to choose a value for $\epsilon$, it is necessary to consider the underlying event since it will affect the rates at small $p_T$. The amount of hadronic activity within a cone around a prompt photon can be seen in Fig. 6.13. With a cutoff at 1-3 GeV/$c$ one would have an $\epsilon$ minimum value at $p_T$ =30 GeV/$c$ of: $\epsilon \geq \frac{p_T{}^{UE}}{p_T{}^{\gamma}} = 0.03 - 0.1$.

We studied the isolation of NLO photons in p–p collisions at $\sqrt{s}$=14 TeV and 10 TeV and the isolated fraction for these p–p colliding energies with isolation cuts $R = 0.4$ and $\epsilon$= 0.1. Using these parameters, about 40% of the total inclusive photons at 5 GeV/$c$ are isolated. This fraction increases to about 80% at 30 GeV/$c$ before saturating for larger $p_T$ .

### 6.5.3 Isolation parameter selection in fully reconstructed events

In this section we study the dependence of the prompt photon selection and the jet cluster rejection on the isolation parameters. Jet-jet events with a $\pi^0$ with $p_T > 5$ GeV/$c$ in the acceptance of the EMCal with and without quenching were superimposed on Pb–Pb events. Quenching was implemented with the quenching weights approach, explained in detail in the ALICE PPR Vol 2 [13]. The goal is to determine the optimum set of isolation cut parameter to enhance the selection of prompt photons (signal) over the selection of photons or clusters within jets (background).

In Figure 6.14, we compare the isolated fraction (isolated yield divided by total yield; spectra

integrated for $p_T > 10$ GeV/$c$) of prompt photons and jet clusters (all clusters in jets produced in jet-jet events) for different choices of the cone radius R and the $p_T^{th}$ parameters for p–p collisions and quenched Pb–Pb collisions. The dependence of the ratio of isolated prompt photons to isolated jet clusters is also shown in Figure 6.14.

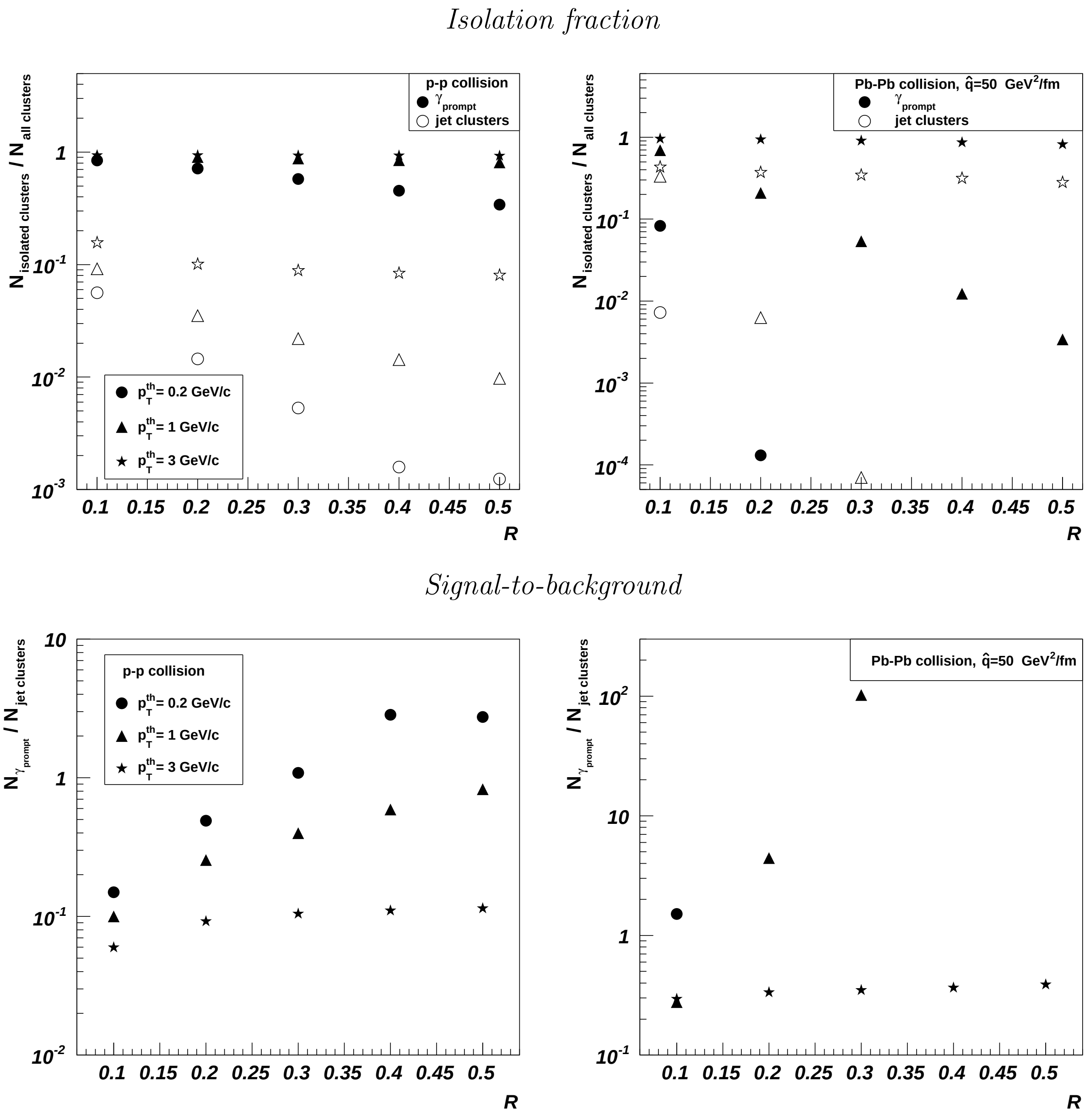


**Figure 6.14:** Isolated fraction (top) of prompt photons (full) and jet clusters (open) and the ratio of isolated photon to isolated jet clusters (bottom), calculated from yields integrated over $p_T > 10$ GeV/$c$, as a function of $R$ for different $p_T^{th}$. Results are shown for PYTHIA p–p collisions at $\sqrt{s}$= 14 TeV (left), PYTHIA+HIJING Pb–Pb collisions at $\sqrt{s_{\rm NN}}$= 5.5 TeV with quenching (right) (similar trends are observed without quenching but with prompt photon to jet cluster ratio 10 times smaller).

For p–p collisions it is observed that the efficiency for prompt photon selection is highest, when the cone radius is small and $p_T^{th}$ is high. The efficiency exceeds 80% for $p_T^{th}$ larger than 1 GeV/$c$. The dependence on the cone radius is small for $p_T^{th} > 2$ GeV/$c$. The highest signal-to-background (S/B) is observed with $p_T^{th} = 0.2$ GeV/$c$ and $R = 0.4$, but the efficiency for photon selection with such a low threshold is small, about 40%. The optimized compromise between isolation and purity is reached with $p_T^{th} = 0.5$ GeV/$c$ and $R$= 0.4. Here the prompt photon identification efficiency is of the order of 70% and the S/B is close to unity.

For Pb–Pb collisions, it is similarly observed that the efficiency for prompt photon selection is high, when the cone radius is small and $p_T^{th}$ is high, i.e. larger than 2 GeV/$c$, for both quenched and unquenched jets. The efficiency exceeds 50% and the dependence on the cone radius is small for $p_T^{th} > 3$ GeV/$c$. Although the signal to background is highest with decreasing $p_T^{th}$ and increasing cone radius, the prompt photon selection efficiency decreases drastically. The optimized parameter set for Pb–Pb collisions is $p_T^{th} = 2$ GeV/$c$ and $R = 0.3$. Here the prompt photon identification efficiency is of the order of 60% and S/B = 0.1 for unquenched jets and unity for quenched jets.

## 6.6 Photon identification with shower shape and isolation cuts

By combining particle identification and isolation cuts the prompt photon selection and jet cluster rejection can be further improved. The ratio of the photon (signal) and the jet cluster (bakcground) spectra and isolation purity (fraction of measured clusters to be really prompt photons) with various isolation cuts is shown in Fig. 6.15. It is observed that the signal-to-background ratio exceeds unity, and purity is close to unity for $p_T > 20$ GeV/$c$ both in p–p collisions and in Pb–Pb collisions with quenching.

By applying the efficiencies from the Bayesian PID and the isolation cuts we can calculate corrected spectra with their corresponding statistical error and systematic error. The correction factor $\xi$ is calculated in the following way

$$N_{measured} = N_\gamma \epsilon_\gamma^{PID+ICM} + N_{hadrons}\epsilon_{hadrons}^{PID+ICM} = N_\gamma(\epsilon_\gamma + \frac{N_{hadrons}}{N_\gamma}\epsilon_{hadrons}^{PID+ICM}) = N_\gamma \xi, \quad (6.3)$$

where $N_{measured}$ is the total number of identified isolated clusters (sum of prompt and jet clusters), $N_\gamma$ is the prompt photon yield, $N_{hadrons}$ is the total number of jet clusters, and $\epsilon$ is the identification and isolation efficiency for correctly identifying a photon or falsely identifying a hadron as a photon, respectively. The correction factor $\xi$ is obtained from the simulation results shown in Fig. 6.15. The prompt photon spectra extracted using Eq. 6.3 for p–p collisions and for Pb–Pb collisions is shown in Fig. 6.16. The $R_{AA}$ calculated from the corrected p–p and Pb–Pb spectra is also shown in Fig. 6.16 . There is no suppression observed since the simulated photon spectra are not supressed. The magnitude of the systematic and

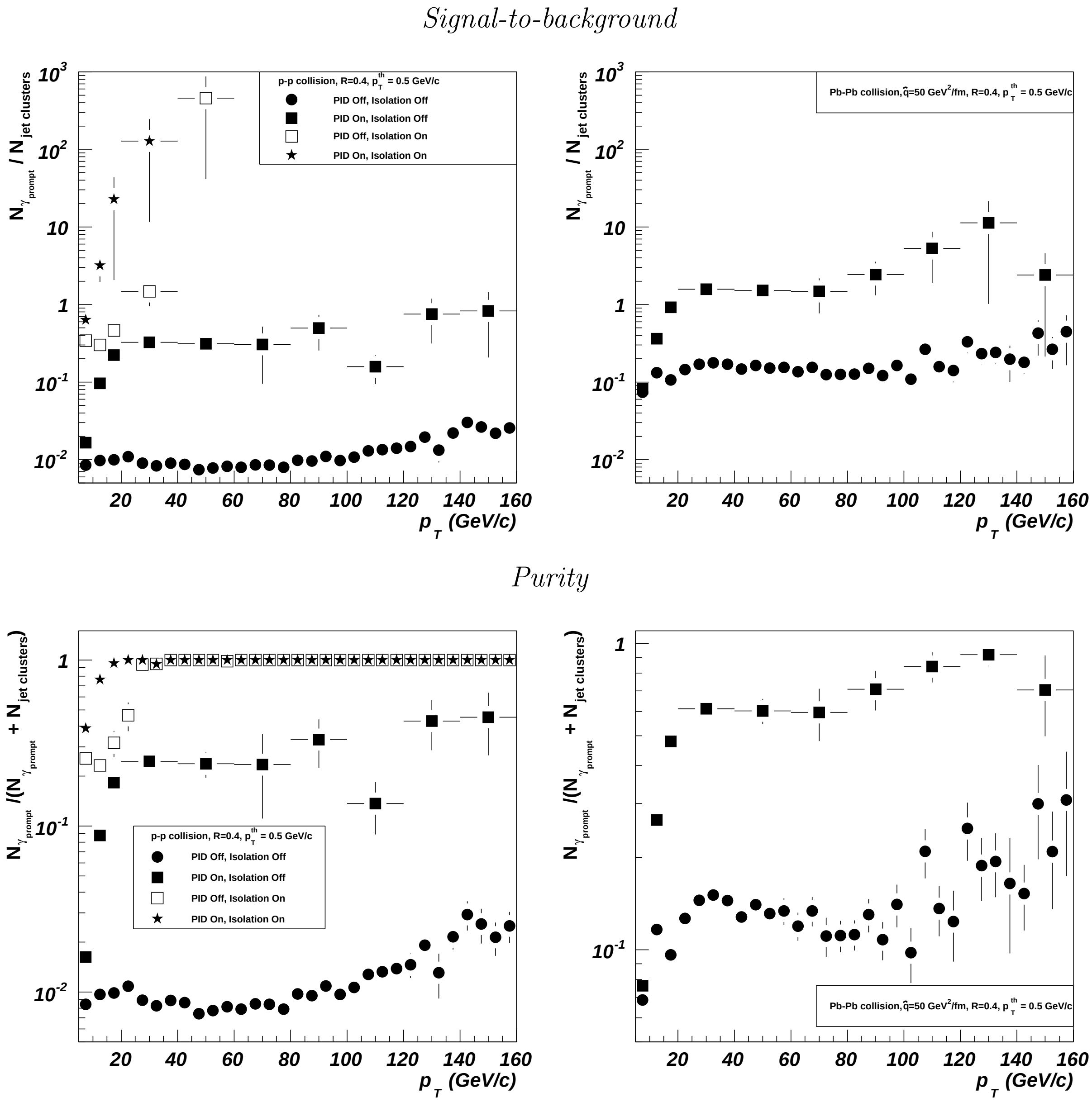


**Figure 6.15:** Ratio of isolated prompt photons to isolated jet clusters (top) and ratio of isolated prompt photons to isolated jet clusters plus isolated prompt photons (bottom, purity) as a function of the particle $p_T$ for PYTHIA p–p collisions at $\sqrt{s} = 14$ TeV (left), PYTHIA+HIJING Pb–Pb collisions at $\sqrt{s_{\rm NN}} = 5.5$ TeV with quenching (right, similar without quenching, but ratios 10 times lower). The isolation parameters used are $p_T^{th} = 0.5$ GeV/$c$ and $R = 0.4$ for p–p collisions and $p_T^{th} = 2$ GeV/$c$ and $R = 0.3$ for Pb–Pb collisions.

statistical errors indicate that any suppression due to isospin or cold nuclear matter effects, as in Fig. 6.1 [8], could be observed in the $p_T$ range from 20 to 80 GeV/$c$ with good accuracy.

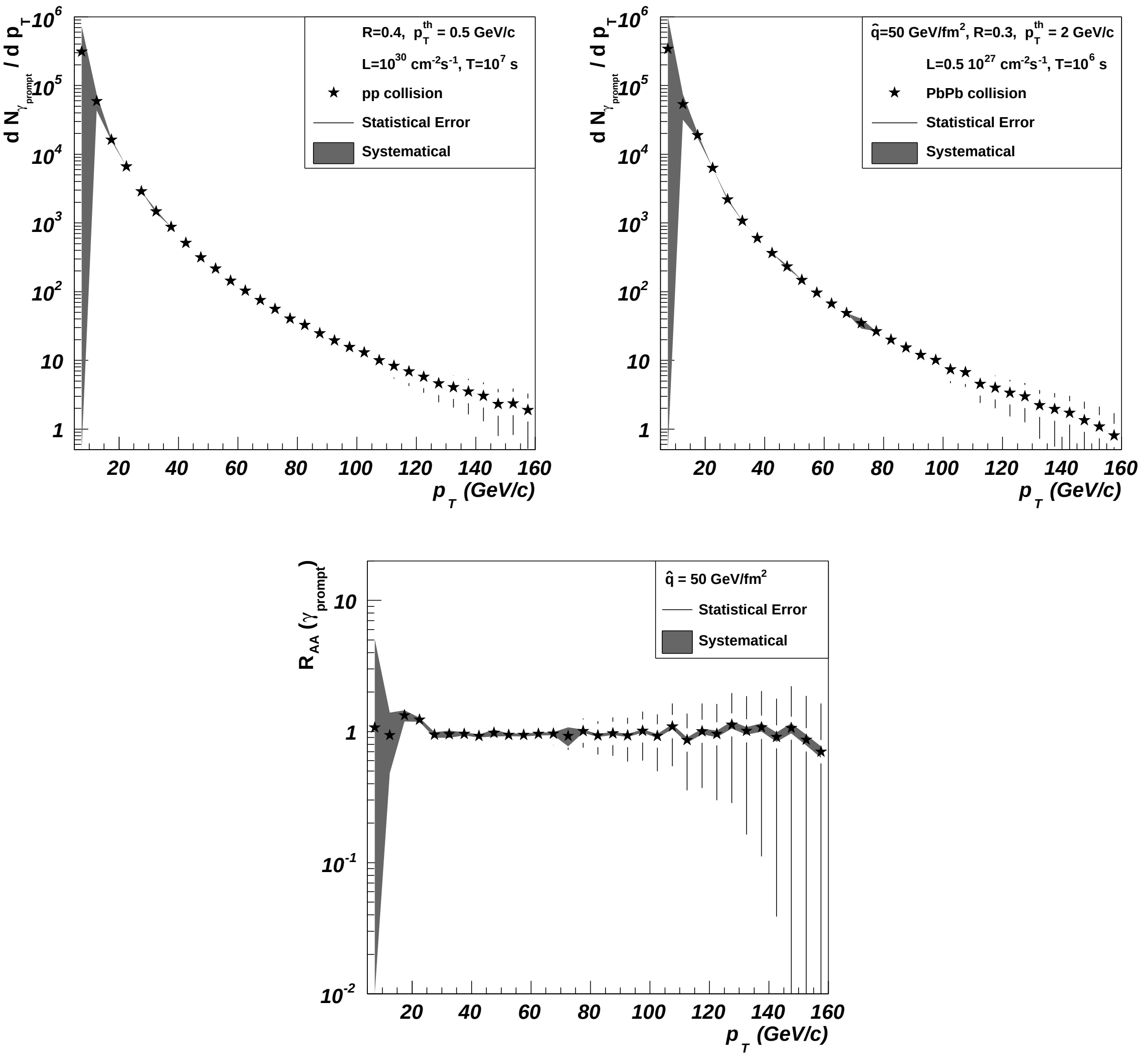


**Figure 6.16:** Top: Corrected spectra of prompt photons to be measured, calculated with Eq. 6.3 with the corresponding systematical and statistical error of a year of data taking. p–p collisions at $\sqrt{s}$= 14 TeV (Left), PYTHIA+HIJING Pb–Pb collisions at $\sqrt{s_{\mathrm{NN}}}$= 5.5 TeV without quenching (Right). Bottom: Ratio of the corrected spectra of prompt photons Pb–Pb over p–p collisions, corrected by the different production yield of prompt photons in p–p collisions at $\sqrt{s}$= 5.5 TeV and $\sqrt{s}$= 14 TeV and binary scaling. The isolation parameters used are $p_T^{th}$ = 0.5 GeV/$c$ and $R$ = 0.4 for p–p collisions and $p_T^{th}$ = 2 GeV/$c$ and $R$ = 0.3 for Pb–Pb collisions.

## 6.7 Photon correlations

The $\gamma$+jet coincidence channel provides the most precise technique for jet quenching [18] studies. Since the photon does not interact with the colored medium, it provides the least biased measurement of the energy and direction of the recoiling jet, essential information for the study of the jet fragmentation modification due to the medium. The main processes contributing to $\gamma$-jet events are Compton scattering and quark annihilation, see Eq. 6.1, with Compton scattering contributing more than 90% to the cross section. Thus, a $\gamma$ tagged jet is predominantly a quark-jet, whereas dijets consist of more than 60% of gluon-jets. Measurements of both, $\gamma$-jet and dijets, are complementary and can provide new insight into the interaction of quarks or gluons with the medium. The kinematic reach of the $\gamma$-jet measurement is however limited due to the small cross section of electromagnetic processes. In practice, we will perform the analysis of $\gamma$+jet events and $\gamma$-hadron correlations in Pb–Pb collisions at $\sqrt{s_{\rm NN}}$= 5.5 TeV from $E_\gamma$ = 20 GeV (below this energy the decay photon contamination is too large, see previous section) up to a maximum energy of about $E_\gamma$ = 50 GeV dictated by limited yield (considering a threshold of about 1000 events per year, see Fig. 6.3-right). The goal is to identify the medium induced modifications to the jet fragmentation function. Standard jet reconstruction algorithms, like the ones used in Sec. 5 will be complemented by photon tagged jet studies to pursue the medium modifications in the fragmentation function. Several $\gamma$-tagged algorithms were developed in Refs. [15, 19] for analysis of the photon-jet events in p–p and Pb–Pb collisions, demonstrating the feasibility of such analyses with the ALICE detectors. These measurements will be complemented by photon-hadron correlations [20, 21]. In the following we discuss correlations of prompt photons detected in the EMCal with jets or hadrons in the tracking detectors of ALICE.

### 6.7.1 Photon correlation with jets

In the $\gamma$-jet analysis the photon is detected in the EMCal and the jet in the TPC. The energy of the jet can not be fully reconstructed (see Fig. 6.17) but this is not necessary in this approach as it is determined by the energy of the photon. The approach is explained in detail elsewhere [15]; it is based on the algorithm developed for the PHOS detector of ALICE [19].

The method progresses as follows: first, we identify and isolate a prompt photon in the EMCAL, as shown in previous sections; second, we select the highest transverse momentum particle in the opposite direction of the photon (in azimuth), in a window of $\pi \pm 0.5$ rad, (almost 100% of jets in the TPC acceptance fall in this window), and with energy of at least $0.1E_\gamma$ to avoid fake jet reconstruction; third, using the highest momentum particle as a seed we determine the jet axis in an iterative process according to the UA1 cone finder. The jet energy is reconstructed summing the energy of all particles in a cone with radius $R_c$ around the jet axis. In the analysis, we use only jets with energy exceeding 20% of the

photon energy in order to avoid jets that fall only partially in the acceptance or fake jets due to the background. Then, we use a cone radius around the jet axis $R_s$, being $R_s > R_c$ (see next paragraphs for explanation), to obtain the jet fragmentation function.

In the case of Pb–Pb collisions, a large background from non-jet particles complicates the analysis. This background is evaluated outside of the jet cone. In the following, we present preliminary results which demonstrate the feasibility of this measurement based on the statistical uncertainties due to the out-of-cone background. The calculations are done for photon energies $\mathrm{E}_\gamma > 30$ GeV, where the contamination of the prompt photon sample with isolated $\pi^0$ or other hadrons is small, and can be neglected to first order.

Fig. 6.17-left shows the distribution of the reconstructed jet energy (expressed as a fraction of $E_\gamma$) using the TPC only in p–p events with different jet cone $R_c$ values. The distribution peaks at about half of the initial photon energy because of the missing neutral particle contribution to the jet energy.

Fig. 6.17-right shows the ratio of the reconstructed jet energy over the background fluctuations, evaluated as RMS of the background energy distribution, from central HIJING Pb–Pb simulations as a function of the cone radius $R_c$. This ratio is highest for the smallest cone radii, but the correction factor to the reconstructed energy will also be largest for small cone radii. A compromise between these two effects is to use $R_c = 0.25$ for 30 GeV jets. Larger cone sizes (see jet chapter) can be used if the neutral jet energy is measured as well (i.e. for EMCAl+TPC jets). Since we only use the cone radius to find the jet and determine the jet axis, the large correction factor to the reconstructed jet energy does not play a role in this analysis. In order to determine the fragmentation function we use a larger cone around the jet axis to investigate the particle momenta within the jet and we use the photon-energy to calculate their fractional momentum.

In order to determine the accuracy of a fragmentation function measurement based on the away-side jet in $\gamma$-jet events we have first simulated $\gamma$-jets in p–p collisions at $\sqrt{s} = 5.5$ TeV (with the full detector response) generated with PYTHIA using a realistic parton spectrum with $p_T$ -hard in the range $30 < \hat{p}_T < 100$ GeV/$c$. Due to the sharply falling exponential spectrum the jet distribution has an average energy of about 44 GeV. To unravel the effect of jet quenching we apply the PYQUEN (PYthia QUENching) [16] event generator, which is based on an accumulated energy loss approach, i.e. the method has no path length dependence. The gluon radiation associated with each parton scattering in the longitudinally expanding quark-gluon fluid is simply added. The PYQUEN energy loss is parametrized through a modified radiation spectrum $dE/dl$ which depends on the impact parameter ($b$), the initial temperature of the plasma ($T_0$), and the formation time of the plasma ($\tau_0$) among others. We used PYQUEN default parameters and initial conditions which were set according to an estimation for LHC heavy-ion energies, ($\tau_0 = 0.1$ fm/$c$, $T_0 = 1$ GeV/$c$ and b $< 5$ fm corresponding to a value of $\hat{q} \approx 30$ GeV$^2$/fm). We also simulate the heavy-ion background based on quenched HIJING events ($b < 5$ fm, $\sqrt{s_{\mathrm{NN}}} = 5.5$ TeV) which are then merged accordingly with the PYTHIA and PYQUEN events.

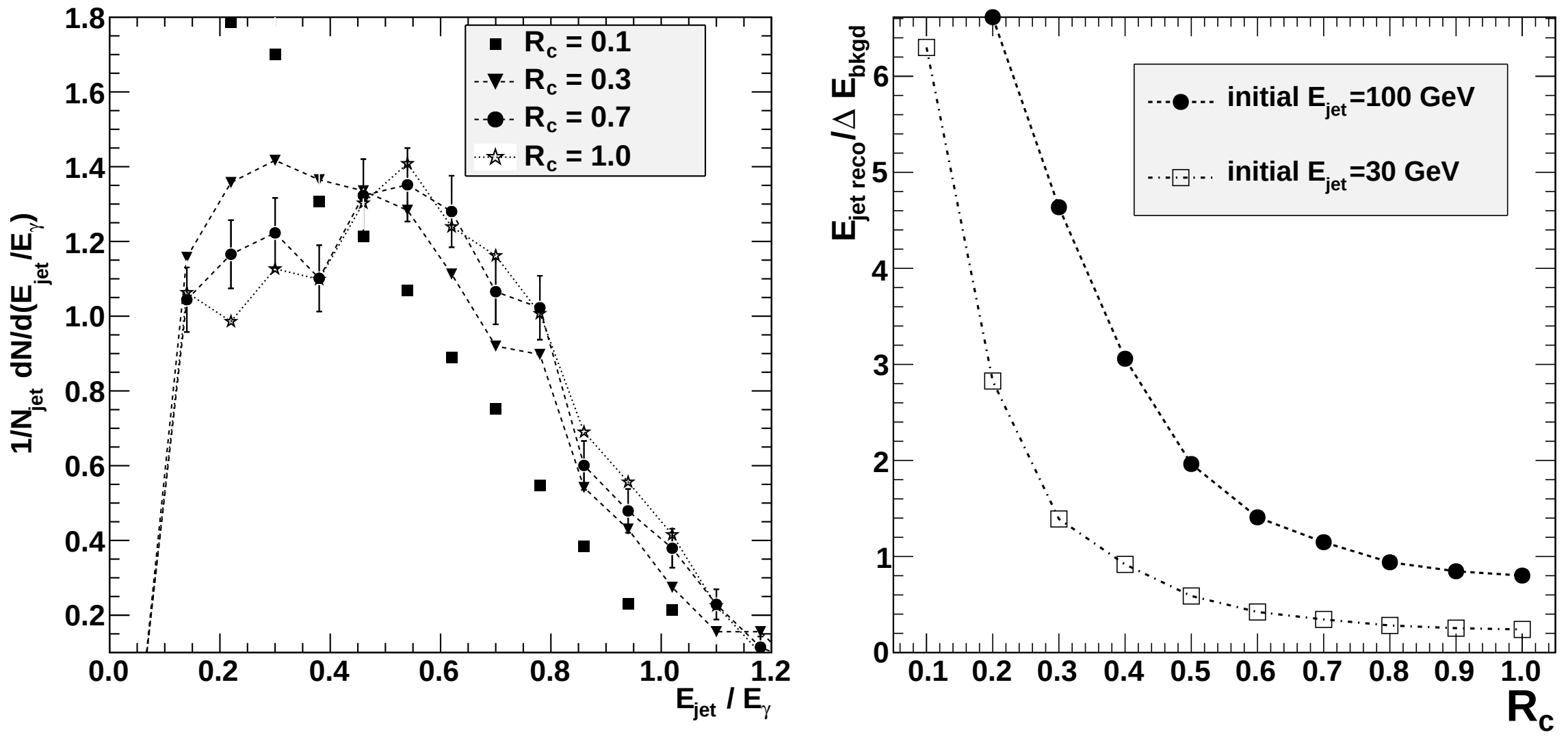


**Figure 6.17:** Left panel: Distribution of reconstructed jets as a function of the reconstructed jet energy expressed as a fraction of the $\gamma$ energy $E_{\rm jet}/E_\gamma$ for various values of the jet cone opening angle $R_c$ for $\gamma$-jets (PYTHIA, $\hat{p}_T$ = 100 GeV/$c$) in p–p collisions at $\sqrt{s}$ = 14 TeV. For clarity, the error bars are shown only in the case of $R_c = 0.7$. For other radii, the uncertainties are similar. The error bars include only statistical errors. Right panel: Ratio of the jet energy over the background fluctuation in Pb–Pb collisions as a function of the cone size (for PYTHIA $\gamma$-jet events of 30 and 100 GeV embedded in central HIJING Pb–Pb events at $\sqrt{s_{\rm NN}}$= 5.5 TeV.

To study the medium modification we plot the fragmentation function, i.e. the distribution of hadrons in the jet as a function of the scaling variable $\xi = \ln(1/x)$ with $x = E_{had}/E_\gamma$. In order to determine a fragmentation function accurately, the opening cone angle $R_c$ should be as large as possible. However, large cone radii will have large background contaminations in heavy-ion collisions. Since our primary goal is a relative measurement of the modification of the fragmentation function distributions compared to vacuum jets, this effect can be revealed even if smaller cone radii are used and the jet is not fully reconstructed.

The contribution of the underlying event is based on determination of a 'background' fragmentation function outside the jet cone in quenched HIJING events. By using regions in the same rapidity bin as the jet, with a large azimuthal window but separated from the jet cone by twice the cone size an unbiased background distribution can be extracted. This distribution is then subtracted from the measured fragmentation function in the cone. Figure 6.18 shows the raw fragmentation function and 'background' fragmentation function using a cone of size $R_s = 0.6$. One should note that effects such as anisotropic flow have not yet been taken into account and will contribute to the background, although the effect is likely small for central collisions.

Fig. 6.19 shows the fragmentation functions for p–p and quenched jets in Pb–Pb collisions

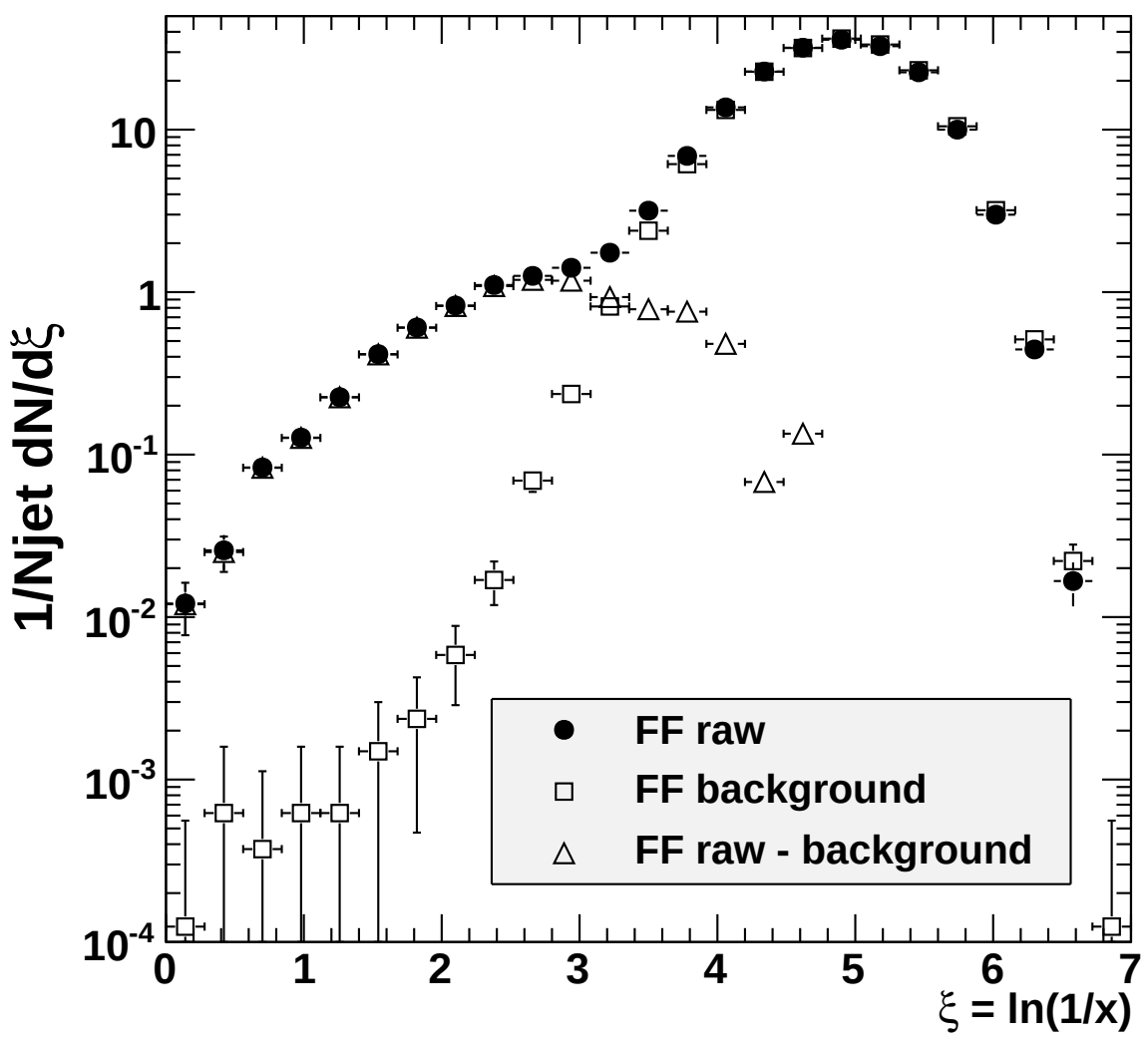


**Figure 6.18:** Fragmentation functions for $\gamma$-jet events in Pb–Pb collisions (PYTHIA jets embedded in central HIJING) for jets with $p_T$ greater than 30 GeV/$c$ with a cone size of $R_s = 0.6$. Shown are the fragmentation function determined in the jet cone (black dots), the 'background' fragmentation function determined outside the jet cone(triangles) and corrected jet fragmentation function (jet-background, squares).

based on the aforementioned method with $R_s = 0.6$ around the pre-determined away-side jet axis. The uncertainty in the fragmentation function due to the underlying background subtraction becomes too large above $\xi = 4$, which corresponds to a lower $p_T$ cut-off of roughly 500 MeV/$c$ for 30 GeV jets. Since there is considerable uncertainty in the quenching strength, as well as in the background level, further studies with varying jet algorithms and quenching scenarios are warranted.

Further improvements in the determination of the fragmentation function in $\gamma$-jet events can be expected for the case of a photon in the PHOS detector and the recoil jet reconstructed in the EMCal [19]. In this case the jet energy reconstruction and the direction is considerably more reliable, as shown in the jet reconstruction section. Considering the photon rates shown on Fig. 6.3, we expect that the PHOS, which has approximately seven times less acceptance than the EMCal, will measure prompt photons for energies close to $E_\gamma = 20$ GeV. Here the number of $\gamma$ (PHOS) + jet (EMCal) coincidences could be of the order of 1000 events[3]. The studies in Ref. [19] suggest that a measurement of the fragmentation function modification in the configuration $\gamma$ in PHOS + jet in EMCal can be achieved in the range $0.1 \pm 0.05 < x < 0.6 \pm 0.05$ or equivalently $0.6 \pm 0.1 < \xi < 1.5 \pm 0.3$ with a 5% sensitivity.

[3]In Ref. [19]. It was found that the probability to match a photon in PHOS with a jet in EMCAL is about 30% and in the TPC about 40% with a similar algorithm.

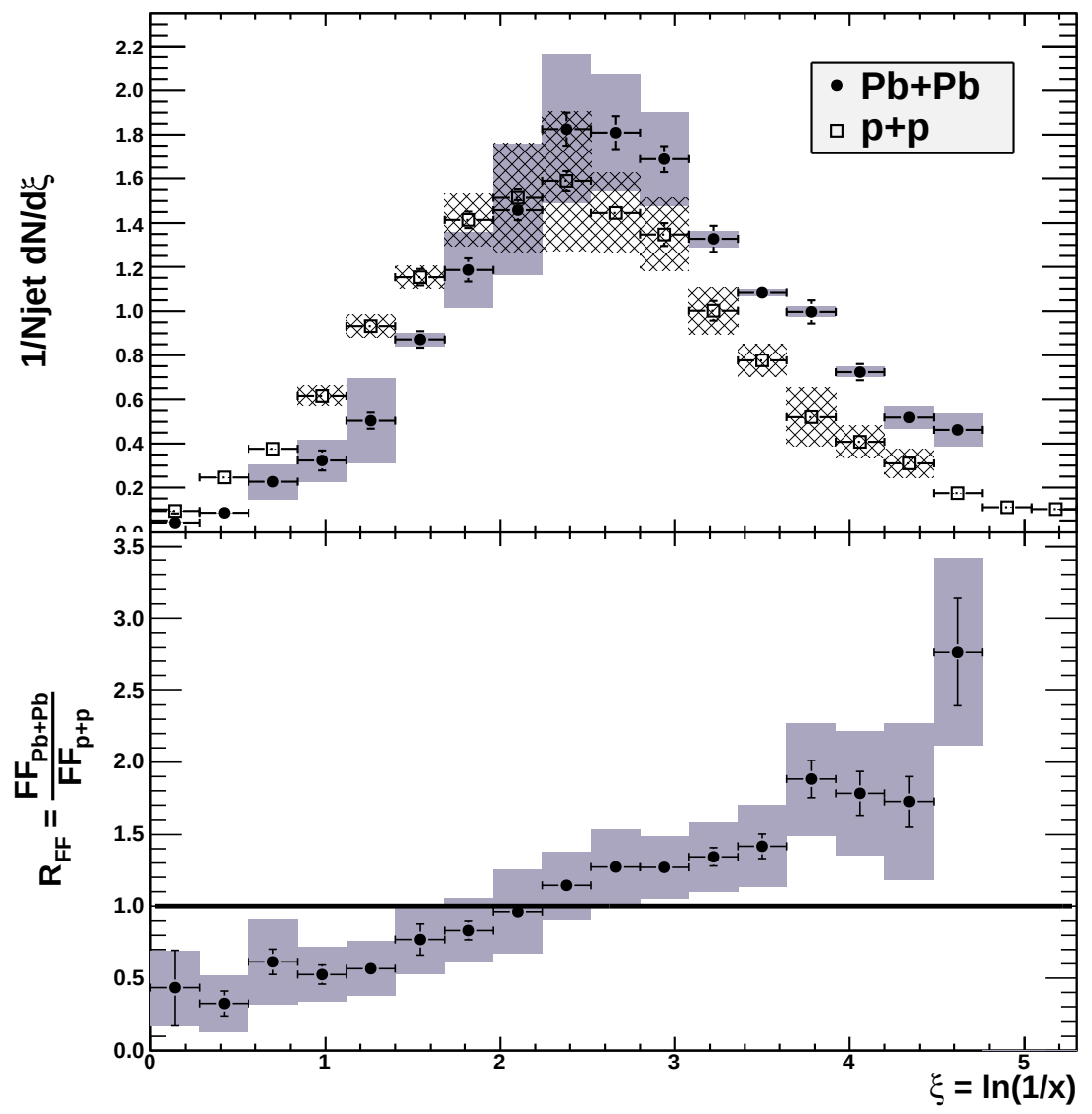


**Figure 6.19:** Fragmentation functions determined for Pb–Pb (black dot) and p–p collisions (open squares) at $\sqrt{s} = 5.5$ TeV as a function of the scaling variable $\xi$. The jets were reconstructed using a cone opening $R_s = 0.6$. The grey area represents the systematic uncertainties due to the background subtraction. Statistical errors are based on 5000 $\gamma$-jet events.

## 6.7.2 Photon correlation with hadrons

In this section we present the results of a first study aimed to measure the modification of the jet fragmentation in nuclear collisions using prompt photon–hadron correlations, a method that does not require the reconstruction of the jet. This method is based on studies of Refs. [21, 22] , and it rests on the fact that under given kinematical conditions, the imbalance parameter $x_{\mathrm{E}}$, calculated as:

$$x_{\mathrm{E}} = \left| \frac{p_{x_h} p_{x_\gamma} + p_{y_h} p_{y_\gamma}}{p^2_{T_\gamma}} \right| , \tag{6.4}$$

is in Leading Order (LO) kinematics equivalent to the jet fragmentation variable $x$. The kinematic conditions which verify this equivalence are the following: First, hadrons must be produced from the fragmentation of a hard scattered parton to exclude hadrons from the soft underlying event; second, the photon must be produced directly from a hard partonic process, and not from jet fragmentation, therefore we want isolated photons; and finally, the range over which the equivalence is verified is given by $p^{\mathrm{min}}_{\mathrm{T_h}} / \, p^{\mathrm{min}}_{\mathrm{T}_\gamma} \leq x_E \leq 1$. Therefore to probe the broadest range the photon and hadron momenta cuts must be very asymmetric, $p^{\mathrm{min}}_{\mathrm{T}_\gamma} \gg p^{\mathrm{min}}_{\mathrm{T_h}}$.

In the following, the feasibility to measure azimuthal correlations and $x_E$ distributions in p–p collisions and Pb–Pb at $\sqrt{s} = 5.5$ TeV will be discussed. The steps of the analysis procedure are: first, an isolated prompt photon is identified in the EMCal; then the distribution as a

function of $x_E$ or the relative azimuth angle $\Delta\phi = \phi_\gamma - \phi_h$ is accumulated. Hadrons are required to fall within $\pi/2 < \Delta\phi < 3\pi/2$ to contribute to the $x_E$ distribution (conditional yield (CY)). Unless specified, a miminum $p_T$ cut of 200 MeV/c on hadrons is used and the prompt photon must have $p_T$ larger than 30 GeV/$c$. We have analyzed $\gamma$-jet events generated with PYTHIA and qPYTHIA ($\hat{q}$ =50 GeV$^2$/fm) in p–p collisions at $\sqrt{s} = 5.5$ TeV for 5 to 200 GeV photon energy. No underlying event study for heavy-ion collisions has yet been performed.

The relative azimuthal angle, $\Delta\phi = \phi_\gamma - \phi_h$, between the direct photon and charged hadrons is strongly peaked at $\pi$ radian for $\gamma$–jet events in p–p collisions as shown in Fig. 6.20 . The $\Delta\phi$ distribution between the photon trigger and the charged hadrons broadens the far side peak when quenching effects are introduced via qPYTHIA. The broadening is small, however, and the background of an underlying heavy ion event might shadow this difference.

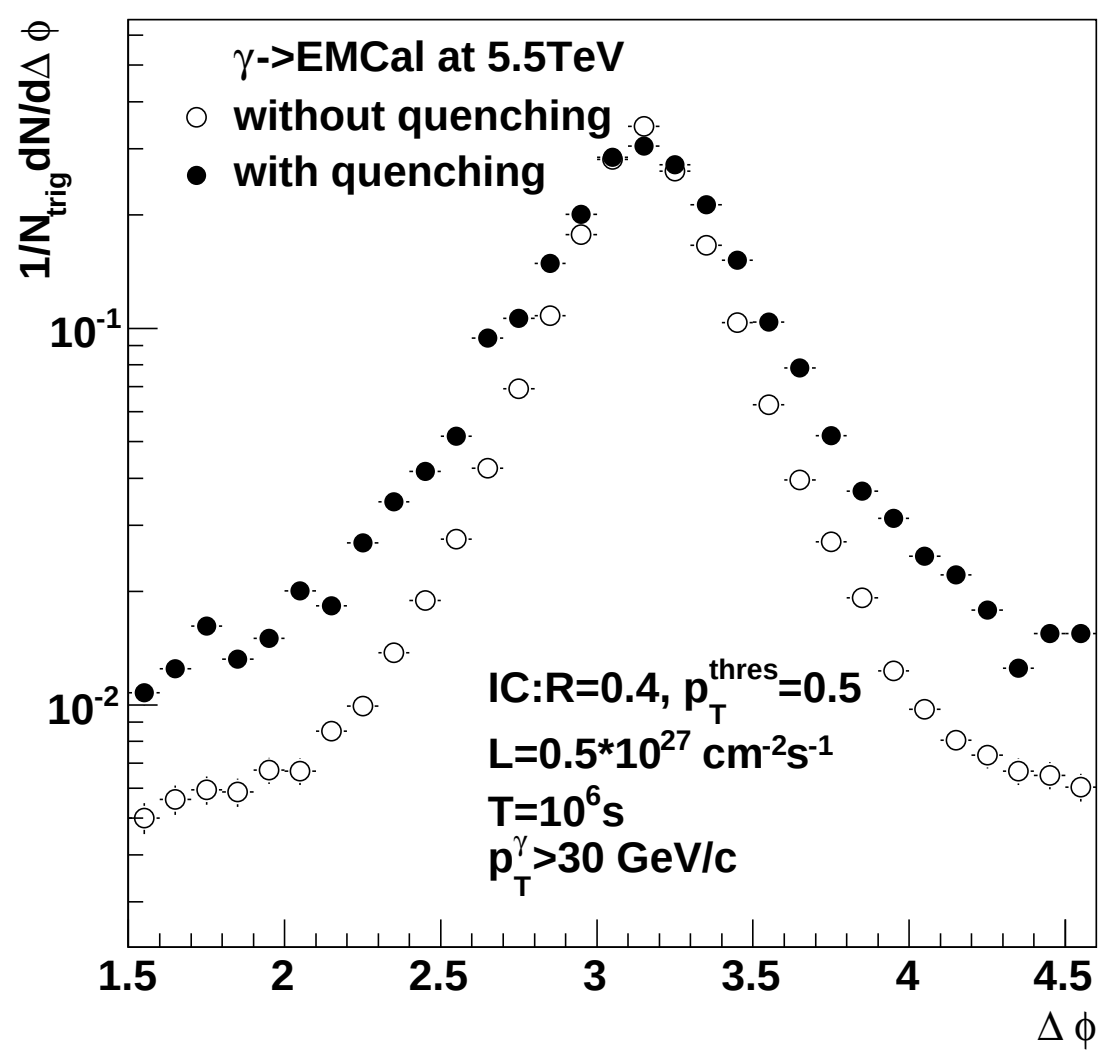


**Figure 6.20:** Relative azimuthal angle distribution $\Delta\phi = \phi_\gamma - \phi_{hadron}$ for $\gamma$-jet events in p–p collisions at $\sqrt{s} = 5.5$ TeV. Photons with $p_{T,\gamma} > 30$ GeV/$c$ are measured and isolated in EMCAL. Associated hadrons have $p_{T,hadron} > 2$ GeV/$c$. Full circles indicate qPYTHIA events and empty circles indicate PYTHIA events. Distributions are normalized to the total number of isolated photons.

In order to obtain an $x_E$ distribution in unquenched p–p collisions at $\sqrt{s} = 5.5$ TeV, the contribution from the underlying event was estimated by correlating the isolated photon with charged hadrons emitted on the same side as the photon, in the azimuthal window $-\pi/2 < \Delta\phi < \pi/2$. Figure 6.21 shows the inclusive and underlying event spectra. The final conditional yield distribution for hadrons, after subtraction of the underlying event contribution, is also shown in Fig. 6.21. The statistical error estimates are based on one month of data taking at LHC.

To quantify the medium modification of the photon triggered hadron distributions in heavy-ion collisions relative to p–p collisions, one can determine the ratio of the conditional yields

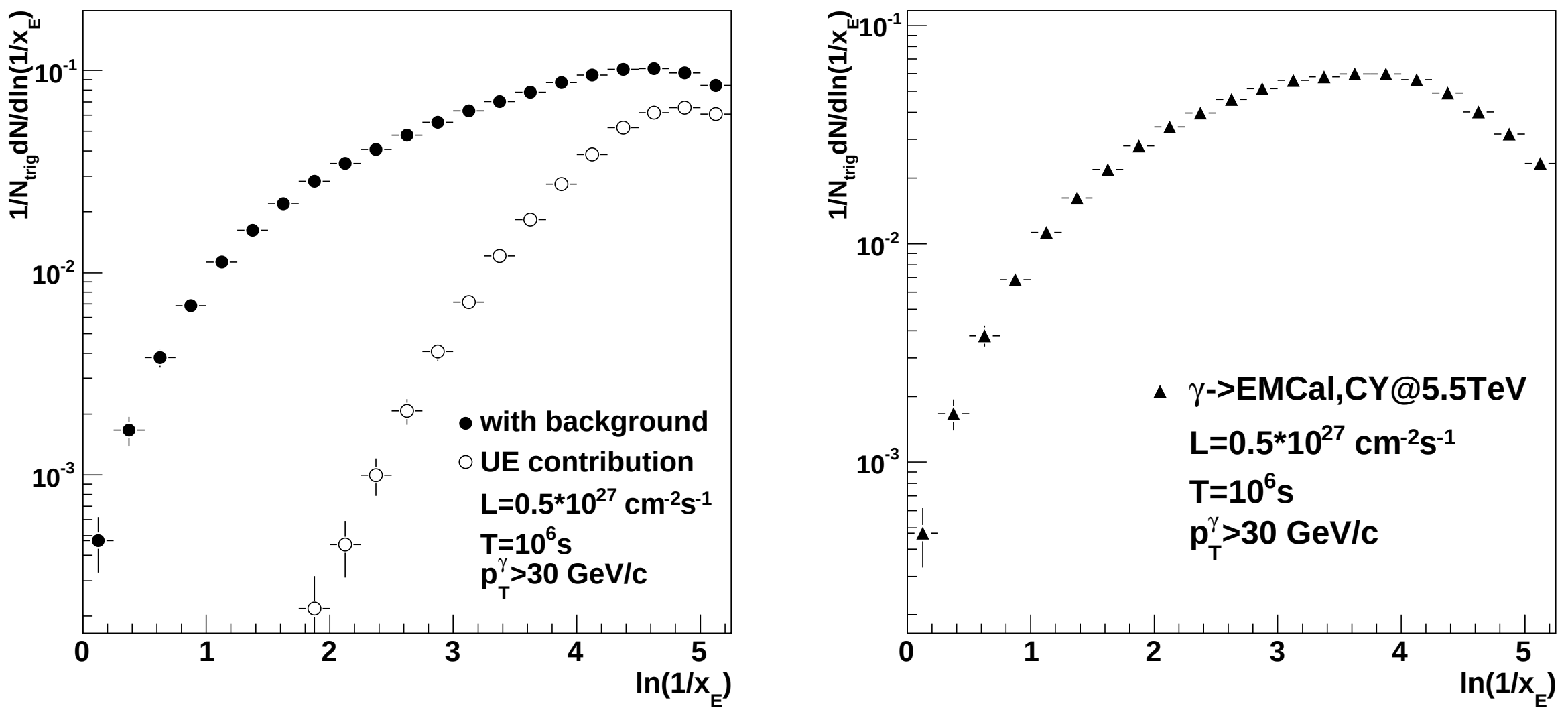


**Figure 6.21:** Photon-charged hadron correlation spectrum using photons identified by EMCal in p–p collisions at $\sqrt{s} = 5.5$ TeV. Left: Inclusive and underlying event distributions. Right: The corrected conditional yield spectrum (inclusive - underlying event). The statistical uncertainty is based on one month of data taking.

in quenched and unquenched jet events. Each distribution is normalized to the number of triggers, i.e. the number of isolated photons $(1/N_{trigger}dN_{hadrons}/d(\ln(1/x_E)))$. The hadron distributions were determined in the azimuthal window $\pi/2 < \Delta\phi < 3\pi/2$ relative to the photon. In this report, we only compare quenched distributions from qPYTHIA to unquenched distributions from PYTHIA, which means that the quenching effect is modeled but the background is based on the underlying event distributions in p–p collisions only. The resulting photon-triggered hadron spectra after the underlying event subtraction are shown in Fig. 6.22-left. The ratio of the PYTHIA over qPYTHIA distributions, labeled $R_{CY}$, is shown in Fig. 6.22-right. The statistical errors are based on the achievable annual yield of hadrons correlated with photons with $p_T$ larger than 30 GeV. An enhancement at low $x_E$ and a suppression at high $x_E$ can be observed. These distributions have to be folded with realistic background distributions, as described in Chapters 5 and 8, in order to obtain final statistical and systematic uncertainties for the heavy-ion environment.

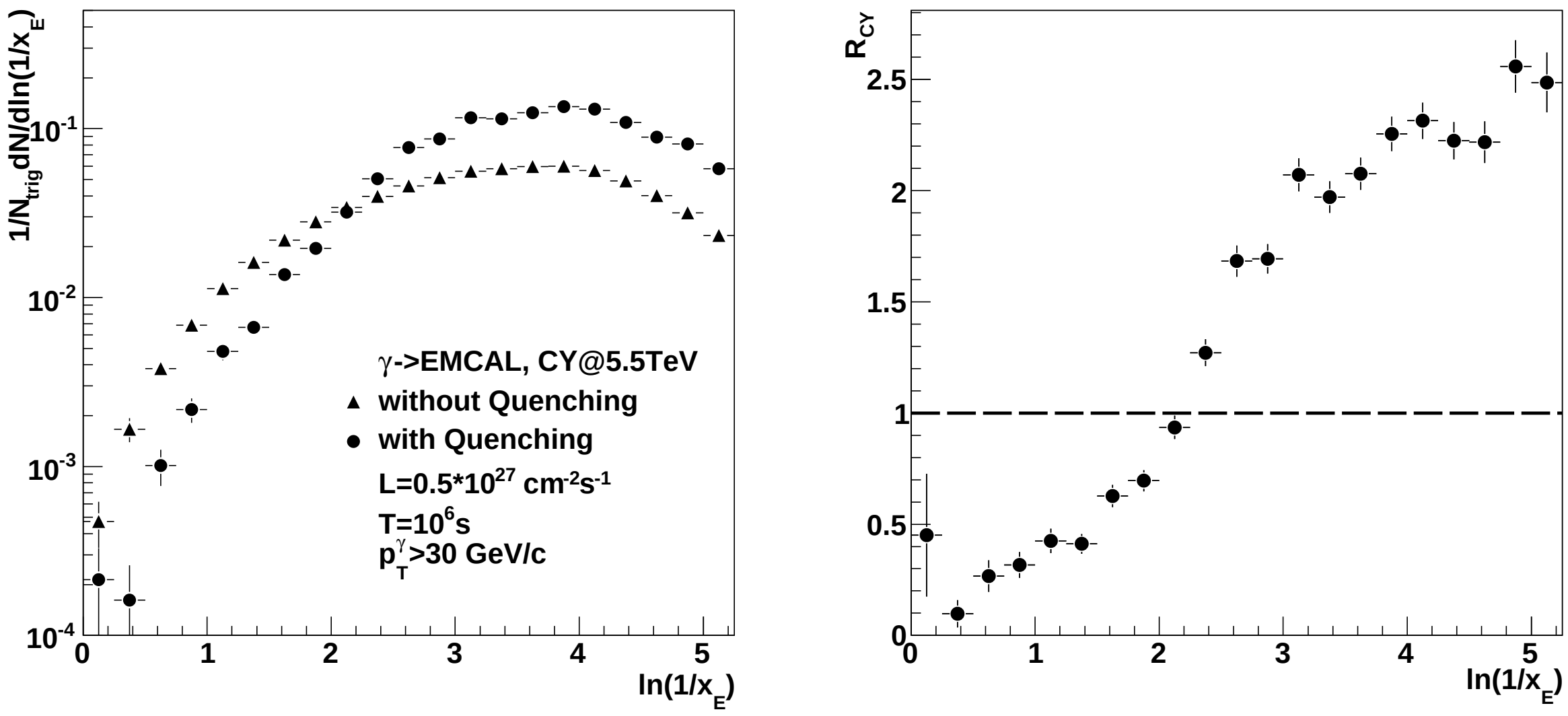


**Figure 6.22:** Left: $\gamma$-hadron correlation distributions in quenched and unquenched PYTHIA events as a function of $\xi$. Right: The ratio of the photon-triggered conditional hadron yields $R_{CY}$ for PYTHIA and qPYTHIA events.

# Chapter 7

# Heavy Flavor Measurements

## 7.1 Introduction

In this chapter we discuss the capabilities of the EMCal to measure heavy flavor production using high $p_T$ electrons.

Heavy flavor hadrons, ie. those containing charm or bottom quarks, will be abundantly produced at the LHC and are important probes of the Quark-Gluon Plasma (QGP) [1]. At high $p_T$, the generic prediction that the energy loss of massive quarks in a colored medium is reduced due to the suppression of forward radiation ("dead cone effect" [2]) has not been validated by measurements at RHIC [3, 4], which find heavy flavor production at high $p_T$ to be suppressed at the same level as light flavor quarks and gluons (see Fig. 7.1). This surprising result has led to significant theoretical and experimental effort to improve our understanding of partonic energy loss. For a recent overview see [5] and references therein.

Theory calculations from different groups of the $R_{\rm AA}$ for non-photonic electrons from charm and bottom are shown in Fig. 7.1. The calculations in Fig. 7.1(right) by DGLV (curve I) and BDMPS (curve II) are based on radiative effects while others (van Hees et al. (curve IV)) consider the collisional (elastic) energy loss of partons. It is clear that in order to approach the level of suppression measured by experiment both channels must be included in the calculation (i.e. curve III). It is essential to establish experimentally the relative strength of charm versus bottom electrons in this $p_T$ range, as it is not well constrained by NLO theory.

Fig. 7.2 (left) shows the predicted $R_{\rm AA}$ of light and heavy quarks vs gluons at RHIC energies. In the $p_T$ range above 10 GeV/c these predictions show that $R_{\rm AA}$ of charm will indeed be similar to $R_{\rm AA}$ of light quarks, but less than that of bottom quarks. An illustration of the reduced energy loss for heavy quarks is shown in Fig. 7.2 (right), where using bottom quarks will allow one to probe "deeper" into the QGP in the initial stages of heavy-ion collisions [6].

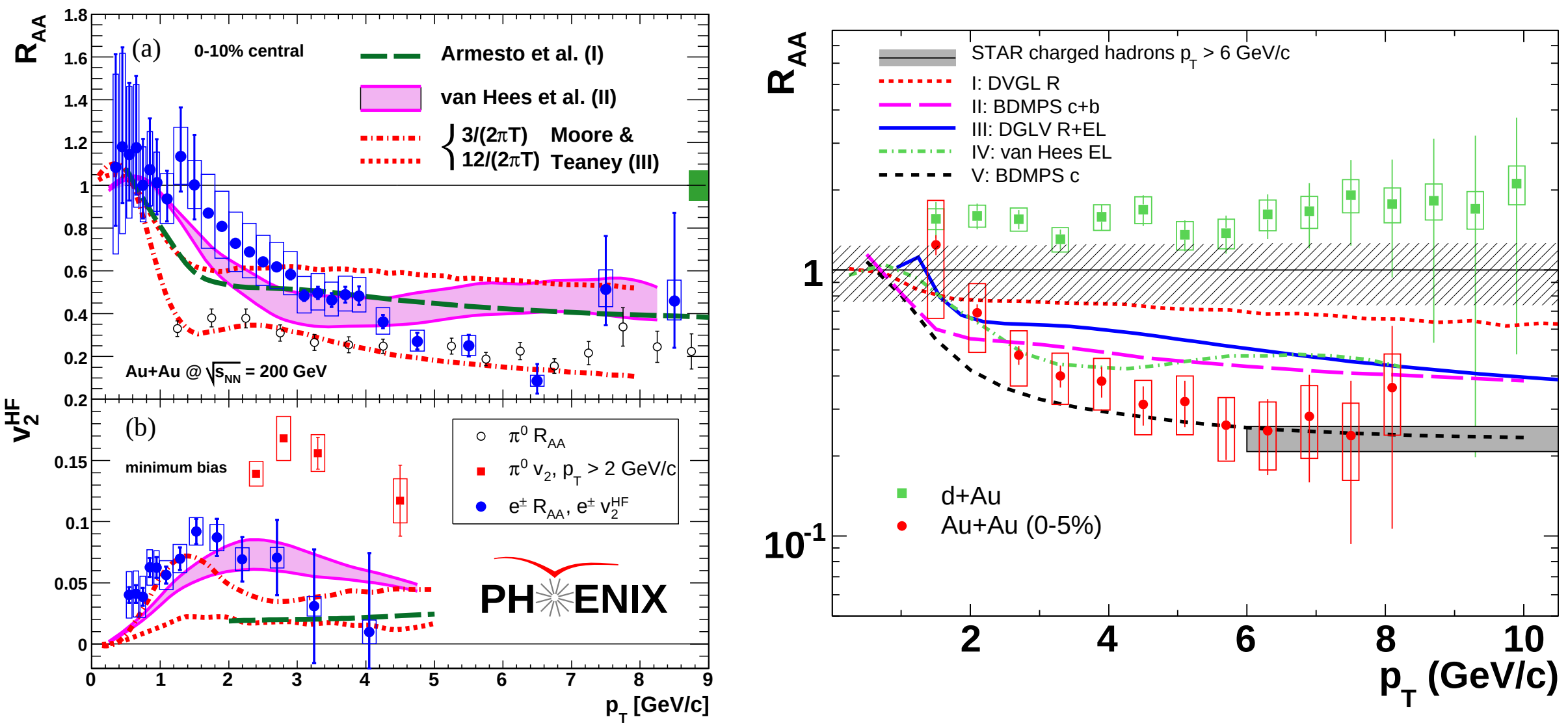


**Figure 7.1:** Measurements of non-photonic electrons in $\sqrt{s_{NN}}$ = 200 GeV p–p and Au–Au collisions Left: from PHENIX [3]. Upper panel shows $R_{\rm AA}$ for central Au–Au collisons, lower panel shows $v_2$ for minimum bias Au–Au collisions. Right: from STAR [4]. $R_{\rm AA}$ for non-photonic electrons in d+Au and Au+Au collisions. See the cited papers for discussion of the theory curves.

Beyond the energy range where the dead cone effect is expected to be significant (i.e. $E_T \gg m$), heavy quark jet production may provide the best tool to probe the color-charge dependence of energy loss [7]. In a more speculative vein, modelling of heavy "quark" propagation in a strongly coupled fluid using the techniques of AdS/CFT has received significant recent attention [8], including a suggestion that the relative suppression of charm vs. bottom quarks is different in pQCD and AdS/CFT, with a magnitude that could be resolvable by experiment [9].

The measurement of heavy flavor production at high $p_T$ will clearly play a central role in the ALICE physics program. While exclusive reconstruction of charm mesons will be carried out via the use of the ITS, measurements of high $p_T$ heavy flavor production require a fast trigger, which is possible only for the semi-leptonic decay mode, with a branching ratio of $\sim$ 10%. The TRD has an efficient electron trigger and good hadron rejection for $p_T < 10$ GeV/c, but at higher $p_T$ additional tools are required for heavy flavor measurements. The EMCal has excellent capabilities for fast triggering and hadron rejection, and provides unique coverage in ALICE for heavy flavor measurements at very high $p_T$.

In this chapter we present backgrounds and systematic uncertainties for representative measurements of high $p_T$ electrons in ALICE using the EMCal. Such measurements rely crucially on the discrimination of electrons from hadrons, and we estimate the hadron suppression required for systematically significant electron measurements. We project the statistical reach and systematic uncertainties for the measurement of the non-photonic electron (NPE) spectrum for central Pb–Pb collisions in one running year.

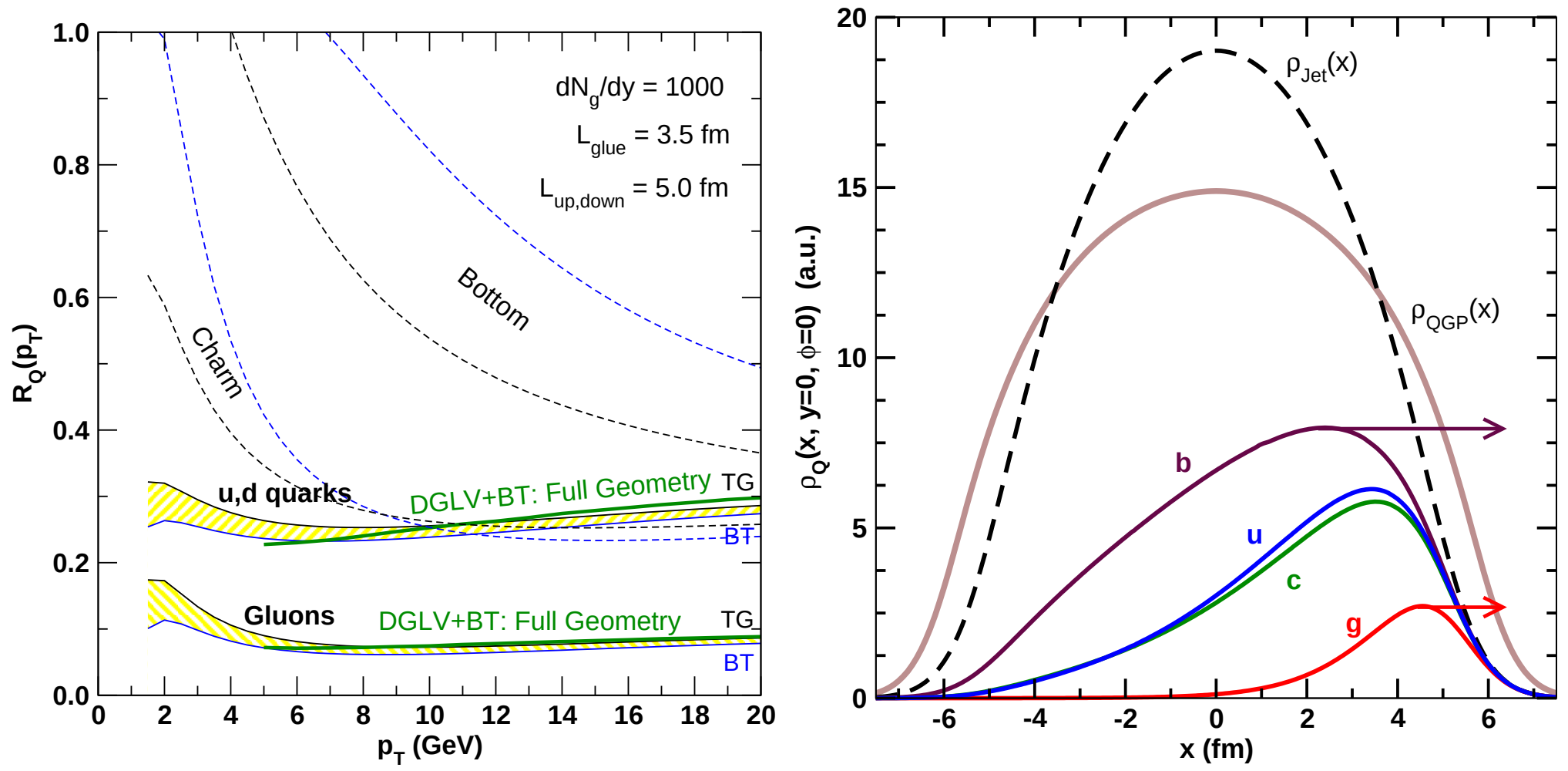


**Figure 7.2:** Left: Nuclear Modification factor ($R_{AA}$) for gluons, light quarks and heavy quarks at RHIC energies. The bands indicate an estimate of the theoretical uncertainties. Right: Transverse coordinate of the production point of surviving quark and gluon jets ($p_T = 15$ GeV/c) as calculated by the DGLV model. Both figures from [6].

The combination of jet-finding and electron identification capabilities in ALICE opens up the interesting avenue of full reconstruction of heavy flavor jets. First studies are presented which aim at "tagging" jets with heavy flavor. Similar studies have been carried out at the Tevatron [10].

## 7.2 High $p_T$ Electron Rates

The primary physics sources of electrons at high $p_T$ are the semi-leptonic decays of charm (C) and B-hadrons (mostly mesons), and the decay of W-bosons. Electrons from C-hadron decay come both from prompt charm production and secondary decay of C-hadrons produced by the hadronic decay of B-hadrons.

In addition, there are significant backgrounds to these sources, due both to other physical processes and detector effects. The physics backgrounds consist primarily of electrons from Dalitz decays of high $p_T$ $\pi^0$ and $\eta$ hadrons in jet fragmentation, while the dominant detector source in ALICE is photon conversion in the ITS.

To obtain the expected annual yields of single electrons in the EMCal, a sample of PYTHIA p-p collisions at 5.5 TeV, triggered on specific subprocesses, were simulated. A large sample of inclusive jet events, with one jet constrained to point towards the EMCal acceptance, was analyzed to evaluate the dominant physics and detector backgrounds. A smaller sample of W-boson events was generated to estimate the contribution from W-decay. The signal events were created using the ALICE standard PYTHIA heavy-flavor settings and requiring a B-jet in the EMCal acceptance and the B-hadron to decay semi-leptonically.

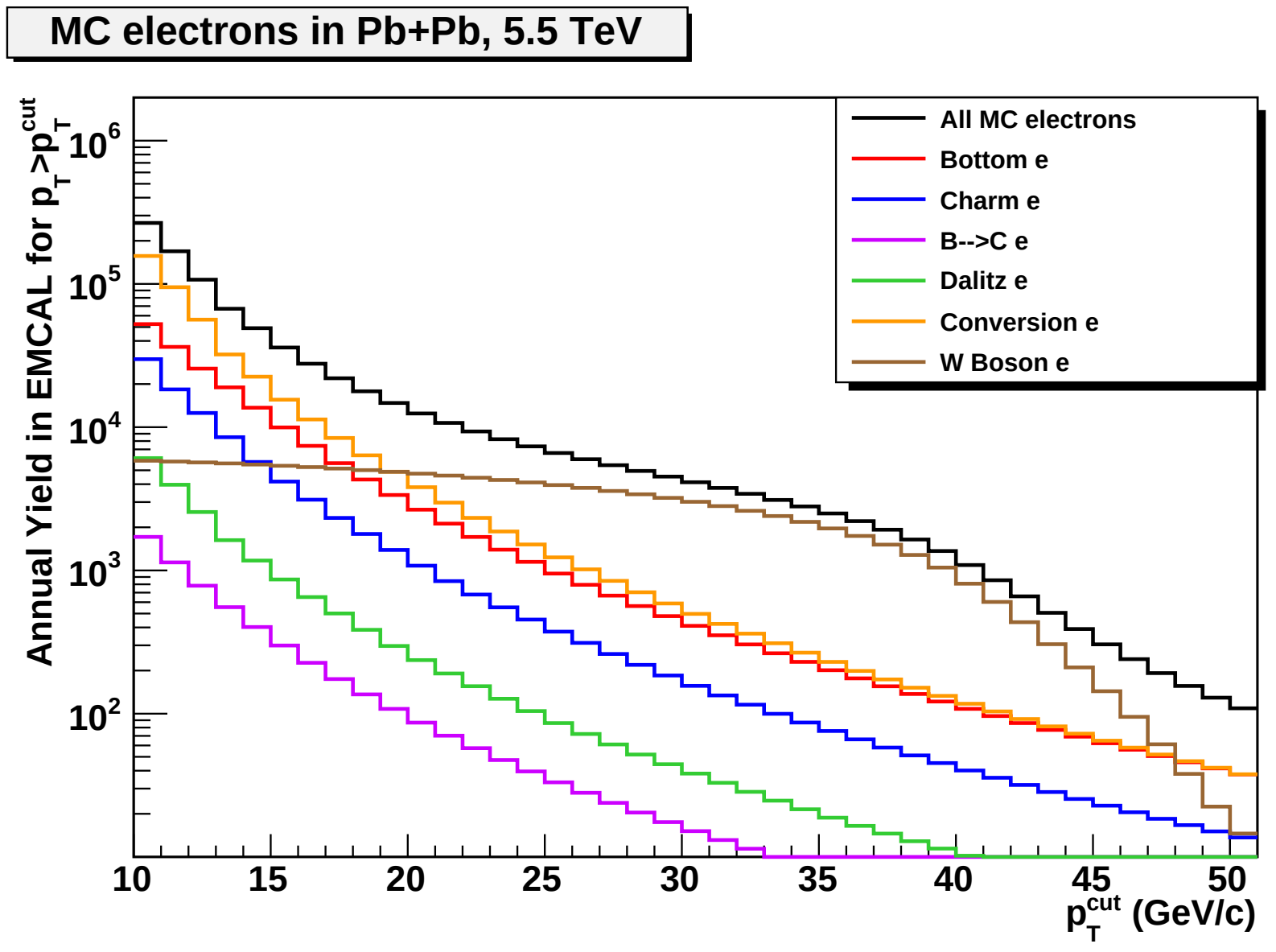


**Figure 7.3:** Rates of MC-Electrons from the various physics/detector sources shown as the integrated annual yield above a $p_T$-cut for a nominal ALICE Pb–Pb year at $\sqrt{s_{\mathrm{NN}}} = 5.5$ TeV.

Fig. 7.3 shows the contributions of the different physics sources to the distribution of electrons in the EMCal acceptance. Yields correspond to the expected Pb–Pb luminosity of 0.5 $\mathrm{mb}^{-1}\mathrm{s}^{-1}$ and one month ($10^6$ s) of Pb–Pb running to obtain an annual yield.

From this figure it is clear that there is a significant rate of bottom electrons to $p_T \sim 50$ GeV/c in the EMCal. The dominant backgrounds to heavy flavor electrons come from the photon conversions and from W-boson decays for $p_T > 20$ GeV/c. The other significant background to measuring B-electrons will be from charged hadrons misidentified as electrons. In order to evaluate the minimum hadron rejection required by the electron identification algorithm, the ratio of charged hadrons to electrons from the same simulation is shown in Fig. 7.4. The ratio for transverse momentum in the range of interest from 10-50 GeV/c is a few hundred, which sets the scale for the hadron rejection requirement.

## 7.3 Electron Identification

The method of identifying electrons in the EMCal relies on the fact that electrons deposit all of their energy in the EMCal as compared to hadrons, which typically leave only a small fraction of their energy in the EMCal. In order to calculate the ratio of EMCal energy to reconstructed track momentum, tracks are matched to EMCal clusters. The procedure takes a reconstructed track and extrapolates it to the EMCal. If the distance between the extrapolated track-position and any cluster-position is less than a given value (in this study about the size of 1 tower) then the track is considered to be matched. In a second loop the

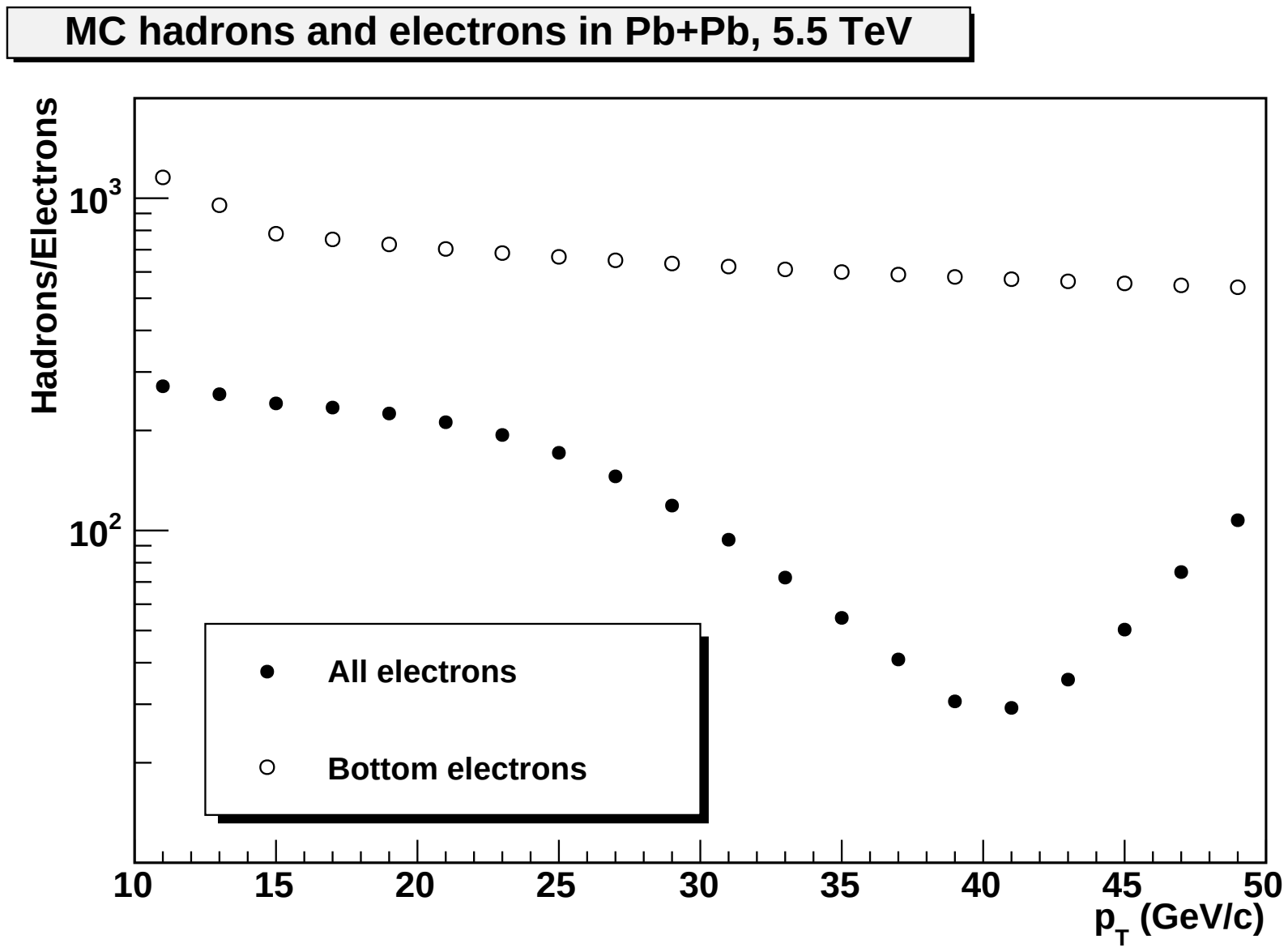


**Figure 7.4:** Ratio of charged hadrons to electrons at the particle level expected in Pb–Pb events. The ratio for all electrons including backgrounds is shown as filled circles, while that for signal electrons from B-decays is shown as the open circles. The dip near 40 GeV/c is due to the W-decay electrons.

matches are compared and only the closest matches are kept, thereby eliminating competing pairs. Fig. 7.5 shows the distance from track projection to the closest cluster from single particle simulations of pions and electrons at two different momenta. These distributions show that a cut value of $\Delta R = \sqrt{\Delta\eta^2 + \Delta\phi^2} < 0.02$ accepts in excess of 99% of the matches.

With the matching algorithm cut established, we can now study the matching efficiency of tracks to the EMCal in p-p collisions. To define the efficiency we require that the TPC-tracks be well reconstructed within the TPC (50 hits minimum) and ITS (3 hits minimum) and extrapolate to the fiducial volume of the EMCal. Efficiencies for track-cluster-matching are displayed in Fig. 7.6. Results for charged particles in a p-p event and two different values of residual-cut $\Delta R$ are shown on the left side. In the right panel of figure 7.6 we determine the matching efficiency for electrons and pions separately. The current algorithm yields $\sim$ 80% matching efficiency for electrons and $\sim$ 60% for pions at $p_T > 5$ GeV/c. This difference is due to the lower probability of reconstructing a pion cluster than an electron cluster.

We use the matched cluster-track pairs, the cluster energy, and the reconstructed track momentum to calculate the E/p ratio. This distribution is shown in Fig. 7.7 for single electron and pion tracks for two different momenta. The normalization of the two distributions is arbitrary and does not reflect the respective ratio of pions to electrons. The effect of particle interactions with the detector material is shown for each species. This was done by comparing the MC-input to the reconstructed momentum. If the particle had lost more than 10% of its input momentum it was considered to have interacted with the detector material and labeled "interacting". If, on the other hand, the reconstructed momentum was more than

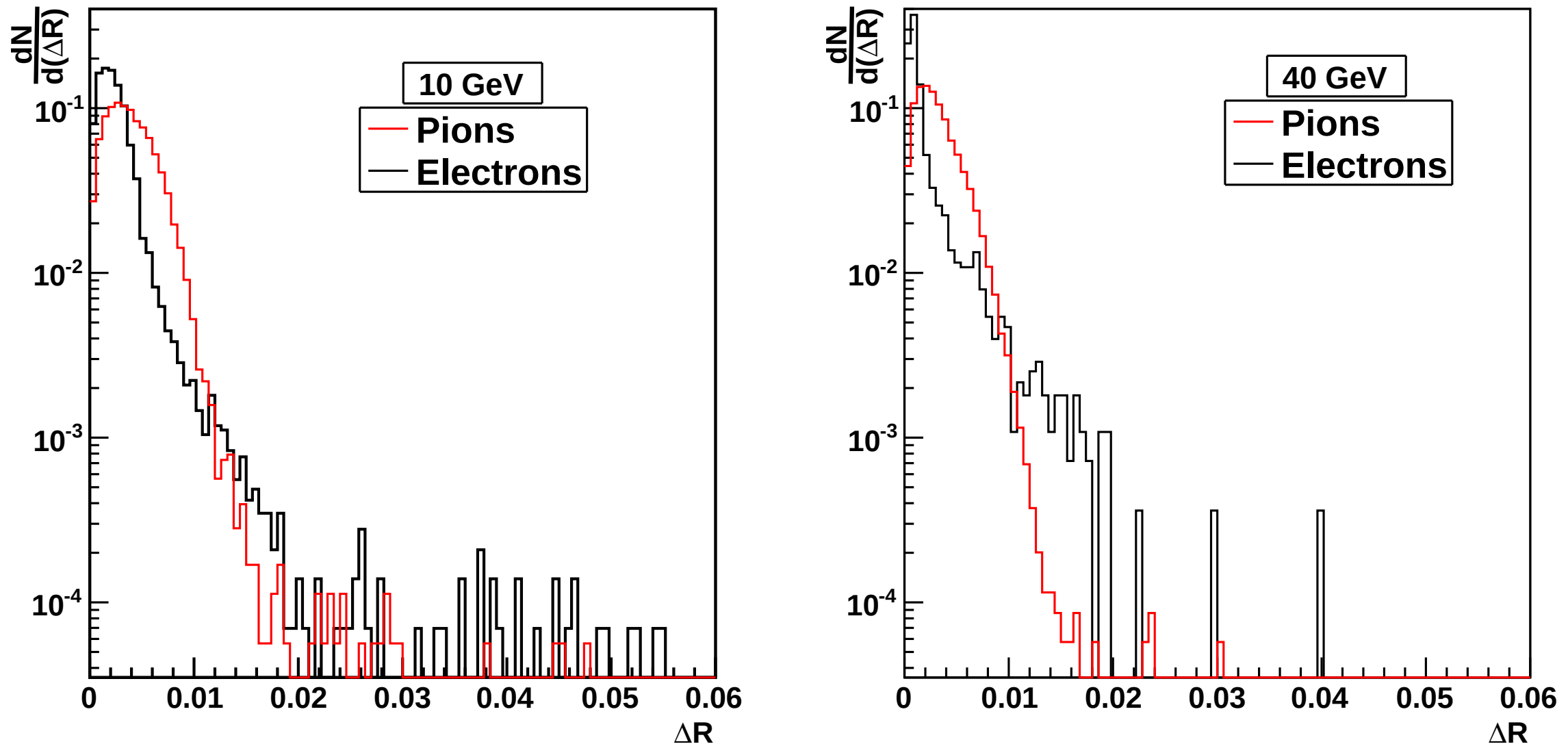


**Figure 7.5:** Track-Cluster residuals as a function of $\Delta R = \sqrt{\Delta\eta^2 + \Delta\phi^2}$. The distributions are from single particle simulations at momenta of 10 GeV/c (left) and 40 GeV/c (right) with full ALICE material. Only track-cluster pairs with single cluster matches are shown.

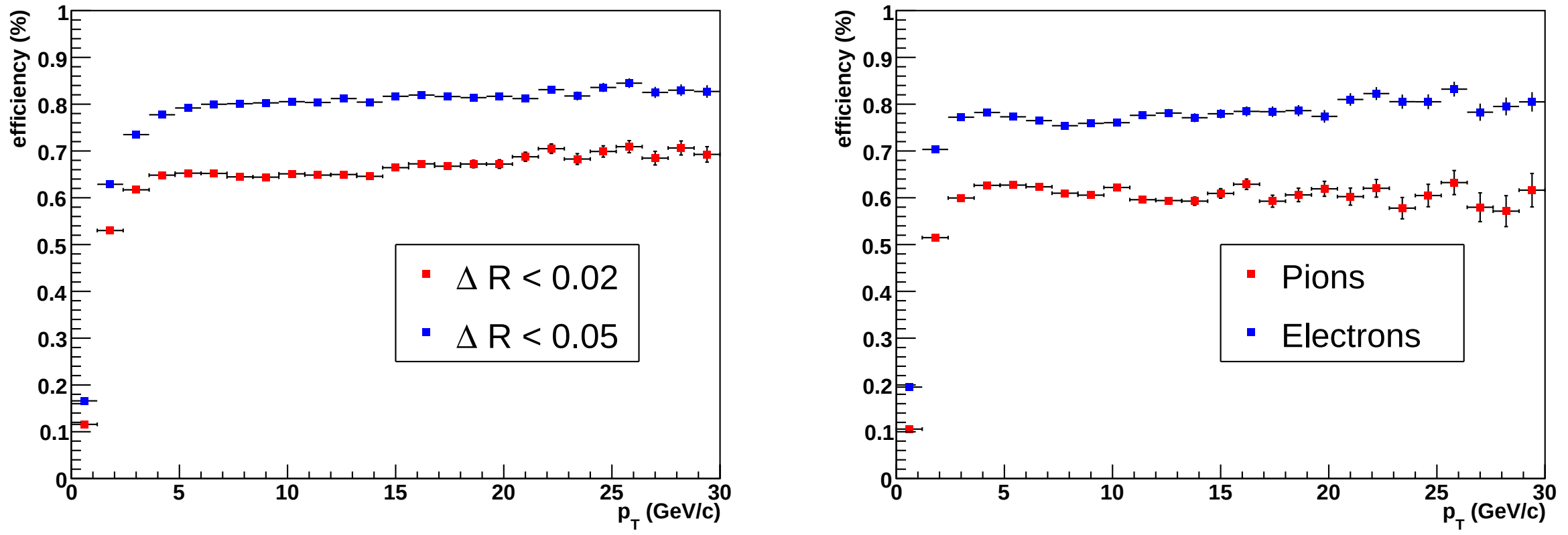


**Figure 7.6:** (left) Track-matching efficiency for charged particles in p-p events shown for two different cuts of $\Delta R = \sqrt{\Delta\eta^2 + \Delta\phi^2}$. (right) Track-matching efficiency for pions vs electrons for $\Delta R = 0.02$.

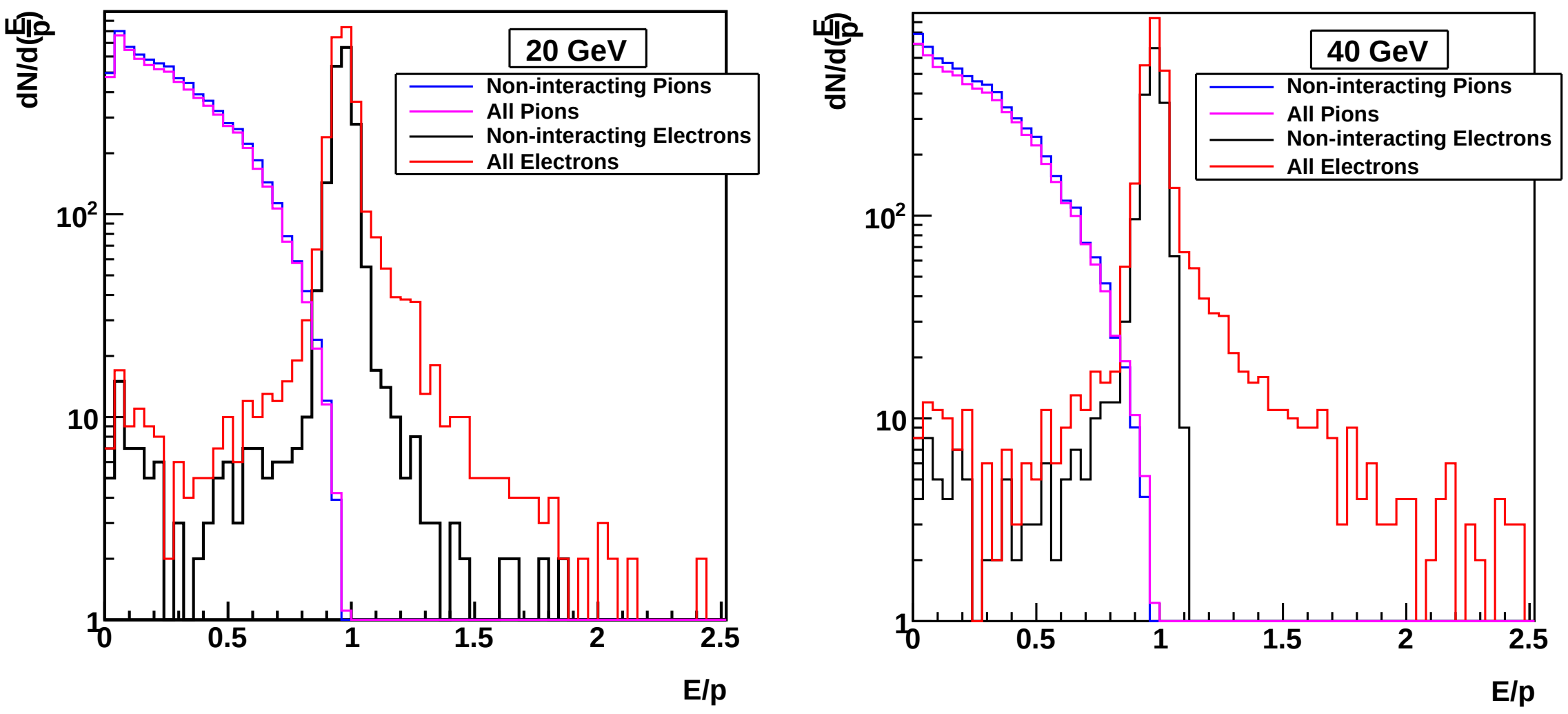


**Figure 7.7:** E/p distributions for single electrons and pions at momenta of 20 GeV/c (left) and 40 GeV/c (right). The normalization of the particle species relative to each other is arbitrary. The lines marked "non-interacting" refer to particles that lose less than 10% of their momentum before arriving at the EMCal surface.

90% of the input momentum then it was considered "non-interacting". Electrons are known to suffer bremsstrahlung in the detector material as can be seen in the slight "tails" of the E/p distribution towards the larger values.

We establish an electron identification criterion by setting a lower limit on the E/p ratio. We define the efficiency, $\epsilon$, as the number of electrons that pass the cut divided by the total number of electrons in our sample. One can then estimate the amount of hadron contamination by integrating the counts above the cut value. This determines the rejection power of the method, defined as $\epsilon^{-1}$ for pions. The actual electron purity in a p-p (Pb-Pb) collision will depend on the relative ratio of pions to electrons in the data. As shown in Fig. 7.4 this ratio was determined to be $\sim$ 100 in simulations.

The hadron rejection power is plotted as a function of track $p_T$ in Fig. 7.8. The efficiency values and error bars were determined using Bayesian statistics. In order to compare to results from the 2007 CERN-SPS test-beam data, we first considered particles that had suffered only little interaction with the material in front of the EMCal, and had lost less than 10% of their momentum with respect to the MC-input. Using this criterion the results for 90% electron efficiency are consistent with or better than the test-beam data. In the presence of an overwhelming hadron background it was necessary to lower the electron efficiency to obtain a purer sample of electrons. At 80% electron efficiency the hadron rejection in this idealized case is above 1000 at all momenta and improves with increasing $p_T$ as expected.

In Fig. 7.8 (right) we compare the simulation results without this cut on momentum loss of particles, reflecting the realistic detector material and obtain results that are significantly

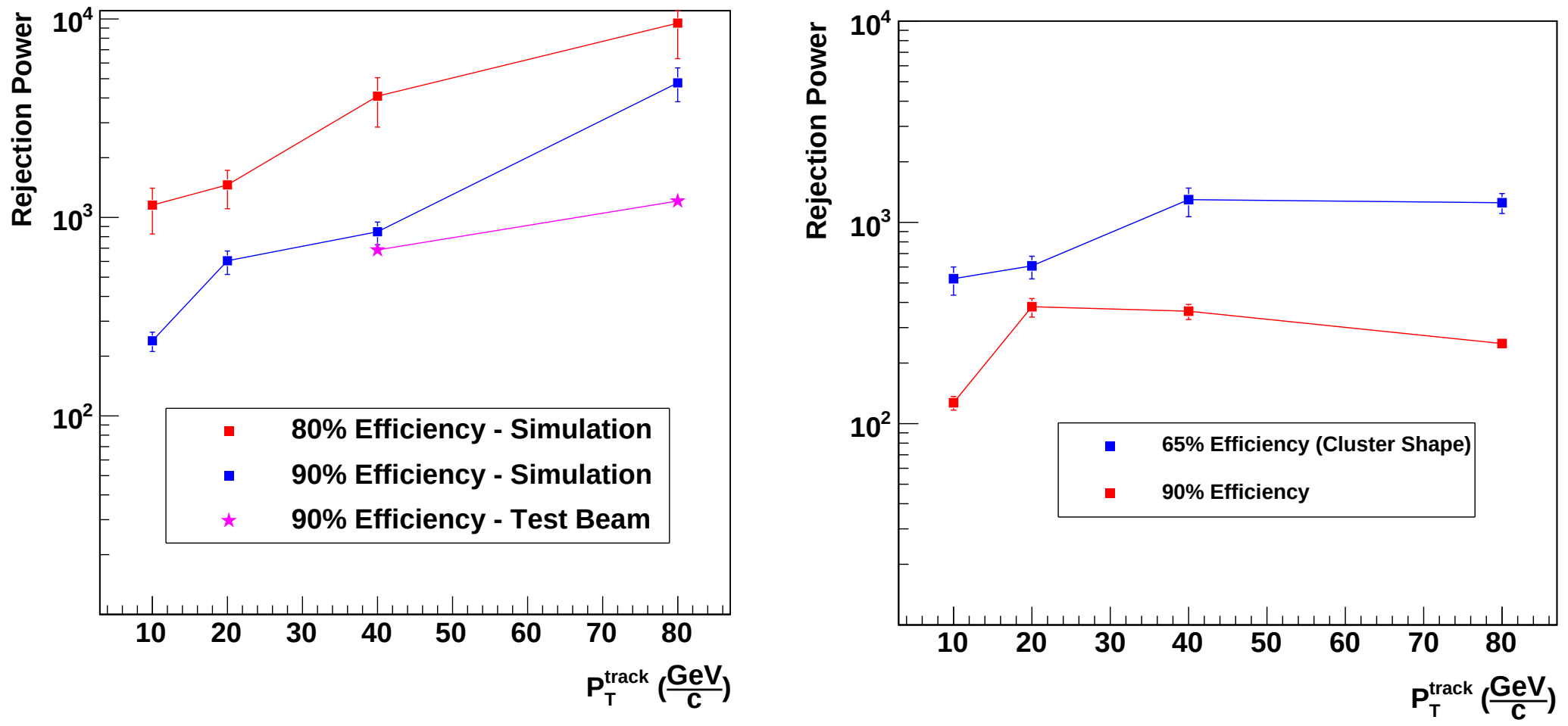


**Figure 7.8:** Hadron rejection factor as a function of electron efficiency and track $p_T$. (left) The two curves show the rejection power for particles that lose less than 10% of their momentum (as compared to input MC), and is therefore an idealized case, comparable to the test-beam setup shown as star symbols from the 2007 CERN SPS-testbeam at the two available energies of 40 and 80 GeV. (right) Pion rejection for 90% and 65% electron efficiency were obtained using all candidates and illustrate the effect of detector material in front of the EMCal (see text). The curve for 65% uses additional cuts on cluster-shape and size to optimize the purity of the electron sample. The tracks used are required to have at least 50 TPC hits and 3 ITS hits.

lower for all $p_T$. In simulations with large hadronic and photonic backgrounds (PYTHIA, HIJING) we determined that an additional cut on EMCal cluster shape and size was necessary to ensure that the electron clusters are correctly identified. This comes at the expense of electron efficiency, which drops to $\sim$ 65%. The hadron rejection power after adding this additional cut is shown as the blue symbols in Fig. 7.8. With these cuts the rejection power reaches 1000 at 40 GeV/c, a factor of $\sim$10 above the required minimum value.

## 7.4 Reconstructed Electron Spectra

Utilizing the PID capabilities of the EMCal we can now reconstruct inclusive and non-photonic electron (NPE) spectra. In order to remove the photonic conversion electrons from the inclusive sample, all candidate electrons are checked against the list of reconstructed secondary vertices (V0) created from charged particles. Those matching a V0 with invariant mass near that of a photon, $\rho^0$, $\omega$ or $\phi$ hadron are considered "photonic" and are subtracted from the inclusive spectrum to obtain the NPE candidate yield.

Fig. 7.9 shows the integrated annual yields in Pb–Pb collisions of reconstructed non-photonic electron candidates (with remaining conversions and mis-identified hadrons) identified with the EMCal for $p_T > {p_T}^{cut}$. The colors indicate the source of the electrons produced in the GEANT simulation (MC-truth). The orange line shows the residual conversion electrons

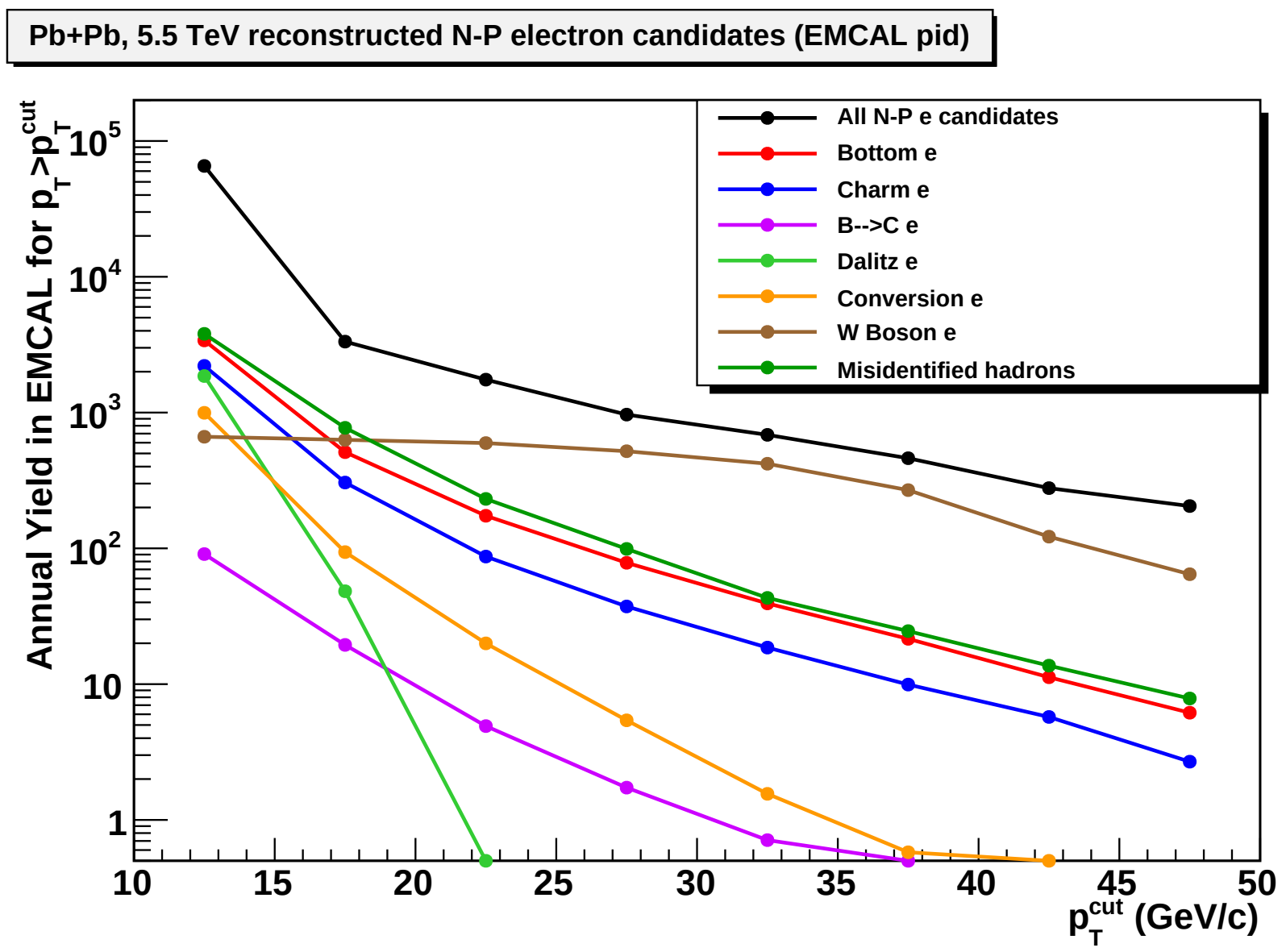


**Figure 7.9:** Integrated annual yields in the EMCal acceptance of reconstructed "non-photonic" electron candidates identified with ALICE tracking and EMCal from various sources for $p_T > {p_T}^{cut}$.

not eliminated by the V0 cut. Further investigation of the ALICE V0 algorithms and their ability to identify photon conversions is underway and we expect that the results should improve with some optimization. The contribution from W-boson decays can be reduced by requiring the electron to match a secondary vertex with decay length consistent with typical B-mesons. Further details on this method are described in section 7.5.1 on b-jet tagging. The contribution due to misidentified hadrons that pass the electron identification cuts is shown in dark green. The various contributions to the reconstructed NPE candidates after cuts are clearly different from the MC-input distributions (Fig. 7.3) as expected and will be further investigated.

To determine the total yield of non-photonic electrons, the estimated contamination from misidentified hadrons is subtracted from the inclusive NPE candidate spectrum and the resulting signal distribution is corrected by the efficiency. The result of these operations is presented in Fig. 7.10. The efficiency corrected spectrum is in agreement with the MC-input distribution within the statistical and systematical uncertainties. The systematic uncertainty bands were determined by varying the track-matching residual criterion and the electron identification cut $E/p$ in several combinations. The decreased efficiency resulting from tighter cuts is compensated by the increased purity and vice versa. With these simulations we obtain point-by-point uncertainties that vary by less than a few percent about an average value of $\sim$40%. Additional systematic effects, in particular in conjunction with the reconstruction of conversion electrons, still need to be evaluated.

While these results are already very promising, there are several additional studies of the tracking efficiency, EMCal cluster reconstruction and track-cluster matching that are cur-

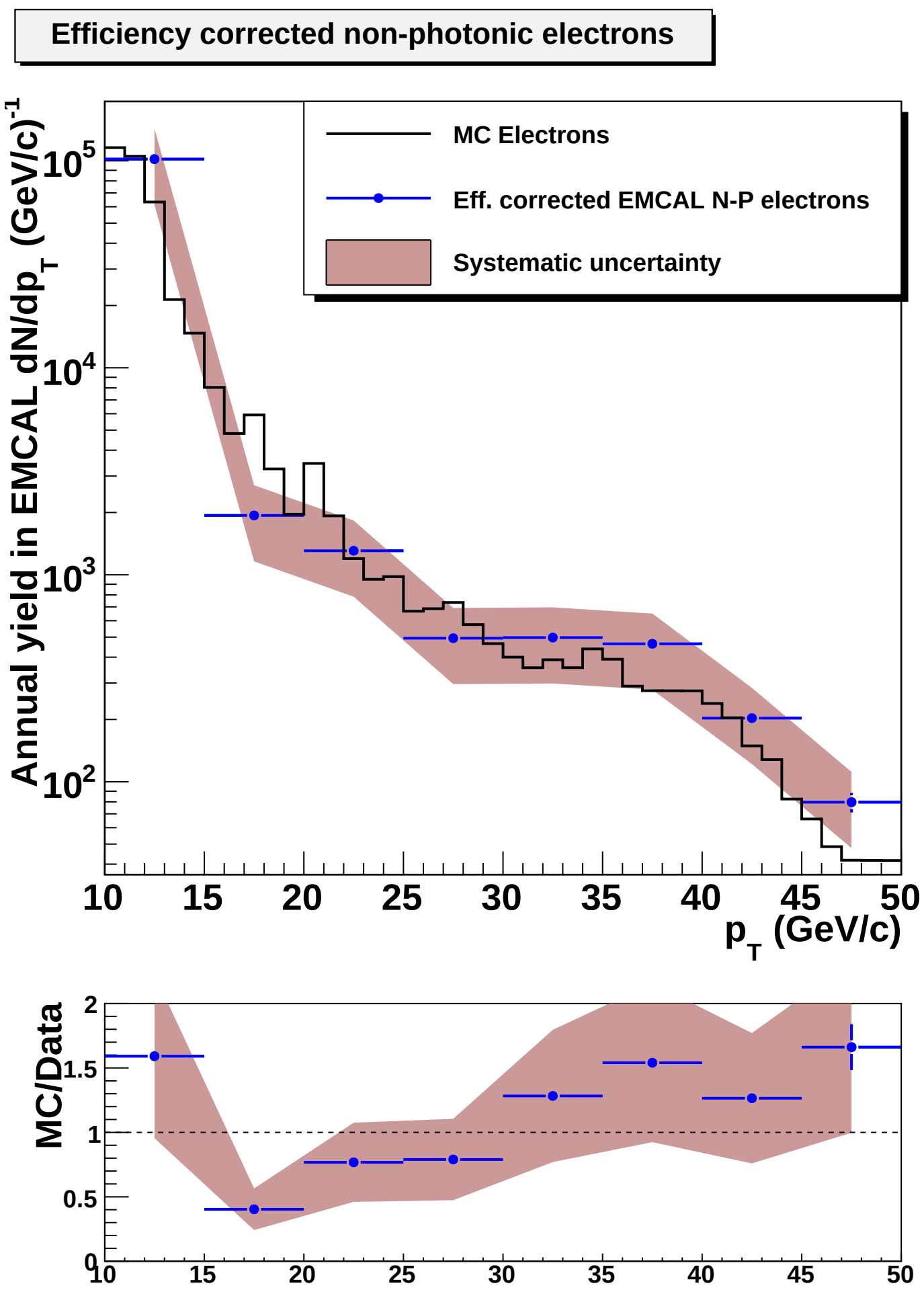


**Figure 7.10:** Efficiency corrected signal of non-photonic electrons for EMCal PID compared to MC truth for bottom and W-decay electrons. Systematic errors due to varying the EMCal electron identification criteria are shown.

rently underway. We expect that the results presented here represent the worst-case scenario which will only improve as the analysis matures. The measurement of NPE in p-p and Pb–Pb collisions will provide a robust measurement of the suppression of non-photonic electrons in Pb–Pb collisions that can be compared to theoretical models. Ultimately, these measurements should enable us to clarify the role of color-charge in partonic energy loss. However, since NPE have contributions from charm and bottom it is important to measure the bottom contribution separately using other techniques described next.

## 7.5 Methods of Beauty-Jet Tagging

A more ambitious undertaking is to identify jets from b-quarks using tagging algorithms. In the following section we will describe two distinct methods of b-tagging that have been

explored by this group. The first method relies on a jet containing a high-$p_T$ electron, whereas the second does not. In Fig. 7.11 the B-hadron $p_T$-spectrum from PYTHIA (at 5.5 TeV) is shown as a function of the electron $p_T$. This figure demonstrates the advantage of selecting high-$p_T$ electrons to preferentially access the higher $p_T$ B-hadron (thereby b-jets) population.

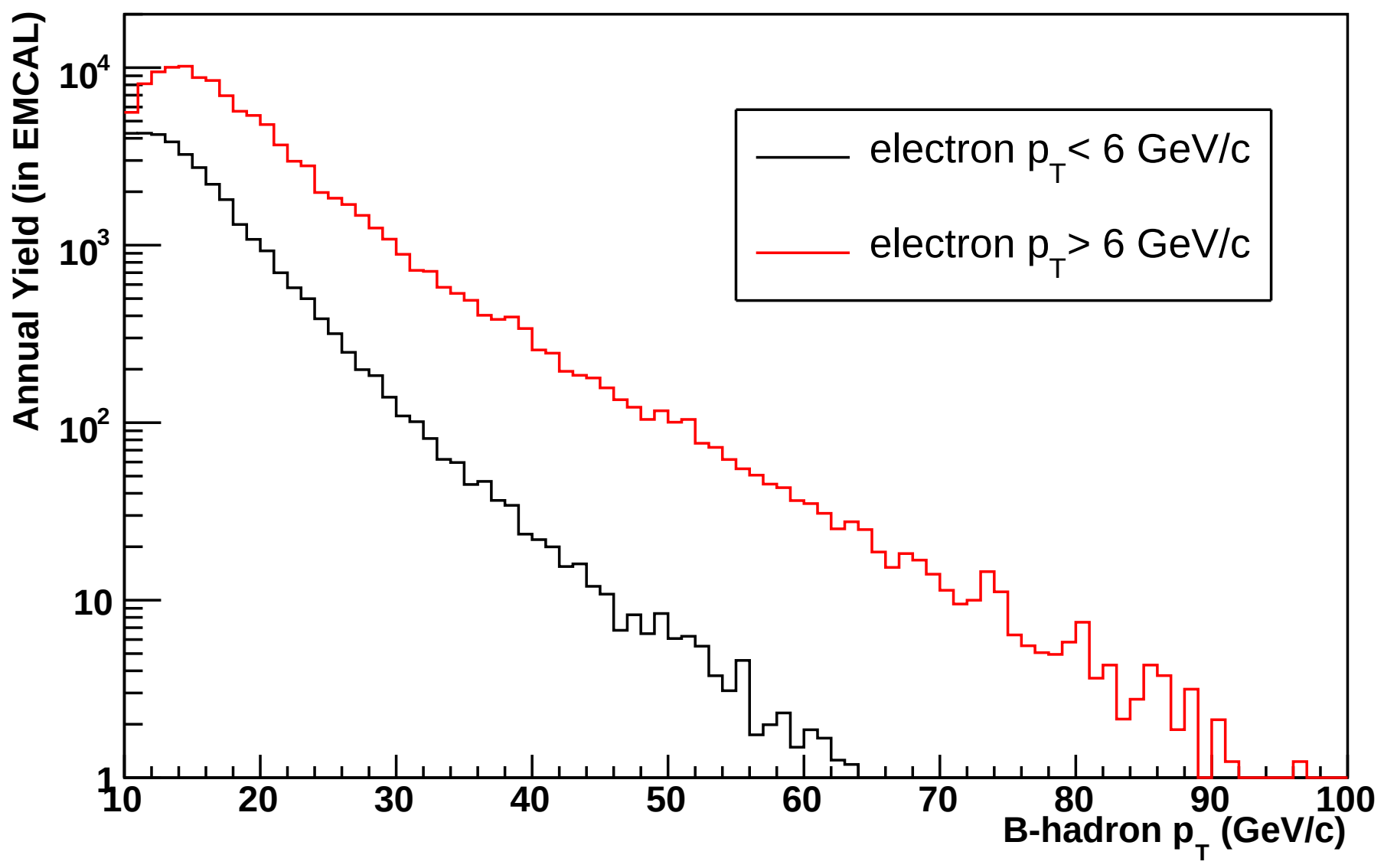


**Figure 7.11:** B-hadron yields (scaled to nominal Pb–Pb year) for selecting electrons in the EMCal with $p_T < 6$ GeV/c (black) and $p_T > 6$ GeV/c (red). Yields are corrected for electron reconstruction and PID efficiency, but not for tagging efficiency.

### 7.5.1 Displaced (Secondary) Vertex Method ("DVM")

This tagging method relies on the reconstruction of displaced secondary vertices from semi-leptonic B-decays, which are typically displaced by a few hundred $\mu$m from the primary vertex. It has been used by CDF to identify bottom contributions in semi-leptonic muon decays [11]. In addition, bottom decays typically produce a large number of charged particles by decaying via charm mesons to lighter hadrons. The highest $p_T$ hadrons are correlated in phase-space and all point back to a common, displaced vertex. By identifying the semi-leptonic displaced vertex consistent with the B-meson lifetime and at least one additional hadron from the decay we have a powerful tool to discriminate non-bottom electrons from bottom electrons.

The method uses a high $p_T$ electron as a seed, then searches for intermediate $p_T$ hadrons (above 1.0 GeV/c) from a common secondary, displaced vertex within a cone of $R = \sqrt{d\eta^2 + d\phi^2} < 1.0$ around the trigger. This R value was successfully used by CDF but still needs to be optimized for the heavy ion environment [11]. A minimum of 4 (out of 6)

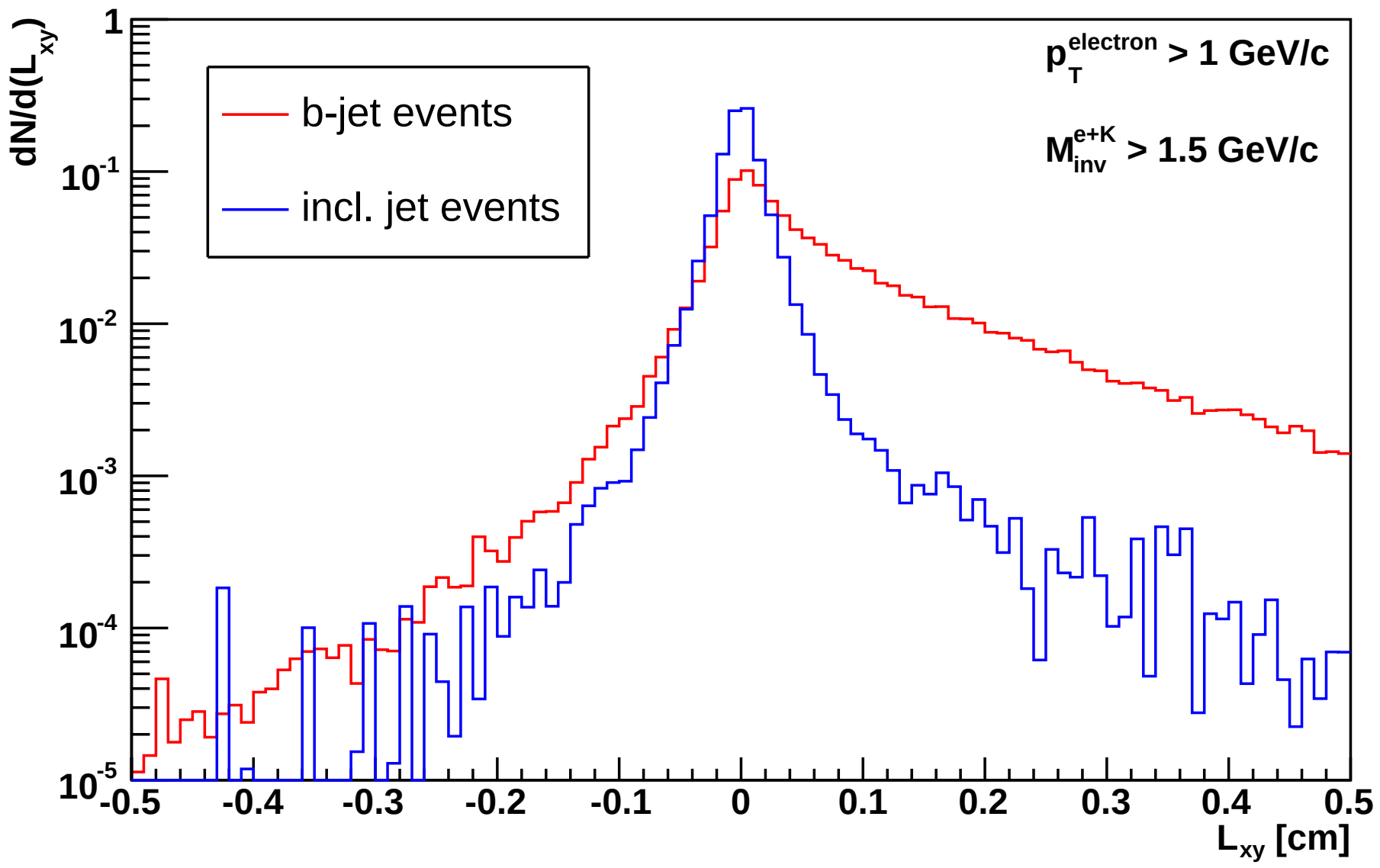


**Figure 7.12:** Signed DCA ($L_{xy}$) distributions from a sample of b-jet events ("Signal") and inclusive jet events ("Background+Signal") at 5.5 TeV. The distributions are normalized to their integral. The inclusive jet events also contain a small contribution of b-jets that was not subtracted.

ITS hits are required on both tracks to ensure sufficient spatial resolution of the secondary vertex. Once a pair is found and its displaced vertex determined, the quantity $L_{xy}$ (in the bending plane) is calculated:

$$L_{xy} = \frac{r \cdot p_e}{|p_e|} = |r| \cdot cos(\theta) \tag{7.1}$$

Where **r** is the vector from the primary vertex to the secondary vertex and **p** is the electron momentum. The distribution of this quantity is symmetric around zero for the background, but strongly biased towards positive values for real decays. Fig. 7.12 shows the distributions of $L_{xy}$ for electrons from inclusive jet ("Background+Signal") and b-jet ("Signal") events. More details can be found in reference [12].

Based on the distribution of $L_{xy}$ we define cuts to select electrons from b-decays. In this study we have adopted preliminary cuts that require at least n (n=1,2,3...) secondary vertices with $L_{xy} > 0.1$cm and $M_{inv}^{e+K} > 1.5$ GeV. Cutting on the invariant mass is a powerful discriminator to reject semi-leptonic vertices from charm. The electron tagging efficiency for different values of n can be measured from simulation and is shown in Fig. 7.13. This efficiency depends strongly on the number of "valid" secondary vertices which were found for a given electron, *i.e.* the number of electron-hadron combinations passing the above mentioned cuts for consistency with a semi-leptonic B-decay.

The next step consists of identifying such tagged electrons within jets. For this we ran a jetfinder $k_T$ algorithm (FASTJET) on our data and then compared the charged jet constituents to our collection of tagged electrons. If a jet contained a tagged electron it was tagged as a b-jet. All jets first had to pass EMCal acceptance criteria in $\eta$, $\phi$ and neutral energy fraction to ensure complete reconstruction of the neutral and charged components.

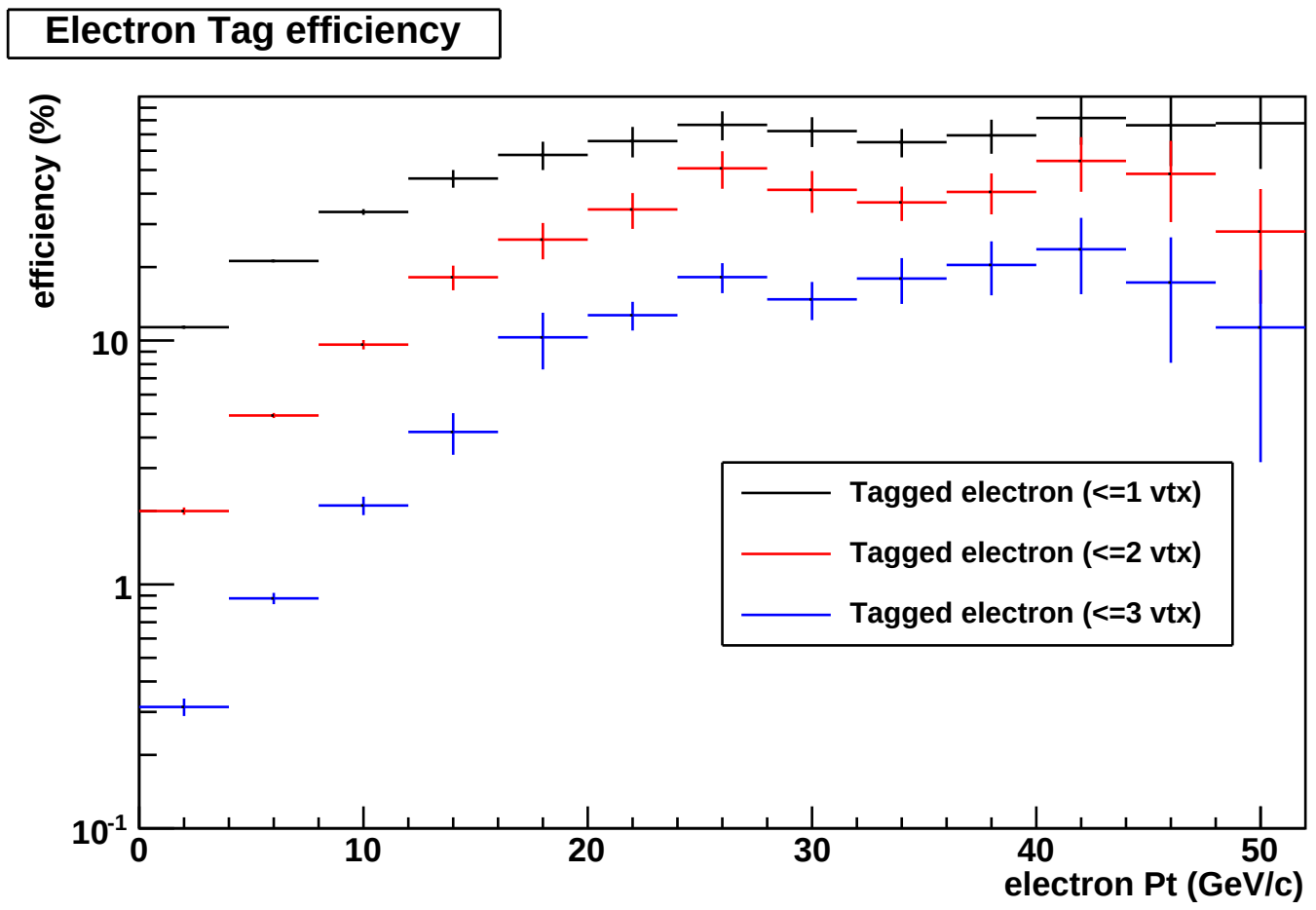

**Figure 7.13:** Efficiency for tagging a bottom-decay electron using the DVM algorithm vs electron $p_T$. Three different levels of cuts for the algorithm are shown requiring 1, 2 or 3 particles originating from the secondary vertex .

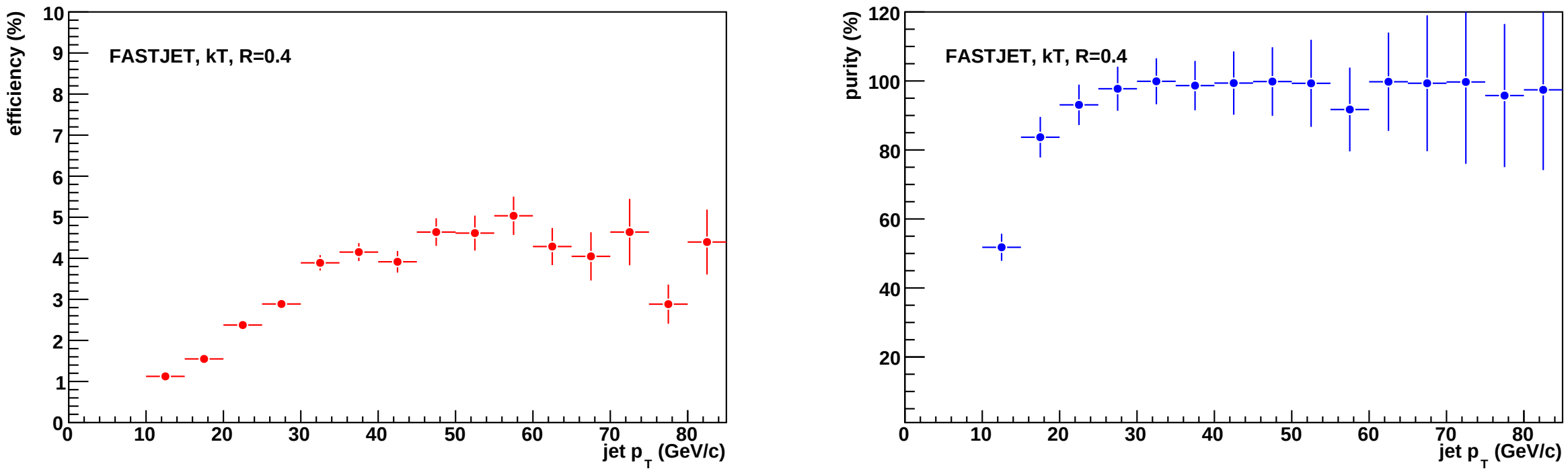

**Figure 7.14:** Tagging efficiency (left) and purity (right) for DVM B-tagging as a function of reconstructed jet $p_T$. These results were obtained requiring at least 2 vertices. Jets were reconstructed using the $k_T$-algorithm of the FASTJET package with R=0.4.

In Fig. 7.14 the current results of the DVM B-tagging algorithm are shown as a function of reconstructed jet $p_T$. With the algorithm configured for highest purity we currently achieve an efficiency of about 5% for $p_T > 40\ GeV/c$ as compared to MC-input B-jets. This efficiency is expected since we require one high $p_T$ electron and two hadrons from the secondary displaced vertex. The purity, defined as well-tagged jets divided by all tagged jets, for this configuration is shown on the right panel of Fig. 7.14 and is 100% within errors above 30 GeV/c. These studies are still preliminary and work is ongoing to optimize the efficiency of the algorithm.

### 7.5.2 Impact Parameter Significance ("IPS")

This alternative B-tagging algorithm is based on the transverse impact parameter (IP), and has been studied by CMS [13]. It uses the fact that B-hadrons have a large decay length as compared to the resolution of the ITS, that is of order ∼50 $\mu$m. Fig. 7.15 depicts how this topology is used. First jets are reconstructed using a standard finder algorithm. Then we calculate the dot-product between the IP vector of all tracks belonging to a given jet with the jet axis vector as shown in Fig. 7.15.

The IP significance is computed by taking the ratio IP/$\sigma_{IP}$, where $\sigma_{IP}$ is the uncertainty on the computed IP. A positive significance larger than 3 indicates that a track has its origin separated from the primary vertex beyond the tracking/vertex resolution. A negative significance would indicate a poorly reconstructed track, or that it is badly associated with the reconstructed jet. The left panel of Fig. 7.16 shows the distribution of the IP significance.

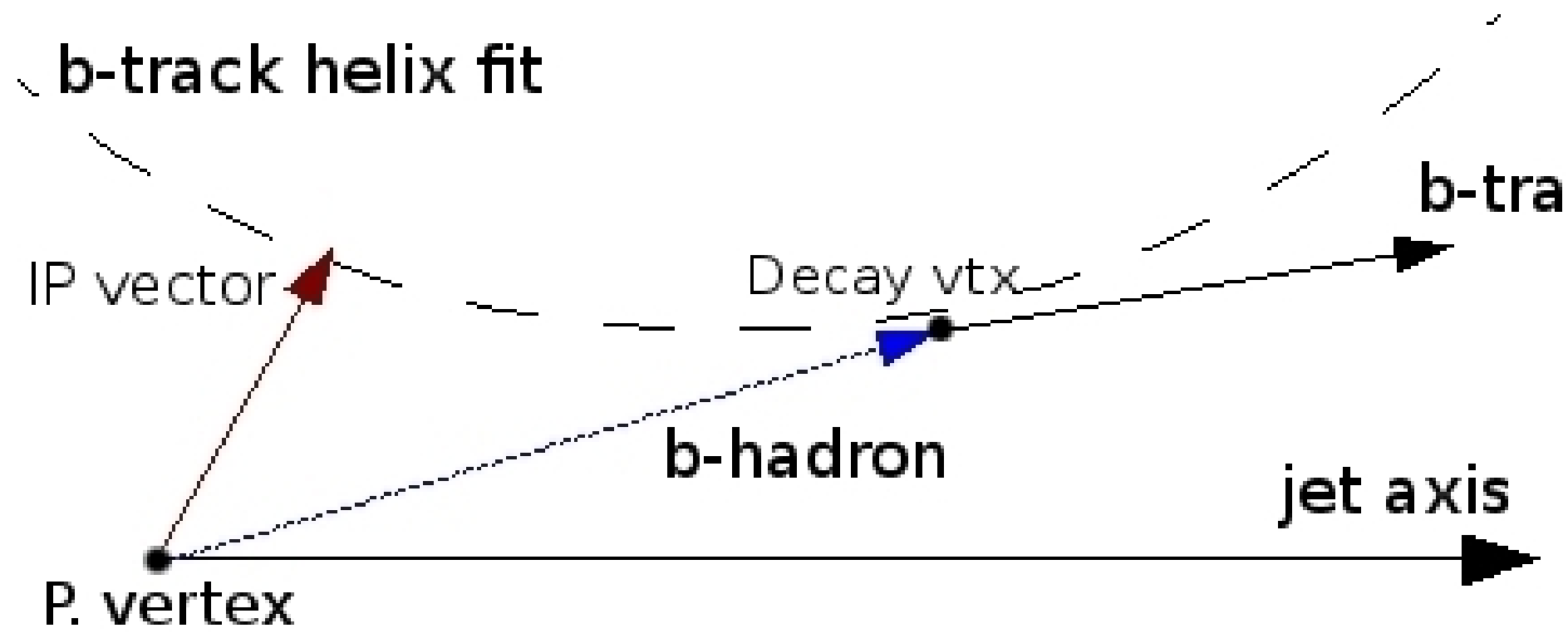

**Figure 7.15:** Topology of the bottom-decay related to the jet axis. The red line corresponds to the transverse IP of the decay track. In this situation the impact parameter is of positive sign.

The black line accounts for jet-tracks, while the blue line is the mirrored distribution, made from the reflection of the negative side of the original distribution onto the positive side. A clear excess on the positive side is evident, accounting for the existence of particles with large decay lengths within the jet cone. This figure was made selecting only those tracks whose vertices are located within 2.5 cm from the beam line (among other standard track quality cuts). The right panel of Fig. 7.16 shows the ratio of the integrals of the IP-significance distribution above a given cut value. The ratio full/reflected increases roughly linearly with the cut on IP-significance.

Once the data are pre-analyzed by a jet finding algorithm (UA1, FastJet, etc), the jet sample is submitted to the tagging. The tagging criteria is based on a positive significance threshold and a minimum number of tracks that pass the significance cut within the reconstructed jet cone. In this performance study, three tagging configurations were tested over PYTHIA inclusive jet and b-jet events at 5.5 TeV. The configurations were: 1 track with significance greater than 4 (1⊗4), 2⊗3 and 3⊗2. The preliminary results of efficiency and purity for these configurations are presented in Fig. 7.17. From these plots one can conclude that the efficiency of the method is sufficiently high to be applicable in ALICE. However, given the current purity of only 5% we need to study the impact of an additional requirement on an identified electron track, although this will be at the expense of a loss in efficiency. This study is continuing.

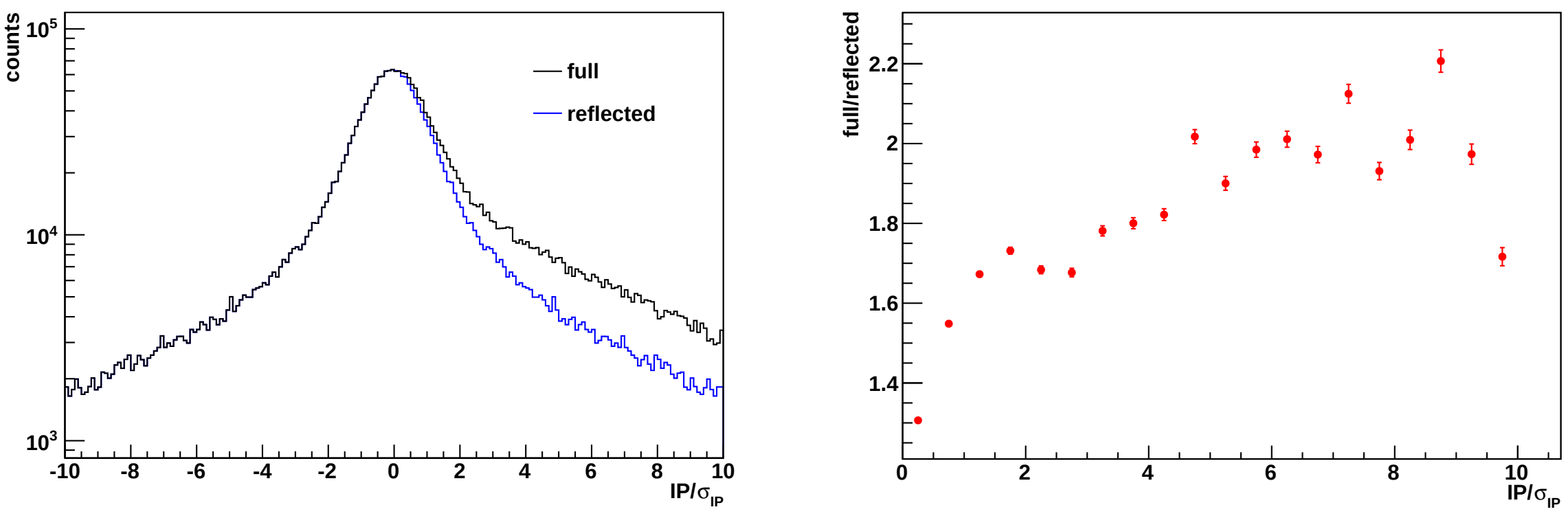

**Figure 7.16:** Left panel: the distribution of jet-track significance in p-p collisions. The excess of the black line over the blue is due to B- and C-hadron decays. The right panel shows the trend of the ratio of the black line over the blue line.

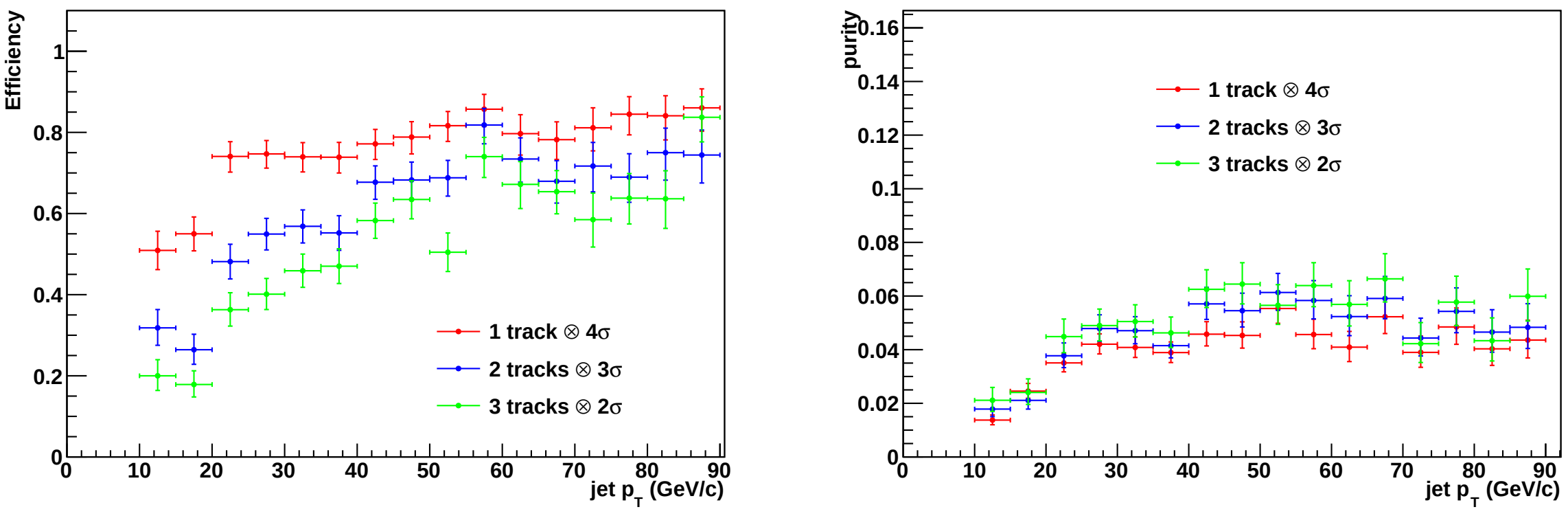

**Figure 7.17:** Efficiency (left) and purity (right) of the 3 tagging configurations of the IPS-method.

## 7.6 Summary

We have shown that the EMCAL possesses excellent electron PID capabilities in the $p_T$ range of 10-50 GeV/c, with hadron rejection factors of the order of several hundreds. This capability will allow the ALICE collaboration to measure non-photonic electrons in Pb–Pb collisions for $p_T$ ranges well above that previously reported at RHIC, and thus gain insight into color-charge and/or quark mass effects of partonic energy loss. Further studies for effectively reducing conversion and W-boson backgrounds are underway. The studies of B-tagging are at a proof-of-principle stage. We have shown two different algorithms that yield good results in p-p collisions and expect to further optimize both algorithms to find an optimal balance between efficiency and purity. The algorithms are complementary and could in the future be combined into a single algorithm. The next step is to evaluate their performance in the presence of the large background produced in heavy ion events.

# Chapter 8

# Particle Identified Measurements in Jets

## 8.1 Introduction

ALICE features a superior suite of particle identification detectors, which enable unique measurements over a very wide range of momenta with very good accuracy.

In the mid-rapidity region PID information is provided by the Inner Tracking System (ITS) and the Time Projection Chamber (TPC) through dE/dx measurements of charged particles, the Time of Flight (TOF) through timing measurements of charged particles, the High Momentum PID (HMPID) through Cerenkov light measurements of charged particles, and finally the Transition Radiation Detector (TRD) and the Electro-magnetic calorimeter (EMCal) through energy deposition analysis of neutral particles and electrons.

Measurements such as production cross sections and various types of correlations can be performed for many species ranging from electrons, charged and neutral pions, kaons, and protons to heavier mesons, baryons, and hadronic resonances. This provides an opportunity to study the flavor composition of jets, in particular measure the fragmentation functions of many particle species, which is of interest as a test of fragmentation models and the impact of the medium on the jets. In previous sections we have discussed the neutral pion and electron capabilities of the EMCal in concert with other ALICE detectors. In this section we will focus on the reconstruction of charged hadrons and hadronic resonances from jets. The hadro-chemistry in jets has recently been the topic of several novel theoretical approaches to questions of medium properties and hadronization.

Sapeta and Wiedemann [1] have suggested that the most likely partonic energy loss mechanism in the dense partonic medium formed in heavy ion collisions at the LHC, namely gluon splitting, will cause a hadron mass dependent modification of the parton fragmentation function. The authors postulate that such a modification would be measurable through high momentum hadronic yields and ratios, i.e. above 8-10 GeV/c, in fully reconstructed jets.

Fries and Liu [2] have suggested that, in the same kinematic region, one could find evidence for flavor conversion for a parton traversing the medium. The partonic matter at the LHC is expected to be gluon dominated, but through heavy flavor or photon tagging as described in the previous chapters, one could identify a sub-set of quark jets in addition to the majority of reconstructed gluon jets. The distinction between quark and gluon jets should enable us to determine the impact of the Casimir (color) factor on the partonic energy loss, and thus study its non-Abelian nature. Surprisingly no evidence for color factor effects in the energy loss have been detected at RHIC, but they could be washed out if the parton can change its flavor due to interactions in the medium. Fries and Liu predict, in particular, that identified suppression factor double ratios such as $R^{\gamma}_{AA}/R^{\pi}_{AA}$ or $R^{p}_{AA}/R^{\pi}_{AA}$ are sensitive to the conversion probabilities. Again, particle identified yield and ratios at high $p_T$ are necessary. Evidence for flavor conversion would alter our understanding of the relation between final hadronic cross sections and the initial parton flavor.

Recent results by the RHIC-STAR collaboration show hints of particle dependent patterns in the suppression of high momentum particles in nucleus-nucleus collisions.

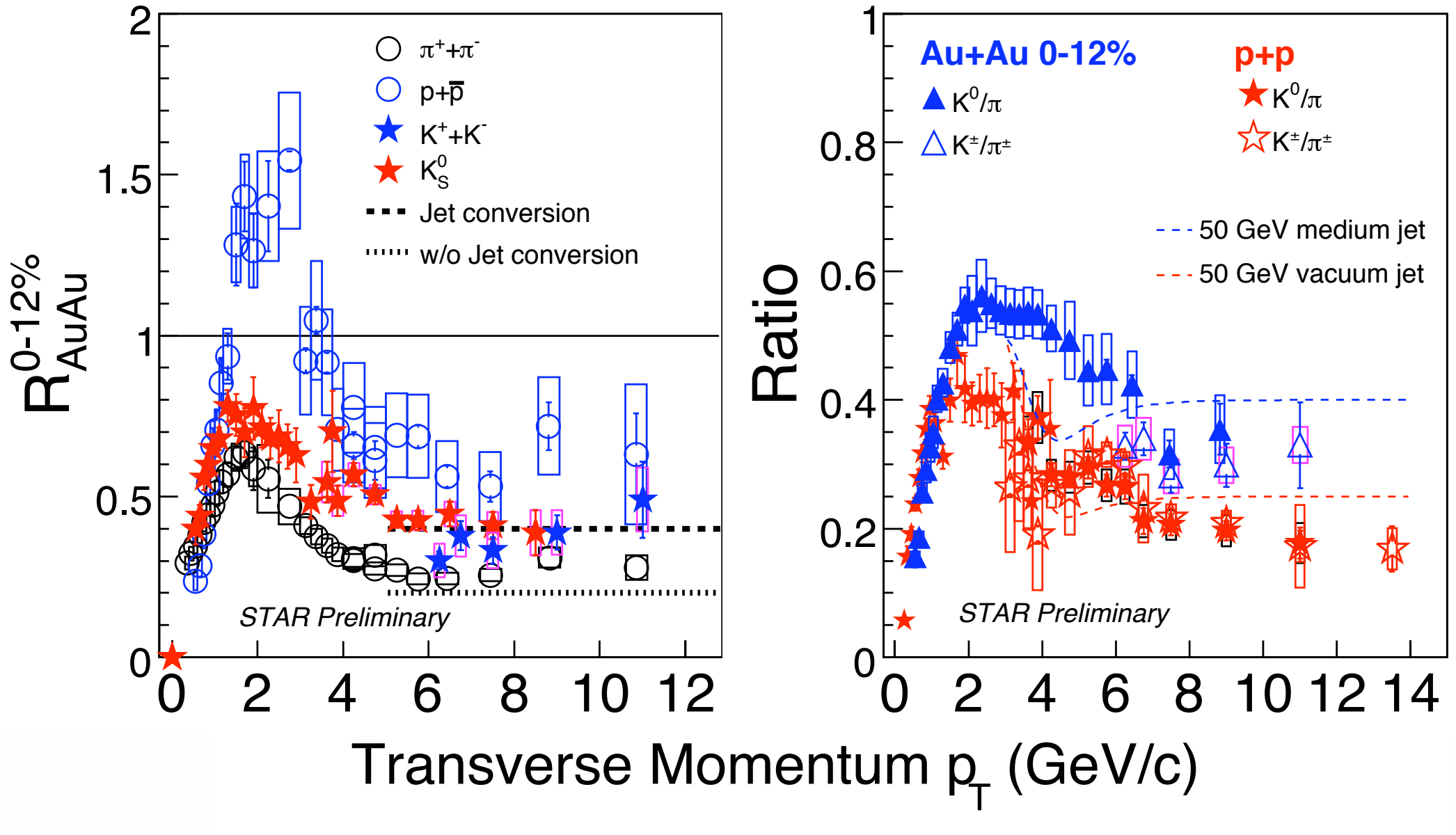

**Figure 8.1:** Left: nuclear suppression factors measured by STAR out to high $p_T$,compared to predictions for kaon suppression taking into account jet conversion [2]. Right: comparison of STAR measurements of K/$\pi$ ratio in p–p and AuAu collisions to predictions at LHC energies taking into account enhanced gluon splitting [1].

Fig. 8.1 (left panel) shows the nuclear suppression factor for pions, kaons and protons out to the fragmentation region ($p_T$ > 6 GeV/c). There is an unexpected difference in the suppression factors for the different species. Fries et al. have speculated that at RHIC energies, other than at LHC energies, the kaons should be enhanced over the pions due to jet conversion. Their predictions with and without jet conversion are shown in the figure. This effect however does not explain the proton-pion difference. That difference was predicted in [1] and Fig. 8.1 (right panel) shows the K/$\pi$ ratio measured by STAR in p–p and Au-Au collisions compared to the calculations by Sapeta and Wiedemann. Although the calculation was performed for LHC energies the discrepancy between the p–p and A-A ratio in the RHIC data in the high $p_T$ fragmentation region is surprisingly well described.

Finally Markert et al. [3] have suggested that high momentum resonances in jets which are formed early in the partonic phase might be used to study chiral symmetry restoration. The possibility for the generation of color singlet pre-hadrons or pre-resonances within the partonic matter is indicated by formation time measurements of hadrons from partonic fragmentation in cold nuclear matter [4]. Due to color transparency [5] the re-interaction cross section of these states is likely small and there is a finite probability that early formed hadronic resonances from the fragmentation process (i.e. in jets) might escape the medium carrying initial production properties.

A study of these predictions requires the identification of hadronic states in jets, a measurement that is uniquely accessible with the ALICE detector. At the LHC, only ALICE features the jet triggering and tagging capabilities, based on the EMCal, and the necessary PID at high momentum. Even though the PID information from the aforementioned detectors overlaps in certain kinematic ranges and can be combined to improve the identification quality, the sole contribution for charged hadrons above 5 GeV/c comes from the energy loss measurement in the TPC in the so-called relativistic rise region of the Bethe-Bloch parametrization of charged particle energy loss (rdE/dx). The method of relativistic dE/dx measurements for high momentum hadron identification has been successfully applied by the STAR collaboration with a setup very similar to the ALICE detector [6, 7].

In the following we will describe the physics performance for the combination of EMCal plus tracking detectors in separate subsections for hadrons and hadronic resonances. Although we focus on measurements suggested in the above theory papers we point out that 'standard' high momentum PID measurements, such as the determination of the high momentum identified particle $v_2$ or the question of flavor equilibration in jets, will be studied. The comparison to specific predictions in this chapter is meant to demonstrate the level of sensitivity that ALICE will provide with respect to certain observables, but my no means constitutes an exhaustive list of all measurements that can be accomplished by ALICE based on particle identification.

Regarding the statistical enhancement of the high $p_T$ particle sample through the EMCal trigger we assume the enhancement factors quoted in the trigger chapter of this document. The yearly number of di-jets with one of the jets triggered in the EMCal in minbias Pb–Pb collisions is about 2 Million for a jet energy of 50 $\pm$10 GeV and 100,000 for a jet energy of 100 $\pm$10 GeV. For all quenched jet simulations in this chapter we use qPYTHIA events with a $\hat{q}$= 50 GeV$^2$/fm.

## 8.2 PID capabilities of the ALICE tracking detectors at very high $p_T$

High momentum PID was discussed in detail in several parts of the original ALICE-PPR [8]. Here we will focus on updates to previous TPC performance benchmarks, which are based on a better description of the detector response and calibration measurements with cosmics and/or test beam. Fig. 8.2 shows the performance of the fully installed TPC based on cosmic ray data. The achieved resolution in the relativistic rise region (5.7%) is very close to the theoretical design value of 5.5% and better than the original detector requirement.

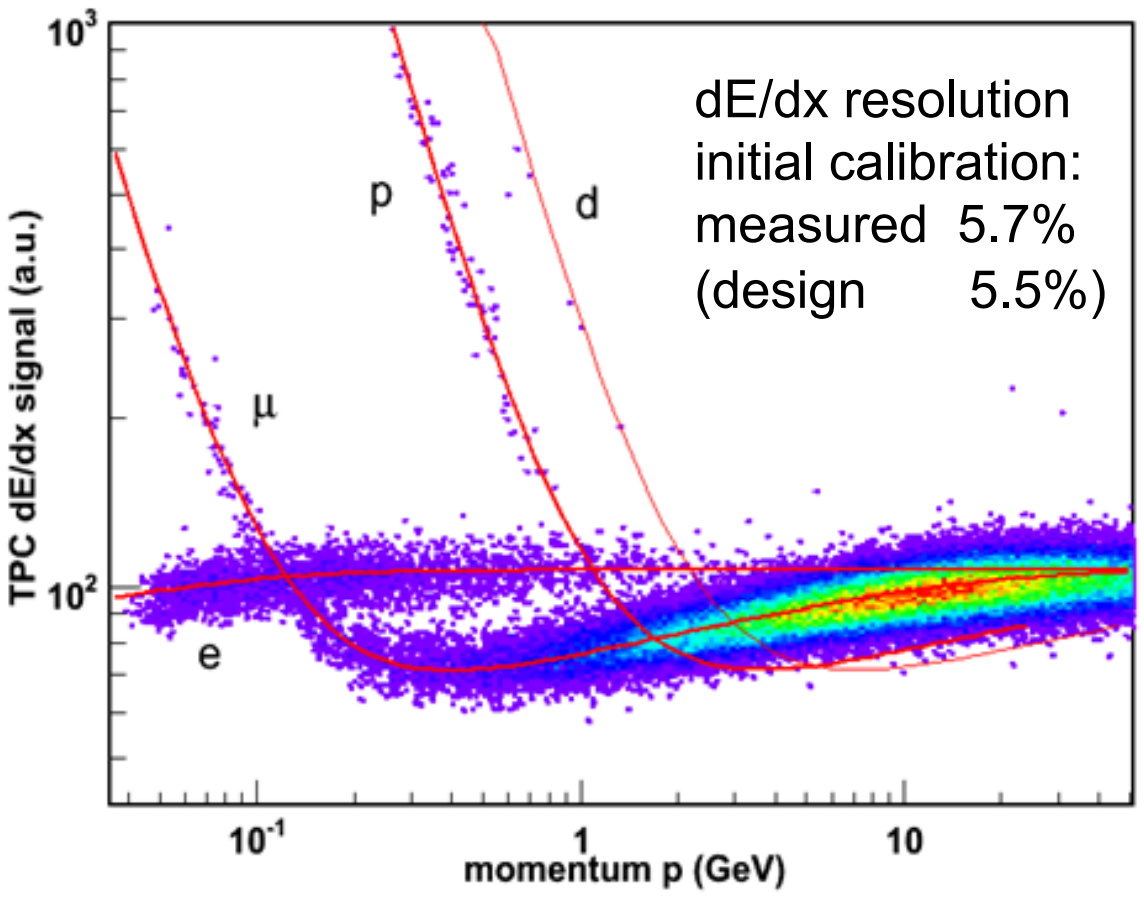

**Figure 8.2:** TPC dE/dx calibration results based on cosmic rays.

The particle species separation quality that can be deduced from these results in the relativistic rise region is summarized in Fig. 8.3. Here we show the separation power between two particles in terms of $\sigma$. The two curves in each plot bracket the range of the anticipated final resolution.

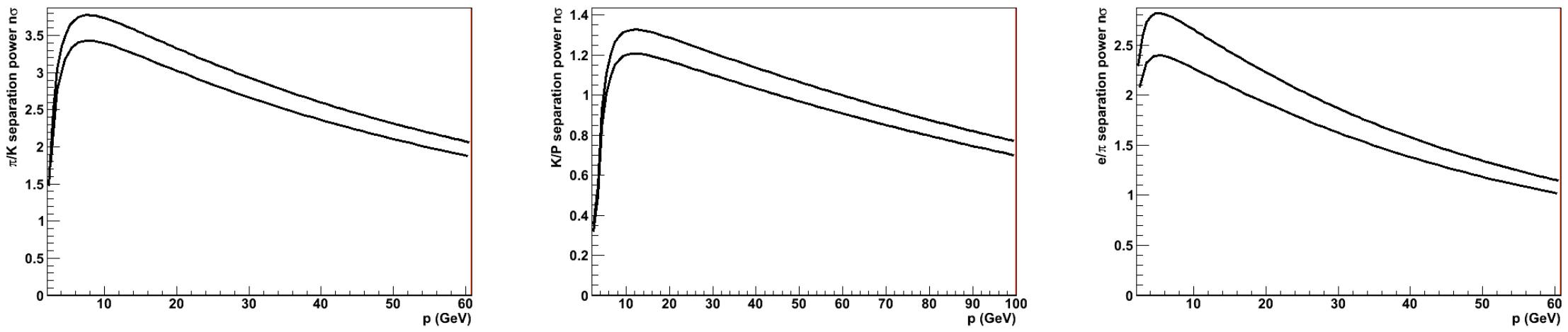

**Figure 8.3:** Anticipated PID separation power in the relativistic rise region based on the TPC test performance for $\pi$/K, K/p, e/$\pi$, respectively.

The other detectors that will contribute at lower momenta are the TOF, the HMPID and the TRD (for e/h separation). In addition, both the EMCal and the TRD, will help with the normalization of the relativistic dE/dx curves of hadrons through identification of high momentum electrons. The e/h separation capabilities in the EMCal are detailed in Chapter 7 of this report.

## 8.3 Momentum spectra and ratios of identified particles in jets

Based on the dE/dx resolution from Fig. 8.2 we can determine a momentum dependent total uncertainty for particle yields by folding the resolution with the detector acceptance as well as the reconstruction efficiencies and momentum resolution that were presented in the original PPR [8].

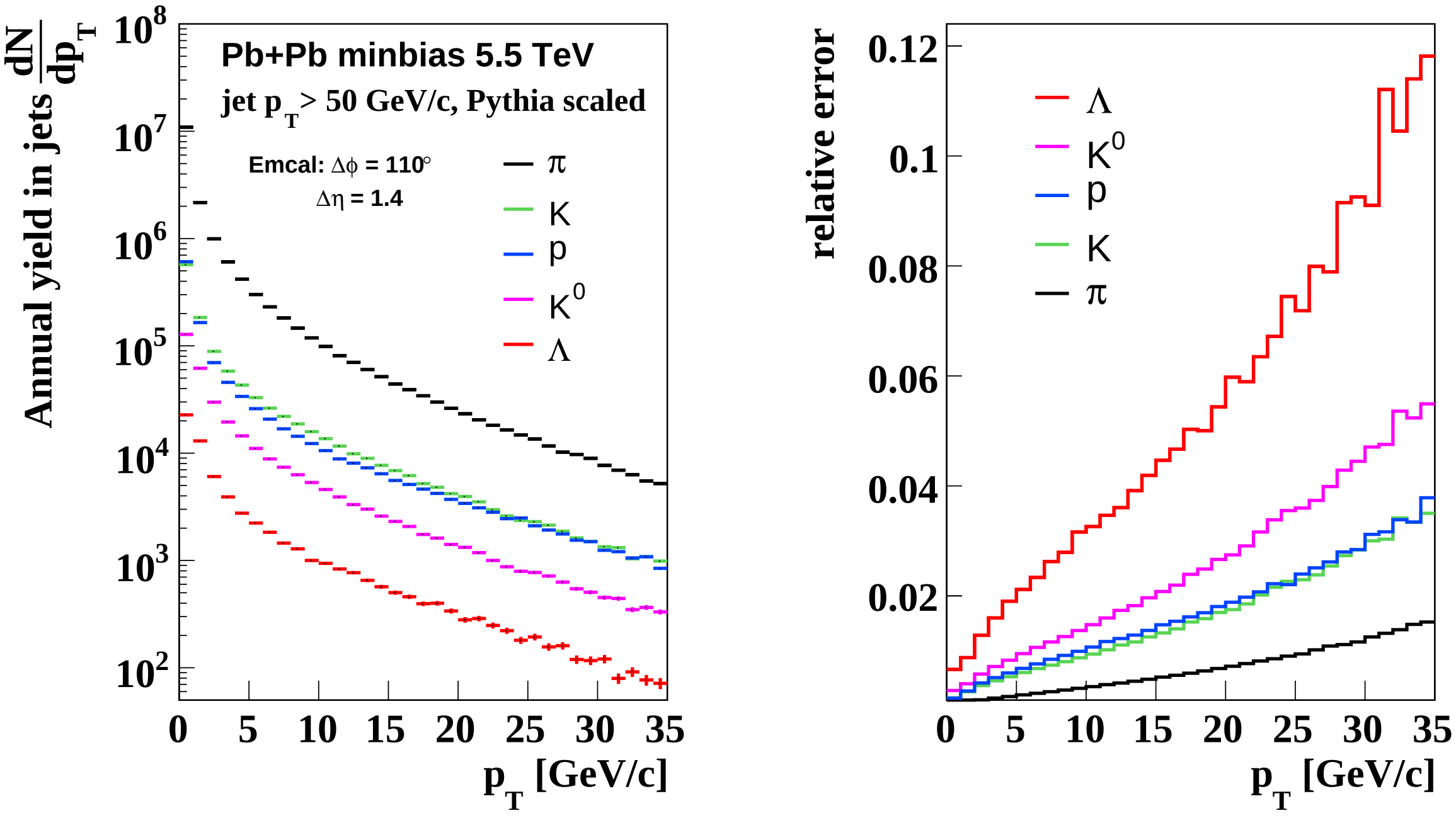

**Figure 8.4:** Left: annual particle yields in reconstructed (unquenched) jets with a jet energy higher than 50 GeV. Right: relative error based on PID separation power, momentum resolution, detector acceptance and reconstruction efficiencies for pions, kaons, protons and Λs.

Fig. 8.4(left) shows the reconstructed annual particle yield in Pb–Pb collisions as a function of transverse momentum for unquenched jets with a jet energy larger than 50 GeV (embedded PYTHIA) as triggered and reconstructed by the EMCal using the anti-$k_T$ algorithm (R=0.4) in the FastJet package (see Chapter: Jet Reconstruction). Around 4 Million Pb–Pb events annually contain an EMCal reconstructed jet above 50 GeV. Fig. 8.4(right) shows the relative

uncertainty as a function of particle species and momentum. A systematic error on the tracking efficiency which was estimated to be 5% in the original ALICE-PPR has not been included.

As can be seen there is good statistics and resolution for the reconstruction of identified particles in jets up to 30 GeV/c. Sapeta and Wiedemann claim in their paper that the in-medium spectrum is considerably modified above 10 GeV/c. In order to compare to their predictions, we show the achievable measurements with error bars (based on Fig. 8.4(right)) in ALICE, using the identified particles from reconstructed unquenched PYTHIA jets in HIJING background, together with MLLA (Modified Leading Log Approximation) predictions for vacuum jets and the Sapeta-Wiedemann prediction for in-medium jets in Fig. 8.5. There is a slight difference in the ratio between PYTHIA and MLLA for vacuum jets, which could be due to either intrinsic differences between the leading order and the modified leading log prediction or the uncertainty in the jet reconstruction. The main conclusion of this study though pertains to the relative difference between the predicted effect and the measurement uncertainty in medium. The model predicts about a factor two increase in the ratios above 10 GeV/c in case the quenching is due to hadron mass dependent gluon-splitting. Our anticipated error bars are well within the accuracy needed to perform a definite measurement.

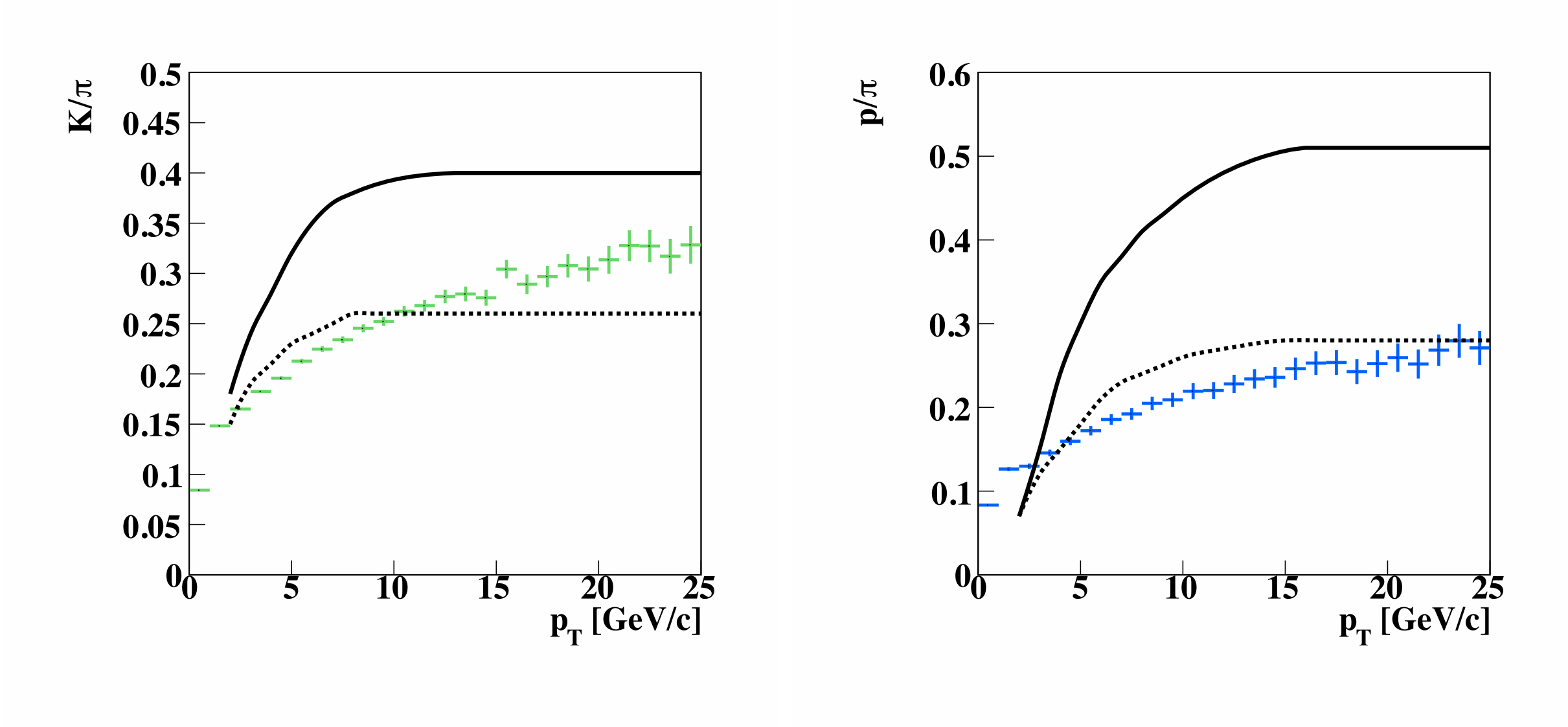

**Figure 8.5:** Achievable uncertainty in particle ratios based on rdE/dx PID information in the TPC for measurements using the annual yield of reconstructed and identified particles in jets in Pb–Pb collisions in ALICE (PYTHIA jets embedded in HIJING). The lines show the estimates by Sapeta and Wiedemann for vacuum jets (dashed) and in-medium jets (solid) [1].

## 8.4 Fragmentation functions of identified particles

The enhanced gluon splitting which is the preferred energy loss mechanism of several quenching models (e.g. JEWEL and qPYTHIA) leads to a softening of the fragmentation function, generically seen as an increase and a slight shift in the hump-back plateau if plotted against the inverse fractional momentum $\xi$ [9]. Based on fragmentation measurements in elementary collisions at lower energies the shape and strength of this plateau is particle species dependent, an effect which has been successfully described by MLLA calculations. The predicted enhanced gluon splitting in the partonic medium in Pb–Pb collisions enhances the species differences even further, as shown in Fig. 8.6 [1].

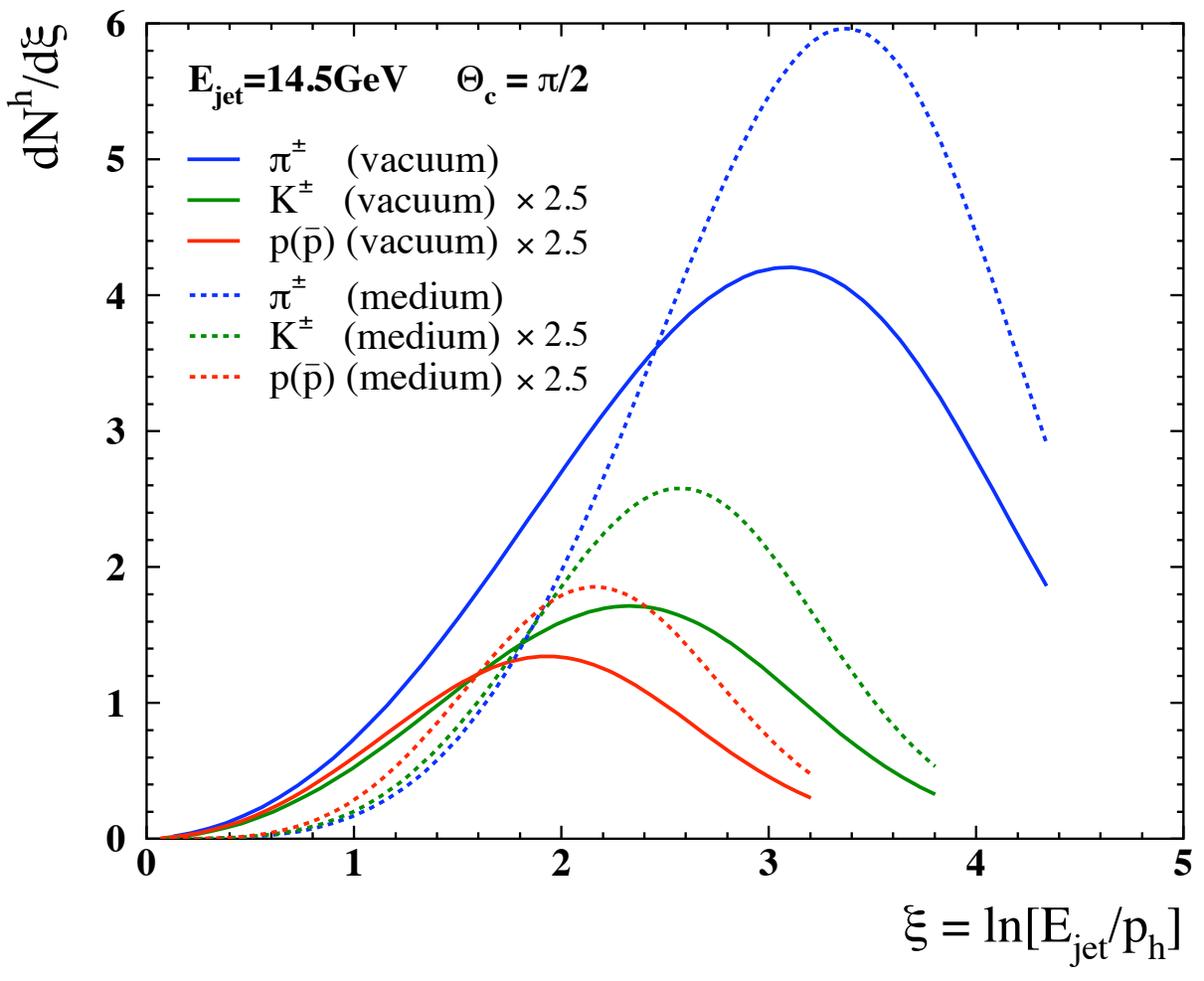

**Figure 8.6:** Prediction of medium modification of identified fragmentation functions based on gluon splitting [1].

Again, the good EMCal jet reconstruction efficiency and triggering capabilities in connection with the PID capabilities of the other ALICE detectors should allow us to make significant measurements of the quenching effect in particle identified fragmentation functions. Unfortunately the event generators presently in use to model quenching do not incorporate a particle species dependent effect. Fig. 8.7 shows that a general softening of the fragmentation function in medium (high $p_T$ suppression and low $p_T$ enhancement) is featured in the PYTHIA over qPYTHIA ratio for each particle species, but the medium modification is not particle specific.

The mass of the particle still shifts the $<\xi>$ though and Fig. 8.8(left) shows a comparison of fragmentation functions for identified particles in reconstructed PYTHIA and qPYTHIA jets with $E_{jet}$=50-60 GeV. A cone algorithm with R=0.4 has been applied to obtain the spectra. The statistical error bar is based on the annual jet yield in Pb–Pb collisions, but no HIJING background has yet been taken into account. The simulated jet energy (MC truth) has been assumed for the reconstructed jet energy, since the correction and unfolding scheme detailed in Chapter 5 indicates that the jet energy can be recovered to high accuracy. Alternatively photon-jets could be used to further constrain the jet energy (see Chapter 6). Fig. 8.8 (right) shows the same comparison for a jet reconstruction using a cone radius of R=1.

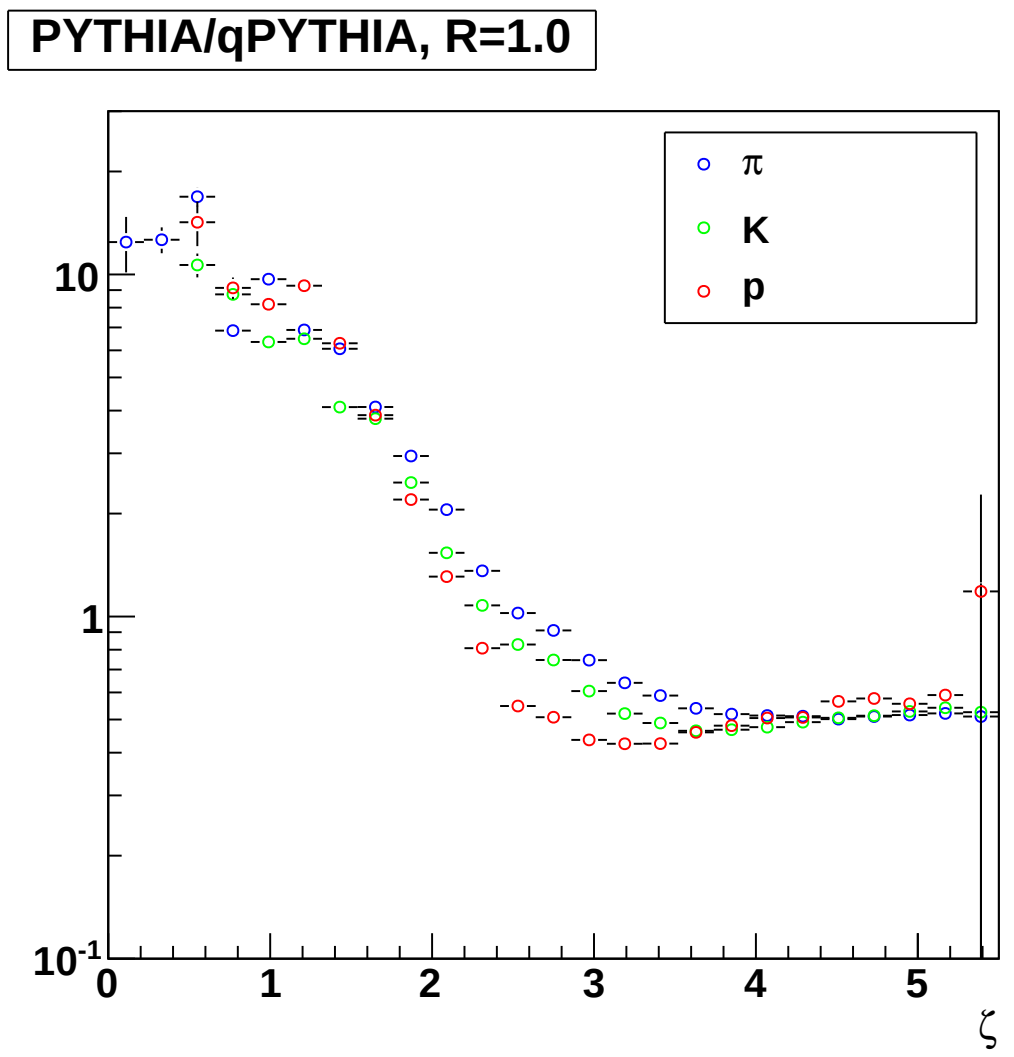

**Figure 8.7:** Comparison of fragmentation functions of identified hadrons for reconstructed jets (cone algorithm with R=1) of 50-60 GeV energy in PYTHIA and qPYTHIA.

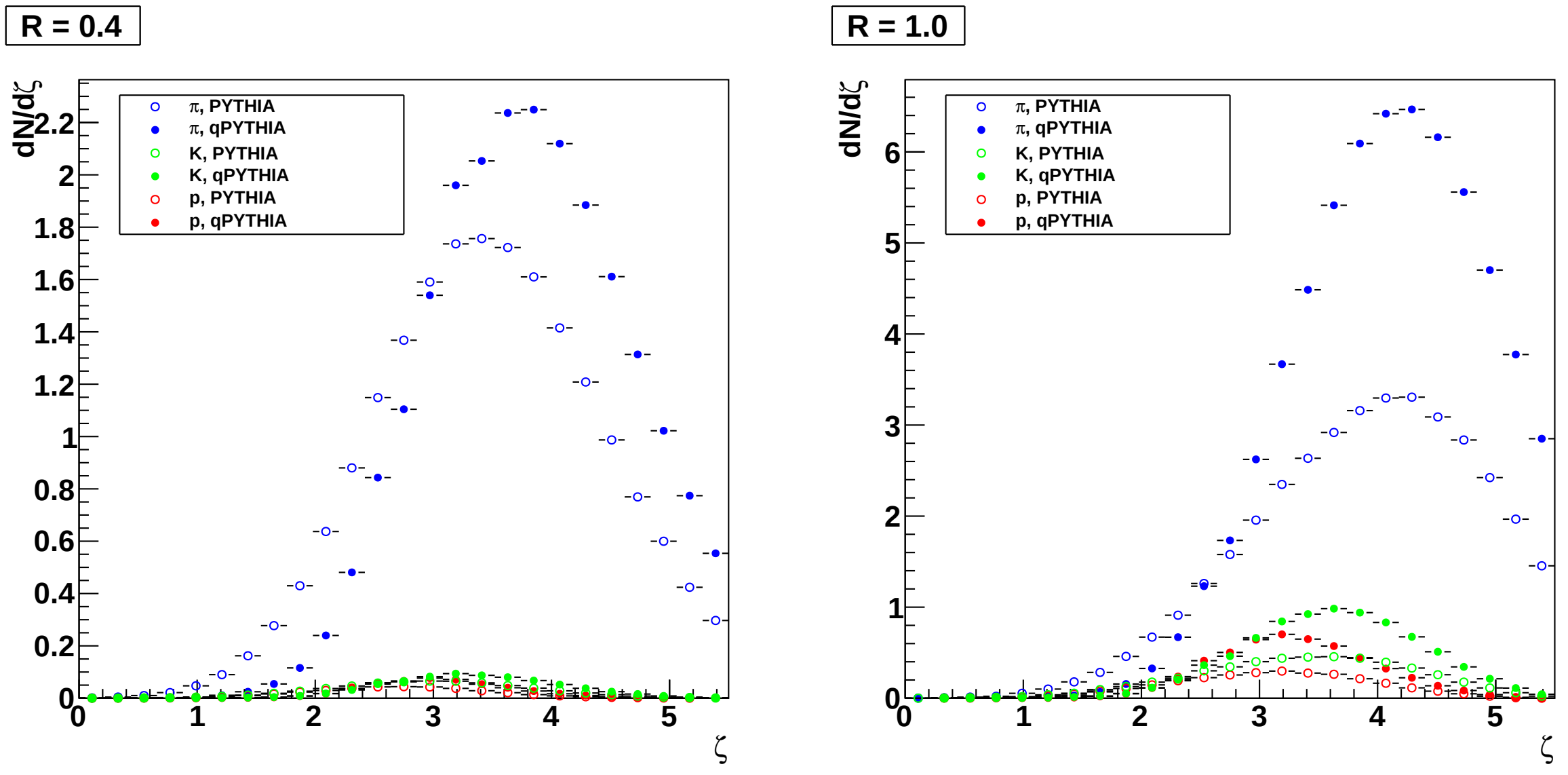

**Figure 8.8:** Comparison of reconstructed fragmentation functions for identified particles from unquenched (PYTHIA) and quenched (qPYTHIA) 50-60 GeV jets using R=0.4 (left) and R=1 (right) in the jet reconstruction.

The difference in the high $\xi$ region between the quenched and the unquenched simulation is significantly smaller for the R=0.4 analysis than the R=1 analysis. One additionally notes the distributions are peaked at lower $\xi$, which indicates that the quenching, as modeled in qPYTHIA, pushes the additional low momentum particles out to higher distances from the jet axis, i.e. the high $\xi$ particles are less contained in the small cone. If confirmed by data, the evolution of the fragmentation function ratio ($FF_{PbPb}/FF_{pp}$) as a function of cone radius (for seeded cone or SIScone algorithms) or resolution parameter (for $k_T$ or anti-$k_T$ recombination algorithms) will carry constraining information on the energy distribution of the quenched jet in the heavy ion environment. Although a R=1 analysis in the EMcal is not possible due to acceptance restrictions studies can be performed up to R=0.7. Upon reconstruction of the corrected jet energy the fragmentation function reliability depends largely on the proper unfolding of the underlying event spectrum for identified particles. As an example, Fig. 8.9 shows the uncertainty in the fragmentation function measurement based on the normalized per jet yield and the anticipated background level in Pb–Pb collisions (from HIJING) for identified positively charged kaons in unquenched reconstructed 50-60 GeV jets. The statistical errors are included and are based on the anticipated annual di-jet yield including particle reconstruction efficiencies. The relative uncertainty quoted here is defined as the inverse of the significance $\sigma$ which is defined as $\sigma$=S/$\sqrt{S+B}$. For example, the annual signal in the $\xi \in [4 - 4.4]$ bin is 12,000 $K^+$, whereas the background, as determined in random cones in HIJING events, is annually on the order of $3\times10^6$ positive kaons. Thus, $\sigma$=7 and the statistical uncertainty is close to 15%.

This uncertainty needs to be compared to any anticipated quenching effect, which for example exceeds a 50% change in qPYTHIA below $\xi$=2 and above $\xi$=3.5. Fig. 8.10 shows the uncertainty for a ratio of kaon fragmentation functions based on qPYTHIA jets in Pb–Pb and PYTHIA jets in p–p. Statistical uncertainties due to the underlying event contributions and the particle reconstruction efficiencies are taken into account. The fragmentation functions are based on the MC truth jet energy, and are in agreement with the functions shown in Fig. 8.8. The uncertainty in the fragmentation function for other hadrons in Pb–Pb collisions is roughly comparable since the relative particle abundance compared to the kaons scales about the same in the jet and the background.

Based on these studies we expect to measure identified fragmentation functions reliably out to $\xi$=4 for 50 GeV jets. Systematic uncertainties, though, both on the jet energy determination and the unfolding procedure on the kaon spectrum, have not yet been taken into account. Both of these operations were discussed in detail in Chapter 5.

If a particle species dependence in the medium modified fragmentation process can be confirmed then the energy loss as well as the hadronization mechanism in medium will be significantly constrained. The same change in the fragmentation function should also already be visible in comparing the integrated particle identified $p_T$-spectra in quenched (Pb–Pb) and unquenched jets (p–p) which were shown in Fig. 8.4.

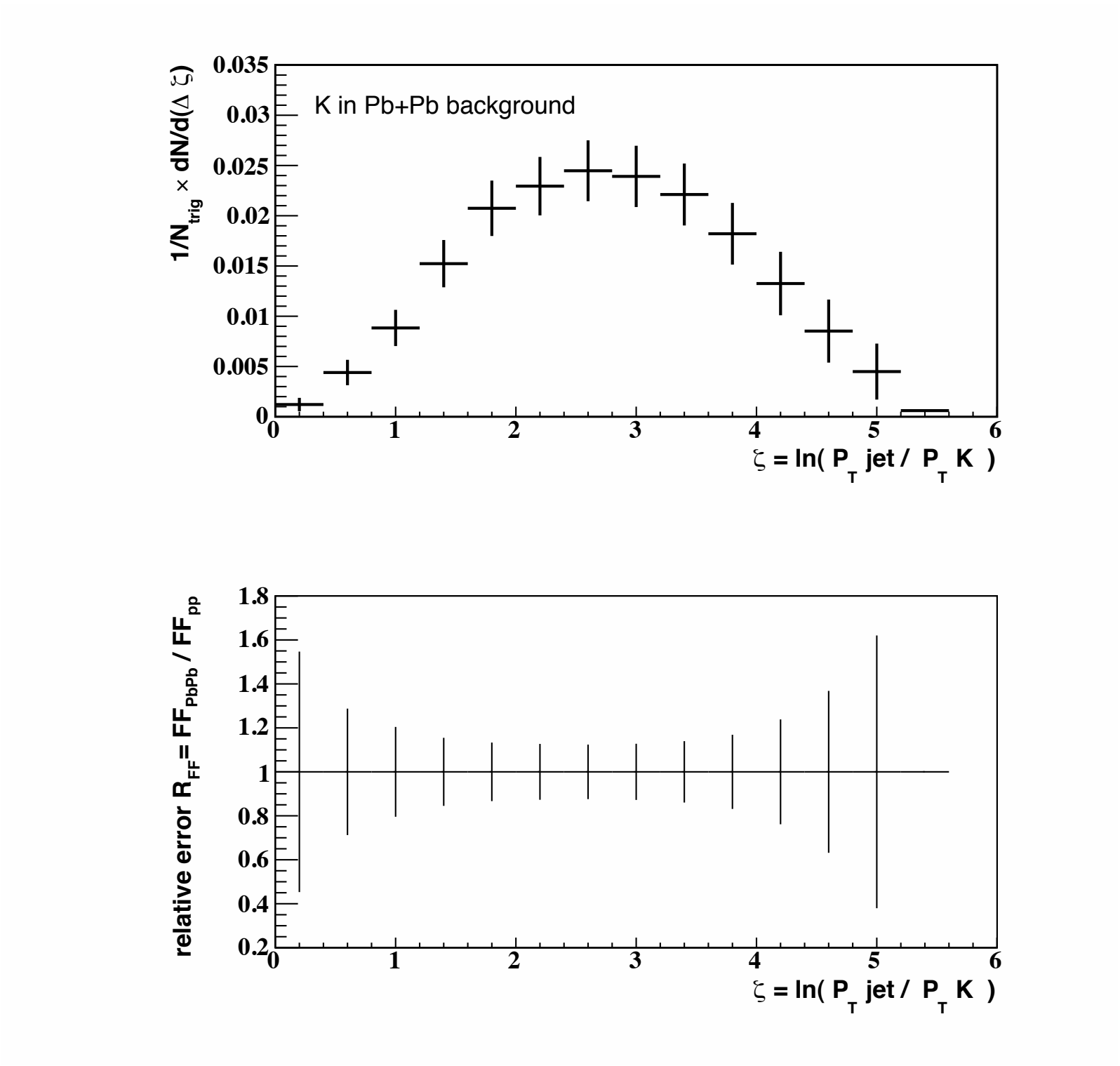

**Figure 8.9:** Top: uncertainty in the kaon fragmentation function measurement from unquenched reconstructed jets ($E_j$=50-60 GeV) in HIJING background. Bottom: relative uncertainty of the Pb–Pb / p–p fragmentation function ratio.

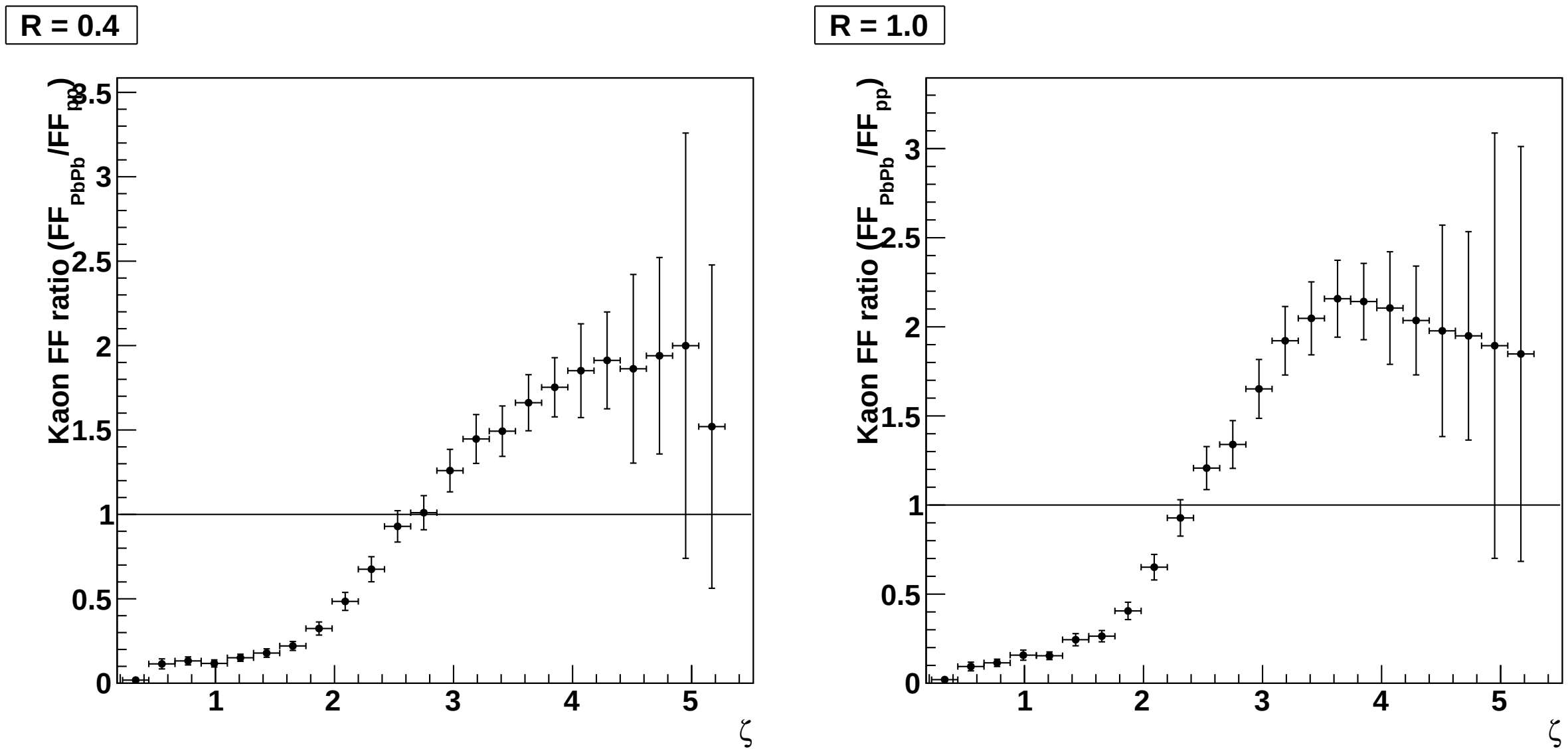

**Figure 8.10:** Achievable uncertainty in the reconstructed kaon fragmentation function ratio (for R=0.4 and R=1), if qPYTHIA is assumed for the fragmenting jet in Pb–Pb and PYTHIA is assumed for the fragmenting jet in p–p.

## 8.5 Momentum spectra and fragmentation functions of identified resonances in jets

The analysis of identified hadrons can be extended to include identified hadronic resonances. STAR has successfully reconstructed a long series of rare hadronic resonances ($\rho$, K*, $\phi$, $\Delta$, $\Lambda^*$, $\Sigma^*$, $\Xi^*$). Based on their lifetime, which ranges from a few fm/c to tens fm/c, these strongly decaying states are particularly sensitive to the evolution, lifetime and properties of the partonic medium. As shown by STAR, the reconstruction of rare hadronic resonances in the large heavy ion background is very challenging and requires superior particle identification capabilities, which are,at the LHC, unique to ALICE. The measurements at RHIC led to specific results on the partonic and hadronic system lifetimes [10] as well as first, inconclusive, attempts to identify chiral symmetry restoration [11, 12]. A novel idea to further constrain the level of medium modifications of resonances will be discussed in the following correlations chapter. Here we first focus on the reconstruction of high momentum resonances in jets. Measurements of the yield of resonant states are interesting on their own since there is only little known, and measured, about resonance production at LHC or Tevatron energies. The simple assumptions of leading order calculations, such as PYTHIA, lead to resonant over non-resonant ratios which approach unity at high momentum. Experimentally the scarce data available from DIS and STAR measurements show that quarks fragment with equal probability into pions and $\rho$ mesons (see e.g. [13, 14]), but a possible over-abundance of resonances at even higher initial energy or resonance momentum still needs to be experimentally verified. Fig. 8.11 shows the anticipated annual statistics for momentum spectra of reconstructed resonances in jets with jet energies higher than 50 GeV (embedded PYTHIA) as triggered and reconstructed by the EMCal using the anti-$k_T$ algorithm (R=0.4) in the FastJet package (see Chapter 5). The figure also shows selected relative uncertainties as a function of particle species and momentum. These uncertainties fold in reconstruction efficiencies which include $2\sigma$ dE/dx cuts. The resonance reconstruction efficiencies are based on the mixed event technique first used by STAR and subsequently applied to the untriggered bulk matter resonance studies detailed in the original ALICE-PPR. Due to statistics limitations the S/B was held constant from 7 GeV/c on out although the ratio is likely to improve. Thus the uncertainties quoted in Fig. 8.11(right) should be considered pessimistic at higher transverse momentum. Table 8.1 summarizes the anticipated annual reconstructed yield of K* and $\phi$-mesons in EMCal triggered 50 and 100 GeV di-jets in Pb–Pb collisions for certain cuts on the resonance transverse momentum.

Based on the good statistics achievable in the equivalent of one year of p–p or Pb–Pb running, invariant mass spectra can be reconstructed out to high $p_T$ in both the heavy ion and the reference system. Fig. 8.12 shows the annual statistical significance in the reconstructed invariant K* mass spectrum for selected high transverse momentum bins in reconstructed 50 $\pm$ 10 GeV jets in p–p collisions. The combinatorial background is about an order of magnitude larger than the signal. Even higher quality spectra can be expected for the $\phi$-meson, as shown in Fig. 8.13.

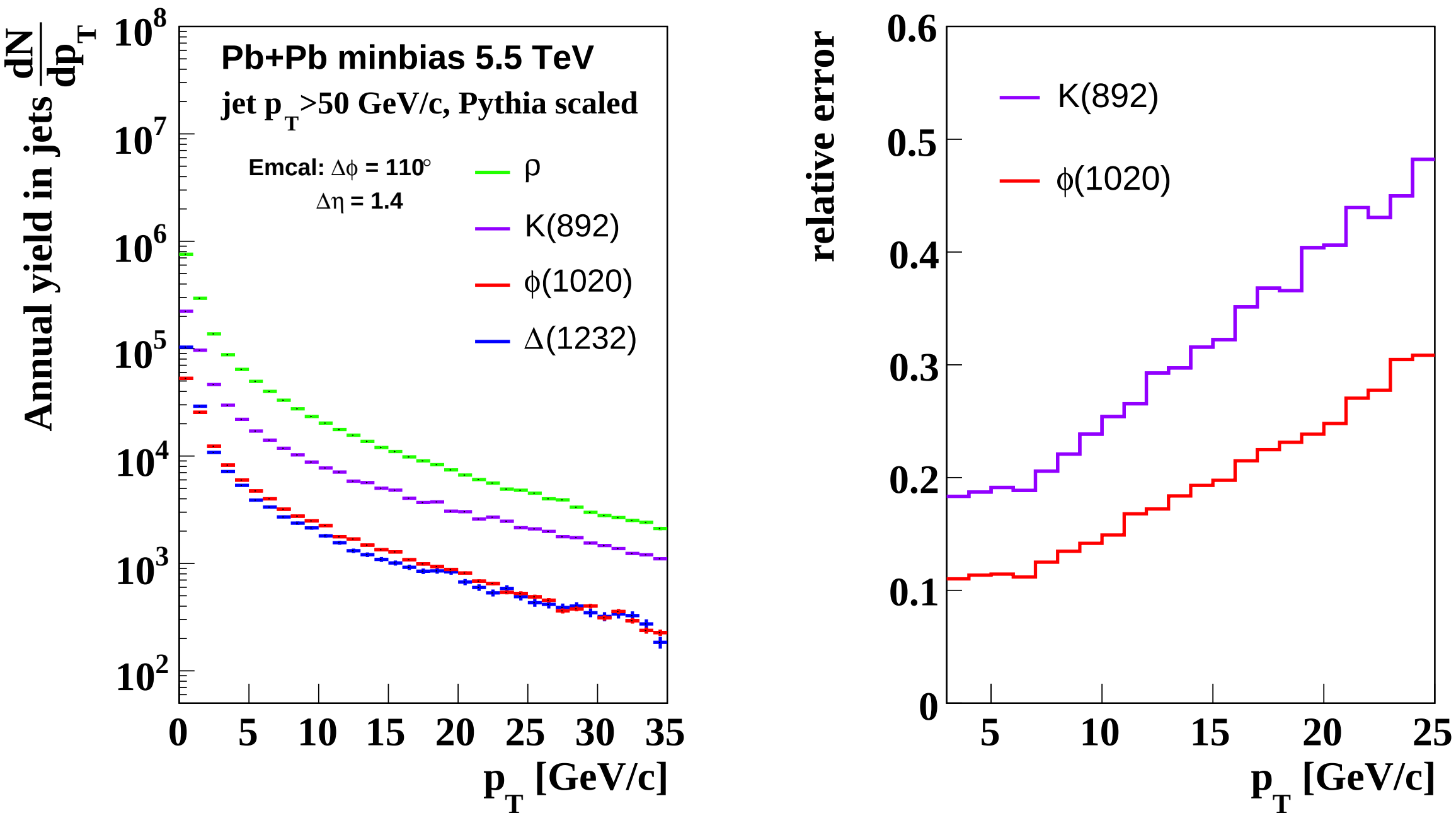

**Figure 8.11:** Left: Annual hadronic resonance yields in reconstructed (unquenched) jets with a jet energy higher than 50 GeV. Right: relative error based on PID separation power, momentum resolution, detector acceptance and resonance reconstruction efficiencies. A systematic error on the tracking efficiency, which was estimated to be 5% in the original ALICE-PPR, has not been included.

**Table 8.1:** Annual yields of reconstructed resonances in jets in Pb–Pb collisions.

| Jet energy plus resonance $p_T$-cut | annual $\phi$-yield | annual K*-yield |
|---|---|---|
| 50 GeV with 1 GeV/c cut | $3 \times 10^5$ | $2.3 \times 10^6$ |
| 50 GeV with 4 GeV/c cut | $1.4 \times 10^5$ | $1.0 \times 10^6$ |
| 100 GeV with 1 GeV/c cut | $1.8 \times 10^4$ | $1.4 \times 10^5$ |
| 100 GeV with 4 GeV/c cut | $1.1 \times 10^4$ | $8.1 \times 10^4$ |

The K* background level in the jet cone can be determined by analyzing the resonance population outside the jet cone (off-axis component). Fig. 8.14 shows the relative strength of the K* background to the K* jet component in proton-proton collisions with a 200 MeV/c $p_T$ cut on the reconstructed resonances. The resulting S/B ratio is about 10:1. The error bar due to combinatorial background subtraction is not shown here, but it is negligible based on the small particle multiplicity in p–p collisions, which is very different from the Pb–Pb distributions that will be discussed in Fig. 8.17.

The relative yield and momentum distribution of the in-cone/out-of-cone resonances can be used to determine the reliable $\xi$-range for a resonance fragmentation function measurement. Fig. 8.15 shows the signal and background contributions to the fragmentation function in p–p collisions. As expected the underlying event has a considerably different momentum distribution than the jet and only affects the measurement at high $\xi$. We estimate that based on the statistical errors the resonance fragmentation functions from 50 GeV jets can be determined reliably out to $\xi$=4.4 ($p_T$=600 MeV/c) in p–p collisions.

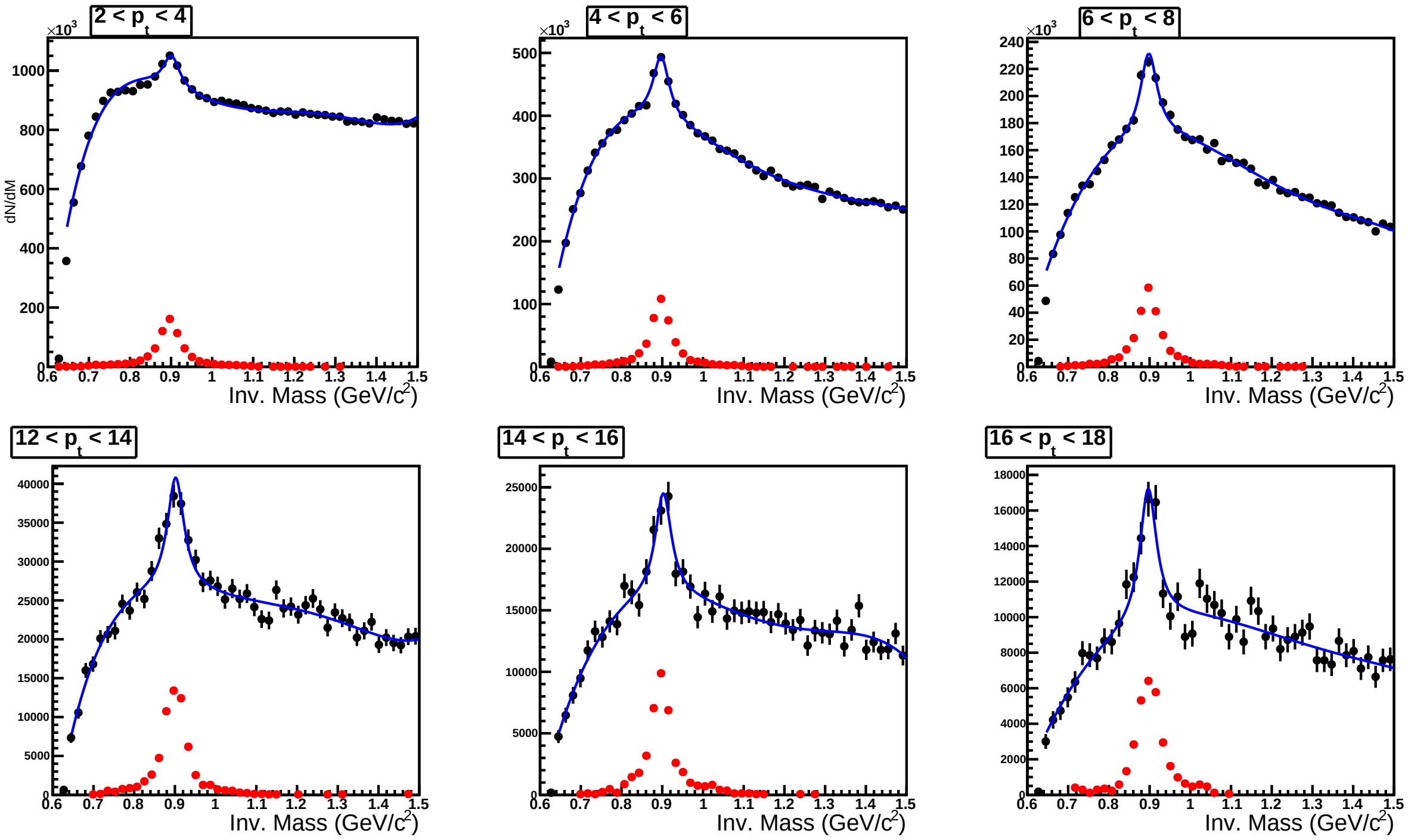

**Figure 8.12:** Invariant mass spectra for K* reconstructed in p–p jets with energies of 50-60 GeV.

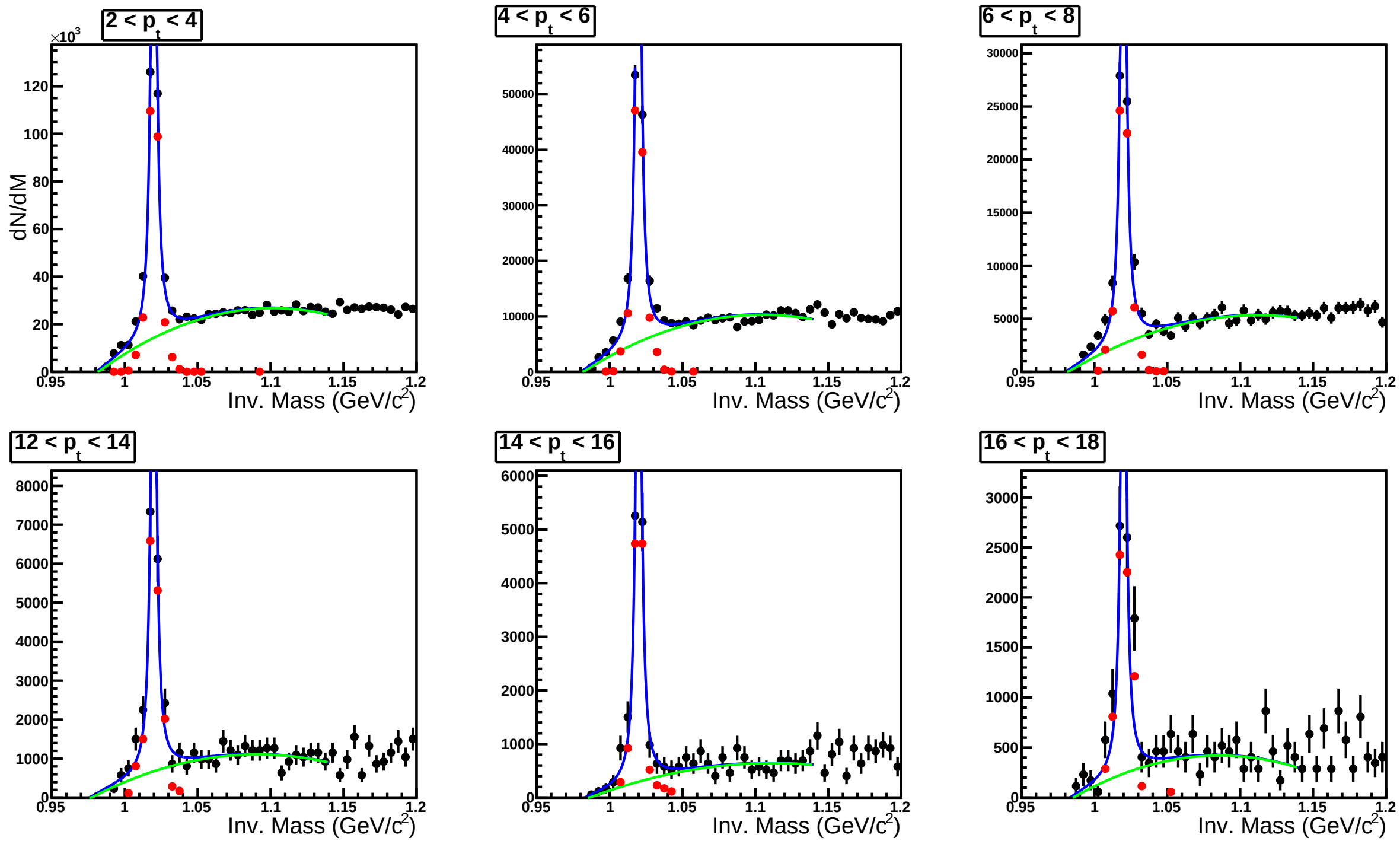

**Figure 8.13:** Invariant mass spectra for $\phi$-mesons reconstructed in p–p jets with energies of 50-60 GeV.

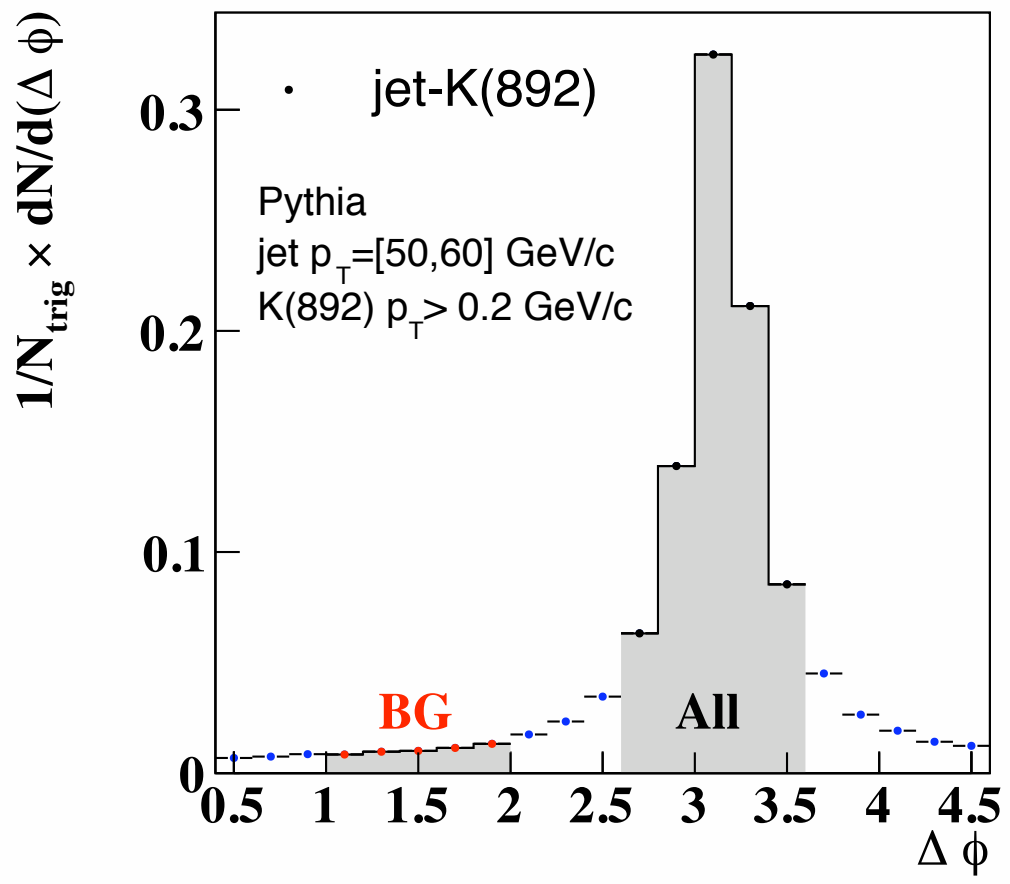

**Figure 8.14:** Jet signal and off-axis background per jet trigger in p–p collisions for K* in jets with energies of 50-60 GeV.

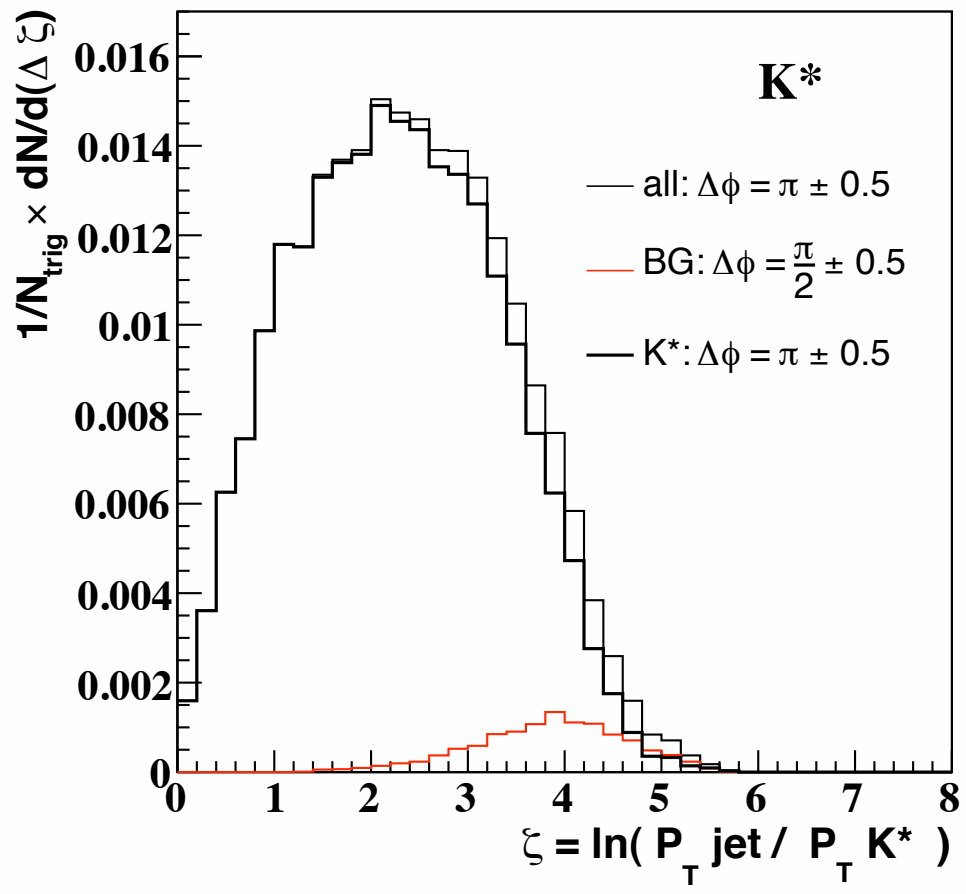

**Figure 8.15:** Contributions to the K* fragmentation function per jet trigger in p–p collisions for PYTHIA jets with 50-60 GeV energy.

In Pb–Pb collisions we do not only expect an increase in the combinatorial and resonance background but also quenching of the away-side (non-triggered) jet. In order to estimate the effect on our measurement we first compare PYTHIA and qPYTHIA fragmentation functions in Fig. 8.16 for K* per reconstructed jet.

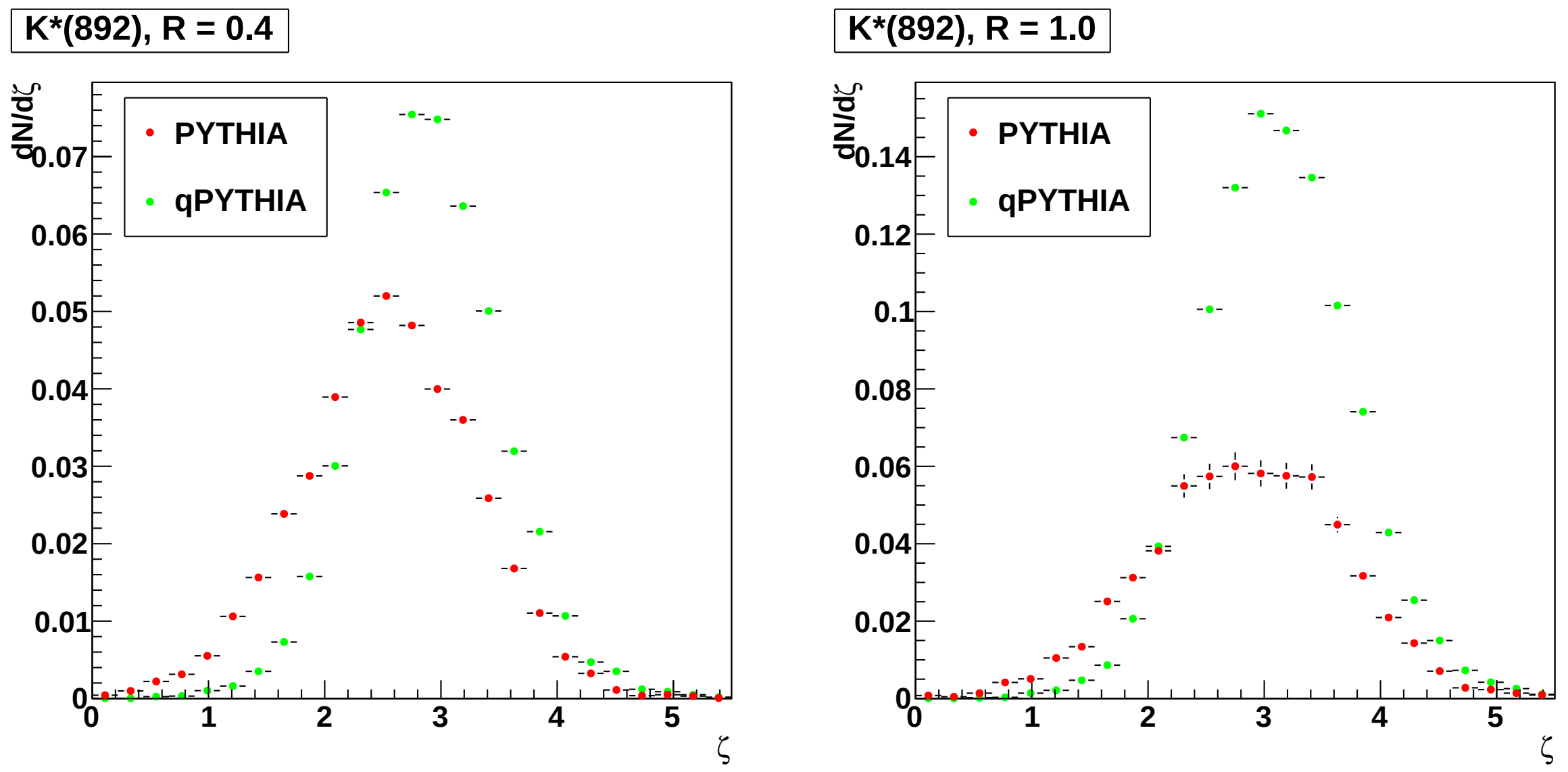

**Figure 8.16:** Reconstructed (R=0.4 and R=1.0 cone algorithm) K* fragmentation function for 50-60 GeV PYTHIA and qPYTHIA jets.

We then embed the K* PYTHIA distributions into the HIJING background (Fig. 8.17) and obtain $p_T$ threshold dependent S/B ratios for K* resonances in jets with energies of 50-60 GeV. The distributions are normalized per jet trigger. The annual statistics can be obtained by multiplying with the triggered di-jet rate which is about 2 Million per year for 50-60 GeV jets in Pb–Pb collisions. We find S/B ratios that are reduced by many order of magnitudes compared to the p–p level. The $p_T$ threshold on the resonance is required to attain manageable levels. For 1 GeV/c and 4 GeV/c cuts the S/B ratios are 2 $\times 10^{-4}$ and 1 $\times 10^{-3}$, respectively, in Pb–Pb collisions.

In these distributions the unfolding method, which subtracts the non-jet resonance contribution, was augmented by a method taking into account the combinatorial background which is based on a mixed event background subtraction performed on $\xi$-binned invariant mass spectra (see e.g. Fig. 8.12). The combinatorial background outweighs the out-of-cone resonance background by at least an order of magnitude even for high $p_T$ cuts ($p_T > 4$ GeV/c) on the resonance spectrum. The error bars shown in Fig. 8.17 reflect the uncertainty based on both background contributions.

In order to obtain a fragmentation function for resonances the method shown in Fig. 8.15 can not be applied since the background, as determined outside the jet cone, is considerably larger (see Fig. 8.17) and has a similar momentum distribution than the jet particles. We therefore have to apply the unfolding method as described in Chapter 5 and in the previous section for the kaon fragmentation function. The resulting uncertainty in the K* fragmentation function determination is shown in Fig. 8.18 for the unquenched jet case in Pb–Pb together with the statistical uncertainty in the Pb–Pb to p–p ratio. By comparing to Fig. 8.16 we can determine that, based on statistical errors only, the difference between PYTHIA and qPYTHIA should be measurable in the $\xi$-range between 0.5 and 1.5 and between 3 and 4. The caveats regarding the systematic uncertainties of the jet energy reconstruction and the background unfolding, which were mentioned in the previous section, apply here as well.

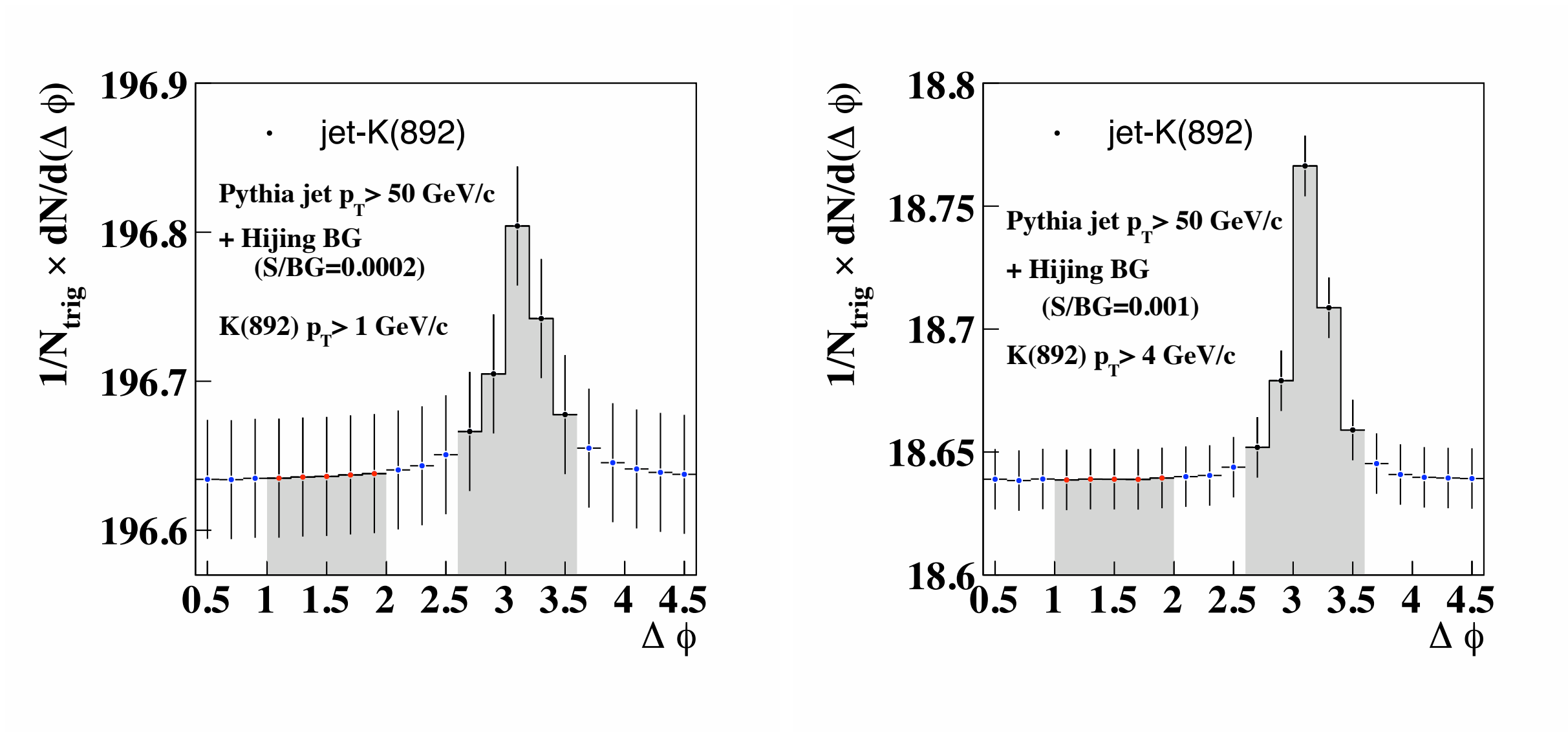

**Figure 8.17:** Signal and background levels for K* in PYTHIA jets embedded in unquenched HIJING background (for 1 GeV/c and 4 GeV/c cuts on the resonance $p_T$, respectively).

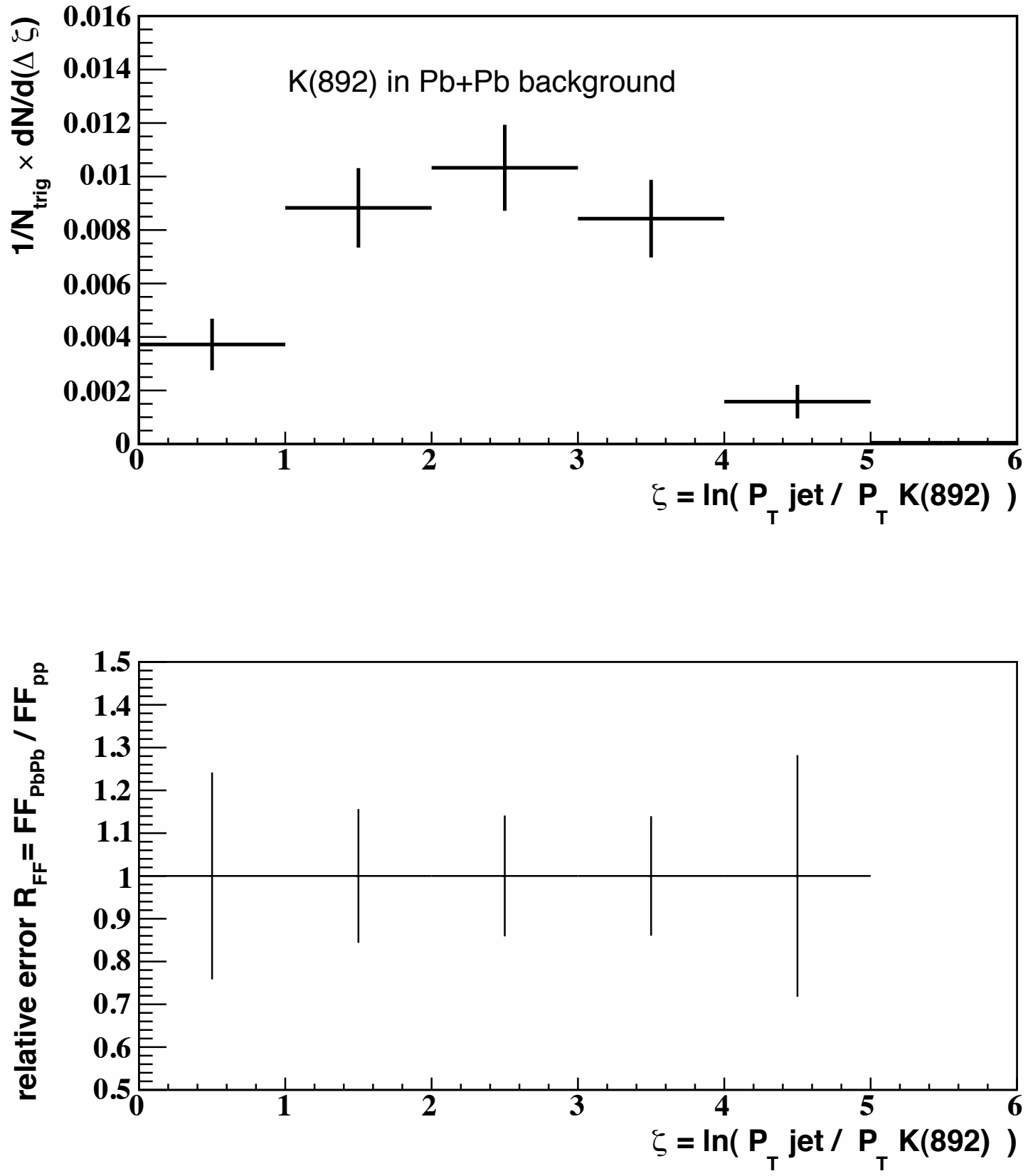

**Figure 8.18:** Top: uncertainty in the K* fragmentation function measurement from unquenched reconstructed jets ($E_j$=50-60 GeV) in HIJING background. Bottom: relative uncertainty of the Pb–Pb / p–p fragmentation function ratio.

## 8.6 Correlation function for associated hadronic resonances

The measurement of high momentum resonances from jets was proposed as an alternate way to measure chirality in the partonic medium [3]. In this paper it is assumed that, in contrast to the standard theory of jet quenching in heavy ion collisions, the fragmentation process actually has a finite probability to occur inside the partonic medium. This can lead to the formation and decay of pre-hadronic resonances, which carry the medium properties upon decay. In order for the chiral effect to not be washed out these resonances have to be produced through fragmentation deep in the partonic fireball and decay prior to hadronization. These conditions suggest a particular optimum $p_T$-range (4-10 GeV/c at the LHC) for a variety of short-lived hadronic resonances. The momentum and configuration space requirements can be matched through specific angular correlation conditions, i.e. an analysis of the $\Delta\phi$-$\Delta\eta$ correlation function is required for associated resonances of a certain momentum in a jet cone with respect to the jet axis. The simplest correlation angle restriction, besides an in-cone vs. out-of-cone correlation comparison, is a so-called quadrant analysis, which is explained in detail in [3], and described schematically in Fig. 8.19. Here the quadrants perpendicular to the triggered jet carry the bulk matter resonance information, the triggered jet is purposefully surface biased, and the away-side quadrant thus carries the modified resonances from the in-medium fragmentation process.

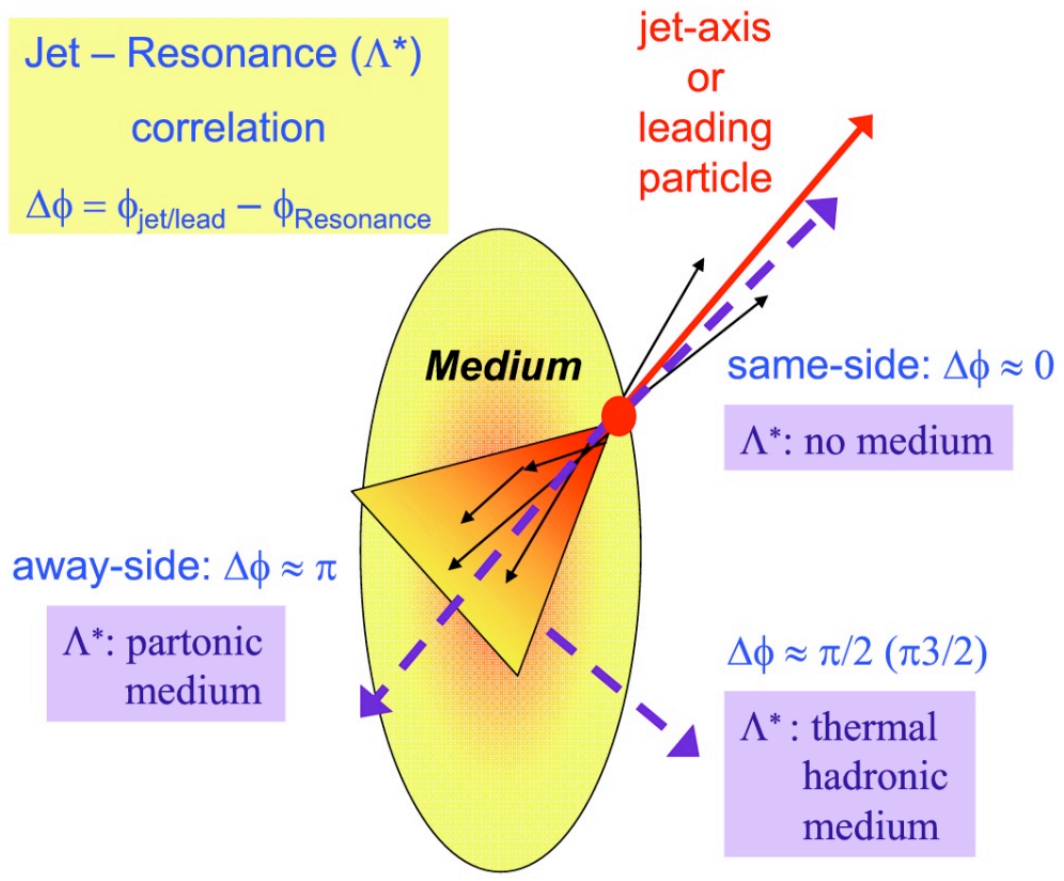

**Figure 8.19:** Schematic description illustrating the quadrant analysis (based on $\Lambda^*$) of jet-resonance correlations: resonances are plotted relative to the same-side jet axis in four 90 degree quadrants.

As an example Fig. 8.20 shows the per jet normalized correlation function of associated $K^*$ mesons for reconstructed jets with energies of 50-60 GeV after combinatorial background subtraction in p–p collisions.

Fig. 8.21(left) shows the $K^*$ invariant mass spectra in the four quadrants with respect to the triggered jet axis in p–p collisions. The corresponding difference in the $K^*$ transverse

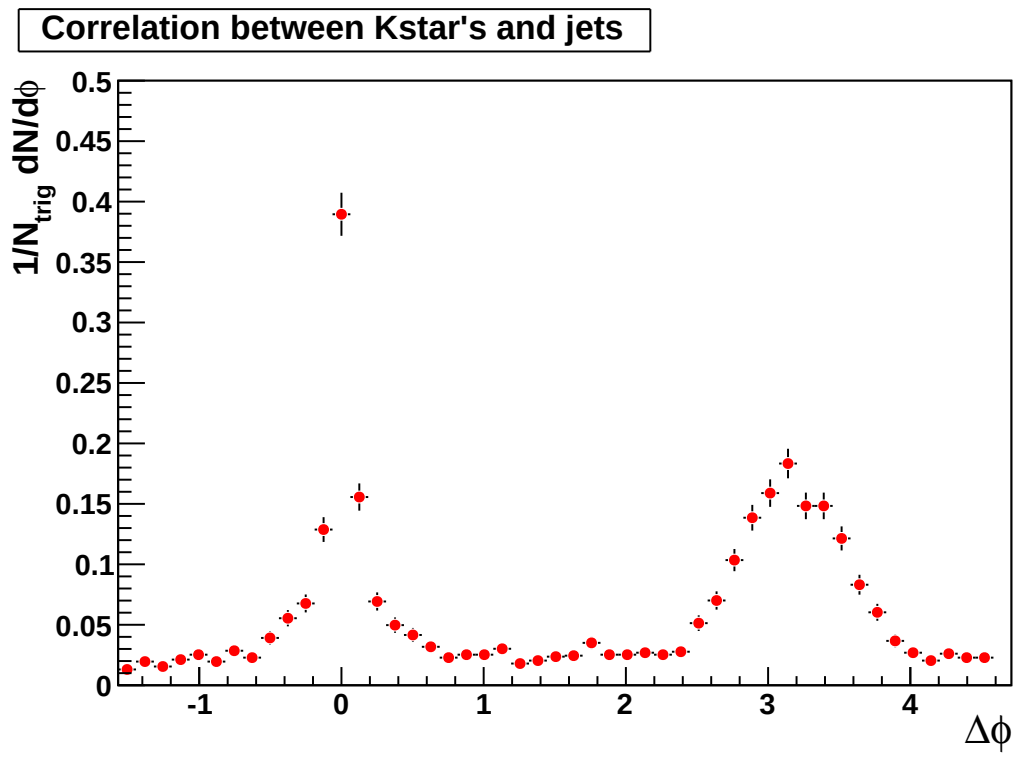

**Figure 8.20:** Associated K* $\Delta\phi$ distribution (relative to the reconstructed jet axis) in events with reconstructed di-jets with jet energies of 50-60 GeV. The y-axis shows the per jet K* yield.

momentum spectra in the jet and underlying event quadrants is shown in Fig. 8.21(right).

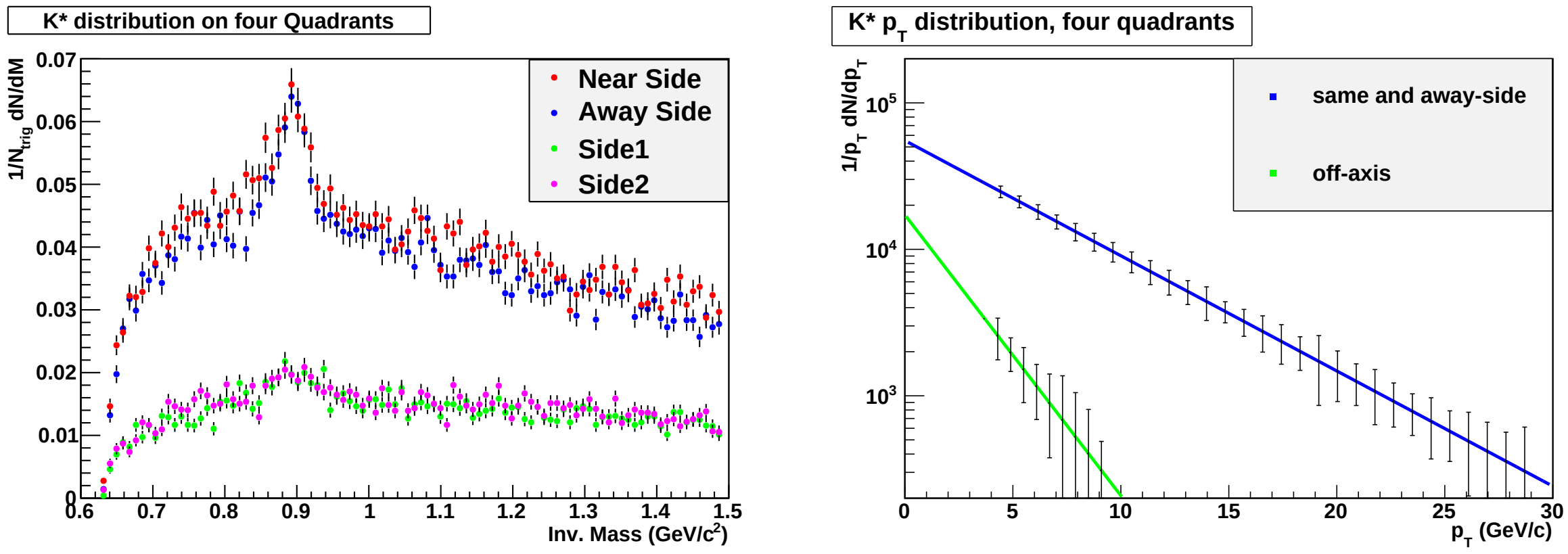

**Figure 8.21:** Left: Invariant mass distributions for K* in jet and underlying event quadrants. Right: Corresponding transverse momentum distributions in jet and underlying event quadrants.

The distributions in the jet quadrants (near-side and away-side) were then folded with HIJING in order to determine the accuracy of the invariant mass distribution in Pb–Pb collisions based on the heavy-ion combinatorial and resonance background. Fig. 8.17(right) showed the estimated signal to background level in case of a 4 GeV/c cut on the resonance. Based on these statistics and the achievable invariant mass resolution shown in Figs. 8.12, 8.13 and 8.21 we conclude that any medium modification (width broadening or mass shift) could be measured to a level of 10 MeV/c$^2$ for K* or $\phi$ and 20 MeV/c$^2$ for $\Lambda$* in the 4-10 GeV/c p$_T$-range.

Finally, the resonant over non-resonant ratios, which were used by STAR to determine the partonic lifetime of the system [10], can be measured differentially as a function of transverse momentum in the $\Delta\phi$ quadrants out to p$_T$ = 10 GeV/c as shown in Fig. 8.22 for unquenched PYTHIA jets. Even in p–p there is a slight difference between the jet resonance production and the underlying event. In Pb–Pb events the differentiation between bulk matter and jet fragments in the different quadrants (in cone vs. out of cone) will enable us to determine the differences between partonic and hadronic rescattering and regeneration of resonances in the same event. An analysis similar to the one in STAR will constrain the partonic and hadronic lifetimes at the LHC and it will enable us to compare the production mechanisms and rates of resonances compared to stable states in the fragmentation process.

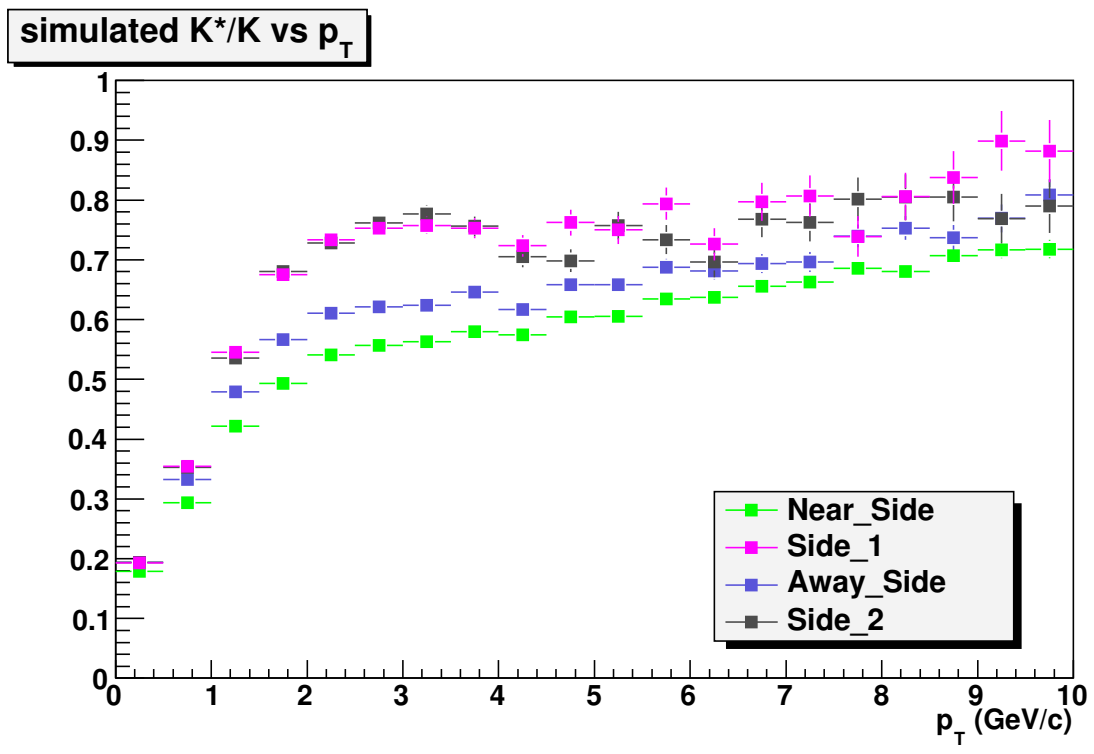

**Figure 8.22:** Resonance over non-resonant ratios inside and outside the reconstructed jet cones.

# Chapter 9

# Summary and Conclusions

This report discusses the contributions of the EMCal to the physics program of ALICE. The highly granular and triggerable EMCal, combined with the tracking and particle identification detectors of ALICE, enables full exploration of the physics of jet interactions with dense matter, utilizing measurements in Pb-Pb, p-p, and p-A collisions at the LHC.

EMCal measurements in combination with other ALICE detectors allow investigation of observables based on high energy jets, photons and electrons, all rare probes requiring efficient triggers. Jet physics relies heavily on precise determination of jet spectra, thus requiring large data sets. Detailed calculations of the ALICE/EMCal Triggering capabilities at Level 1 and for the High Level Trigger and for the various different systems (Pb-Pb, lighter collision systems and p-p) are presented in the report. The triggering capabilities establish ALICE's ability to exploit the kinematic reach in these observables, as made possible by their increased yields at the LHC energy.

The emphasis of the EMCal physics program is on the modification of jets in medium, and the corresponding response of the medium. We have shown that the medium modified jet spectrum can be measured in ALICE with precision ($\sim 10 - 20\%$) over a broad kinematic range. Current theoretical models have larger systematic uncertainties than this, and there is an intensive, community-wide effort to improve the precision of such calculations. We expect that the interplay between the well-controlled jet measurements enabled by the EMCal with the new Monte Carlo generators now under development will result in a much deeper, quantitative understanding of the underlying physics of jet-medium interactions.

The measurement of hadron momentum distributions in the jet, based on charged or identified particles, can be used to further constrain the physics of quenching, by determining both the hadro-chemistry in the jet and the nature of flavor conversion in the medium. Resonance measurements in jets, also shown in this report to be achievable, are of interest because they may also undergo medium modification due to their short lifetime, which is comparable to the lifetime of the medium.

The measurement of inclusive prompt photons provides a crucial benchmark of perturbative QCD calculations for hadronic and nuclear collisions at LHC energies. Prompt photons also play an important role as triggers for photon-jet and photon-hadron coincidence measurements. We have shown in this report that the application of shower-shape analysis in conjunction with isolation cuts in the EMCal promises to provide controlled discrimination of prompt photons from backgrounds, even in the heavy ion environment. While photon-jet and photon-hadron correlation measurements are limited in energy reach due to small cross section, they enable the cleanest measurement of hadron distributions in jets and their medium modification.

A well-controlled comparison of quark and gluon jet quenching would provide important constraints on the physics of quenching. Tagging of quark jets is needed, since at LHC energies the inclusive jet population for $E_T$ up to a few hundred GeV is gluon dominated. A bias towards light quarks jets is achieved in the photon+hadron coincidence measurement, which is dominated by Compton scattering in which the recoiling jet is a quark jet.

High momentum electrons from semi-leptonic heavy quark decays are good tags for quark jets, but must be separated from a competing yield from W decays, as well as hadronic backgrounds. The EMCal is the primary detector in ALICE for hadron suppression beyond $\sim$ 10 GeV/c. Our studies show that the semi-leptonic decay channel, which is dominated by B-meson decay out to $\sim$25 GeV electron energy, can be measured with good systematic control out to $\sim$50 GeV. An additional method of tagging displaced vertex electrons in order to identify B-meson decays has also been explored and provides promise for tagging of B-mesons out to $\sim$80 GeV.

In summary, this report demonstrates that the EMCal adds to ALICE significant capabilities for triggering on and measuring jets, photons and electrons at large transverse momentum. The EMCal coupled with the ALICE detectors for tracking and particle identification enables systematic studies of high momentum identified hadrons and resonances in tagged jets, as well as lower momentum identified particles to investigate the medium response to jet energy deposition. No other experiment at the LHC can match the scope of these studies.